\documentclass[a4paper,12pt,openright,twoside]{book}
\usepackage[utf8]{inputenc}
\usepackage{t1enc}
\usepackage[greek.polutoniko,english]{babel}
\usepackage[outer=2.5cm, inner=2.5cm, top=2.5cm, bottom=2.5cm,headsep=0.5cm]{geometry}
\usepackage{comment}
\usepackage[title]{appendix}
\usepackage{teubner}
\usepackage{enumitem}
\usepackage{hyperref}

\usepackage{lineno}
\usepackage{graphicx}
\usepackage[onehalfspacing]{setspace}
\usepackage{physics}
\usepackage[backend=bibtex,sorting=none,style=numeric-comp,maxbibnames=5,minbibnames=3,firstinits=true]{biblatex}

\makeatletter
\DeclareCiteCommand{\fullcite}
  {\defcounter{maxnames}{\blx@maxbibnames}%
    \usebibmacro{prenote}}
  {\usedriver
     {\DeclareNameAlias{sortname}{default}}
     {\thefield{entrytype}}}
\makeatother

\renewbibmacro{in:}{}
\usepackage{fancyhdr}
\usepackage[nameinlink,capitalize]{cleveref}
\usepackage{ mathrsfs }
\usepackage{slashed}
\usepackage{url}

\usepackage[usestackEOL]{stackengine}

\usepackage{braket}
\usepackage{bibentry,amssymb}
\usepackage{bbold}
\usepackage{xurl}
\usepackage{mathtools}
\usepackage{amsmath}
\usepackage{blindtext}
\usepackage{float}
\usepackage{setspace}
\usepackage{subcaption}
\usepackage{tikz}
\newcommand*\circled[1]{\tikz[baseline=(char.base)]{
    \node[shape=rectangle, color = black, draw, inner sep=2pt, 
        minimum height=12pt] (char) {#1};}}

\counterwithout{footnote}{chapter}

\begin{document}

\pagestyle{fancy}
\fancyhf{}
\fancyhead[LE]{\slshape\nouppercase{\leftmark}}
\fancyhead[RO]{\slshape\nouppercase{\rightmark}}
\fancyfoot[LE,RO]{\thepage}

\renewcommand{\chaptermark}[1]{\markboth{Chapter \thechapter. #1}{}}

\renewcommand{\sectionmark}[1]{\markright{\thesection. #1}}

\frontmatter

\begin{titlepage}
\begin{center}
{\scshape\large PhD Dissertation}\\
\vspace{0.8cm}

\begin{onehalfspacing}
\fontsize{12}{20}\selectfont
{\bfseries\scshape\Large Odderon Exchange in Elastic}\\
{\bfseries\scshape\Large Proton-Proton and Proton-Antiproton}\\
{\bfseries\scshape\Large Scattering at TeV Energies}\\
\end{onehalfspacing}
 \vspace{0.8cm}

{\scshape \large István Szanyi}\\
\vspace{0.6cm}

\begin{figure}[h!]
    \centering
    \includegraphics[scale=0.22]{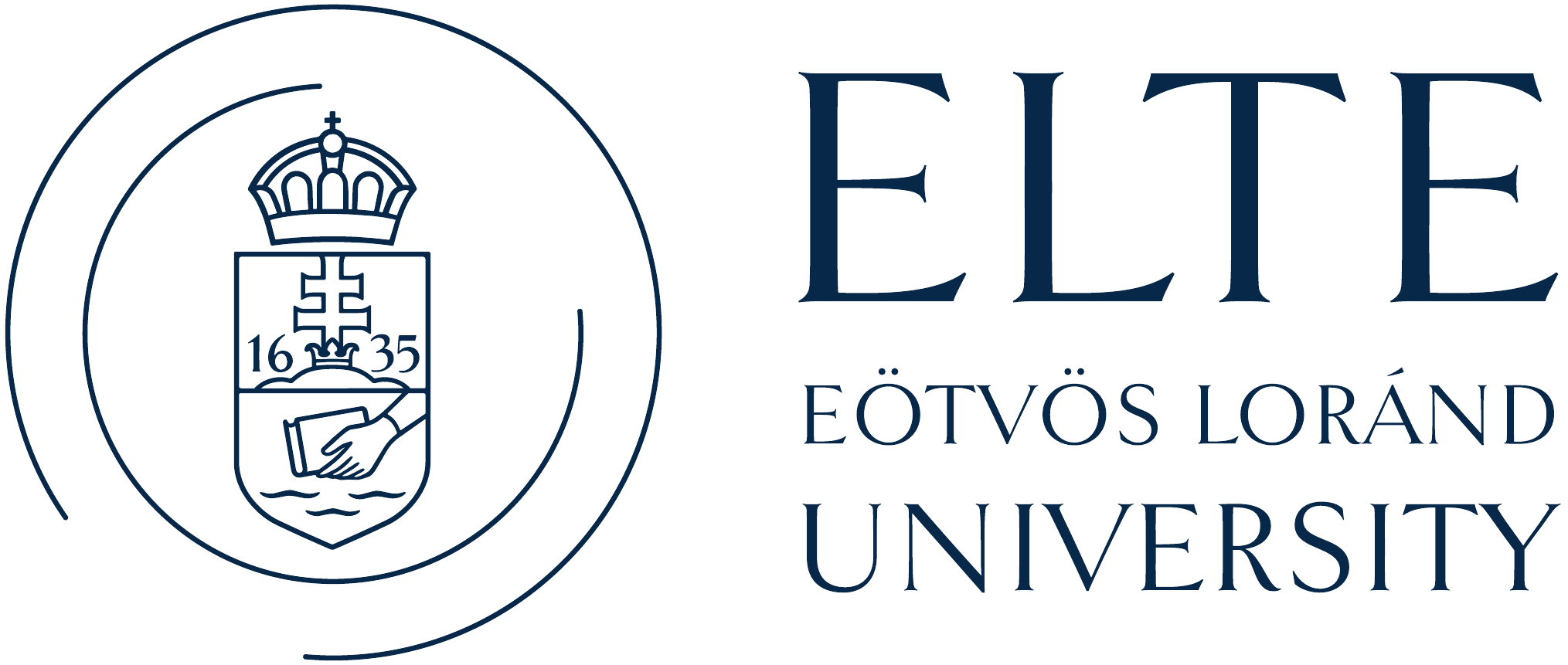}
\end{figure}
\vspace{-0.4cm}
{\scshape  ELTE Eötvös Loránd University\\
  Doctoral School of Physics\\
 Particle and Nuclear Physics Program\\
\vspace{0.5cm}
}

\begin{figure}[h!]
    \centering
    \includegraphics[scale=0.08]{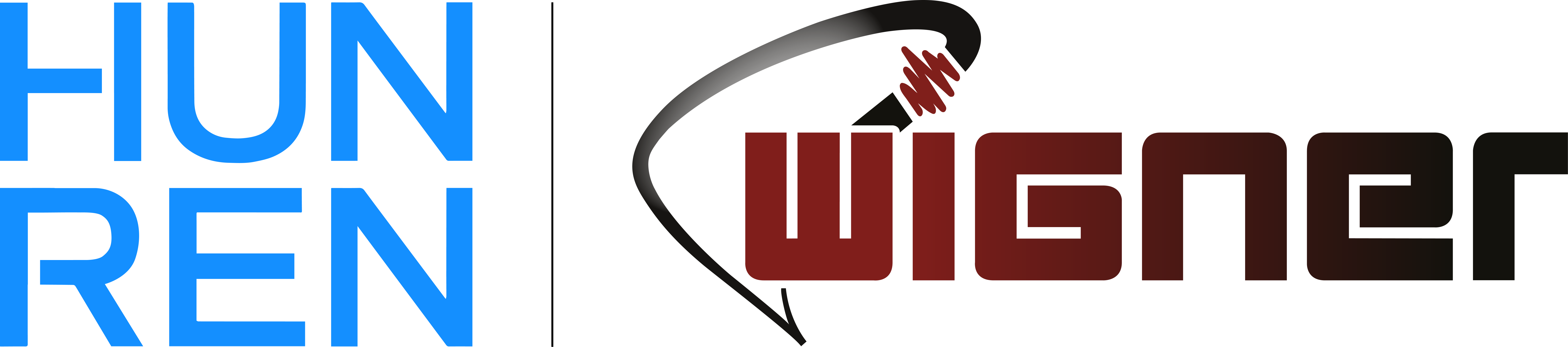}
\end{figure}
\vspace{-0.4cm}
{\scshape  HUN-REN Wigner Research Centre for Physics \\
Institute of Nuclear and Particle Physics\\
Department of Theoretical Physics\\
}

\vspace{0.6cm}
{\scshape  doi: 10.15476/ELTE.2024.155}\\

\vspace{-0.6cm}
\begin{align*}
\text{ Supervisor:}~& \text{{\scshape Tamás Csörgő},}\\
&\text{Member of Academia Europaea}\\[3.5pt]
\text{ Co-supervisor:}~ & \text{Prof. {\scshape Máté Csanád}}\\[3.5pt]
\text{ Head of the doctoral school:}~  & \text{Prof. {\scshape Gergely Palla}}\\[3.5pt]
\text{ Head of the doctoral program:}~  & \text{Prof. {\scshape Zoltán Trócsányi}}\\[0.2cm]
\end{align*}

\vfill
\vspace{-0.7cm}
{ \scshape BUDAPEST\\
2024}
\end{center}

\end{titlepage}

\newpage
\thispagestyle{empty}
\hspace{0cm}
\vfill
\begin{center}
\begin{minipage}{0.85\textwidth}
\centering
{\scshape\large Odderon exchange in elastic proton-proton and proton-antiproton scattering at TeV energies}

A dissertation submitted for the degree of Doctor of Philosophy

~

©  István Szanyi, 2024

~

~

\text{Supervisor:}\\
\text{{Tamás Csörgő}, D.Sc.,}\\
\text{Member of Academia Europaea,}\\
\text{Scientific Advisor at HUN-REN Wigner RCP,}\\
\text{Research Professor at MATE Institute of Technology,}\\
\text{Honorary Professor at ELTE Eötvös Loránd University}

~

\text{Co-supervisor:} \\
\text{{Máté Csanád, D.Sc.,}}\\
\text{Professor at ELTE Eötvös Loránd University}

\end{minipage}

\end{center}
\hspace{0cm}
\vfill
\newpage
\onehalfspacing

\clearpage

\thispagestyle{empty}

\hspace{0cm}
\vfill
\begin{center}
\begin{minipage}{0.7\textwidth}

We do not know what the rules of the game are; all we are allowed to do is to watch the playing. Of course, if we watch long enough, we may eventually
catch on to a few of the rules.

~

{\hspace{6.1cm} --- Richard P. Feynman \cite{feynman2011six}}
\end{minipage}
\end{center}
\vfill
\hspace{0cm}

\thispagestyle{empty}
\clearpage
\thispagestyle{empty}
\clearpage

\tableofcontents

\chapter{Preface}

\markboth{Preface}{Preface}

\vspace{-0.5cm}


Elastic\, proton-proton\, ($pp$)\,\, scattering\,\, measurements\,\, at\, CERN's\, LHC\, and\, elastic \mbox{proton-antiproton ($p\bar{p}$)}\, scattering\, measurements\, at\, FNAL's\, Tevatron,\, performed at \mbox{TeV-scale} center-of-mass energies ($\sqrt{s}$) and over wide ranges of squared four-momentum transfer~($t$), gave new opportunities to study the physics of elastic hadronic processes.
Though the fundamental theory of strong interactions is quantum chromodynamics (QCD), elastic scattering processes in the kinematic domain $|t|\lesssim3$ GeV$^2$, called ``soft'' processes and studied here,  are outside the domain of applicability of perturbation methods used to carry out QCD calculations. Though nonperturbative methods in the framework of QCD are developing, comprehensive theoretical investigations of experimental data on ``soft'' processes can be performed so far only by applying complementary approaches.

One of the basic approaches for high-energy elastic hadronic collisions is the eikonal picture, where the elastic scattering is related to particle formation in the $s$-channel. In the usual terminology, it is said that the infinite number of inelastic open channels at high energies builds up a ``shadow'' leading to ``diffraction'', $i.e.$, elastic and ``quasi-elastic'' scattering.
In this consideration, the amplitude, as a function of the impact parameter, is written as the solution of the $s$-channel unitarity equation in terms of the opacity function whose parametrization is fixed by different theoretical assumptions. 
Successful parametrizations have been obtained based on, \textit{e.g.}, R.~J.~Glauber's diffractive multiple scattering theory.

In the framework of T. Regge's theory of complex angular momenta,  the scattering amplitude in the momentum representation is specified by $t$-channel exchanges of Regge trajectories. A Regge trajectory represents a whole family of particles that have the same quantum numbers but varying spin and mass. A Regge trajectory exchange can be interpreted as an exchange of a virtual particle with continuously varying spin and virtuality. At energies of the order of a few GeV, the mesonic trajectories dominate causing the cross sections to decrease as the energy grows. The pomeron trajectory exchange, characterized by the quantum numbers of vacuum, was invented in the 1960s to account for hadronic cross sections that appeared at first to be constant and were later found to be increasing with energy. Mesons are lying on essentially linear mesonic Regge trajectories. In perturbative QCD, a pomeron exchange is interpreted as an exchange of an even number of interacting gluons. Based on this result, the pomeron trajectory is often regarded as a  Regge trajectory representing bound states of gluons, $i.e$, glueballs composed of an even number of gluons. However, there is no experimental evidence or compelling theoretical argument that the pomeron trajectory is unambiguously a glueball trajectory.

In 1973, L. Lukaszuk and B. Nicolescu proposed the negative spatial and charge parity counterpart of the pomeron, which was later named as odderon. In perturbative QCD, an odderon exchange corresponds to the $t$-channel exchange of a color-neutral gluonic compound consisting of an odd number of gluons feeding the belief that the odderon trajectory is again a glueball trajectory. The odderon satisfies all basic requirements of analyticity and unitarity, and does not vanish relative to the pomeron contribution or vanishes but only slowly with increasing energy.

Elastic $pp$ and $p\bar p$ scattering are related by crossing symmetry. This means that the amplitudes of the two processes, when their variables are continued to complex values, are part of the same analytic function. These amplitudes can have a crossing-even and a crossing-odd part. The crossing-even part of the scattering amplitude, when passing from $pp$ to $p\bar p$ scattering or vice versa, remains the same, while the crossing-odd part changes sign. Pomeron and mesonic exchanges with positive charge parity constitute the \mbox{crossing-even} part of the scattering amplitude. In contrast, the odderon and mesonic exchanges with negative charge parity constitute the crossing-odd part. 

At very high energies, including the TeV energy range probed by the LHC, the contribution of mesonic exchanges is highly suppressed; only the effects of the pomeron and the possible odderon exchange are measurable. Since odderon is a crossing-odd contribution, any statistically significant difference measured between $pp$ and $p\bar p$ elastic scattering in the TeV energy domain indicates the existence of the $t$-channel odderon exchange. For \mbox{48 years,} there was no definitive answer to the question about the existence of the \mbox{$t$-channel odderon.}  The joint analysis of the new $pp$ and the earlier $p\bar p$ elastic scattering data allowed us to do comparative studies of elastic $pp$ and $p\bar p$ scattering in the same kinematic ($s$, $t$) range and reveal the existence of the \mbox{$t$-channel odderon.} Given that the elastic $pp$ and $p\bar p$ scattering data never were measured at the same (or close enough) energies in the TeV region, the main task was to close the energy gap as much as possible, without direct measurement, based on the analysis of already published data. Such a program was carried out using different methods, utilizing model-dependent and model-independent extrapolations as well as the so-called $H(x)$ scaling property of \mbox{elastic $pp$ scattering.}


In this dissertation, I detail my contribution to the discovery of the $t$-channel odderon exchange and related studies. In \cref{chap:intro}, I introduce the basics of high-energy elastic $pp$ and $p\bar p$ scattering; I detail the relation of these processes to the odderon exchange and the methodology I utilized in my work. In \cref{chap:ReBBpbarp}, I analyze elastic $p\bar p$ scattering data by using the Real-extended Bialas--Bzdak (ReBB) model that incorporates R.~J.~Glauber's diffractive multiple scattering theory; this is the generalization of the ReBB model of elastic $pp$ scattering to elastic $p\bar p$ scattering, a crucial step towards the odderon analysis based on the ReBB model. In \cref{chap:oddTD0}, I present a preliminary study of closing the energy gap between $pp$ and $p\bar p$ elastic differential cross section measurements and analyze the odderon effects at $\sqrt{s}=1.96$ TeV utilizing physical models -- the ReBB model and a phenomenological model based on Regge theory -- calibrated to experimental data on $pp$ and $p\bar p$ elastic scattering. These results served as a guide during the joint D0-TOTEM odderon analysis. In \cref{chap:rebbdesc}, I present a refined analysis of elastic $pp$ and $p\bar p$ scattering data with the ReBB model. In \cref{chap:odderon}, I utilize this refined calibration of the ReBB model for the final ReBB model odderon analysis that results in the model-dependent observation of $t$-channel odderon exchange signals with discovery-level statistical significances.  In \cref{chap:Hxvalidity}, first, I study the reason behind the $H(x)$ scaling of high-energy elastic $pp$ scattering data at higher $|t|$ values which was used for analyzing odderon effects, then I identify the $H(x)$ scaling limit of the ReBB model and fit it to experimental data to test against the measurements the conditions required for $H(x)$ scaling. In \cref{chap:levy}, first I analyze low-$|t|$ $pp$ and $p\bar p$ differential cross section data utilizing a newly introduced simple Lévy $\alpha$-stable model, then I formulate the real extended Lévy $\alpha$-stable generalized Bialas--Bzdak (LBB) model which may be used in the future for a more detailed analysis of elastic hadronic scattering and odderon effects.
Finally, in \hyperref[chap:summary]{Chapter Summary}, I shortly recapitulate my main results. This dissertation contains also three appendices. \hyperref[sec:app_multscatt]{\mbox{Appendix A}} details the basic idea behind R.~J.~Glauber's diffractive multiple scattering theory in a non-relativistic setting. \hyperref[sec:app_BBcalc]{\mbox{Appendix B}} details some analytic ReBB model calculations. \hyperref[sec:app_biGaussLevy]{\mbox{Appendix C}}
fixes the notation of bivariate Gaussian and symmetric Lévy $\alpha$-stable distributions and gives some additional details on the latter. 
In \hyperref[chap:TP]{\mbox{Chapter Thesis points}}, I present my main scientific achievements formulated in five points. These five points form the basis of this dissertation. \hyperref[chap:publications]{\mbox{Chapter Publications}} lists my scientific publications.  \hyperref[chap:talks]{\mbox{Chapter Talks}} lists the talks I gave at international scientific events and in the TOTEM Experiment.  The acronyms used for particle accelerators and colliding experiments are expanded in \hyperref[chap:LA]{\mbox{Chapter List of acronyms}}.  \hyperref[chap:bib]{\mbox{Chapter Bibliography}} lists the references. At the end of the dissertation, a one-page summary, both in English and in Hungarian, is included. 

I achieved my scientific results presented in this dissertation at ELTE Eötvös Loránd University and HUN-REN Wigner Research Centre for Physics in Budapest, as well as at MATE Institute of Technology in Gyöngyös. 

\vspace{0.6cm}

\textbf{Acknowledgements}
\vspace{0.2cm}

I want to express my sincere gratitude to everyone who contributed to the successful completion of my dissertation and PhD studies in physics. 

First and foremost, I am very thankful to my supervisor, Tamás Csörgő MAE for carefully guiding my scientific work since my master's studies at Eötvös University. \mbox{I am} happy for his wise advice not only in scientific questions but also in life in general. \mbox{Under} his guidance, I continued my theoretical research and embarked on my journey into experimental physics, first as a member of the CERN LHC TOTEM Collaboration and then as a member of the CERN LHC CMS Collaboration. I am grateful to him for providing me with such great opportunities. I express my gratitude to my co-supervisor, \mbox{Prof. Máté Csanád}, for his help and guidance during my studies at Eötvös Loránd University and for providing me the opportunity to gain some teaching \mbox{experience, which I enjoyed.}

I thank my co-authors Dr. Tamás Novák, Dr. Roman Pasechnik, and András Ster. Their collaborative effort, among others, led to comprehensive research concerning the \mbox{$t$-channel} odderon observation using the $H(x)$ scaling property of elastic $pp$ scattering. I also thank the members of the CERN LHC TOTEM and FNAL Tevatron D0 Collaborations with whom I worked on the odderon search. I thank Dr. Frigyes Nemes, who provided valuable technical help concerning coding issues in the initial phase of my research. \mbox{I thank} Prof. Christophe Royon and Prof. Wesley Metzger for the valuable discussions. I especially thank Prof. Royon, who read through my dissertation and provided valuable feedback and suggestions for corrections. 
I thank Dr. Gábor Kasza for his technical help in shaping the style of my dissertation in LaTeX.

I thank Prof. Ferenc Siklér for his comments and questions during my dissertation's pre-defense refereeing stage, which helped to improve the final text. I especially thank Dr. Matteo Giordano for the numerous essential remarks and illuminating comments during my dissertation's pre-defense refereeing stage, which substantially contributed to the refinement of my dissertation.

I am very grateful to Prof. László Jenkovszky for introducing me to the realm of research in the field of high-energy particle diffraction during my bachelor's studies at Uzhhorod National University. Under his guidance and mentorship, I wrote my first scientific papers and gave my first talks at international conferences.

I thank my fiancée, Daniella, for her unwavering support and for making my life happier. Her love gave me the strength and persistence to write this dissertation. \mbox{I thank} my parents for their love and support throughout my life. 
I am grateful to my grandparents, who always supported me and my studies.
My sister and little brother deserve my wholehearted thanks as well for always being there for me. I especially thank my sister, Iza, whose expertise as a graphic designer contributed to producing nice figures in the introductory chapter of this dissertation.


I express my gratitude to my secondary school teachers in Nagydobrony, specifically to my physics and mathematics teachers, whose dedicated efforts played a significant role in shaping my path toward becoming a physicist. I thank all my instructors at Uzhhorod National University and at Eötvös Loránd University who imparted valuable knowledge to me. Finally, I want to express my gratitude to all my colleagues, fellow students, and friends who helped, supported, and encouraged me throughout \mbox{my studies and scientific activities.}
\vspace{0.3cm}

My research was supported by the \'UNKP-18-2
New National Excellence Program of the Ministry of Human Capacities, ÚNKP-21-3 and \'UNKP-22-3 New National Excellence Program of the Hungarian Ministry for Innovation and
Technology from the source of the National Research, Development and Innovation Fund, \'UNKP-23-3 New National Excellence Program of the Ministry for Culture and Innovation from the source of the National Research, Development and Innovation Fund; by the NKFIH grants \mbox{No. FK-123842,} \mbox{FK-123959,} K133046, 2020-2.2.1-ED-2021-00181, and K147557; by the Research Excellence Programme of the Hungarian University of Agriculture and Life Sciences; and by the Márton Áron Szakkollégium program.

~
~
~
~

{\noindent Budapest} \hfill István Szanyi

\noindent September, 2024

\clearpage
\thispagestyle{empty}

\mainmatter

\chapter{High-energy elastic $\mathbf{\rm{pp}}$ and $\mathbf{\rm{p\bar p}}$ scattering}\label{chap:intro}

This dissertation deals with elastic proton-proton and proton-antiproton scattering at high energies and small
momentum transfers and focuses on the search for the $t$-channel odderon exchange. The phenomena under consideration belong to the field of elastic particle diffraction or -- more loosely -- ``soft physics''. In this introductory chapter, I briefly review the basic theoretical tools for studying elastic diffractive processes, the status of odderon search, and the methods I applied in my research. 

In \cref{sec:kinem}, I detail the kinematics of the processes studied. 
In \cref{sec:unit} and \cref{sec:crossing}, I introduce unitarity and crossing symmetry focusing on elastic proton-proton and proton-antiproton scattering. I outline the measurable quantities in \cref{sec:measurables}. In \cref{sec:amplitude}, I detail the two basic and complementary approaches of elastic scattering of hadrons: the $s$-channel eikonal approach and the $t$-channel Regge approach. \cref{sec:oddintro} is dedicated to the odderon, where I briefly outline the theory behind the odderon and the strategy for odderon search in elastic scattering. In \cref{sec:ReBBmodel}, I detail the \mbox{Bialas--Bzdak model} and its improved version, the Real extended \mbox{Bialas--Bzdak} model, which I utilized during my study. In \cref{sec:Hx}, I present the so-called $H(x)$ scaling of elastic $pp$ scattering. In \cref{sec:fittingmethod}, I detail the data fitting methods and the statistical tools I applied in my work. Finally, in \cref {sec:introsum}, I briefly summarize the goals of my work and the methods I used.

\newpage

\section{Kinematics}\label{sec:kinem}

\begin{figure}[hbt!]
	\centering
\includegraphics[width=0.7\linewidth]{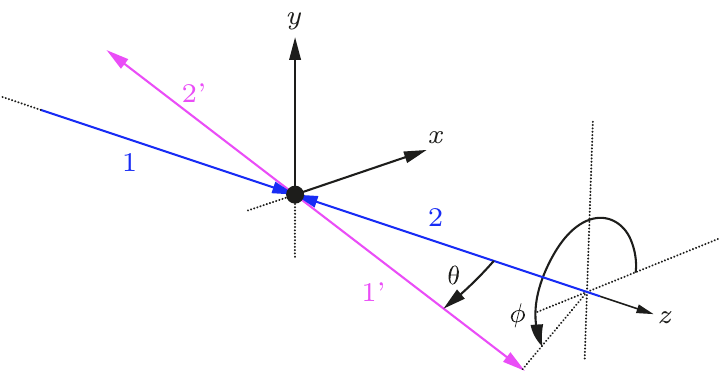}
	\caption{Elastic scattering of two protons in center-of-mass frame. $1$ and $2$ are the incoming protons, $1'$ and $2'$ are the outgoing protons. The arrows represent the protons' momenta. $\theta$ is the polar coordinate or scattering angle while $\phi$ is the azimuthal angle.
 }
	\label{fig:kinematics_cm}
\end{figure}

Elastic proton-proton ($pp$) and proton-antiproton ($p\bar p$) scattering data are measured in colliding experiments where bunches of particles are accelerated and thrown against each other. The setup is such that only one pair of particles collide at a time. If an elastic collision happens -- unless spin polarisation is also measured -- what we observe experimentally is only a change in the direction of motion of the incoming particles. To study the collisions, it is convenient to choose a center of mass frame (c.m.) and quantify the change in the direction of motion by the scattering angle $\theta$ and the azimuthal angle $\phi$ (see \cref{fig:kinematics_cm}). Given that the scattering process is azimuthally symmetric about the beam axis, the interesting physics is encoded in the scattering angle.
\vspace{-5mm}
\vspace{-5mm}

The physics of elastic $pp$ and $p\bar p$ scattering can be conveniently investigated utilizing the Lorentz-invariant Mandelstam variables, $s$, $t$, and $u$. Let us consider a general \mbox{two-particle} to two-particle scattering process, $1+2\rightarrow 1' + 2'$, where each particle has the same mass $m$ and energy $E$. In the c.m. frame of the colliding particles, the four-momenta of the incoming protons are\footnote{Except for Appendix A, throughout in this dissertation, in the formulas, I use the natural units where the speed of light, $c$, and the reduced Planck constant, $\hbar$, are set equal to unity: $c=\hbar=1$.} 
\begin{equation}\label{eq:pin}
    p_1 = (E, \vec{p}),~~~p_2=(E, -\vec{p}),
\end{equation}
and the four-momenta of the outgoing protons are
\begin{equation}\label{eq:pout}
    p'_1 = (E, \vec{p}\,'),~~~p'_2 = (E, -\vec{p}\,').
\end{equation}
In \cref{eq:pin} and \cref{eq:pout}, $\pm\vec p$ and $\pm\vec{p}\,'$ are the particles three momenta before and after the collision, respectively. 
Then, the Mandelstam variables are \cite{Barone:2002cv}
\begin{equation}\label{eq:s}
s=(p_1+p_2)^2=(p'_1+p'_2)^2 = 4E^2,
\end{equation}
\vspace{-12mm}
\begin{align}\label{eq:t}
t&=(p_1-p'_1)^2=(p_2-p'_2)^2 = -2\vec{p}^{\,\, 2} \, (1-\cos\theta),
\end{align}
and
\begin{equation}\label{eq:u}
u=(p_1-p'_2)^2=(p_2-p'_1)^2 = -2\vec{p}^{\,\, 2} \, (1+\cos\theta).
\end{equation}  
$s$ is the squared total c.m. energy, $t$ is the squared four-momentum transfer, and $u$ is the squared four-momentum transfer for the case when particles $1'$ and $2'$ are interchanged. Note that these variables satisfy the relation
\vspace{-2mm}
\begin{equation}\label{eq:stu}
s+t+u=4m^2.
\end{equation}

\begin{figure}[hbt!]
	\centering
\includegraphics[width=0.99\linewidth]{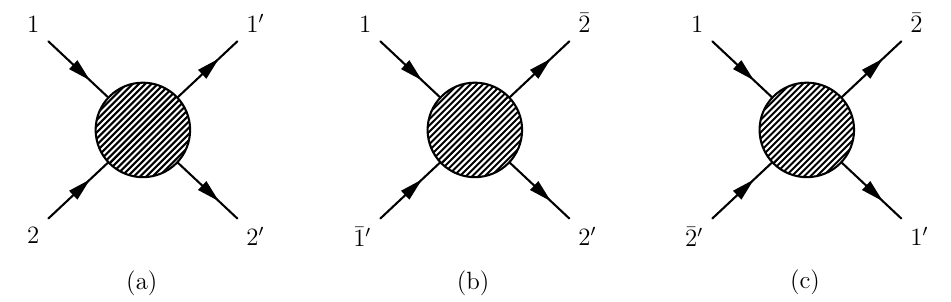}
	\caption{(a) $s$-channel, (b) $t$-channel, and (c) $u$-channel processes. The interactions of the particles are represented by the shaded circles. 
 }
	\label{fig:channles}
\end{figure}

\vspace{-3mm}
For the physical description, it is useful to define $s$-, $t$-, and $u$-channel scattering processes transforming into one another by crossing. Crossed processes are obtained by viewing an incoming (outgoing) particle as an outgoing (incoming) antiparticle with an opposite sign of four-momentum. That is,  if the $s$-channel  or direct channel process is given as $1+2\rightarrow 1' + 2'$,
the corresponding crossed $t$- and $u$-channel processes are $ 1+\bar 1'\rightarrow \bar 2 + 2'$ and $1+\bar 2'\rightarrow \bar 2 + 1'$,
respectively (see \cref{fig:channles}). The bar above the number denotes the antiparticle of the given particle. 

\begin{figure}[hbt!]
	\centering
\includegraphics[width=0.75\linewidth]{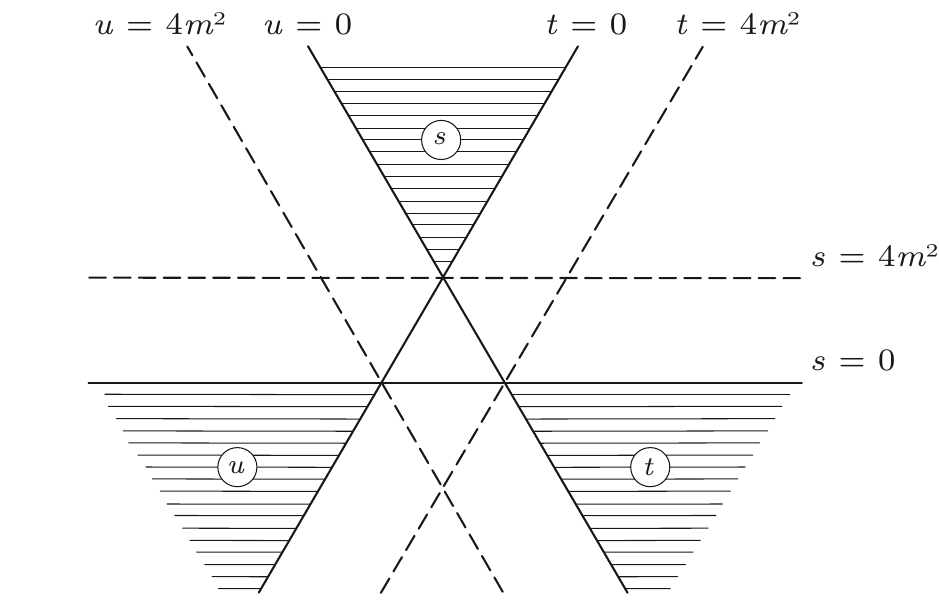}
	\caption{Mandelstam plot for equal mass scattering. Physical regions in the different channels are shown shaded.
 }
	\label{fig:mandelplot}
\end{figure}

In an $s$-channel process, the variable $s$ is the squared total c.m. energy, while the variables $t$ and $u$ are squared four-momentum transfers. In a $t$-channel process, the role of $s$ and $t$ is interchanged: $t$ is the total c.m. energy, while $s$ is the squared four-momentum transfer. 
Analogously, in a $u$-channel process, the role of $s$ and $u$ is interchanged. The physical domains of the Mandelstam variables are represented in \cref{fig:mandelplot}. These are 
\vspace{-5mm}
\begin{eqnarray}
    s\geq 4 m^2,~~~t\leq0,~~~u\leq 0,~~~\text{for the $s$-channel proccess}; \nonumber \\ \nonumber t\geq 4 m^2,~~~s\leq0,~~~u\leq 0,~~~\text{for the $t$-channel proccess}; \\\nonumber u\geq 4 m^2,~~~t\leq0,~~~s\leq 0,~~~\text{for the $u$-channel proccess.}
\end{eqnarray}
\vspace{-5mm}

\vspace{-5mm}
Since protons are spin-$\frac{1}{2}$ particles, it is possible to do experiments with spin-polarized beams and/or measure the spin polarization of the scattered particles. In this work, I analyzed data from experiments where neither the spin orientation of the initial state particles is predetermined, nor the spin direction of the final state particles is measured. As it is the common practice in this case, I simply neglected the spin degree of freedom of the colliding particles during the analysis.

\section{Unitarity}\label{sec:unit}

The scattering of subatomic particles can be described relying on quantum mechanical principles. A complete quantum-mechanical theory of particle interactions would allow us to calculate the probability of getting a particular final state from a given initial state. Scattering happens only when particles interact. Before the interaction, there is an initial state $\ket{i}$ consisting of two free particles. After the interaction, there is a final state $\ket{f}$ consisting of, in general, $n$ free particles. In the case of elastic scattering, the particle content of the final state is the same as that of the initial state. Initial and final states are called asymptotic states, formally at times $t=\mp\infty$. The duration of the interaction is relatively short, and compared to that, the initial state is considered to be in the distant past while the observations on the final state are considered to be made in the distant future. 

The scattering matrix or $S$-matrix is a time-evolution operator that evolves the initial state at $t=-\infty$ to the final state at $t=+\infty$. In formulas \cite{Barone:2002cv}:
\begin{equation}
    S \ket{i} = \ket{f}.
\end{equation}

\noindent Then the probability of the transition from the state $\ket{i}$ to the state $\ket{f}$ is given as 
\begin{equation}\label{eq:prob}
    P_{fi}=|\mel{f}{S}{i}|^2.
\end{equation}

The unitarity of $S$,
\begin{equation}\label{eq:unit}
    S^\dagger S=SS^\dagger=\mathbb{1},
\end{equation}
expresses the conservation of probability: the probability of all possible outcomes of the scattering process must add up to 1,
\begin{equation}\label{eq:elements}
    \sum_j\big|\mel{f_j}{S}{i}\big|^2=1,
\end{equation} 
where $\{\ket{f_j}\}$ is the set of all the possible final states of the scattering process from an initial state $\ket{i}$. 

The $S$-matrix elements between initial and final states can be written as
\begin{equation}
    \mel{f}{S}{i}=\delta_{if}+i\left(2\pi\right)^4 \delta^4\big({\textstyle\sum} p_i-{\textstyle\sum} p_f\big)\mathcal{M}_{i\rightarrow f},
\end{equation}
where $\delta_{if}=\braket{f|i}$, ${\textstyle\sum} p_i$ is the sum of four-momenta of the initial state particles,  ${\textstyle\sum} p_f$ is the sum of four-momenta of the final state particles, and $\mathcal{M}_{i\rightarrow f}$
is called the scattering amplitude for the transition from the state $\ket{i}$ to the state $\ket{f}$. When no interaction happens, the $S$ is simply the identity matrix.

\begin{figure}[hbt!]
	\centering
\includegraphics[width=0.92\linewidth]{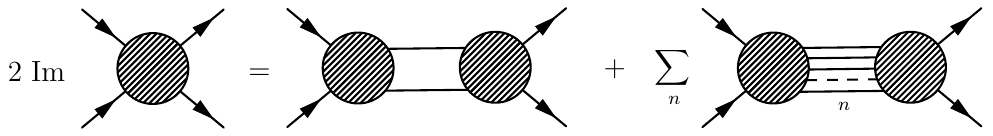}
	\caption{Graphical representation of the unitarity equation \cref{eq:optth0}.}
	\label{fig:unitarity}
\end{figure}

In the case of two-particle to two-particle elastic scattering, when initial and final states are identical, \textit{i.e.}, $\ket{i}=\ket{f}$, unitarity of the $S$-matrix leads to the relation,
\begin{equation}\label{eq:optth0}
    2~{\rm Im }~\mathcal{M}_{\rm el}= \int d\Pi_2 \mathcal{M}_{\rm el}\mathcal{M}_{\rm el}^*+\sum_{n}\int d\Pi_n ~\mathcal{M}_{i\rightarrow n}\mathcal{M}_{i\rightarrow n}^*,
\end{equation}
where $\mathcal{M}_{\rm el}=\mathcal{M}_{i\rightarrow i}$ is the elastic scattering amplitude, $\mathcal{M}_{i\rightarrow n}$ are the inelastic scattering amplitudes for the scattering of the two-particle initial state to an $n$-particle final state, and $d\Pi_n$ is the Lorentz-invariant $n$-particle phase space. A scattering amplitude is associated with a graph that illustrates the corresponding scattering process from a given initial state to a given final state. The graphical representation of the unitarity relation \cref{eq:optth0} is shown on \cref{fig:unitarity}.

\cref{eq:optth0} means that the imaginary part of the elastic scattering amplitude emerges from the contribution of all possible intermediate states that the initial state can get into. The first term on the right-hand side of \cref{eq:optth0} is the elastic or two-particle intermediate state contribution to the imaginary part of the elastic scattering amplitude. The second term on the right-hand side of \cref{eq:optth0} is the inelastic contribution to the elastic scattering amplitude given by the sum over all possible $n$-particle inelastic intermediate states.



Assuming spherically symmetric interactions, the relativistic, $s$-channel elastic scattering amplitude can be expanded to partial wave series and written as \cite{Collins:1977jy} 
\begin{equation}\label{eq:realpartialexp}
    \mathcal{M}_{\rm el}(s,t) = 16\pi \sum_{\ell=0}^\infty(2\ell+1)f_\ell(s)P_\ell(\cos\theta_s(s,t)),
\end{equation}
where $f_\ell(s)$ is the partial-wave scattering amplitude of angular momentum $\ell$, $P_\ell$ is the Legendre polynomial of degree 
$\ell$, and $\theta_s$ is the $s$-channel scattering angle (see \cref{fig:kinematics_cm}). For equal-mass scattering the cosine of the $s$-channel scattering angle is \cite{Collins:1977jy}
\begin{equation}\label{eq:cosths}
    \cos\theta_s(s,t) = 1+\frac{2t}{s-4m^2}.
\end{equation}
The partial wave amplitude in the high-energy limit can be expressed as 
\begin{equation}\label{eq:pwamprel}
    f_\ell(s) = \frac{i}{2}\left(1-S_\ell(s)\right)
\end{equation}
with 
\begin{equation}\label{eq:complexpf}
    S_\ell(s) = e^{2i\delta_\ell(s)},
\end{equation}
where $\delta_\ell(s)$ is the phase shift of the $\ell$-th wave and $S_\ell(s)$ is the element of the $S$-matrix in a state of definite orbital angular momentum $\ell$ \cite{Barone:2002cv}. The $\delta_\ell(s)$ phase shift contains all the dynamical information of the scattering process.

Denoting collectively all the quantum numbers that characterize the initial and final states of the scattering process by $\lambda$ and $\lambda'$, respectively, the $S$-matrix element $S_\ell$ develops additional indices: $S^{\lambda\lambda'}_\ell$. In case of elastic scattering $\lambda = \lambda'$. Setting $S_\ell^{\lambda\lambda}\equiv S_\ell $, the unitarity restriction of the $S$-matrix implies that \cite{Barone:2002cv}
\begin{equation}\label{eq:prtialSunit}
  \sum_{\lambda'} |S_\ell^{\lambda\lambda'}|^2 =|S_\ell|^2 + \sum_{\lambda'\neq \lambda}|S_\ell^{\lambda\lambda'}|^2 = 1,
\end{equation}
where $|S_\ell|^2$ is the contribution from elastic scattering, while $\sum_{\lambda'\neq \lambda}|S_\ell^{\lambda\lambda'}|^2$ is the contribution from inelastic scattering.
For elastic collisions $|S_\ell(s)|=1$ and thus $S_\ell(s)$ can be written as a complex phase factor given in \cref{eq:complexpf} 
with real valued $\delta_\ell(s)$ phase shifts. When inelastic channels are open, $|S_\ell(s)|\leq1$ and the phase shift becomes a complex number and thus \cite{Barone:2002cv}:
\begin{equation}
S_\ell(s)=\eta_\ell(s)e^{2i{\rm Re}\delta_l(s)},
\end{equation}

\noindent where $\eta_\ell(s)=e^{-2{\rm Im}\delta_l(s)}$ is called the absorption (or inelasticity) coefficients \cite{Barone:2002cv, Collins:1977jy}. The unitarity constraints $|S_\ell(s)|\leq1$ imply:
\begin{equation}\label{eq:phaseshuni}
{\rm Im} \delta_l(s)\geq0.    
\end{equation}
Requiring the condition \cref{eq:phaseshuni} to hold,  the unitarity relation \cref{eq:optth0} can be rewritten for the relativistic partial wave amplitudes\footnote{In Ref.~\cite{Barone:2002cv} the unitarity relation for the partial wave amplitudes is written out in the non-relativistic case. Using the correspondence between the non-relativistic and relativistic partial wave amplitudes, $a_\ell(k)\longleftrightarrow\frac{2}{\sqrt{s}}f_\ell(s)$, where $k\underset{s\to\infty}{\simeq}\frac{1}{2}\sqrt{s}$, here I write the unitarity relation for the partial wave amplitudes in the relativistic case.}  as \cite{Barone:2002cv}
\begin{equation}\label{eq:pwunit}
    4{\rm Im}f_\ell(s)=4|f_\ell(s)|^2 +\left(1-e^{-4{\rm Im}\delta_\ell(s)}\right).
\end{equation}

\section{Crossing symmetry}\label{sec:crossing}

Unitarity of the scattering matrix, as well as Lorentz invariance and analyticity of the scattering amplitudes, are the general features of a physical description of scattering \mbox{processes \cite{Barone:2002cv, Schwartz:2014sze}.} The analyticity property means that the scattering amplitudes are analytic functions of the kinematic variables when these are continued to complex values.\footnote{The only singularities of the scattering amplitudes are those required by the unitarity of the scattering matrix. These singularities are simple poles and branch points on the real axis of the variables. A simple pole corresponds to the exchange of a physical particle, while a branch point corresponds to the exchange of two or more physical particles. See more details in Refs. \cite{Barone:2002cv,Collins:1977jy}.} A consequence of analyticity is crossing symmetry: the scattering amplitude of all three crossed-channel processes is a single analytic function that can be analytically continued to different domains of the variables.  

The $p\bar p$ scattering amplitude, $\mathcal{M}^{p\bar p}_{\rm el}$, can be obtained by 
crossing the $s$-channel $pp$ scattering amplitude, $\mathcal{M}^{pp}_{\rm el}$, to the $u$-channel \cite{Ewerz:2003xi,Hu:2008zze, Block:2006hy}:
\begin{equation} \label{eq:ppcross}
    \mathcal{M}^{p\bar p}_{\rm el}(s,t,u) = \mathcal{M}^{pp}_{\rm el}(u,t,s).
\end{equation}
\cref{eq:ppcross} corresponds to the analytical continuation of the $s$-channel scattering amplitude outside the $s$-channel physical domain to obtain the $u$-channel scattering amplitude in the physical domain of the $u$-channel. In other words, $\mathcal{M}^{p\bar p}_{\rm el}$ is obtained from $\mathcal{M}^{p p}_{\rm el}$ by interchanging the roles of the variables $s$ and $u$.

Because of the relation \cref{eq:stu} only two of the three Mandelstam variables are independent. For high-energy and small-scattering-angle elastic scattering, analyzed in this work, the scattering amplitude is advantageously written as a function of \mbox{$s$ and $t$.} $\mathcal{M}^{p p}_{\rm el}(s,t)$ and $\mathcal{M}^{p \bar p}_{\rm el}(s,t)$ are parts of the same function, $\mathcal{M}_{\rm el}(s,t)$, analytic in $s$ and $t$ regarded as complex variables. The physical amplitude used to calculate measurable quantities is recovered in the limit $s,t\rightarrow$ real \cite{Barone:2002cv,Block:2006hy}.

If $\mathcal{M}^{pp}_{\rm el}(s,t)$ and $\mathcal{M}^{p\bar p}_{\rm el}(s,t)$ are the physical $pp$ and $p\bar p$ elastic scattering amplitudes, one can introduce even–under–crossing and odd–under–crossing amplitude components as \cite{Barone:2002cv}
\begin{equation}\label{eq:ampl_M_even}
\mathcal{M}^+=   \frac{1}{2} \left(\mathcal{M}^{p\bar p}_{\rm el}+\mathcal{M}^{pp}_{\rm el}\right)
\end{equation}
and
\begin{equation}\label{eq:ampl_M_odd}
\mathcal{M}^-=  \frac{1}{2} \left(\mathcal{M}^{p\bar p}_{\rm el}-\mathcal{M}^{pp}_{\rm el}\right),
\end{equation}
respectively. $\mathcal{M}^-$ changes sign under crossing from $s$-channel to $u$-channel while $\mathcal{M}^+$ does not: the amplitude component $\mathcal{M}^+$ is the same for $pp$ and $p\bar p$ scattering while 
the amplitude component $\mathcal{M}^-$ changes sign when passing from $pp$ scattering to $p\bar p$ scattering. 
In terms of crossing even and crossing odd components, the $pp$ and $p\bar p$ elastic scattering amplitudes are
\begin{equation}\label{eq:M_ampl_pp}
\mathcal{M}^{pp}_{\rm el} = \mathcal{M}^+ - \mathcal{M}^-
\end{equation}
and
\begin{equation}\label{eq:M_ampl_pbarp}
\mathcal{M}^{p\bar p}_{\rm el}=  \mathcal{M}^+ + \mathcal{M}^-,
\end{equation}
respectively.



\section{Observables}\label{sec:measurables}

In scattering experiments, the main quantity to be measured is the cross section $\sigma$ of the studied process. The cross section is the measure of the probability that the given scattering process takes place, and it has the unit of squared length usually given in barns: 1 b = 10$^{-28}$ m$^2$. In a collider experiment, the performance of the particle accelerator is quantified by the luminosity, $\mathcal{L}$. The higher the luminosity, the higher the number of possible scattering events. The unit of luminosity is inverse squared length times inverse time, usually b$^{-1}$s$^{-1}$.

Basically, when the experiment is performed, the rate of scattering events $R$ in units of inverse time is measured, the luminosity of the particle accelerator is determined, and the cross section of the scattering process is extracted: $\sigma = R/\mathcal{L}$.
While the event rate depends on the details of the experimental setup, the cross section of the process is independent of the details of the experimental setup. 

When scattering events with prescribed final-state features are counted, we use the notion of the differential cross section. Denoting the continuous kinematic variable which specify the final state of the scattering process by $\xi$, the differential cross section denoted as $d\sigma/d\xi$ gives the $\xi$-distribution of the scattering events. Then the cross section in some finite interval of $\xi$ is calculated by integration. The differential cross section can be understood as a measure of the interaction probability for two particles as a function of a continuous variable characterizing the final state.

The cross section of a scattering process is calculated from the scattering amplitude. The interactions that practically determine the elastic scattering of protons are the electromagnetic and the strong interactions. 
Consequently, the amplitude has a Coulomb part and a nuclear part. The interference between Coulomb and nuclear amplitudes is generally expressed by a phase factor and the  full elastic amplitude is written in the \mbox{form \cite{Kundrat:1993sv,Leader:2011vwq}}
\begin{equation}
    \mathcal{M}_{\rm el} = \mathcal{M}_{\rm el}^C + \mathcal{M}_{\rm el}^Ne^{\pm i\delta_C},
\end{equation}
where $\delta_C$ is called the Coulomb phase.\footnote{The phase factor can multiply either the nuclear amplitude or the Coulomb amplitude \cite{Block:2006hy,Kopeliovich:2023xtu}.} The phase factor accounts for the effect of mixed electromagnetic and strong interaction mediator \mbox{exchanges \cite{Block:2006hy}.}

The nuclear component plays a significant role at all kinematically allowed value of $t$ while Coulomb effects are dominant only at very low-$|t|$ values  \cite{Kundrat:1993sv,TOTEM:2017sdy}, typically \mbox{$|t|\lesssim$ 0.01 GeV$^2$.} In this work, I analyze data in a kinematic range where the pure Coulomb and Coulomb-nuclear interference effects are negligible within experimental errors. Thus, in what follows, the scattering amplitude means the purely nuclear component, \mbox{$\mathcal{M}_{\rm el}\equiv\mathcal{M}^N_{\rm el}$}, which describes the effects of the strong interactions between protons.

The differential cross section for a high-energy two-particle to two-particle  azimuthally symmetric scattering process in c.m. frame ignoring the spin degree of freedom of the particles is \cite{Barone:2002cv} 
\begin{equation}\label{eq:dsigma_2}
    \frac{d\sigma_{\rm el}}{dt} (s,t) = \frac{1}{16\pi s^2} \big|\mathcal{M_{\rm el}}(s,t)\big|^2.
\end{equation}
The differential cross section at a given $\sqrt{s}$ energy measures the probability of elastic interaction with squared four-momentum transfer $t$ that results in the scattering of the particles to the angle $\theta$ (see \cref{eq:t}).

The differential cross section at $t=0$ is called the optical point parameter and denoted as 
\begin{equation}\label{eq:optpoint}
    a(s)=\frac{d\sigma_{\rm el}}{dt}(s,t)\Big|_{t=0}.
\end{equation}

The $t$-dependent or local nuclear slope of the differential cross section is defined as
\begin{equation}\label{eq:Bst}
    B(s,t) = \frac{d}{dt} \ln \frac{d\sigma_{\rm el}(s,t)}{dt}.
\end{equation}  
The low-$|t|$ limit of $B(s,t)$,
\begin{equation}\label{eq:Bst0}
    B_0(s)=\lim_{t\to 0}B(s,t),
\end{equation}
is called the nuclear slope parameter or simply slope parameter. The ${t\to 0}$ limit is taken within the low-$|t|$ region free from observable electromagnetic effects.

The elastic cross section is calculated by integrating the differential cross section for all possible squared four-momentum values 
\begin{equation}\label{eq:elastic_cross_section}
\sigma_{\rm el}(s)=\int\limits_{t_{\rm min}}^0 dt~\frac{d\sigma_{\rm el}}{dt}(s,t).
\end{equation}
At high energies when $s\to\infty$, $t_{min}\to-\infty$, however, the essential contribution comes from the domain $-1$ GeV$^2$ $\lesssim t \leq 0$ . The elastic cross section measures the probability of the elastic interaction at a given c.m. energy.

The total cross section of a scattering process with two particles in the initial state can be related to the imaginary part of the elastic scattering amplitude at $t=0$ by the unitarity of the $S$-matrix. The physical situation when the initial and final states of the scattering process are identical is forward elastic scattering where $t=0$.
In the high energy limit, \cref{eq:optth0} leads to the relation \cite{Barone:2002cv}:
\begin{equation}\label{eq:optth}
    \sigma_{\rm tot}(s) = \frac{1}{s} {\rm Im }\mathcal{M}_{\rm el}(s,t=0).
\end{equation}
\cref{eq:optth} is referred to as the optical theorem \cite{Schwartz:2014sze, Barone:2002cv}. The total cross section measures the total interaction probability or interaction strength at a given c.m. energy.

The ratio of the real to imaginary parts of the nuclear elastic scattering amplitude, 
\begin{equation}\label{eq:rhost}
\rho(s,t)= \frac{{\rm Re}\, \mathcal{M}_{\rm el}(s,t)}{{\rm Im}\, \mathcal{M}_{\rm el}(s,t)}, 
\end{equation}
can be measured only at very low-$|t|$ by observing the interference of the nuclear amplitude in the low-$|t|$ region with the Coulomb amplitude\footnote{The elastic nuclear differential cross section in the forward direction depends on $\rho_0$ quadratically. Measuring the interference between the Coulomb and the nuclear amplitude at low $|t|$ values, one can determine both the sign and the value of $\rho_0$ \cite{Pancheri:2016yel} (see \cref{eq:dsdt-exp} and \cref{eq:norma0}).} \cite{Barone:2002cv}. The low-$|t|$ limit of $\rho(s,t)$ is
\begin{equation}\label{eq:rhos}
    \rho_0(s) \, = \, \lim_{t\rightarrow 0} \rho(s,t).
\end{equation}
\cref{eq:rhos} is generally called the $\rho$ parameter. In this work, I call \cref{eq:rhos} as $\rho_0$ parameter to distinguish it from \cref{eq:rhost}.

Defining the phase of the nuclear amplitude, $\zeta(s,t)$, as \cite{Dremin:2012ke} 
\begin{equation}
    \mathcal{M}_{\rm el}(s,t) = i |\mathcal{M}_{\rm el}(s,t)|{\rm e}^{-i\zeta(s,t)},
\end{equation}
the relation between this phase and the $\rho(s,t)$ is
\begin{equation}\label{eq:rhosz}
    \rho(s,t) \, = \, \tan \zeta(s,t),\,\,\text{with} \,\,\zeta(s,t)=\frac{\pi}{2}-\arg\left(\mathcal{M}_{\rm el}(s,t)\right).
\end{equation}

The inelastic scattering cross section, $\sigma_{\rm in}$, that measures the probability of inelastic scattering at a given energy is defined as \cite{Barone:2002cv}
\begin{equation}\label{eq:inelastic_cross_section}
\sigma_{\rm in}(s) =  \sigma_{\rm tot}(s)-\sigma_{\rm el}(s) .
\end{equation}

By measuring $d\sigma_{\rm el}/dt$ we know the absolute value of the elastic nuclear scattering amplitude. The measurement of $\sigma_{\rm tot}$ reveals the imaginary part of the elastic nuclear scattering amplitude at $t=0$, and by measuring $\rho_0$ we acquire knowledge about the real part of the elastic nuclear scattering amplitude at $t=0$. 



\section{Scattering amplitude}\label{sec:amplitude}

\begin{figure}[hbt!]
	\centering
\includegraphics[width=0.4\linewidth]{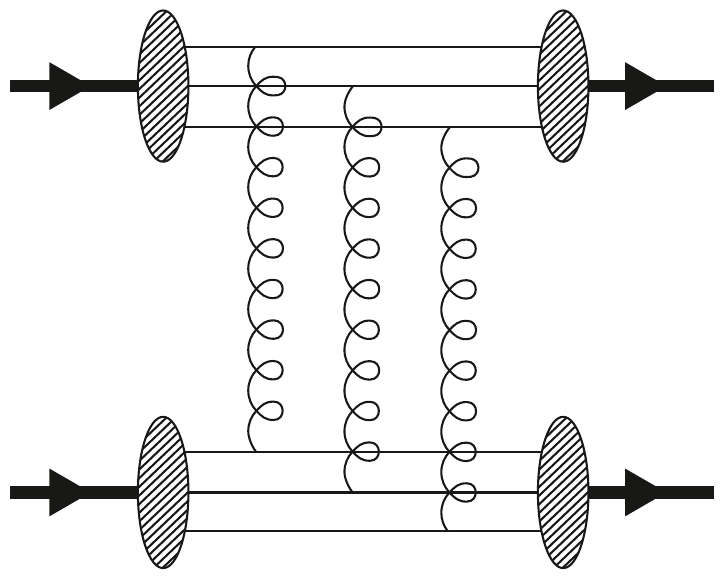}
	\caption{Three-gluon exchange between the valence quarks of two colliding protons.}
	\label{fig:threegluon}
\end{figure}

Perturbative methods can be applied in the fundamental theory of
the strong force, quantum chromodynamics (QCD), to describe the elastic scattering of protons at relatively high $|t|$ values ($|t|\gtrsim$ 3 GeV$^2$) in the framework of the three gluon-exchange model \mbox{\cite{Barone:2002cv,donnachie_dosch_landshoff_nachtmann_2002}.} The interaction at this $|t|$ region can be interpreted as the exchange of three gluons,
which couple to the valence quarks of the protons as illustrated in \cref{fig:threegluon}. The resulting differential cross section has an energy-independent power law decreasing behavior \mbox{in $|t|$ \cite{Donnachie:1979yu,Donnachie:1996rq},}
\begin{equation}
    \frac{d\sigma_{\rm el}}{dt}\sim |t|^{-8}.
\end{equation}
The amplitude of the three-gluon exchange has an opposite sign for $pp$ and $p\bar p$ scattering. As detailed in \cref{sec:oddintro}, the three-gluon exchange is interpreted as a QCD model of the odderon.

To derive a concrete form of the high-energy scattering amplitude that describes the effects of the strong interactions between protons in the non-perturbative squared four-momentum transfer domain $|t|\lesssim$ 3 GeV$^2$, one can use other approaches 
than field \mbox{theory \cite{Matthiae:1994uw}}. In \cref{sec:eik}, I detail the $s$-channel eikonal approach, while in  \cref{sec:Regge}, I discuss the $t$-channel Regge pole approach. In $t$-channel models, the $s$-dependence is almost unambiguous, while the $t$-dependence is less constrained and almost arbitrary. In $s$-channel models, the $t$-dependence is strongly suggested, while the $s$-dependence is rather arbitrary \cite{Barone:2002cv}.  




\subsection{Eikonal picture: diffraction}\label{sec:eik}

The quantum mechanical description of high-energy small-angle elastic scattering is obtained by applying the eikonal approximation\footnote{The eikonal approximation originates in the application of Maxwell's electromagnetic theory to geometrical optics \cite{fried1990basics}. The word \textit{eikonal} was used first by the German mathematician and astronomer Heinrich Bruns \cite{bruns1895eikonal} and can be considered as the German version of the Greek word  \textit{\textgreek{εἰκών}} that means likeness, icon or image.}. The resulting formulas are analogous to those obtained using Kirchhoff's theory of Fraunhofer diffraction of light on absorbing and refracting obstacles based on Maxwell's equations \cite{Barone:2002cv,Glauber:1987sf}. The intensity of the diffracted light at small
angles and large wave numbers has a forward peak and a rapid decrease followed by secondary maxima. A similar pattern is observed in the angular distributions of high-energy elastic hadron-hadron and hadron-nucleon processes \cite{Barone:2002cv,Czyz:1969jg}. This fact allows the use of optical nomenclature when discussing high-energy, small-angle elastic scattering.

\begin{figure}[hbt!]
	\centering
\includegraphics[width=0.4\linewidth]{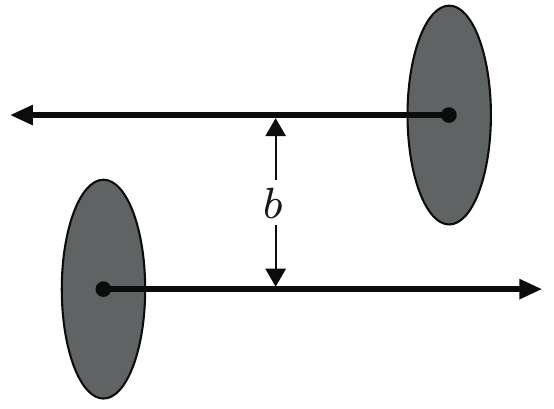}
	\caption{Schematic representation of the collision of two hadrons with impact \mbox{parameter $b$.}}
	\label{fig:impact}
\end{figure}

The eikonal representation of the relativistic high-energy, small-angle elastic scattering amplitude can be obtained from its partial wave expansion (see \hyperref[sec:app_multscatt]{Appendix A} for the eikonal amplitude in the non-relativistic case derived from the Schrödinger equation). The eikonal approach is an approximation for describing low angle ($\theta_s,t\rightarrow0$) scattering at high energies ($s\rightarrow \infty$). At such conditions, many partial waves contribute, especially those for large $\ell$ \cite{Barone:2002cv}.  For large values of $\ell$ and $s$ one can write the semi-classical correspondence \cite{Barone:2002cv,Glauber:2019roq,Block:2006hy}
\begin{equation}\label{eq:sccorr}
    \frac{\sqrt{s}}{2}b\longleftrightarrow\ell+\frac{1}{2},
\end{equation}
where $b=|\vec b|$ is the impact parameter, and $\vec b$ is the two-dimensional impact parameter vector in the plane transverse to the direction of the momenta of the incoming particles. The impact parameter $b$ has the semiclassical interpretation as the minimal distance of approach between the two colliding \mbox{particles \cite{Matthiae:1994uw}.}  Two hadrons colliding at very high energies semiclassically can be imagined as the interaction of two Lorentz-contracted objects, disks of hadronic matter flying through each other \cite{ Barone:2002cv,Dremin:2012ke} (see \cref{fig:impact}).

Since at high energies many partial waves are contributing, $\ell$ can be regarded as a continuous variable and the expanded amplitude of \cref{eq:realpartialexp} can be rewritten \mbox{as \cite{Barone:2002cv,Block:2006hy,Glauber:2019roq}}
\begin{equation}\label{eq:relPWtoeik0}
    \mathcal{M}_{\rm el}(s,t)=4\pi i s \int db b J_0(qb)\Gamma(s,b),\,\,\,q=\sqrt{-t}\,,
\end{equation}
where $J_0$ is the zeroth order Bessel function of the first kind and $\Gamma(s,b)$ is called the profile function. $q$ is the absolute value of $\vec q$, the momentum transfer vector, which is a two-dimensional vector in the plane transverse to the direction of the momenta of the incoming particles. In formulas, $q=|\,\vec q\,|$.
 For azimuthally symmetric processes \cref{eq:relPWtoeik0} is equivalent to a Fourier transform\footnote{Because of the azimuthal symmetry, the $\vec b$-dependence can be written as a $b$-dependence. 
In this dissertation, when $s$ is regarded as a variable, I write $b$-dependence; otherwise, I write $\vec b$-dependence. This convention is beneficial for a clear presentation of the formulas.}:
\begin{equation}\label{eq:relPWtoeik}
    \mathcal{M}_{\rm el}(s,t) = 2 i s \int d^2\vec b e^{i\vec q\cdot\vec b }\Gamma(s,b).
\end{equation}

The profile function, characterizing the ``obstacle'' that produces diffraction pattern, is given as \cite{Barone:2002cv} 
\begin{equation}\label{eq:profile}
    \Gamma(s,b)=1-e^{i\chi(s,b)},
\end{equation}
where $\chi(s,b)$ is called the eikonal function. In the high-energy limit, $\chi(s,b)$ can be expressed in terms of  the phase shift $\delta_\ell(s)$ of the $\ell$-th wave (see the discussion in \cref{sec:unit}) as
\begin{equation}\label{eq:eiktops}
    \chi(s,b) = 2\delta_\ell(s)|_{\ell = \sqrt{s}b/2}.
\end{equation}

Instead of the eikonal function, it is common to use the opacity function \cite{Barone:2002cv}: 
\begin{equation}\label{eq:optoeik}
    \Omega(s,b) = -i\chi(s,b).
\end{equation}
The elastic scattering amplitude in impact parameter representation is defined as
\begin{equation} \label{eq:impamp}
    \widetilde T_{\rm el} (s,b) = i \Gamma(s,b).
\end{equation}
Using \cref{eq:profile} and \cref{eq:optoeik}, the amplitude \cref{eq:impamp} can be rewritten in terms of the opacity function as
\begin{equation}\label{eq:impact_ampl_eik_sol}
    \widetilde T_{\rm el} (s, b) = i \left(1-e^{-\Omega(s,b)}\right).
\end{equation}

Using \cref{eq:relPWtoeik0}, \cref{eq:relPWtoeik}, and \cref{eq:impamp}, one can introduce the elastic scattering amplitude in momentum representation as
\begin{equation}\label{eq:relPWtoeik_2}
    T_{\rm el}(s,t) = 2\pi \int db b J_0(qb)\widetilde T_{\rm el} (s,b) = \int d^2\vec b e^{i\vec q\cdot\vec b }\widetilde T_{\rm el} (s,b)
\end{equation}
so that 
\begin{equation}\label{eq:MtoT}
\mathcal{M}_{\rm el}(s,t)=2sT_{\rm el}(s,t).
\end{equation}
In case, $T_{\rm el}(s,t)$ is known, $\widetilde T_{\rm el} (s,b)$ can be recovered by an inverse Fourier transform:
\begin{equation}\label{eq:invFourier}
    \widetilde T_{\rm el} (s,b) = \frac{1}{(2\pi)^2}\int d^2\vec q e^{-i\vec q\cdot\vec b } T_{\rm el} (s,q).
\end{equation}
The formulas for the elastic differential cross section and total cross section in terms of $ T_{\rm el} (s,t)$ read
\begin{equation}\label{eq:Tdsigma}
    \frac{d\sigma_{\rm el}}{dt} (s,t) = \frac{1}{4\pi} \big|T_{\rm el}(s,t)\big|^2,
\end{equation}
\begin{equation}\label{eq:Ttot}
    \sigma_{\rm tot}(s) = 2{\rm Im }T_{\rm el}(s,t=0).
\end{equation}

The unitarity relation \cref{eq:optth0} for high-energy scattering can be rewritten in a useful form in the impact parameter space. 
Based on \cref{eq:pwamprel}, \cref{eq:complexpf}, \cref{eq:profile}, \cref{eq:eiktops}, and \cref{eq:impamp}, in the high energy limit,
\begin{equation}\label{eq:relPWtoprof}
    f_\ell(s)\big|_{\ell=\sqrt{s}b/2}  = \frac{1}{2}\widetilde T_{\rm el} (s,b).
\end{equation}
Thus, $\widetilde T_{\rm el}(s,b)$, up to a constant factor, is the correspondent of $f_\ell(s)$ in the impact parameter space. Utilizing this correspondence, as well as \cref{eq:eiktops}, \cref{eq:optoeik}, \cref{eq:pwunit}, and considering the constraints for the phase shifts that follow from unitarity, \cref{eq:phaseshuni}, the unitarity relation \cref{eq:optth0} in the impact parameter space in the high energy-limit takes the form
\begin{equation}\label{eq:unitb_2}
    2{\rm Im} \widetilde T_{\rm el} (s,b) = \big| \widetilde T_{\rm el} (s,b)\big|^2 + \tilde\sigma_{\rm in}(s, b),
\end{equation}
where 
\begin{equation}\label{eq:shadowprof}
     \tilde\sigma_{\rm in}(s, b) = 1-\big|e^{-\Omega(s, b)}\big|^2= 1 - e^{-2\text{Re}\Omega(s,b)}
\end{equation}
with the constraint
  \begin{equation}\label{eq:unitcond}
     {\rm Re} \Omega(s,b)\geq 0\,\,\, \left({\rm or~equivalently}~ {\rm Im} \chi(s,b)\geq 0\right).
 \end{equation}
$\sigma_{\rm in}(s, b)$ is called the shadow profile function or inelastic overlap function \cite{Barone:2002cv}. The consequence of the  constraint \cref{eq:unitcond} that follows from the unitarity constraints on the partial wave amplitudes is
\begin{equation}
    0\leq\tilde\sigma_{\rm in}(s, b)\leq1.
\end{equation}
$\tilde\sigma_{\rm in}(s, b)$ gives the probability of inelastic scattering (absorption) associated with each value of the impact parameter. $\tilde\sigma_{\rm in}(s, b)$ corresponds to the second term on the right-hand side of \cref{eq:optth0} that gives the contribution of all inelastic intermediate states and determines how absorptive the interaction region is.

Calculating the elastic cross section by \cref{eq:elastic_cross_section} and the total cross section by \cref{eq:optth} with the amplitude of \cref{eq:realpartialexp}, in the high-energy low-scattering-angle limit, we find that \cite{Barone:2002cv}
\begin{equation}\label{eq:sigelPW}
    \sigma_{\rm el}(s) = \int d^2\vec b |\widetilde T_{\rm el} (s,b)|^2,
\end{equation}
\begin{equation}\label{eq:sigtotPW}
    \sigma_{\rm tot}(s) = 2\int d^2\vec b {\rm Im }\widetilde T_{\rm el} (s,b).
\end{equation}
Considering \cref{eq:sigelPW}, \cref{eq:sigtotPW}, and the definition of the inelastic cross section,  \cref{eq:inelastic_cross_section}, it follows form \cref{eq:unitb_2}  that  the quantity
\begin{equation}\label{eq:unitb_2a}
     \tilde\sigma_{\rm in}(s, b) = 2{\rm Im} \widetilde T_{\rm el} (s,b) - \big| \widetilde T_{\rm el} (s,b)\big|^2,
\end{equation}
after integrating over the impact-parameter plane, gives the inelastic cross section: 
\begin{equation}\label{eq:siginelPW}
    \sigma_{\rm in}(s) = \int d^2\vec b \tilde\sigma_{\rm in}(s, b).
\end{equation}




Note that the characteristics of the ``obstacle'', producing a diffraction pattern,  at a given c.m. energy and impact parameter $b$ is specified by the profile function $\Gamma(s,b)$ given in terms of the opacity function $\Omega(s,b)$. It follows from \cref{eq:shadowprof}, that the scattering is completely absorptive (refractive) at a given energy and impact parameter if $\Omega(s,b)$ is purely real (imaginary). 

Let us investigate now the unitarity equation of \cref{eq:optth0}. At high c.m. energies, more particle production channels contribute, and the number of intermediate states, $n$, tends to infinity. Hence, at high c.m. energies, the imaginary part of the forward elastic scattering amplitude is a sum over infinitely many positive terms. Since no such condition exists for the real part, the scattering amplitude should be predominantly \mbox{imaginary \cite{Barone:2002cv}.} This imaginary part decreases rapidly away from the forward direction because the phases of the extremely large number of terms in the right-hand side of \cref{eq:optth0} vary \mbox{randomly \cite{Barone:2002cv}}. These conclusions are supported by the measurements. At low-$t$, the $pp$ and $p\bar p$ elastic differential cross section show a sharp forward peak which is a typical feature of Fraunhofer diffraction \cite{Barone:2002cv,Matthiae:1994uw}.  This forward peak is generally called diffractive cone.

\begin{figure}[hbt!]
	\centering
\includegraphics[width=0.8\linewidth]{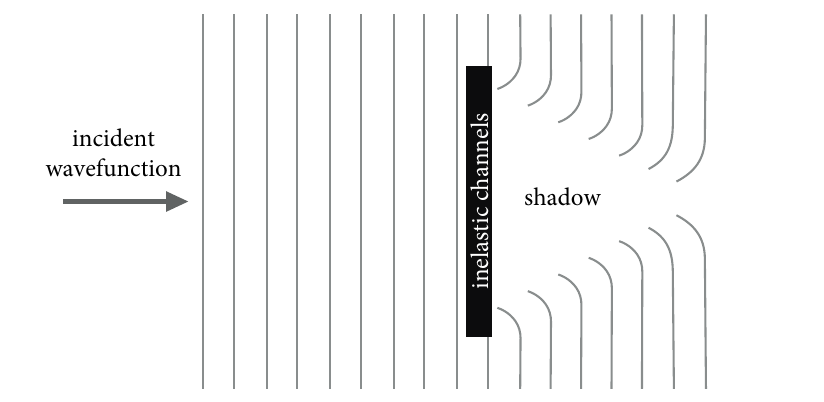}
	\caption{Schematic representation of high-energy particle diffraction.}
	\label{fig:diffraction}
\end{figure}

Given the elastic scattering amplitude is dominantly imaginary, it follows that $\Omega(s,b)$ is dominantly real and elastic scattering is dominantly absorptive scattering. Absorptive scattering results from the propagation of the attenuated incoming wave function, and this attenuation happens 
because of the absorption of the incident hadronic wave in some region of the impact parameter due to many open inelastic channels. In the usual terminology, it is said that the infinite number of inelastic open channels at high energies builds up a ``shadow'' leading to ``diffraction'', $i.e.$, elastic and ``quasi-elastic'' \mbox{scattering \cite{Barone:2002cv,Matthiae:1994uw}} (see \cref{fig:diffraction}). Diffracted waves add up coherently in the forward direction, giving rise to a sharp forward peak \cite{Matthiae:1994uw}, the diffractive cone.  \cref{eq:unitb_2} shows that if inelastic scattering happens, $i.e.$, $\tilde\sigma_{\rm in}(s, b)>0$, then $Im \widetilde T_{\rm el} (s,b)>0$ and $\big| \widetilde T_{\rm el} (s,b)\big|^2\neq 0$. This means that inelastic scattering is always accompanied by elastic scattering. 

Many diffraction models were constructed in the past with specific assumptions on the shape of the $b$ distribution of the opacity function, $\Omega(s,b)$. For a short review, see Refs.~\cite{Barone:2002cv,Matthiae:1994uw}. The ReBB model \cite{Nemes:2015iia}, used in this work, is detailed in \cref{sec:ReBBmodel}, where the $b$ distribution of $\tilde\sigma_{\rm in}(s,b)$ is constructed based on R. J. Glauber's diffractive 
multiple scattering theory and the functional form of $\Omega(s,b)$ is chosen in terms of $\tilde\sigma_{\rm in}(s,b)$.





\subsection{Regge pole approach and the pomeron}\label{sec:Regge}

The Regge pole approach is a great tool to explain high-energy and small-angle hadronic scattering. In this approach, the high-energy behavior of the relativistic scattering amplitude is determined by its singularities in the complex angular momentum plane \cite{Barone:2002cv}. 

In 1959, Tullio Regge continued the angular momentum to complex values in non-relativistic quantum mechanics \cite{Regge:1959mz,Regge:1960zc}. In 1961, Chew, Frautschi \cite{Chew:1961ev} and Gribov \cite{Gribov:1961ex} investigated the high-energy relativistic scattering amplitude with complex values of angular momentum. By continuing the angular momentum to the complex plane, they obtained a representation of the relativistic elastic scattering amplitude, $\mathcal{M}_{\rm el}(s,t)$, valid in all channels, $s$, $t$, and $u$. It turned out that the leading complex angular momentum singularity of the partial wave amplitude in a given channel determines the asymptotic behavior in the crossed channels. In the following, I briefly summarize the idea of complex angular momentum in relativistic scattering theory based mainly \mbox{on Refs.~\cite{Barone:2002cv,Collins:1977jy}.}

The relativistic, $t$-channel scattering amplitude expanded to partial wave series \mbox{reads \cite{Collins:1977jy}} 
\begin{equation}\label{eq:realpartialexp_tch}
    \mathcal{M}_{\rm el}(s,t) = 16\pi \sum_{\ell=0}^\infty(2\ell+1)f_\ell(t)P_\ell(\cos\theta_t(s,t)),
\end{equation}
where $f_\ell(t)$ is the relativistic, $t$-channel partial-wave scattering amplitude of angular momentum $\ell$, $P_\ell$ is the Legendre polynomial of degree 
$\ell$, and $\theta_t$ is the $t$-channel scattering angle. For equal-mass scattering the cosine of the $t$-channel scattering angle is \cite{Collins:1977jy}
\begin{equation}\label{eq:cosths}
    \cos\theta_t(s,t) = 1+\frac{2s}{t-4m^2}.
\end{equation}
$f_\ell(t)$ can be analytically continued to complex $\ell$ values to obtain an interpolating function, $f(\ell,t)$, which reduces to $f_\ell(t)$ for real integer $\ell$. With some additional assumptions on $f(\ell,t)$ \cite{Barone:2002cv}, this continuation can be uniquely done by defining  two different amplitudes $f^+_\ell(t)$ and $f^-_\ell(t)$, which coincide with $f_\ell(t)$ for even and odd integer values of \mbox{$\ell$, respectively, $i.e.$,}
\begin{equation}
    f^+_\ell(t) = f_\ell(t)\,\,\,{\rm for}~\ell=0,~2,~4,~...
\end{equation}
and
\begin{equation}
    f^-_\ell(t) = f_\ell(t)\,\,\,{\rm for}~\ell=1,~3,~5,~...~.
\end{equation}
This corresponds to introducing a new quantum number, the signature $\xi$, which can take two values: $\xi=\pm 1$. The amplitude $f^+_\ell(t)$ has a positive signature while the amplitude $f^-_\ell(t)$ has a negative signature. In terms of the definite signature partial wave amplitudes $f^\xi_\ell(t)$ the definite signature elastic scattering amplitude is given by the partial wave expansion 
\begin{equation}\label{eq:realpartialexp_t}
    \mathcal{M}^\xi(s,t) = 16\pi \sum_{\ell=0}^\infty(2\ell+1)f^\xi_\ell(t)P_\ell(z_t),
\end{equation}
where $z_t \equiv   z_t(s,t) = \cos\theta_t(s,t)$.
The full scattering amplitude
is given in terms of positive-signature, crossing-even and negative-signature, crossing-odd
amplitude components:
\begin{equation}\label{eq:fullampl}
    \mathcal{M}_{\rm el}(s,t)=\frac{1}{2}\left[ \mathcal{M}^+(z_t,t)+\mathcal{M}^+(-z_t,t)+\mathcal{M}^-(z_t,t)-\mathcal{M}^-(-z_t,t)\right].
\end{equation}
Since $P_\ell(-z_t)=(-1)^\ell P_\ell(z_t)$ and the crossing operation yields $z_t \to - z_t$, only even(odd)-signature amplitude contributes to the crossing-even(odd) part of the physical scattering amplitude.

The scattering amplitude with a definite signature is obtained by a contour integral of the definite signature partial wave amplitude, $f^\xi(\ell,t)$, in the complex $\ell$ plane:

\begin{equation}\label{eq:WStransform}
    \mathcal{M}^\xi(s,t) = -\frac{16\pi}{2i}\int_C\left(2\ell+1\right)f^\xi(\ell,t)\frac{P_\ell(-z_t)}{\sin \pi \ell}{\rm d}\ell,
\end{equation}
where the contour $C$ embraces the positive integers and zero, but avoids any singularities of $f^\xi(\ell,t)$. \cref{eq:WStransform} is called a Watson-Sommerfeld transform of the partial wave amplitude $f^\xi(\ell,t)$. Provided that $f^\xi(\ell,t)$ is analytic in $\ell$, \cref{eq:WStransform} is equivalent to \cref{eq:realpartialexp_t}.

More than one pole of the same signature can contribute to the scattering amplitude. Let us suppose that in the domain ${\rm Re}\ell>-\frac{1}{2}$, $f^\xi(\ell,t)$ has singularities, simple poles at $\ell=\alpha_{\xi n}(t)$, $n=1,~2,...$, which have the form:  
\begin{equation}
f^\xi(\ell,t) \underset{\ell \to \alpha_{\xi n}(t)}{\sim}  \frac{\beta_{\xi n}(t)}{\ell - \alpha_{\xi n}(t)}   ,
\end{equation}
where $\alpha_{\xi n}(t)$ are called Regge trajectories and give the location of the signature $\xi$ poles of the partial wave amplitude, and $\beta_{\xi n}(t)$ are the residues at the corresponding poles. Then, we can rewrite
\cref{eq:WStransform} as
\begin{align}\label{eq:WStransform_re}
    \mathcal{M}^\xi(s,t) =& -\sum_{n}16\pi^2\left(2\alpha_{\xi n}(t)+1\right)\beta_{\xi n}(t)\frac{P_{\alpha_{\xi n}(t)}(-z_t)}{\sin \pi \alpha_{\xi n}(t)} \\\nonumber&-\frac{16\pi}{2i}\int_{C'}\left(2\ell+1\right)f^\xi(\ell,t)\frac{P_\ell(-z_t)}{\sin \pi \ell}{\rm d}\ell,
\end{align}
where the integration contour $C'=(-1/2-i\infty,-1/2+i\infty)$. For $s \to \infty$, the second term of \cref{eq:WStransform_re}, called ``background integral'', vanishes because of the asymptotic $z$ behavior of $P_\ell(z)$. The first term of \cref{eq:WStransform_re} is the contribution of signature $\xi$ Regge poles, $i.e.$, simple poles in the complex $\ell$ plane, to the scattering amplitude. For $s \to \infty$, the signature $\xi$ scattering amplitude is\footnote{The powers of $s$ should be normalized to some fixed value $s_0$ that sets the scale beyond which the asymptotic expressions are valid \cite{Barone:2002cv}. Typically $s_0\simeq1$ GeV$^2$. For simplicity, this $s_0$ is omitted, but one should keep in mind that any asymptotic formula is a function of $s/s_0$.}
\begin{align}\label{eq:signampl}
    \mathcal{M}^\xi(s,t) \underset{s\to\infty}{\simeq} -\sum_{n}\beta_{\xi n}(t)\frac{s^{\alpha_{\xi n}(t)}}{\sin \pi \alpha_{\xi n}(t)}
\end{align}
and the leading behavior is determined by the pole with the largest ${\rm Re}\,\alpha_{\xi n}(t)$. Thus, the leading angular momentum singularities of the partial wave amplitude in the $t$-channel determine the asymptotic behavior in the crossed $s$-channel. The full amplitude based on \cref{eq:fullampl} is given by the sum over the contributions by definite-signature Regge poles,
\begin{align}\label{eq:signamplfull}
    \mathcal{M}_{\rm el}(s,t) \underset{s\to\infty}{\simeq} -\sum_{\xi}\sum_{n}\beta_{\xi n}(t)\frac{1+\xi e^{-i\pi\alpha_{\xi n}(t)}}{\sin \pi \alpha_{\xi n}(t)}s^{\alpha_{\xi n}(t)},
\end{align}
where the factor
\begin{align}\label{eq:signfact}
    -\frac{1+\xi e^{-i\pi\alpha_{\xi n}(t)}}{\sin \pi \alpha_{\xi n}(t)}
\end{align}

\noindent is called the signature factor. Furthermore, in \cref{eq:signamplfull}, the irrelevant $t$-dependent factors are incorporated into the residues $\beta_{\xi n}(t)$.

The shape of the residues $\beta_{\xi n}(t)$ and the trajectories $\alpha_{\xi n}(t)$ are not fixed. Bounds on the asymptotic and threshold behavior of the trajectory exist, but in most phenomenological applications, exponential residues and linear trajectories are used \cite{Jenkovszky:1987yd} as they are quite well suited to the experimental data. There are also theoretical arguments for linear
mesonic Regge trajectories based on a string picture of $q\bar q$ interactions (see, e.g., Ref.~\cite{Greensite:2003bk}).

Single particle intermediate states 
give simple poles in the amplitude in the corresponding energy variable. In the $t$-channel, the creation of physical particles (strong interaction resonances or bound states) gives rise to poles at some values of $t$. In the crossed $s$ or $u$ channel, the same poles are still there at those $t$ values, but now they appear as poles in squared four-momentum transfer at unphysical values. The scattering and the spectroscopy domains of $t$ are connected by extrapolation \cite{Phillips:875552}. Pole locations extrapolated from the spectroscopy domain to the scattering domain through the Regge trajectory give rise to particle exchanges that mediate the strong interaction between hadrons. This particle exchange is referred to as $t$-channel exchange. Thus, we say that the $s$-channel asymptotic behavior of the high energy scattering amplitude is determined by the exchanges of families of particles in the crossed $t$-channel with quantum numbers such that they may be formed in the crossed $t$-channel. Particles in a family have the same quantum numbers except for the spin ($J$) and the mass $m$.  

\begin{figure}[!hbt]
\centering
\includegraphics[width=0.95\linewidth]{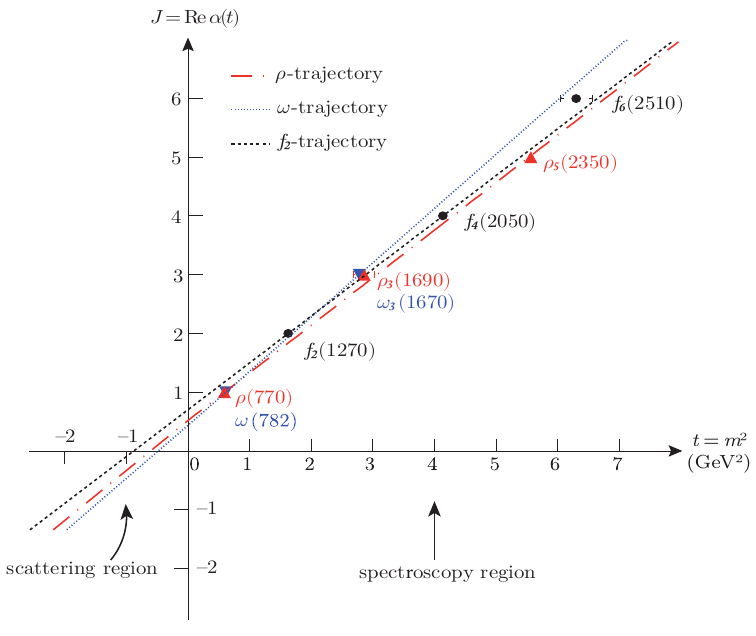}
 \vspace{-5mm}
 \caption{Chew-Frautschi	plot for the $\rho$, $\omega$ and $f_2$-trajectory with the corresponding experimentally measured mesonic spectra. Lines are calculated by fitting a linear function to each family of particles.}
	\label{fig:chew_frautschi}
 \vspace{-2mm}
\end{figure}

For real and positive values of $t$ Regge poles represent bound states and resonances of increasing angular momentum. A Regge trajectory interpolates such resonances with spin $J={\rm Re}\alpha(t)$ and squared mass $m^2=t$. The plot of the particles' spin versus their squared masses is called Chew-Frautschi	plot. This plot is shown for the $\rho$-, $\omega$- and \mbox{$f_2$-trajectory}
in \cref{fig:chew_frautschi}. One can see that mesonic trajectories are essentially linear; they can be approximated with a real linear function\footnote{Note that linear trajectories are only an approximation to reality. Non-linear (complex) trajectories are required by $t$-channel unitarity and preferred by the experimental data \cite{Jenkovszky:1987yd,Jenkovszky:2017efs,Szanyi:2019kkn}. }
\begin{equation}\label{eq:lintra}
    \alpha(t)=\alpha(0)+\alpha't,
\end{equation}
where $\alpha(0)$ is the intercept of the trajectory and $\alpha'$ is the slope of the trajectory. 
There are also similar linear baryonic trajectories leading to the conclusion that the experimentally observed strongly interacting composite particles lie on essentially linear Regge trajectories.

We say that the $s$-channel asymptotic behavior of the scattering amplitude is determined by Regge pole exchanges, reggeon exchanges, or Regge trajectory exchanges corresponding to exchanges of whole families of particles. A trajectory exchange can be interpreted as an exchange of a virtual particle with continuously \mbox{varying spin and virtuality \cite{Jenkovszky:2018itd}.}

The denominators of the terms in \cref{eq:signamplfull} vanish whenever $\ell=\alpha(t)$ crosses an integer. Because of the signature factors, the numerators of the terms in \cref{eq:signamplfull} also vanish, but only at every other integer value of $\ell$. Consequently, a positive signature $\xi=1$ trajectory interpolates between even angular momentum states, whereas a negative signature $\xi=-1$ trajectory interpolates between odd angular momentum states. 

Experimentally the trajectory can be found from high-energy scattering data ($t<0$) or by measuring the hadronic resonances ($t>0$). This way, Regge trajectories anticipate duality by relating low- and high-energy phenomena \cite{Jenkovszky:1987yd}. 


\cref{eq:signampl} implies that the contribution of a single definite-signature Regge pole to the total cross section, \cref{eq:optth},  is 
\begin{equation}\label{eq:reggetot}
    \sigma_{\rm tot}(s) \underset{s\to\infty}{\simeq} s^{\alpha_\xi(0)-1},
\end{equation}
whereas to the differential cross section, \cref{eq:dsigma_2}, is
\begin{equation}\label{eq:reggediff}
    \frac{d\sigma_{\rm el}}{dt}(s,t)\underset{s\to\infty}{\simeq}s^{2\alpha_\xi(0)-2}{\rm e}^{-B|t|}
\end{equation}
with 
\begin{equation}\label{eq:reggeslope}
    B=B_0+2\alpha'_\xi\ln s.
\end{equation}
\cref{eq:reggediff} with \cref{eq:reggeslope} implies that the width of the forward peak decreases with increasing energy: the exponential fall-off of the $t$-distribution has a slope parameter that increases with energy. This effect, called shrinkage of the diffractive peak or diffractive cone, is observed experimentally \cite{TOTEM:2017asr}. The shrinkage, interpreted as an increase of the interaction radius, $R_{\rm int}\underset{s\to\infty}{\simeq}\sqrt{\alpha'\ln s}$, is a non-trivial achievement of Regge theory \cite{Barone:2002cv}.

We also see on \cref{fig:chew_frautschi} that, for all the leading meson trajectories, $\alpha(0)\lesssim 0.8$. Other mesonic trajectories, such as those interpolating strange particles, have even lower intercepts. It follows from \cref{eq:signamplfull} that the smaller the intercept, the less important the contribution is as $s$ grows. This means that the contribution of the mesonic trajectories is highly suppressed at very high energies, including the TeV energy range probed by the LHC.

The total hadronic cross sections, as a function of the c.m. energy, below a few GeV, usually have a complex structure with peaks due to the formation of excited hadronic states. At higher energies, hadron-hadron total cross sections have smooth behaviors: these total cross sections, as a function of the c.m. energy, are rather flat around \mbox{$\sqrt s\sim$ (10$-$20) GeV$^2$} and increase at higher energies \cite{TOTEM:2017asr}. As discussed above, the exchange of mesonic Regge poles with $\alpha(0)<1$ leads to total cross sections decreasing with energy. To explain constant total cross sections, in 1961, Chew, Frautschi \cite{Chew:1961ev} and Gribov \cite{Gribov:1961ex} introduced a Regge trajectory with intercept $\alpha(0)=1$ and the quantum numbers of vacuum. This reggeon, after I. Ya. Pomeranchuk, was named pomeron. 

In the 1970s, it was experimentally observed that the hadronic cross sections rise with energy, implying that the intercept of the pomeron trajectory should be slightly above 1. From fitting elastic scattering data, one finds \cite{Barone:2002cv,donnachie_dosch_landshoff_nachtmann_2002, Matthiae:1994uw,Donnachie:1992ny} that the pomeron intercept is $\alpha_P(0)\simeq 1.08$ while the pomeron slope \mbox{$\alpha_P'\simeq 0.25$ GeV$^{-2}$}.


Thus, in the TeV energy domain, the pomeron exchange gives the dominant contribution in elastic hadron-hadron scattering. Pomeron has the quantum numbers of vacuum,
$$\mathbb{P}:~~P=+1,~C=+1,~G=+1,~I=0,~\xi=+1\,,$$ where $P$ is the spatial parity, $C$ is the charge parity, $G$ is the $G$-parity, $I$ is the isospin, and $\xi$ is the signature. Since $\xi=+1$, the pomeron-dominated high energy forward elastic scattering amplitude behaves\footnote{The origin of the prefactor $i$ in \cref{eq:reggepom} is the presence of the signature factor. This signature factor (see \cref{eq:signfact}) is equal to $i$ in the limit $\alpha_\mathbb{P}\to1$ \cite{Barone:2002cv}.} as 
\begin{equation}\label{eq:reggepom}
    \mathcal{M}_{\rm el}(s,t=0) \underset{s\to\infty}{\simeq} i\beta_\mathbb{P}(0)s^{\alpha_\mathbb{P}(0)}
\end{equation}
and is purely imaginary in harmony with the conclusions drawn in \cref{sec:eik} based on the optical analogy. This way, scattering dominated by pomeron exchange is called diffractive scattering or diffraction \cite{Barone:2002cv}. The leading mesonic trajectories contributing to $pp$ and $p\bar p$ scattering but highly suppressed at TeV and higher energies are $f_2$, $a_2$, $\omega$, and $\rho$. Their quantum numbers are $$f_2:~~P=+1,~C=+1,~G=+1,~I=0,~\xi=+1\,,$$ $$a_2:~~P=+1,~C=+1,~G=-1,~I=1,~\xi=+1\,,$$ $$\omega:~~P=-1,~C=-1,~G=-1,~I=0,~\xi=-1\,,$$ $$\rho:~~P=-1,~C=-1,~G=+1,~I=1,~\xi=-1\,.$$

The Froissart-Martin theorem, which can be proved using analyticity and \mbox{unitarity \cite{Matthiae:1994uw},} states that the total cross section can not grow faster than $\ln^2s$ at asymptotic energies: $\sigma_{\rm tot}\leq C \ln^2s$ as $s\to\infty$, where $C$ is a constant and $C\gtrsim60$ mb \cite{Barone:2002cv}. At asymptotically high energies, $\alpha_P(0)> 1$ violates unitarity since, in this case, the growth of the total cross section exceeds the Froissart-Martin asymptotic bound.  However, this violation would happen only at very huge energies, $\sqrt s\sim10^{27}$ GeV \cite{Matthiae:1994uw}. One can argue that the presently available energies are far not asymptotic, and later on, a presently unknown mechanism will set in and unitarize the cross section \cite{Barone:2002cv}. It is also known that simple poles are not the only singularities in the complex angular momentum plane \cite{Barone:2002cv}. Besides simple poles, there are more complicated singularities, such as higher-order poles (multiple poles) or cuts giving logarithmic corrections. A cut singularity corresponds to the exchange of two or more reggeons. Regge cuts are important to understand $s$-channel unitarity in the framework of Regge theory. Interpreting \cref{eq:signampl} as a Born approximation, higher order terms corresponding to multiple reggeon exchanges might take care of restoring unitarity \cite{Barone:2002cv, Ewerz:2003xi,Collins:1977jy,Matthiae:1994uw,Covolan:1992mf}. 



The ultimate goal would be to derive Regge theory from QCD, the microscopic theory of the strong interactions, and to determine the positions of Regge singularities
from first principles \cite{Ewerz:2003xi}. The first steps were made by Low \cite{Low:1975sv} and Nussinov \cite{Nussinov:1975mw}, describing the pomeron as a colorless state of two gluons. Balitzkii, Fadin, Kuraev, and \mbox{Lipatov \cite{Kuraev:1977fs,Balitsky:1978ic},} in a perturbative approach, described the pomeron exchange as the exchange of a colorless state of two interacting gluons in the $t$-channel. This description is called the BFKL pomeron and motivates the interpretation of the pomeron trajectory as a glueball\footnote{Besides mesons and baryons, QCD allows the formation of bound states of gluons called glueballs. Glueballs are composite particles solely consisting of gluons without valence quarks: a glueball's valence degrees of freedom are gluons \cite{Mathieu:2008me}. Direct channel glueballs are hard to observe; there has been no definite experimental evidence for their existence so far. Pure gluonic states are also likely to mix with mesonic states carrying the same quantum numbers. See Refs.~\cite{Mathieu:2008me, Athenodorou:2023ntf} for more details on glueballs and their mixing with mesonic states.} trajectory: as mesonic trajectories interpolate between mesonic states, the pomeron trajectory, in the $t>0$ domain, may interpolate between glueball states composed of an even number of gluons. However, there has been no experimental observation of direct channel glueballs and there is no compelling theoretical argument that the pomeron trajectory must be a glueball trajectory~\cite{donnachie_dosch_landshoff_nachtmann_2002}. It
is not clear how the perturbative ``hard'' pomeron relates to the ``soft'' pomeron that describes elastic scattering in the non-perturbative kinematic domain\footnote{In Ref.~\cite{Szanyi:2019kkn}, based on a Regge phenomenological model fitted to elastic $pp$ and $p\bar p$ scattering data, I and my co-authors made predictions for the masses and decay widths of states lying on the leading pomeron trajectory with quantum numbers $J=2,~4,...$ and $P=C=+1$ . Our predictions give lower masses than lattice QCD calculations on glueball masses with the same quantum numbers \cite{Llanes-Estrada:2021evz}.}.
See Refs.~\cite{Barone:2002cv,donnachie_dosch_landshoff_nachtmann_2002,forshaw1997quantum} for more details on QCD pomeron.

Recently, central exclusive production of charged-hadron pairs resulting from the dominant double-pomeron exchange was studied in $pp$ collisions at $\sqrt{s}=13$ TeV by the CMS and TOTEM Collaborations in Ref.~\cite{TOTEM:2024aso}. A rich structure of interactions related to double-pomeron exchange was observed, and various physical quantities related to pomeron physics were fine-tuned and determined.


\section{The odderon}\label{sec:oddintro}

The theoretically and experimentally rather well-established pomeron may not be the only contribution that can survive at very high, such as TeV c.m. energies. Another candidate is the odderon, the $P=C=-1$ counterpart of the pomeron. Contrary to the pomeron, the odderon, the main topic of this thesis, although rather well-established theoretically, was not supported by experimental evidence for almost half a century after its invention. In \cref{sec:oddth}, I introduce the concept of the odderon in more detail. Then, in \cref{sec:oddsearch}, I detail the strategy of odderon search in elastic $pp$ and $p\bar p$ scattering I utilized in my work. 

\subsection{Theory behind the odderon}\label{sec:oddth}

The odderon was proposed by Lukaszuk and Nicolescu in 1973 \cite{Lukaszuk:1973nt} in terms of Regge language. In Regge theory, the odderon is a Regge singularity in the odd-signature, \mbox{$\xi=-1$,} scattering amplitude, $\mathcal{M}^-$, in the complex angular momentum plane near \mbox{$\ell=1$ \cite{Ewerz:2003xi,donnachie_dosch_landshoff_nachtmann_2002}.} The odderon trajectory intercept is close to unity: $\alpha_\mathbb{O}(t)\simeq 1$. Theoretically, the odderon satisfies all the basic requirements of analyticity and unitarity; it does not vanish relative to the Pomeron contribution or vanishes but only slowly with a small power of $s$ or a power of $\ln s$.   

Assuming (i) asymptotically vanishing odd-under-crossing scattering amplitude, $\mathcal{M}^-$  and (ii)  asymptotically finite total cross sections, the original or strong Pomeranchuk theorem states that the difference between the particle-particle and particle-antiparticle total cross sections is vanishing at asymptotic energies \cite{Barone:2002cv, Matthiae:1994uw, Ewerz:2003xi}: $
    \sigma_{\rm tot}^{p p}-\sigma_{\rm tot}^{p\bar p} \underset{s\to\infty}{\to} 0.
$ However, the original Pomeranchuk theorem can be generalized to a weaker form including the possibility of asymptotically rising cross sections and non-vanishing $\mathcal{M}^-$. The generalized or weak Pomeranchuk theorem states that the ratio of particle-particle and particle-antiparticle total cross sections is asymptotically unity \cite{Barone:2002cv, Matthiae:1994uw, Ewerz:2003xi}:
\begin{equation}
    \frac{\sigma_{\rm tot}^{p\bar p}}{\sigma_{\rm tot}^{p p}}\underset{s\to\infty}{\to}1.
\end{equation}
One can use the Froissart-Martin theorem to show that $|\sigma_{\rm tot}^{p p}-\sigma_{\rm tot}^{p\bar p}|$ can not grow faster than $\ln s$ at asymptotic energies \cite{Matthiae:1994uw, Ewerz:2003xi}:
\begin{equation}
        |\sigma_{\rm tot}^{p p}-\sigma_{\rm tot}^{p\bar p}| \underset{s\to\infty}{\leq} C'\ln s,
\end{equation}
where $C'$ is a constant. This means that asymptotically, though bounded, the particle-particle and particle-antiparticle total cross section difference is allowed by general principles to increase with energy and even diverge. This also shows that general principles do not forbid an odderon contribution.

The name odderon was introduced in Ref.~\cite{Joynson:1975az} for a simple pole of $\mathcal{M}^-$ in the complex angular momentum plane at $\ell=1$, but now used for any type of singularity that may be present at very high energies and corresponds to an exchange with odderon quantum numbers:
$$\mathbb{O}:~~P=-1,~C=-1,~G=-1,~I=0,~\xi=-1\,.$$

The concept of odderon is well established also in perturbative QCD \cite{Ewerz:2003xi}. The simplest picture of the odderon exchange in QCD is a colorless exchange of three interacting gluons with negative charge parity in the $t$-channel. This picture was established by
Bartels \cite{Bartels:1980pe}, Jaroszewicz \cite{Jaroszewicz:1980mq}, and Kwiecinski and Praszalowicz \cite{Kwiecinski:1980wb} in 1980. Almost twenty years later, the odderon trajectory intercept was predicted to be unity in QCD \cite{Braun:1998fs,Bartels:1999yt}. The odderon in QCD with running coupling was studied in Ref.~\cite{Bartels:2019qho}, and the results hint at a physical odderon with intercept one and a small $t$-slope. For an extensive review on odderon in QCD see Ref.~\cite{Ewerz:2003xi}. 

The perturbative QCD picture motivates the interpretation of the odderon trajectory as a glueball trajectory that, in the $t>0$ domain, may interpolate between glueball states composed of an odd number of gluons. However, such an interpretation is again not compelling theoretically because of the unclear relation between the ``soft'' non-perturbative and ``hard'' perturbative kinematic domains\footnote{In Ref.~\cite{Szanyi:2019kkn}, based on a Regge phenomenological model fitted to elastic $pp$ and $p\bar p$ scattering data, I and my co-authors made predictions for the masses and decay widths of states lying on the leading odderon trajectory with quantum numbers $J=3,~5,...$ and $P=C=-1$. Our predictions give lower masses than lattice QCD calculations or calculations based on a field theoretical Coulomb gauge QCD model on glueball masses with the same quantum numbers \cite{Llanes-Estrada:2021evz}.}.

\vspace{-0.5cm}
\subsection{Strategy of odderon search in elastic scattering}\label{sec:oddsearch}

Analyzing proton-proton and proton-antiproton scattering at TeV energies in a common kinematic domain, one can search for the $t$-channel odderon exchange \cite{Ewerz:2003xi}. The details are explained below.  

As detailed in \cref{sec:Regge}, for mesonic trajectories, $\alpha(0)<1$, hence the contributions of mesonic exchanges, with both even and odd signatures, virtually vanish in the TeV c.m. energy domain (see \cref{eq:reggetot} and \cref{eq:reggediff}). Thus, the nonvanishing exchanges in this c.m. energy domain are the pomeron and the odderon. The even signature $\xi=+1$ pomeron amplitude has the same sign in $pp$ and $p\bar p$ elastic scattering amplitudes not resulting any difference between $pp$ and $p\bar p$ collisions. However, the odd signature $\xi=-1$ odderon amplitude has a different sign in $pp$ and $p\bar p$ elastic scattering amplitudes generating differences between $pp$ and $p\bar p$ collisions. Thus, in the TeV c.m. energy domain 
the crossing-even part of the scattering amplitude is associated with the pomeron exchange, $\mathcal{M}^+\equiv\mathcal{M}^\mathbb{P}$, and the crossing-odd part of the scattering amplitude is associated with the odderon exchange $\mathcal{M}^-\equiv\mathcal{M}^\mathbb{O}$. Consequently, based on \cref{eq:M_ampl_pp} and \cref{eq:M_ampl_pbarp}, we can write the $pp$ and $p\bar p$ scattering amplitudes as 
\begin{equation}\label{eq:M_ampl_pp_re}
\mathcal{M}^{pp}_{\rm el} = \mathcal{M}^\mathbb{P} - \mathcal{M}^\mathbb{O}
\end{equation}
and
\begin{equation}\label{eq:M_ampl_pbarp_re}
\mathcal{M}^{p\bar p}_{\rm el}=  \mathcal{M}^\mathbb{P} + \mathcal{M}^\mathbb{O},
\end{equation}
respectively.  Now, the pomeron and odderon amplitudes can be expressed as 
\begin{equation}\label{eq:ampl_M_P}
\mathcal{M}^\mathbb{P}=   \frac{1}{2} \left(\mathcal{M}^{p\bar p}_{\rm el}+\mathcal{M}^{pp}_{\rm el}\right)
\end{equation}
and
\begin{equation}\label{eq:ampl_M_O}
\mathcal{M}^\mathbb{O}=  \frac{1}{2} \left(\mathcal{M}^{p\bar p}_{\rm el}-\mathcal{M}^{pp}_{\rm el}\right),
\end{equation}
respectively. Since observables of $pp$ ($p\bar p$) scattering are calculated from the  $pp$ ($p\bar p$) scattering amplitude, \cref{eq:ampl_M_O} implies that any difference which is observed between $pp$ and $p\bar p$ observables in the TeV energy region is a manifestation of the odderon exchange. Without the odderon, $pp$ and $p\bar p$ observables are the same in the TeV c.m. energy domain.

If the $pp$ and $p\bar p$ differential cross sections are not the same at the same  c.m. energy in a TeV energy domain, then the Odderon contribution to the scattering amplitude cannot be vanishing:
     \begin{equation}\label{eq:odderon-dsdt}
        \frac{d\sigma^{pp}_{\rm el}}{dt} (s,t) \neq  \frac{d\sigma^{p\bar p}_{\rm el}}{dt} (s,t) \,\,\, \mbox{ \rm for}\,\, \sqrt{s}\ge 1 \,\, \mbox{\rm TeV}
        \implies 
                \mathcal{M}^\mathbb{O}(s,t) \neq 0  \, .
    \end{equation}
At $\sqrt{s}\simeq1$ TeV and at higher c.m. energies, the contribution of the mesonic exchanges virtually vanishes, as discussed above. 
Similar conditions hold with the $t=0$ observables (see \cref{sec:measurables}): any of 
\begin{equation}
a^{pp}(s)  \neq  a^{p\bar p}(s) , 
\end{equation}
\begin{equation}
B_0^{pp}(s)  \neq B_0^{p\bar p}(s),
\end{equation}
\begin{equation}
\rho_0^{pp}(s)  \neq  \rho_0^{p\bar p}(s), 
\end{equation}
\begin{equation}
\sigma_{\rm el}^{pp}(s)  \neq  \sigma_{\rm el}^{p\bar p}(s),
\end{equation}
\begin{equation}
\sigma_{\rm in}^{pp}(s)  \neq  \sigma_{\rm in}^{p\bar p}(s), 
\end{equation}
\begin{equation}
\sigma_{\rm tot}^{\rm pp}(s)  \neq  \sigma_{\rm tot}^{p\bar p}(s),
\end{equation}
for $\sqrt{s}\ge 1\,\,\mbox{\rm TeV}$ implies that $\mathcal{M}^\mathbb{O}(s,t) \neq 0$ and/or $\mathcal{M}^\mathbb{O}(s,0) \neq 0$. Note that not all the $t=0$ observables are independent (see \cref{sec:measurables}). 

In my work, I used \cref{eq:odderon-dsdt} to investigate the statistical significance of a possible odderon exchange in $pp$ and $p\bar p$ elastic scattering. Due to the dominating contribution of the pomeron, the odderon had remained elusive for almost five decades. 

An experimental indication of a $C$-odd contribution in elastic $pp$ and $p\bar p$ scattering was observed in 1985 at CERN ISR by comparing the elastic $pp$ and $p\bar p$ differential cross sections in a $|t|$ region including the diffractive minimum, \mbox{0.5 GeV$^2$ $<|t|<$ 4.0 GeV$^2$,} measured at \mbox{$\sqrt s = 53$ GeV} \cite{Breakstone:1985pe}. The qualitative difference observed between the $t$-distribution of $pp$ and $p\bar p$ scattering is that, while in $pp$ scattering, we see a prominent minimum-maximum structure, the minimum is rather filled in in $p\bar p$ scattering.  The statistical significance of the observation of a $C$-odd contribution was 3.35$\sigma$ at \mbox{$\sqrt s = 53$ GeV.} This is below the standard discovery limit of $5\sigma$. Moreover, at \mbox{$\sqrt s = 53$ GeV,} mesonic exchanges are expected to play a significant role, rendering the odderon search at the ISR rather inconclusive.

Later experiments at CERN SPS at $\sqrt s = 546$ GeV \cite{UA4:1985oqn} and 630 GeV \cite{UA4:1986cgb} in the 1980s, and the D0 experiment at FNAL Tevatron at $\sqrt s = 1.96$ TeV in 2012 \cite{D0:2012erd} observed the lack of the diffractive minimum in $p\bar p$ scattering confirming the observation made at CERN ISR. In the $t$-distribution of elastic $p\bar p$ scattering, only a shoulder-like structure is observed even at TeV energies, and no minimum is seen.

The elastic $pp$ differential cross section at $\sqrt{s}=7$ TeV was measured by the TOTEM Collaboration at CERN LHC. The results were published in 2011 \cite{TOTEM:2011vxg} showing that the prominent minimum-maximum structure is seen in the $t$-distribution of elastic $p p$ scattering at TeV energies as observed at CERN ISR in the 1970s in the energy range $23~{\rm GeV}\lesssim\sqrt s\lesssim63~{\rm GeV}$  \cite{Nagy:1978iw}. These observations were confirmed by the TOTEM Collaboration at $\sqrt{s}=13$ TeV \cite{TOTEM:2018hki} in 2019, at $\sqrt{s}=2.76$ TeV in 2020 \cite{TOTEM:2018psk}, and later also at $\sqrt{s}=8$ TeV \cite{TOTEM:2020zzr,TOTEM:2021imi}.

In Ref.~\cite{TOTEM:2018psk}, the TOTEM Collaboration noted that ``{\it Under the condition that the effects due to the energy difference between TOTEM and D0 can be neglected, the result provides evidence for a colorless 3-gluon
bound state exchange in the $t$-channel of the proton-proton elastic scattering}''. In other words, if the effects due to the energy difference between the $\sqrt{s} = 2.76$ TeV $pp$ TOTEM and $\sqrt{s}= 1.96$ TeV $p\bar p$ D0 measurements can be neglected, the direct comparison of these differential cross section data provides conditional evidence for the odderon exchange in the $t$-channel. To overcome the lack of $pp$ and $p\bar p$ data at the same c.m. energies, the idea was to extrapolate the TOTEM measurements at $\sqrt{s}=$ 2.76 TeV, 7 TeV, 8 TeV, and 13 TeV down to $\sqrt{s}=1.96$ TeV.  

An increase of $\sigma_{\rm tot}(s)$, associated with 
a decrease of $\rho_0(s)$ as a function of $s$ first identified 
at $\sqrt{s} = 13$ TeV by the TOTEM Collaboration \cite{TOTEM:2017sdy}, also indicated a possible odderon effect. However, the identification of odderon effects at $t=0$ in $\rho_0(s)$ is disputed in the literature \cite{Gotsman:2018buo,Pancheri:2018yhd}.

Given that the elastic $pp$ and $p\bar p$ scattering data have not been measured at the same (or close enough) energies in the TeV region so far, the main task was to close the energy gap as much as possible at present, without direct measurement, based on the analysis of already published data.

After new measurements at LHC on elastic $pp$ scattering were published, the odderon effects were investigated in elastic scattering in several papers but without reporting a statistically significant odderon observation. Let me briefly mention some of these studies. The authors of Ref.~\cite{Lengyel:2012sw} concluded about indirect odderon evidence by analyzing elastic scattering data using a dipole Regge model for the pomeron and the odderon. The authors of Ref.~\cite{Donnachie:2013xia} used another Regge approach and the triple-gluon exchange model to analyze the data. 
In Ref.~\cite{Ster:2015esa}, the $(s,t)$-dependent contributions of an odderon exchange were studied in the framework of the Phillips-Barger model~\cite{Phillips:1974dnz}. A similar study was performed in Ref.~\cite{Goncalves:2018nsp}. In Ref.~\cite{Csorgo:2018uyp}, the Odderon features have been identified and qualitatively described in a model-independent way using the L\'evy imaging technique. New analyses of elastic $pp$ and $p\bar p$ elastic scattering data were performed in \mbox{Refs.~\cite{Broniowski:2018xbg,Szanyi:2018pew}} using a dipole Regge model for the pomeron and the odderon concluding, by explicit calculations, that (i) the relative contributions of mesonic reggeon exchanges are negligible at LHC energies and (ii) the odderon has an important role in the minimum-maximum region of the \mbox{$t$-distribution} and beyond at the higher $|t|$ region as well as at $t=0$ in describing $\rho_0(s)$ at LHC energies. Odderon effects were investigated in the framework of a two-channel eikonal model in Ref.~\cite{Khoze:2017swe,Khoze:2018kna}. 
Several other Regge parameterizations were used to describe 
the LHC data reasonably well \mbox{in Refs.~\cite{Selyugin:2018uob, Broilo:2018els,Broilo:2018qqs}.} Statements about the maximal nature of the odderon effect were made in Refs.~\cite{Khoze:2018kna,Martynov:2018nyb,Martynov:2018sga,Shabelski:2018jfq,Bence:2021uji}. Authors of Ref.~\cite{Troshin:2018ihb} studied the $s$-dependence of $\rho_0$ and its connection to the $s$-evolution of $\sigma_{\rm tot}(s)$. 
 QCD-based results on odderon features were obtained, $e.g.$, in \mbox{Refs. \cite{Gotsman:2018buo,Gotsman:2020mkd,Hagiwara:2020mqb,Contreras:2020lrh}.}

Note also that the search for the odderon need not be restricted to elastic scattering. See details in Refs.~\cite{Ewerz:2003xi,donnachie_dosch_landshoff_nachtmann_2002}. 

Measurements at LHC on elastic $pp$ scattering combined with data on elastic $p\bar p$ scattering measured at SPS and Tevatron finally allowed to shed light on the existence of the $t$-channel odderon. Based on the fact that the mesonic reggeon exchange contributions to the elastic scattering at $\sqrt{s} =$ 1.96 TeV and at higher energies are negligibly small, the first three journal papers, with my significant contribution, were published in 2021 reporting statistically significant, discovery-level $t$-channel odderon observations using different methods \cite{Csorgo:2019ewn,Csorgo:2020wmw,TOTEM:2020zzr}. 
An anonymously refereed/peer-reviewed ISMD 2019 conference proceedings reporting discovery-level odderon observation was published by our Hungarian-Swedish team even in 2020 \cite{Csorgo:2020msw}. 
The model-independent results of \mbox{Refs.~\cite{Csorgo:2020msw,Csorgo:2019ewn}} are based on the $H(x)$  scaling behavior of elastic $pp$ scattering data. This $H(x)$  scaling is introduced in \cref{sec:Hx} and further investigated in \cref{chap:Hxvalidity} of this dissertation. The model-dependent results of Ref.~\cite{Csorgo:2020wmw}, which was later extended by the study of Ref.~\cite{Szanyi:2022ezh}, were achieved by utilizing the ReBB model of Ref.~\cite{Nemes:2015iia} introduced in \cref{sec:ReBBmodel}. These model-dependent results are detailed in \cref{chap:rebbdesc} and \cref{chap:odderon} of this dissertation. My contribution 
to the results of the joint analysis of the TOTEM and D0 Collaborations published in Ref.~\cite{TOTEM:2020zzr} is detailed in \cref{chap:oddTD0}.

\section{The ReBB model}\label{sec:ReBBmodel}

A. Bialas and A. Bzdak, in 2007, published models for
elastic $pp$ scattering \cite{Bialas:2006qf,Bialas:2006kw}, the BB
models for short. The BB models are based on (i) the idea that the nucleon is composed of constituent quarks and (ii) R. J. Glauber's multiple scattering approach: all possible single and multiple binary inelastic collisions of the constituents is considered in a way that constituent backscattering is prohibited. The elastic
scattering amplitude is then calculated based on the
unitarity relation neglecting the sub-dominant real
part of the scattering amplitude. This approximation resulted in a completely vanishing $pp$ differential cross section at the minimum of the $t$-distribution.

First, in 2014 in Ref.~\cite{CsorgO:2013kua}, then in 2015 in Ref.~\cite{Nemes:2015iia}, the BB model was extended: a real part of the scattering amplitude was added. By these extensions, the model described reasonably the $pp$ elastic differential cross section data also in the region of the minimum structure. In Ref.~\cite{Nemes:2015iia}, the real part was added in a unitary manner leading to the Real extended Bialas--Bzdak model, the ReBB model for short.

In Ref.~\cite{Bialas:2006qf} two variants of the BB model are introduced: the $p=(q,d)$ version, where the constituent diquark is treated as a single object, and the $p=(q,(q,q))$ version, where the constituent diquark is decomposed into two constituent quarks. In both versions, the constituent quark forms a color triplet, and the constituent diquark forms a color antitriplet \cite{Bialas:2006kw}. It was shown in Refs.~\cite{Nemes:2012cp,CsorgO:2013kua,Nemes:2015iia} that the $p=(q,(q,q))$ version gives too many minima for the $pp$ differential cross section, while according to the measured data, there is only a single minimum. This result is supported by the finding in Ref.~\cite{Czyz:1969jg} that the number of diffractive minima increases as either of the colliding composite objects develops a more complex internal structure. However, so far, there is no deeper explanation neither for the preferred quark-diquark structure nor for the preference for the quark flavors entering into the diquark. Though a proton is more naturally imagined as a bound state of three quarks, the two-component quark-diquark structure is simply favored by the experimental data on the $pp$ differential cross section showing a single minimum-maximum structure. The relation of the $(q,d)$ structure of the proton to QCD is presently a research topic. In the following, I discuss only the $p=(q,d)$ version of the model which quite well agrees with the experimental findings and which I used in my research. 

In \cref{sec:obb}, I discuss the original BB model in detail, and then in \cref{sec:ebb}, I present the ReBB model.



\vspace{-0.5cm}
\subsection{Original Bialas--Bzdak model}\label{sec:obb}

In the case of high-energy small-angle scattering the real part of the elastic scattering amplitude is small. 
Neglecting the real part of the scattering amplitude, ${\rm Re}\,\widetilde T_{\rm el} (s,b)=0$, it follows from \cref{eq:unitb_2} that the imaginary part of the scattering amplitude can be written in terms of $\tilde\sigma_{\rm in}(s,b)$ as
 \begin{equation} \label{eq:imaplel}
    {\rm Im}\,\widetilde T_{\rm el} (s,b) = 1-\sqrt{1-\tilde\sigma_{\rm in}(s,b)}.
\end{equation}
In this case the opacity function $\Omega(s,b)$ is purely real (see \cref{eq:impact_ampl_eik_sol}) and reads  
\begin{equation}\label{eq:re_omega}
\text{Re}\,\Omega(s,b)=-\frac{1}{2}\ln\left[1-\tilde\sigma_{\rm in}(s,b)\right]\,,
\end{equation}
\begin{equation}
\text{Im}\,\Omega(s,b)=0\,.
\end{equation}

In Ref.~\cite{Bialas:2006qf} the authors consider the proton as a bound state of a quark and a diquark and construct $\tilde\sigma_{\rm in}(\,\vec b\,)$, the total probability for the occurrence of an inelastic event in the collision of two protons at an impact parameter $b$, based on the multiple scattering idea. The multiple scattering idea emerges from the generalization of the eikonal approach for collisions of composite objects as discussed in the non-relativistic case in \hyperref[sec:app_multscatt]{Appendix A} of this dissertation.

\begin{figure}[b!]
	\centering
 \vspace{-5mm}
\includegraphics[width=0.55\linewidth]{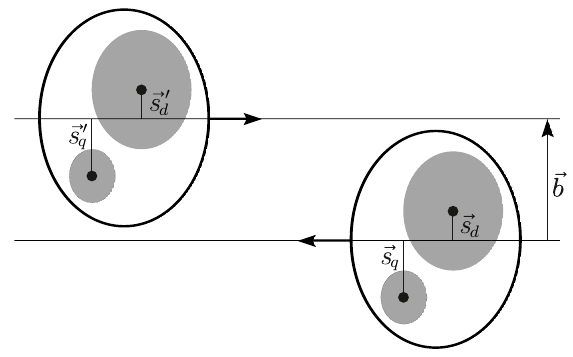}
\vspace{-1mm}
 \caption{Geometry of the collision of two protons in the quark-diquark model and definition of the coordinates used in the cross section formulae.}
	\label{fig:BBcoord}
\end{figure}

Let us introduce the inelastic differential cross section for the collision of two constituents $a$ and $b$ whose relative transverse position is $\vec s$. This inelastic differential cross section is denoted as\footnote{Note that $\sigma_{\rm in}^{ab}(\vec s\,)$ is only a shorthand notation for $\frac{d^2\sigma_{\rm in}^{ab}(\vec s)}{d^2\vec s}$.}
\begin{equation}
    \frac{d^2\sigma_{\rm in}^{ab}(\vec s\,)}{d^2\vec s}\equiv\sigma_{\rm in}^{ab}(\vec s\,)
\end{equation}
and gives the probability of the collision of constituent $a$ in one proton and constituent $b$ in the other as a function of their relative transverse position $\vec s$. Then, by following the same ideas as in the course of the construction of \cref{eq:Gamma_comp} in \hyperref[sec:app_multscatt]{Appendix A}, the sum of the probabilities of all possible single and multiple binary collisions of the constituents is
\begin{align}\label{eq:totalsum}
    \sigma_{\rm in}(\vec b, \vec s_q, \vec s_d, \vec s_q^{\,\prime},\vec s_d^{\,\prime}) = 1- \prod_{a\in\{q,d\}}\prod_{b\in\{q,d\}} \left[1- \sigma_{\rm in}^{ab}(\vec b-\vec s_a + \vec s_b^{\,\prime})\right].
\end{align}
where $\vec s_q$, $\vec s_d$, $\vec s_q^{\,\prime}$, and $\vec s_d^{\,\prime}$ are the transverse positions of the quarks and diquarks inside the two colliding protons. \cref{eq:totalsum} is actually considered for a case when the quarks and diquarks are frozen at their positions. The transverse coordinates of the constituents in the two colliding protons are illustrated in \cref{fig:BBcoord}. \cref{eq:totalsum} says that the collision of two protons is inelastic if at least one constituent-constituent collision is inelastic. 
Expanding \cref{eq:totalsum}, we get
\begin{align}\label{eq:elprob}
\sigma_{\rm in}(\vec b,\vec s_{q},\vec s_{d}%
,\vec s_{q}^{\,\prime},\vec s_{d}^{\,\prime})&=1-\left[1-\sigma^{qq}_{\rm in}(\vec s_{q},\vec s_{q}^{\,\prime}%
;\vec b)\right]\left[1-\sigma^{qd}_{\rm in}(\vec s_{q},\vec s_{d}^{\,\prime}%
;\vec b)\right]\times\\\nonumber &\times\left[1-\sigma^{dq}_{\rm in}(\vec s_{q}^{\,\prime},\vec s_{d}%
;\vec b)\right]\left[1-\sigma^{dd}_{\rm in}(\vec s_{d},\vec s_{d}^{\,\prime}%
;\vec b)\right],
\end{align}
where I used the shorthand notation
\begin{equation}
    \sigma_{\rm in}^{ab}(\vec s_{a},\vec s_{b}^{\,\prime}%
;\vec b)\equiv\sigma^{ab}_{\rm in}(\vec b+\vec s_b^{~\prime}-\vec s_a),\,\,\,a,b\in\{q,d\}.
\end{equation}

Expanding \cref{eq:elprob} we get the single scattering contributions of the constituents as well as the multiple -- double, triple, and quadruple -- scattering contributions:
\begin{align}\label{eq:sig_b_s}
\sigma_{\rm in}&=\sigma^{qq}_{\rm in}+\sigma^{qd}_{\rm in}+\sigma^{dq}_{\rm in}+\sigma^{dd}_{\rm in}\nonumber \\&-\left[\sigma^{qq}_{\rm in}\sigma^{qd}_{\rm in} +\sigma^{qq}_{\rm in}\sigma^{dq}_{\rm in}+\sigma^{qd}_{\rm in}\sigma^{dq}_{\rm in}+\sigma^{qq}_{\rm in}\sigma^{dd}_{\rm in}+\sigma^{qd}_{\rm in}\sigma^{dd}_{\rm in}+\sigma^{dq}_{\rm in}\sigma^{dd}_{\rm in}\right] \nonumber\\ 
&
+\sigma^{qq}_{\rm in}\sigma^{qd}_{\rm in}\sigma^{dq}_{\rm in}+\sigma^{qq}_{\rm in}\sigma^{qd}_{\rm in}\sigma^{dd}_{\rm in}+\sigma^{qq}_{\rm in}\sigma^{dq}_{\rm in}\sigma^{dd}_{\rm in}+\sigma^{dd}_{\rm in}\sigma^{qd}_{\rm in}\sigma^{dq}_{\rm in}\nonumber \\&-\sigma^{qq}_{\rm in}\sigma^{qd}_{\rm in}\sigma^{dq}_{\rm in}\sigma^{dd}_{\rm in},
\end{align}
where the arguments are implicit. The first-order single scattering term sums the probabilities of the cases when only one constituent-constituent inelastic collision happens.  The second-order double scattering term sums the probabilities of the cases when altogether two inelastic constituent-constituent collisions happen, $i.e.$, when a constituent suffers two binary collisions or when two constituents both suffer a single binary
collision. The third-order triple scattering term sums the probabilities of the cases when altogether three inelastic constituent-constituent collisions happen, $i.e.$ when there is a constituent that suffers two binary collisions, and there is another one that suffers only one binary collision. The highest order quadruple scattering term gives the probability that both constituents of both protons suffer two binary collisions.

Let us now introduce the transverse density distribution\footnote{The transverse density distribution is called thickness in the field of heavy ion physics \cite{wong1994,Bialas:2006kw}.} of the proton $D(\vec s_q,\vec s_d)$ as 
\begin{equation}
    D(\vec s_q,\vec s_d) =\int \rho(\vec s_q,\vec s_d, z_q,z_d)dz_qdz_d,
\end{equation}
where $\rho$ is the three-dimensional density distribution of the proton given as a function of the quark and diquark positions; $z_q$ and $z_d$ are the $z$-components of the three-dimensional quark and diquark position vectors. $D(\vec s_q,\vec s_d)d ^2 \vec s_q d ^2\vec s_d$ is the probability of finding the quark in surface element $d^2\vec s_q$ and the diquark in the surface element $d^2\vec s_q$. Thus $D(\vec s_q,\vec s_d)$ is interpreted as the transverse distribution of constituents inside the proton and normalized to give 
\begin{equation}\label{eq:densnorm}
    \int D(\vec s_q,\vec s_d) d ^2 \vec s_q d ^2\vec s_d = 1.
\end{equation}

Then using $D(\vec s_q,\vec s_d)$ and \cref{eq:totalsum}, the probability for inelastic collision of two protons at an impact parameter $\vec b$ is 
\begin{equation}\label{eq:tilde_sigma_inel}
\tilde\sigma_{\rm in}(\vec b)=\int d^{2}\vec s_{q}d^{2}\vec s_{q}^{\,\prime}d^{2}\vec s_{d}d^{2}\vec s_{d}^{\,\prime
}D(\vec s_{q},\vec s_{d})D(\vec s_{q}^{\,\prime},\vec s_{d}^{\,\prime})\sigma_{\rm in}(\vec b, \vec s_q, \vec s_d, \vec s_q^{\,\prime},\vec s_d^{\,\prime}).
\end{equation}
\cref{eq:tilde_sigma_inel} corresponds to an averaging of constituent-constituent inelastic scattering probabilities over the constituents' positions in the two colliding protons.

The transverse density of the proton and the inelastic differential cross sections for the constituent-constituent collisions are not known. The next step is to choose the functional form of these quantities with free parameters to be fitted to the elastic $pp$ scattering data.

Requiring that the transverse density of the proton  (i) satisfies \cref{eq:densnorm}, (ii) has a Gaussian form, (iii) depends only on the absolute value squared  of the constituents' transverse position vectors, and the center-of-mass of the quark-diquark system  is fixed, the authors of Ref.~\cite{Bialas:2006qf} define  
\begin{equation}
D\left({\vec s}_q,{\vec s}_d\right)=\frac{1+\lambda^2}{R_{qd}^2\,\pi}e^{-(s_q^2+s_d^2)/R_{qd}^2}\delta^2({\vec s}_d+\lambda{\vec s}_q).
\label{eq:quark_diquark_distribution}
\end{equation}
where the Dirac delta guarantees the constraint on the center-of-mass of the system introducing position correlation for the constituents,
\begin{equation}
{\vec s}_d=
-\lambda\,{\vec s}_q,
\label{eq:poscorr}
\end{equation} 
and $\lambda=m_q/m_d$ is the ratio of the quark and diquark masses. The presence of the two-dimensional Dirac $\delta$ function in \cref{eq:quark_diquark_distribution} reduces the dimension of the integral in \cref{eq:tilde_sigma_inel} from eight to four. $R_{qd}$ can be interpreted as the typical value of the square root of the sum of the constituents' squared transverse position vectors. Because of the position correlation, \cref{eq:poscorr}, $R_{qd}$ can also be interpreted as a measure of the quark-diquark distance inside the proton. One can go even further and based on \cref{eq:poscorr} derive the relation 
\begin{equation}\label{eq:relation}
    (\vec s_q -\vec s_d)^2 = \frac{(\lambda-1)^2}{1+\lambda^2}\left(\vec s_q^{\,2}+\vec s_d^{\,2}\right). 
\end{equation}
\cref{eq:relation} can be used to relate $R_{qd}$ to $d_{qd}$, the relative distance between the constituents in a single proton. Since $\lambda=1/2$, 
\begin{equation}\label{eq:dr}
    d_{qd} = R_{qd}/\sqrt{5}.
\end{equation}

\begin{figure}[!hbt]
\centering
\includegraphics[width=0.4\linewidth]{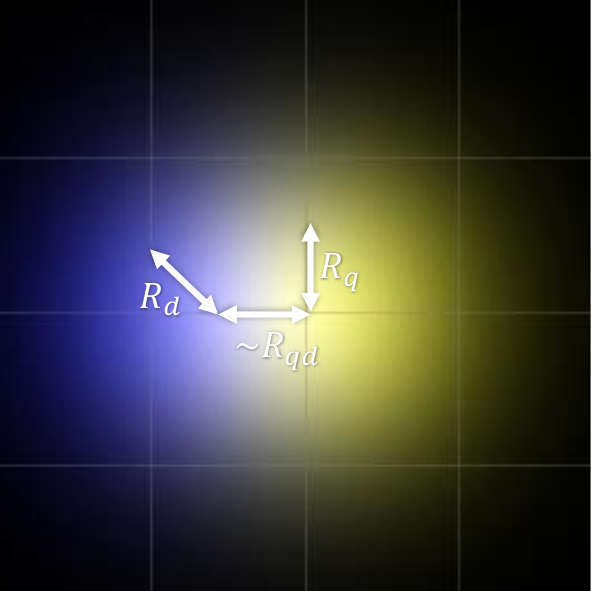}
	\caption{Schematic view of the proton in the $p=(q,d)$ Bialas--Bzdak model \cite{Csorgo:2013usx}.}
	\label{fig:proton_BB}
\end{figure}

The constituent-constituent differential cross sections are parameterized again by using Gaussian forms:
\begin{equation}\label{eq:inelastic_cross_sections}
     \sigma^{ab}_{\rm in}(\vec s) = A_{ab}e^{-\vec s^{\,2}/S^2_{ab} }
 \end{equation}
with $S^2_{ab}=R_a^2+R_b^2$ and $a,b\in\{q,d\}$. $R_q$ and $R_d$, respectively, are interpreted as the typical radii of the quark and the diquark. The quark and the diquark constituents of the proton can be imagined as parton clouds characterized by Gaussian parton distributions with scale parameters $R_q$ and $R_d$; the distance between them is again characterized by a Gaussian scale parameter $R_{qd}$ (see \cref{fig:proton_BB}). It was found in previous \mbox{studies \cite{Bialas:2006qf,Nemes:2012cp, Nemes:2015iia}} that $R_d>R_q, R_{qd}$. This implies that the constituent clouds overlap. Furthermore, it was found in Ref.~\cite{Nemes:2015iia} that the energy evolution trends of the scale parameters are rising with energy logarithmically, implying that the constituents of the proton get bigger, and the distance between the constituents gets larger.

The total inelastic cross sections of the constituent-constituent collisions are given as
\begin{equation}\label{eq:totalinelastic}
\sigma_{\text{in}}^{ab,\text{tot}}=\int{\sigma_{ab}\left({\vec s}\,\right)}\,\text{\rm d}^2\vec s= \pi A_{ab}S_{ab}^2\,.
\end{equation}
Assuming that (i) the diquark contains twice as many partons as the quark and (ii) the colliding constituents do not shadow each other, we have:
\begin{equation}\label{eq:ratiosforsigma}
\sigma_{\text{in}}^{qq,\text{tot}}:\sigma_{\text{in}}^{qd,\text{tot}}:\sigma_{\text{in}}^{dd,\text{tot}}=1:2:4\,. 
\end{equation}

\noindent By this assumption the parameters $A_{qd}$ and $A_{dd}$ can be expressed via $A_{qq}$,
\begin{equation}\label{eq:Apars}
A_{qd}=A_{qq}\frac{4R_q^2}{R_q^2+R_d^2}\,,\;A_{dd}=A_{qq}\frac{4R_q^2}{R_d^2}\,,
\end{equation}
reducing the number of the model's free parameters by two. Thus, five free parameters remain in the model: $A_{qq}$, $\lambda$,  $R_{q}$, $R_{d}$, and $R_{qd}$.

Substituting \cref{eq:tilde_sigma_inel} into \cref{eq:imaplel} and passing into momentum space by computing the integral \cref{eq:relPWtoeik_2} numerically one can calculate the differential cross section by \cref{eq:Tdsigma}. The integral given by \cref{eq:tilde_sigma_inel} is calculated analytically with the methods described in Refs.~\cite{Nemes:2015iia,Bialas:2006qf} and summarized in \hyperref[sec:app_BBcalc]{Appendix B}.

It was shown in Ref.~\cite{Bialas:2006qf} that the original BB model reproduces qualitatively the $pp$ differential cross section data up to $-t=3$ GeV$^2$ by adjusting the model parameters to (i) the total inelastic cross section, (ii) the slope of the elastic cross section at $t=0$, (iii)~the position of the diffractive minimum and (iv) the height of the maximum of the $pp$ differential cross section. The authors of Ref.~\cite{Nemes:2012cp} showed that the $p=(q,d)$ version of the BB model reproduces quantitatively the $pp$ differential cross section data at ISR energies in the squared four-momentum transfer range $0.4$ GeV$^2$ $\leq -t \leq$ 2.5 GeV$^2$ by excluding from the $\chi^2$ fit three data points in the dip range. This exclusion is done because the BB model with a purely imaginary scattering amplitude is singular at around the dip position.

In Refs.~\cite{Bialas:2006kw,Bialas:2007eg} the quark-diquark distribution and the constituent-constituent cross sections as determined from the analysis of the elastic $pp$ data were used quite successfully to describe the particle production in heavy-ion collisions. 

We know that the proton contains three valence quarks; thus, it would be natural to consider the proton as a bound state of three constituent quarks. The multiple scattering calculations, assuming that the proton is a bound state of three uncorrelated constituent quarks, were already performed in the 1970s \cite{bialas1977elastic}, but the result was not compatible with the measurements. It turns out that correlations between constituent quarks play an important role \cite{Bialas:2006qf}: in the BB model, a correlation between the constituent quarks is introduced by combining two of the quarks into one object, a diquark.

\subsection{Extended Bialas--Bzdak model}\label{sec:ebb}

According to the findings of Ref.~\cite{Nemes:2012cp} the original BB model does not describe the LHC TOTEM 7 TeV elastic $pp$ differential cross section data in a statistically acceptable manner even if a few data points are left out in the minimum region during the fit. The scattering amplitude in the BB model was first extended by a real part in Ref.~\cite{CsorgO:2013kua} by introducing a purely imaginary factor in the formula of the inelastic cross section, \textit{i.e.}, by the change
\begin{equation}
   \tilde\sigma_{\rm in}(s, b) \to \left(1-i\alpha(s)\right)\tilde\sigma_{\rm in}(s, b), 
\end{equation}
where $\alpha$ was a new parameter to be determined from the analysis of data. This modification somewhat improved the situation in describing the LHC TOTEM 7 TeV data but did not solve the issue completely. Later, in Ref.~\cite{Nemes:2015iia}, the model was extended by introducing a non-zero imaginary component for the opacity function, ${\rm Im}\Omega(s,b)$, giving rise to the Real-extended Bialas--Bzdak (ReBB) model. In the following, I detail this ReBB model.

It follows from \cref{eq:impact_ampl_eik_sol} that the scattering amplitude has a real part if the opacity function $\Omega(s,b)$ has an imaginary part. In Ref.~\cite{Nemes:2015iia}, several options were investigated for the functional form of ${\rm Im}\Omega(s,b)$, however, it was concluded that the choice which suits the best to the experimental data on $pp$ elastic differential cross section is
\begin{equation}\label{eq:im_omega}
\text{Im}\,\Omega(s,b)=-\alpha_R(s)\tilde\sigma_{\rm in}(s,b)\,,
\end{equation}
where $\alpha_R(s)$ is a free parameter to be fitted to the data measured at a given energy. The real part of the opacity function, as given by \cref{eq:re_omega}, was unaltered. Thus, it is clear that ${\rm Im}\Omega(s,b)\neq0$ not only generates the real part of the scattering amplitude but also modifies the imaginary part of the scattering amplitude.

The physical interpretation of \cref{eq:im_omega} is the following \cite{Nemes:2015iia}: at small four-momentum transfers protons can collide inelastically due to parts of the proton scattering elastically but into different directions, not parallel to one another. As we will see in \cref{chap:oddTD0} and \cref{chap:odderon}, the odderon effects in the ReBB model are characterized by a single parameter, which is the opacity parameter $\alpha_R$ introduced in \cref{eq:im_omega}.

Substituting \cref{eq:re_omega} and \cref{eq:im_omega} into \cref{eq:impact_ampl_eik_sol} we obtain the ReBB model elastic scattering amplitude in $b$ space:
\begin{equation}\label{eq:ReBB_b_ampl}
\widetilde{T}_{\rm el}(s,b)=i\left(1-e^{i\, \alpha_R\, \tilde\sigma_{\rm in}(s,b)}\sqrt{1-\tilde\sigma_{\rm in}(s,b)}\right).
\end{equation}
\cref{eq:ReBB_b_ampl} is a particular solution of the unitarity equation of \cref{eq:unitb_2} in terms of $\tilde\sigma_{\rm in}(s,b)$ for ${\rm Im}\widetilde{T}_{\rm el}(s,b)\neq0$ and ${\rm Re}\widetilde{T}_{\rm el}(s,b)\neq0$. The computation of the momentum space amplitude is performed numerically\footnote{To get the values of the ReBB model scale parameters, $R_q$, $R_d$ and $R_{qd}$, in fm (femtometer) instead of natural units, the conversion \mbox{$\hbar c /$GeV = 0.1973~fm} is used to get 1 GeV$^{-1}$ =  0.1973 fm, since in natural units $\hbar = c = 1$. To get the total cross section in units of mb (millibarn) and the differential cross section in units of mb/GeV$^2$, the conversion \mbox{$\hbar^2c^2$/GeV$^2$~=~0.3894~mb} is applied to get \mbox{1 GeV$^{-2} = 0.3894$ mb.} The conversion between fm$^2$ and mb is: 1 fm$^2$ = 10 mb.} using \cref{eq:relPWtoeik_2}.

In conclusion, the ReBB model has an additional free parameter as compared to the original BB model: this is the opacity parameter, $\alpha_R$, which controls the imaginary part of the opacity function that generates the real part of the scattering amplitude. Setting $\alpha_R=0$, we get back the original BB model. It was shown in Ref.~\cite{Nemes:2015iia} that the typical value of $\alpha_R$ is rather small, the order of $10^{-2}-10^{-1}$, in agreement with the known fact that the real part of the scattering amplitude is small (see \cref{sec:eik}).

It was shown in Ref.~\cite{Nemes:2015iia} that, during the fitting procedure, the parameters $\lambda$ and $A_{qq}$ are better kept fixed at values 1/2 and 1, respectively. $\lambda = 1/2$ means that the mass of the diquark is twice as large as that of the constituent quark. 
$A_{qq} = 1$ means that the head-on $qq$ collisions are inelastic with a 100\% probability as it follows from Eq.~(\ref{eq:inelastic_cross_sections}). Keeping these parameters free improves the fit quality from the statistical point of view ($i.e.$, the $CL$ values are higher); however, the two additional fit parameters introduce more correlations and thus fluctuations in the values of the fit parameters.

Fixing the values of $\lambda$ and $A_{qq}$ the authors of Ref.~\cite{Nemes:2015iia} managed to determine the energy dependencies of the scale parameters, $R_q$, $R_d$, and $R_{qd}$, as well as the energy dependence of the opacity parameter, $\alpha_R$. The energy dependencies of these parameters were determined in Ref.~\cite{Nemes:2015iia} for $pp$ elastic scattering from the fits to the ISR and the LHC TOTEM 7 TeV differential cross section data. It was found that the energy dependencies of the parameters $R_q$, $R_d$, $R_{qd}$, and $\alpha_R$ are compatible with the linear-logarithmic shape
\begin{equation}\label{eq:linlog}
    P(s)=p_0+p_1\ln{(s/s_0)},\,\,\,s_0=1~{\rm GeV}^2,
\end{equation}
with $p_0$ and $p_1$ values that give a strictly increasing behavior. 

The ``impact picture'' of elastic hadronic scattering \cite{Matthiae:1994uw,Barone:2002cv} motivates the asymptotic $\ln^2(s)$ behavior of the total cross section and consequently the $\ln(s)$ behavior of the size parameter as given by \cref{eq:linlog}. The $\ln^2(s/s_0)$ behavior of hadronic total cross sections is supported also by lattice QCD calculations \cite{Giordano:2012mn, Giordano:2013iga}. A faster growth than $\ln^2(s)$ for the hadronic total cross sections would violate the Froissart-Martin theorem, $i.e.$, unitarity (see the discussion in \cref{sec:oddintro}).

In \cref{chap:ReBBpbarp}, I use the ReBB model to describe $p\bar p$ elastic scattering differential cross section data from ISR energies up to Tevatron energies. In \cref{chap:oddTD0} and \cref{chap:rebbdesc}, I use the ReBB model to analyze jointly elastic $pp$ and $p\bar p$ scattering data. 
Then in \cref{chap:oddTD0} and \cref{chap:odderon}, I use the ReBB model to compare $pp$ and $p\bar p$ elastic differential cross sections in the same kinematics domain and conclude about the statistical significance of the odderon exchange. While, in \cref{chap:oddTD0}, a preliminary analysis is presented,  \cref{chap:odderon} details the final odderon analysis utilizing the ReBB model. In \cref{chap:levy}, after demonstrating the power of Lévy $\alpha$-stable distributions in describing elastic scattering data, I generalize the Gaussian distributions in the ReBB model to Lévy $\alpha$-stable distributions.

\section{H(x) scaling of elastic $\rm pp$ scattering}\label{sec:Hx}

An interesting scaling property of elastic $pp$ scattering was discovered in Ref.~\cite{Csorgo:2019ewn}. This is the so-called $H(x)$ scaling. The $H(x)$ scaling function can be derived as follows.

Measurements show that the low-$|t|$ $pp$ and $p\bar p$ differential cross sections can be approximated with an exponential shape \cite{Barone:2002cv},
\begin{equation}\label{eq:dsdt-exp}
    \frac{d\sigma_{\rm el}}{dt}(s,t) = a(s)  e^{B_0(s) t},
\end{equation}
where the normalization parameter, called optical point, can be written as

\begin{equation}\label{eq:norma0}
    a(s)= \frac{1}{4\pi}\left({\rm Re}T_{\rm el}^2(s,0)+{\rm Im}T_{\rm el}^2(s,0)\right) = \frac{1+\rho_0^2(s)}{16\pi} \sigma_{\rm tot}^2(s).
\end{equation}
Substituting \cref{eq:dsdt-exp} into \cref{eq:elastic_cross_section} we get 
\begin{equation}
    \sigma_{\rm el} (s) = \frac{a(s)}{B_0(s)}
\end{equation} 
and
\begin{equation}\label{eq:norma}
    a(s) = B_0(s)  \sigma_{\rm el}(s)  .
\end{equation}
Thus, it is evident that the energy dependence from the elastic differential cross section, given in \cref{eq:dsdt-exp}, can be scaled out if one simultaneously rescales (i) the horizontal axis by the slope parameter $B_0(s)$ and (ii) the vertical axis by the product $B_0(s) \sigma_{\rm el}(s)$. This observation leads to the $H(x)$ scaling function,
\begin{equation}\label{eq:HX_0}
    H(x)=\frac{1}{B_0(s) \sigma_{\rm el}(s)} \frac{d\sigma_{\rm el}}{d t}  \bigg|_{t=-x/B_0(s)}\equiv e^ {-x},
\end{equation}
first introduced in Ref.~\cite{Csorgo:2019ewn}. One can see that $H(0)=1$ and $\int_{0}^{\infty} dx H(x)=1$.

\begin{figure} [hbt!]
	\centering
	\subfloat[\label{fig:1}]{%
		\includegraphics[width=0.8\linewidth]{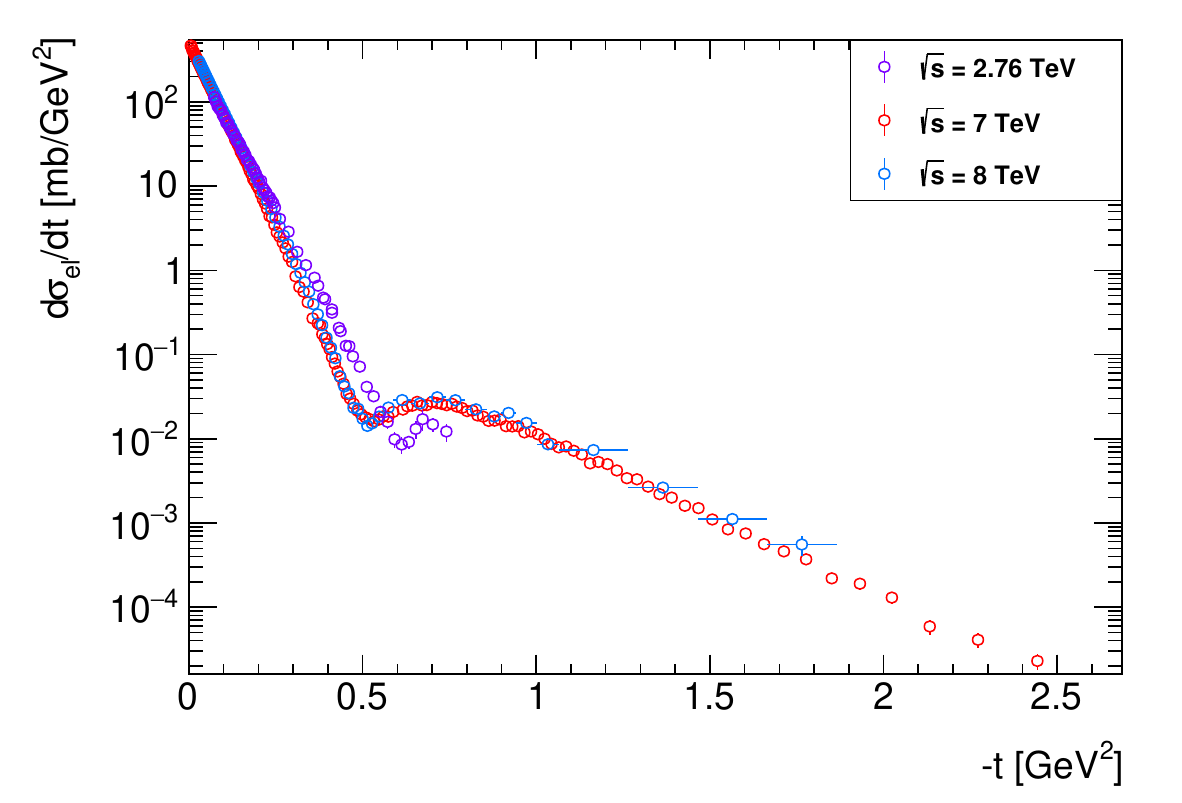}%
	}\hfill
	\subfloat[\label{fig:2}]{%
		\includegraphics[width=0.8\linewidth]{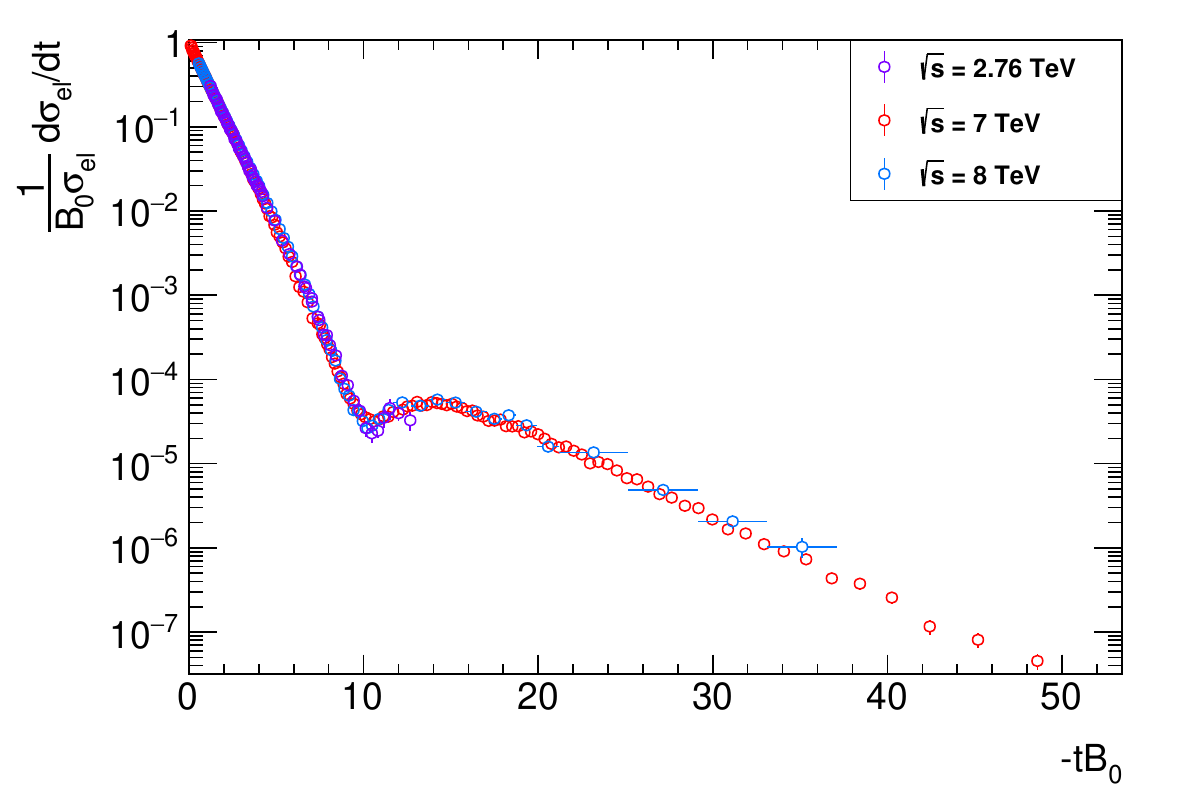}%
	}
	\caption{(a) LHC TOTEM 2.76 TeV, 7 TeV, and 8 TeV data on $pp$  $d\sigma_{\rm el}/dt$  and \mbox{(b) the calculated} $pp$ $H(x)$ scaling functions using these data.}
	\label{fig:exp_hx_pp}
\end{figure}

Since \cref{eq:dsdt-exp} is valid at low-$|t|$, one would expect that an $H(x)$-scaling behavior of the data is observable only in the low-$|t|$ region. Surprisingly, this is not the case. \cref{fig:exp_hx_pp} shows the scaling behaviour of  the LHC TOTEM elastic $pp$ differential cross section data measured at $\sqrt{s}=$ 2.76 TeV, 7 TeV and 8 TeV. One can see that the LHC data show a data-collapsing feature not only at low-$|t|$ values but also at the dip-bump region and even at large-$|t|$ \cite{Csorgo:2019ewn, Csorgo:2023rzm}. The $\sigma_{\rm el}$ and $B_0$ values utilized to calculate the experimental $H(x)$ scaling functions are summarized in \cref{tab:sigel_B0_vals}. The $H(x)$  scaling was successfully utilized to search for odderon effects in the TeV energy range in Ref.~\cite{Csorgo:2019ewn}. 

\begin{figure} [hbt!]
	\centering
	\subfloat[\label{fig:1}]{%
		\includegraphics[width=0.8\linewidth]{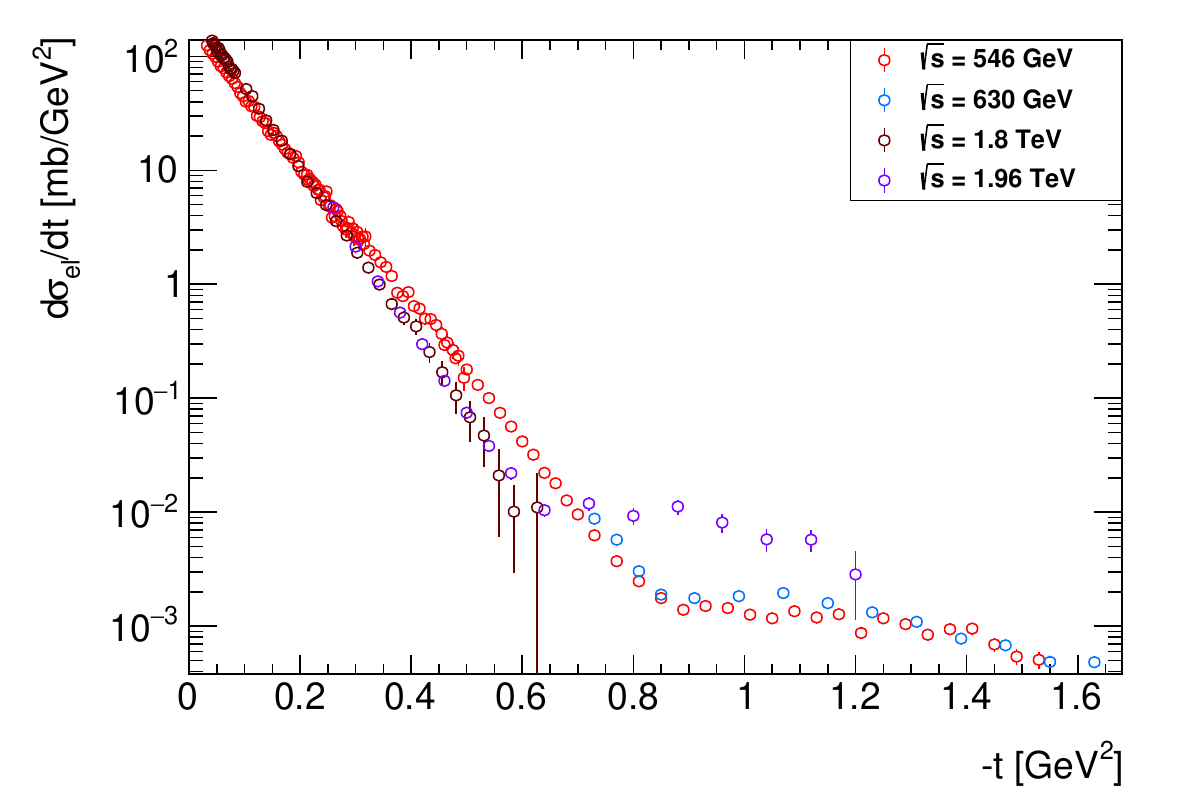}%
	}\hfill
	\subfloat[\label{fig:2}]{%
		\includegraphics[width=0.8\linewidth]{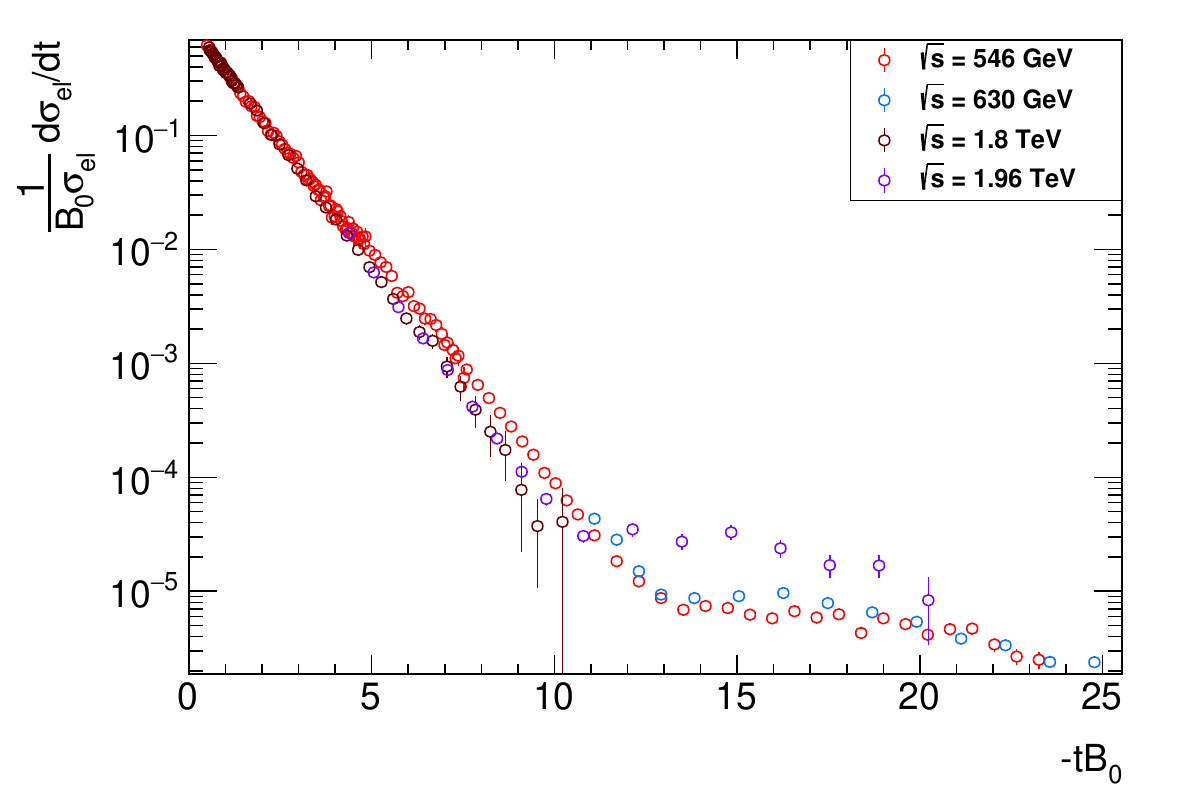}%
	}
	\caption{(a) SPS UA4 546 GeV TeV, 630 GeV, TEVATRON E710 1.8 TeV, and Tevatron D0 1.96 TeV data on $p\bar p$  $d\sigma_{\rm el}/dt$  and \mbox{(b) the calculated} $p\bar p$ $H(x)$ scaling functions using these data. Since at $\sqrt{s}$ = 630 GeV $\sigma_{\rm el}$ and $B_0$ are not measured, to calculate the experimental $H(x)$ scaling function at this energy I utilized the $\sigma_{\rm el}$ and $B_0$ values measured at $\sqrt{s}$ = 546 GeV since this was the closest energy where measurement was done.}
	\label{fig:exp_hx_pbarp}
 \vspace{-5mm}
\end{figure}

In \cref{chap:Hxvalidity}, I provide an interpretation for the observed scaling of the data at higher $|t|$ values, including the region of the diffractive minimum-maximum structure, and I test model-dependently the $H(x)$  scaling against the differential cross section data in the TeV energy domain. I also discuss the possible reasons for scaling violation in $pp$ scattering at \mbox{$\sqrt{s}=$ 13 TeV} and in $p\bar p$ scattering in general. In elastic $p\bar p$ scattering $H(x)$  scaling is present only in the lower-$|t|$ range (see \cref{fig:exp_hx_pbarp}; the $\sigma_{\rm el}$ and $B_0$ values utilized to calculate the experimental $H(x)$ scaling functions are summarized in \cref{tab:sigel_B0_vals}). The observation that elastic $pp$ scattering has an $H(x)$ scaling property in the higher-$|t|$ domain and such a scaling is absent in elastic $p\bar p$ scattering is itself an odderon effect.


\begin{table*}[htb]
\begin{center}
\vspace{-1mm}
\begin{tabular}{l l l}
$\sqrt{s}$ (TeV)  &\,\,\,\, $\sigma_{\rm el}$ (mb)  &\,\,\, $B_0$ (GeV$^{-2}$) \\
\hline
 2.76 ($pp$)          &\,\,\,\, 21.8  $\pm$ $1.4 $  ~\cite{Nemes:2017gut,TOTEM:2018psk}
                     & \,\,\, 17.1  $\pm$ $0.3$ ~\cite{TOTEM:2018psk}     \\
 7.0 ($pp$)          &\,\,\,\, 25.43 $\pm$ $0.03$  ~\cite{TOTEM:2013lle}  
 & \,\,\, 19.89 $\pm$ $0.03$  ~\cite{TOTEM:2013lle} \\
 8.0 ($pp$)          &\,\,\,\, 27.1 $\pm$ $1.4$ ~\cite{TOTEM:2012oyl}
                      & \,\,\, 19.9 $\pm$ $0.3$ ~\cite{TOTEM:2012oyl} \\
 0.546 ($p\bar{p}$)  &\,\,\,\ 13.3 $\pm$ 0.6 \cite{UA4:1984skz}  & \,\,\, 15.2 $\pm$ 0.2 \cite{UA4:1984uui}\\
 1.8 ($p\bar{p}$)  &\,\,\,\, 16.6 $\pm$ 1.6 \cite{E-710:1990xdw} & \,\,\, 16.3 $\pm$ 0.5 \cite{E-710:1990xdw}\\
 1.96 ($p\bar{p}$)  &\,\,\,\, 20.2  $\pm$ $1.7 $   
 & \,\,\, 16.86 $\pm$ $0.1$  ~\cite{D0:2012erd} \\
\hline
\end{tabular}
\end{center}
\vspace{-0.5cm}
\caption {Elastic cross section $\sigma_{\rm el}$ and nuclear slope parameter $B_0$ values at different energies for $pp$ and $p\bar p$ scattering utilized to calculate the experimental $H(x)$ scaling functions. The value and error of the $\sigma_{\rm el}$ at $\sqrt{s} = 1.96$ TeV 
is obtained from a low $-t$ exponential fit to the data of Ref.~\cite{D0:2012erd}. 
}
\label{tab:sigel_B0_vals}
\end{table*}


\vspace{-5mm}
\section{Data fitting: models versus measurements} \label{sec:fittingmethod}

In my work, I analyzed measurements on elastic $pp$ and $p\bar p$ scattering by fitting the parameters of different physical models to the data. I utilized the method of least squares or $\chi^2$ fitting technique \cite{ParticleDataGroup:2022pth,Block:2006hy}. This method can be beneficially used for fitting $N$ measurements at known points $x_i$, which are Gaussian
distributed about their measured values $d_i$ and have error estimates $\sigma_i$ corresponding to standard deviations of the fluctuations. 

Assume that we have a physical model $m(x,\vec p)$ with $n$ free parameters put into an $n$-dimensional vector $\vec p$, and we want to use this model to describe experimental data. In the standard $\chi^2$ fitting procedure the standard $\chi^2$ function, 
\begin{equation}\label{eq:chi_trad0}
    \chi^2(\vec p\,) = \sum_{i=1}^N \frac{\left(d_i - m(x_i,\vec p)\right)^2}{\sigma_i^2},
\end{equation}
is minimized by solving a set of $n$ equations
\begin{eqnarray}
    \frac{\partial \chi^2(\vec p\,)}{\partial p_j} =0,\,\,\,j=1,...,n.
\end{eqnarray}

\noindent There is an equation for each parameter to get a set of parameter values that minimize the $\chi^2(\vec p\,)$ function. When the errors are Gaussian distributed, $\chi^2(\vec p\,)$ is a random variable that follows a $\chi^2$-distribution with number of degrees of freedom $NDF = N-n$. 

The $\chi^2$ minimization procedure returns the best-fit parameter values of the model with statistically meaningful error estimates. For minimizing the $\chi^2$ function, I used the Minuit package \cite{James:2004xla}, and for estimating the fit parameter errors, I used the MINOS algorithm of Minuit.
MINOS calculates parameter errors taking into account both parameter correlations and non-linearities of the model to be fitted. The MINOS error for a given parameter is defined as the change in the value of that parameter which causes the $\chi^2$ at the minimum, $\chi^2_{\rm min}$, to increase by 1.

Before the operation of MINOS, a good minimum is found and the error matrix is calculated using Minuit's MIGRAD algorithm. This error matrix is the inverse of the matrix of second derivatives of the $\chi^2(\vec p\,)$ and takes into account all the parameter correlations, but not the non-linearities. The parameter errors then are given as the square roots of the diagonal elements of the error matrix. Minuit calculates the eigenvalues of the error matrix to decide if the matrix is positive-definite. A physical $\chi^2(\vec p\,)$ function is quadratic in the neighborhood of the minimum, and the error matrix is positive-definite. An accurate error matrix means that MIGRAD converged and a local minimum has been found. I considered a fit acceptable if the error matrix was accurate and MINOS errors were successfully determined.

I quantified the level of agreement between the data and the hypothesis (physical model) by the confidence level:
\begin{equation}
CL\equiv CL(\chi^2,NDF) = 1-F(\chi^2,NDF),
\end{equation}
where $F(\chi^2,NDF)$ is the cumulative distribution function of the $\chi^2$ probability density distribution. The $CL$ is a function of the $\chi^2$ and the $NDF$ and is, therefore, itself a random variable. If the fitting function is the true representation of the data, then the $CL$ will be uniformly distributed between 0 and 1.  The  $CL$ value represents our ``confidence'' in the statement that the fitting function is \mbox{the true representation of the data.}

When searching for new phenomena, we try to reject a hypothesis represented by a physical model describing a given process.  In experimental analyses, it is customary to choose a number for the lower $CL$ limit, like 1\%, 0.5\%, or 0.1\% for rejecting a fit. 
I chose 0.1\% for the lower $CL$ limit from the beginning of the analysis. 
An agreement between the model and the data with $CL\geq99.9$\% is rather unlikely and may indicate that the experimental errors are overestimated. Thus, in my analysis, I accepted a model description when $0.1\%\leq CL<99.9\%$ and rejected it when $CL<0.1\%$ or $CL\geq99.9$\%.

The $CL$ value can be converted into an equivalent significance $\delta$. This $\delta$ significance is 
defined so that outside $\pm\delta$ from the mean of Gaussian distribution, the area of the tails of the distribution is equal to $CL$ \cite{ParticleDataGroup:2022pth}. The $CL<0.1$\% corresponds to a statistical significance, $\delta>3.29\sigma$ for the difference between the model and the data. If the difference between the model and the data has a significance $\delta\leq3.29\sigma$, the difference is considered to be a non-significant difference. If the difference between the model has a significance $\delta>3.29\sigma$ but $\delta<5.0\sigma$, the difference is considered to be an indication that the model assumption is inconsistent with the data. The lower limit of a discovery level statistical significance in high-energy physics is $\delta=5\sigma$ corresponding to a $CL$ value of $5.7\times10^{-5}\%$. Thus, if the difference between the model has a significance $\delta\geq5.0\sigma$, the model assumption is considered inconsistent with the data, the null hypothesis is rejected, and a new effect contradicting the assumptions of the used model is discovered.

\cref{eq:chi_trad0} is not the only possible form of the $\chi^2$ function. In the earlier ReBB model study of elastic $pp$ scattering \cite{Nemes:2015iia}, in addition to the statistical uncertainty, the $t$-independent systematic uncertainty of the data was considered during the $\chi^2$ fit. This $t$-independent systematic uncertainty follows from the luminosity measurement. The $\chi^2$ function, minimized in the course of fitting the model to the $d\sigma_{\rm el}/dt$ data, was the following:
\begin{equation}\label{eq:nemeschi}
    \chi^2 = \sum_{i=1}^N \frac{\left(d_i - \gamma\, m_i(\vec p\,)\right)^2}{\sigma_i^2} + \frac{(1-\gamma)^2}{\sigma^2_{\rm lumi}},
\end{equation}
where $d_i$ is the $i$th data point, $\sigma_i$ is its statistical error, $m_i(\vec p\,)$ is the value of the model corresponding to the $i$th data point, $\sigma^2_{\rm lumi}$ is the $t$-independent systematic uncertainty following from the luminosity measurement, and $\gamma$ is a normalization parameter to be optimized. I utilized a similar $\chi^2$ formula in \cref{chap:oddTD0} of this dissertation.

In my work, I utilized also another, more advanced form of the $\chi^2$ formula, which handles the $t$-dependent statistical and systematic errors as well as the $t$-independent systematic errors. This form of the $\chi^2$ function was derived by the PHENIX Collaboration (see the derivation in Ref.~\cite{PHENIX:2008ove}) for the case when the covariance matrix of errors is not available for a dataset but the available experimental errors can be separated into three different types: 
\begin{itemize}
    \item type \textit{a}: point-to-point varying uncorrelated systematic and statistical errors ($\sigma_a$);
    \item type \textit{b}: point-to-point varying and 100\% correlated systematic errors ($\sigma_b$);
     \item type \textit{c}: point-independent, overall correlated systematic uncertainties ($\sigma_c$ in percent) that scale all the data points up and down by the same factor.
\end{itemize}
The corresponding $\chi^2$ function reads
\begin{equation}\label{eq:phenix_o}
    \chi^2 = \left(\sum_{i=1}^N \frac{\left(d_i + \epsilon_b \tilde\sigma_{bi} +\epsilon_cd_i\sigma_c- m_i(\vec p\,)\right)^2}{\tilde\sigma_i^2}\right) + \epsilon_b^2 + \epsilon_c^2,
\end{equation}
where
\begin{equation}
    \tilde\sigma_i = \sigma_{ai}\frac{d_i + \epsilon_b \sigma_{bi} +\epsilon_cd_i\sigma_c}{d_i}
\end{equation}
is the uncertainty scaled by the shift in $d_i$ to keep the fractional uncertainty unchanged and $\epsilon_l$ is the fraction of the type $l\in\{b,c\}$ error fitted to the data. The use of \cref{eq:phenix_o} allows us to consider during the fit not only statistical but also systematic errors of the data in a proper way, $i.e.$, in a way equivalent to the use of the covariance matrix of experimental errors. When fitting a model, $e.g.$, to the $pp$ and $p\bar{p}$ $d\sigma_{\rm el}/dt$ data, the $\epsilon_b$ allows for a slope variation for the model based on the type $b$ errors of the data points, while $\epsilon_c$ allows for a normalization variation based on the type $c$ error of the data. Both $\epsilon_b$ and $\epsilon_c$ have a known central value, 0, and a known standard deviation, 1. Hence, they must be considered not only as fit parameters but also as new data points. Thus, these parameters do not affect the $NDF$: $NDF=(N_d+N_\epsilon) - (N_p+N_\epsilon)$, where $N_d$ is the number of data points fitted, $N_\epsilon$ is the number of different $\epsilon$ parameters in the used $\chi^2$ formula, and $N_p$ is the number of free parameters of the model utilized to fit the data. Of course, in case the covariance matrix of errors is available, one does not need to use \cref{eq:phenix_o} and the parameters $\epsilon_b$ and $\epsilon_c$. However, the covariance matrix of errors is not publicly available for the data I analyzed. 

The $\chi^2$ formula of \cref{eq:phenix_o} was validated by evaluating the $\chi^2$ from a full covariance matrix fit of the $\sqrt{s}$ = 13 TeV TOTEM differential cross section data using the Lévy expansion method of Ref.~\cite{Csorgo:2018uyp}. The fit with \cref{eq:phenix_o} and that with the full covariance matrix resulted in the same minimum within one standard deviation of the fit \mbox{parameters \cite{Csorgo:2020wmw}.} This shows that the use of \cref{eq:phenix_o} is indeed equivalent to the use/diagonalization of the covariance matrix of statistical and systematic errors. 

In \cite{Csorgo:2019ewn}, the $\chi^2$ formula of \cref{eq:phenix_o} was rewritten in a form to treat data-data comparison. In \cref{chap:rebbdesc} of this thesis, I present a version of \cref{eq:phenix_o} that can be applied for data-model comparison where the data (i) consist of sub-datasets corresponding to several separately measured kinematic ranges and (ii) have both vertical and horizontal errors.

\section{Summary of goals and used methods}\label{sec:introsum}


The main goal of my work was to search for the signal of odderon exchange by comparing pp and $\rm{p\bar p}$ elastic differential cross sections \mbox{(${\rm d}\sigma_{\rm el}/{\rm d}t$-s)} in the same kinematic \mbox{($s$, $t$)} range in the TeV energy region where the contribution of mesonic Regge exchanges is highly suppressed. 
To overcome the lack of experimental data at the same $s$ and $t$ domain, 
I applied model extrapolations. I used the $p=(q,d)$~version of the ReBB model, which is formulated based on R.~J.~Glauber's diffractive multiple scattering theory as detailed above in \cref{sec:ReBBmodel}. I used also a phenomenological model based on Regge \mbox{theory \cite{Broniowski:2018xbg,Szanyi:2018pew}} detailed in \cref{sec:reggemodel}. The obtained results are detailed in \cref{chap:ReBBpbarp}, \cref{chap:oddTD0}, \cref{chap:rebbdesc}, and \cref{chap:odderon}.

Aiming to provide a physical interpretation of the $H(x)$ scaling, I was looking for a class of models of elastic pp scattering that manifests this scaling. Aiming to check if the scaling property of the data \cite{Csorgo:2019ewn} is a valid tool for odderon search, I tested, model dependently, the $H(x)$ scaling by identifying the $H(x)$ scaling limit of the ReBB model and fitting it to the data on $pp$ and $p\bar p$ ${\rm d}\sigma_{\rm el}/{\rm d}t$. The results are detailed in \cref{chap:Hxvalidity}.

To fit the model and compare the calculated curves to experimental data, I utilized different forms of the $\chi^2$ function introduced above and detailed further in the text below. For minimizing the $\chi^2$ functions, I used the Minuit package \cite{James:2004xla}, and for estimating the fit parameter errors, I used the MINOS algorithm of Minuit. I performed the numerical computations using C++ programming language, CERN's ROOT data analysis framework, and Wolfram Mathematica.

After the odderon search was completed, aiming to achieve a quantitative description in a wider kinematic range 
to the data that includes also the strongly non-exponentially behaving low-$|t|$ domain of the $pp$ differential cross section, I worked on improving models of hadronic elastic scattering using, in the impact parameter space, Lévy \mbox{$\alpha$-stable} distributions instead of the Gaussian distributions. The details are given in \cref{chap:levy}.


\newpage
\thispagestyle{empty}

\chapter{Elastic $\rm p\bar p$ scattering and the ReBB model}\label{chap:ReBBpbarp}

The ReBB model was proposed to describe elastic $pp$ scattering data. The generalization of the ReBB model to elastic $p\bar p$ scattering is a crucial step towards the odderon analysis based on the ReBB model. In this Chapter, I generalize the ReBB model to elastic $p\bar p$ scattering by fitting its parameters to all the available experimental elastic $p\bar p$ differential cross section data, which includes the c.m. energy range of 31~GeV~$\leq\sqrt{s}\leq 1.96$~TeV. 
\mbox{I exclude} only the $-t$ region of the Coulomb-nuclear interference. 

In \cref{sec:rebbpbarp}, I present the results of the ReBB model description to elastic $p\bar p$ scattering data. Then, in \cref{sec:rebb_pbarp_en_evolution},
I discuss that the energy dependencies of the ReBB model parameters for $p\bar p$ scattering are consistent with linear-logarithmic shapes in the analyzed c.m. energy range. 

This chapter is based on  Ref.~\cite{Csorgo:2020wmw}.

\newpage

\section{ReBB model versus elastic $\rm p\bar p$ scattering data}\label{sec:rebbpbarp}

The ReBB model was introduced in Ref.~\cite{Nemes:2015iia}, and it was used to describe elastic $pp$ differential cross section data at ISR energies and at the LHC energy of $\sqrt s$ = 7 TeV. In my work, I generalized the ReBB model to describe elastic $p\bar p$ scattering by fitting the parameters of the model to elastic $p\bar p$ differential cross section data. This generalization is supported by the fact that the $t$-distribution of elastic $pp$ scattering is very similar to that of elastic $p\bar p$ scattering except for the minimum-maximum structure, characteristic for elastic $pp$ scattering, and a shoulder structure, characteristic for $p\bar p$ scattering.

Let me summarize the $p\bar p$ elastic differential cross section measurements starting from the era of the CERN ISR accelerator.  There are low-$|t|$  measurements at ISR at \mbox{$\sqrt{s}=31$ GeV,} \mbox{53 GeV} and 62 GeV \cite{AMES-BOLOGNA-CERN-DORTMUND-HEIDELBERG-WARSAW:1984yby}, and a high-$|t|$ measurement at \mbox{$\sqrt{s}=53$ GeV \cite{Breakstone:1985pe}.} The UA4 Collaboration at CERN SPS measured low-$|t|$ data at $\sqrt s = $ 540 GeV \cite{UA4:1983mlb} and 546 GeV \cite{UA4:1984uui}, and high-$|t|$ data at $\sqrt{s}=546$ GeV \cite{UA4:1985oqn} and $630$ GeV \cite{UA4:1986cgb}. The E-710 Collaboration at FNAL Tevatron measured low-$|t|$ data at $\sqrt{s}=1.8$ TeV \cite{E-710:1990vqb}. The D0 Collaboration at FNAL Tevatron measured high-$|t|$ data at $\sqrt{s}=1.96$ TeV \cite{D0:2012erd}.  

The ReBB model parameters fitted to these $p\bar p$ elastic scattering data are summarized in \cref{tab:rebbfit_parameters_ISR}
and \cref{tab:rebbfit_parameters_SPSTEV}. The parameter values and their uncertainties are rounded to two decimal places (the same procedure was applied in previous studies \cite{Nemes:2015iia}; this is not the standard practice, but reliably measuring length in such experiments with errors smaller than 0.01 fm is hardly possible). The fits are done using the traditional $\chi^2$ function given by \cref{eq:chi_trad0}. The above-cited data sources give point-to-point varying statistical or combined statistical and systematic uncertainties and overall normalization errors. The latter is neglected by \cref{eq:chi_trad0}. All fits, summarised in \cref{tab:rebbfit_parameters_ISR}
and \cref{tab:rebbfit_parameters_SPSTEV}, are successful: $CL\geq$ 0.1\% is satisfied, the error matrix is accurate, and MINOS errors are successfully determined. Parameter $\lambda$ is fixed at $1/2$, and parameter $A_{qq}$ is fixed at $1$ as it was done in Ref.~\cite{Nemes:2015iia}.

\begin{table}[h]\small
	\centering
	\begin{tabular}{ccccc} \hline\hline
		$\sqrt{s}$ [GeV]   & 31            & 53            & 53			   & 62				\\ \hline
		$|t|$ [GeV$^{2}$]  & (0.05,0.85)   & (0.11,0.85)   & (0.523,3.52)  &(0.13,0.85) \\ 
		$\chi^{2}/NDF$	   & 17.34/18      &   26.59/20    & 34.53/23	   &14.30/19   \\ 
		CL [\%] 	       & 49.99         & 14.71         & 5.79          & 76.59	 \\ 
		$R_{q}$ [$\rm fm$] 	   & 0.28$\pm$0.01 & 0.27$\pm$0.01 & 0.22$\pm$0.03 & 0.27$\pm$0.01   \\ 
		$R_{d}$ [$\rm fm$] 	   & 0.75$\pm$0.03 & 0.73$\pm$0.02 & 0.71$\pm$0.04 & 0.72$\pm$0.05 	\\ 
		$R_{qd}$ [$\rm fm$]    & 0.36$\pm$0.06 & 0.38$\pm$0.02 & 0.42$\pm$0.03 & 0.39$\pm$0.07   \\
		$\alpha_R$ 	       & 0.27$\pm$0.08 & 0.0$\pm$0.4 & 0.11$\pm$0.01 & 0.19$\pm$0.09    \\ \hline\hline
	\end{tabular}
	\caption{Values of the ReBB model parameters fitted to the proton-antiproton differential cross section data at $\sqrt{s}=$ 31 GeV, 53 GeV, and 62 GeV. The parameter values and their uncertainties are rounded to two decimal places.}
	\label{tab:rebbfit_parameters_ISR}
\end{table} 

\begin{table}[h]\small
	\centering
	\begin{tabular}{ccccc} \hline\hline
		$\sqrt{s}$ [GeV]    & 540 \& 546           & 630	        & 1800	        &  1960			\\ \hline
		$|t|$ [GeV$^{2}$]   &(0.033,1.53)   & (0.73,2.13)  & (0.034,0.627)  &  (0.26,1.2) \\
		$\chi^{2}/NDF$	    &  151.43/117   & 29.74/15      & 29.18/47      &  13.93/13    \\ 
		CL [\%] 	        & 1.77          & 1.29		    &  98.07        &  37.91	 \\ 
		$R_{q}$ [$\rm fm$] 	& 0.35$\pm$0.01 & 0.25$\pm$0.04 & 0.38$\pm$0.01 &  0.39$\pm$0.03	\\ 
		$R_{d}$ [$\rm fm$] 	& 0.83$\pm$0.01 & 0.69$\pm$0.08 & 0.87$\pm$0.02 &  0.87$\pm$0.04	\\ 
		$R_{qd}$ [$\rm fm$] & 0.28$\pm$0.01 & 0.50$\pm$0.09 & 0.33$\pm$0.04 &  0.3$\pm$0.1	\\
		$\alpha_R$ 	        & 0.12$\pm$0.01 & 0.16$\pm$0.03	& 0.13$\pm$0.04 &  0.16$\pm$0.01\\ \hline\hline
	\end{tabular}
	\caption{Values of the ReBB model parameters fitted to the proton-antiproton differential cross section data at $\sqrt{s}=$ 540 GeV, 546 GeV, 630 GeV, 1.8 TeV, and 1.96 TeV. The parameter values and their uncertainties are rounded to two decimal places.}
	\label{tab:rebbfit_parameters_SPSTEV}
\end{table}

\begin{figure}[!hbt] 
\centering
\includegraphics[scale=0.65]{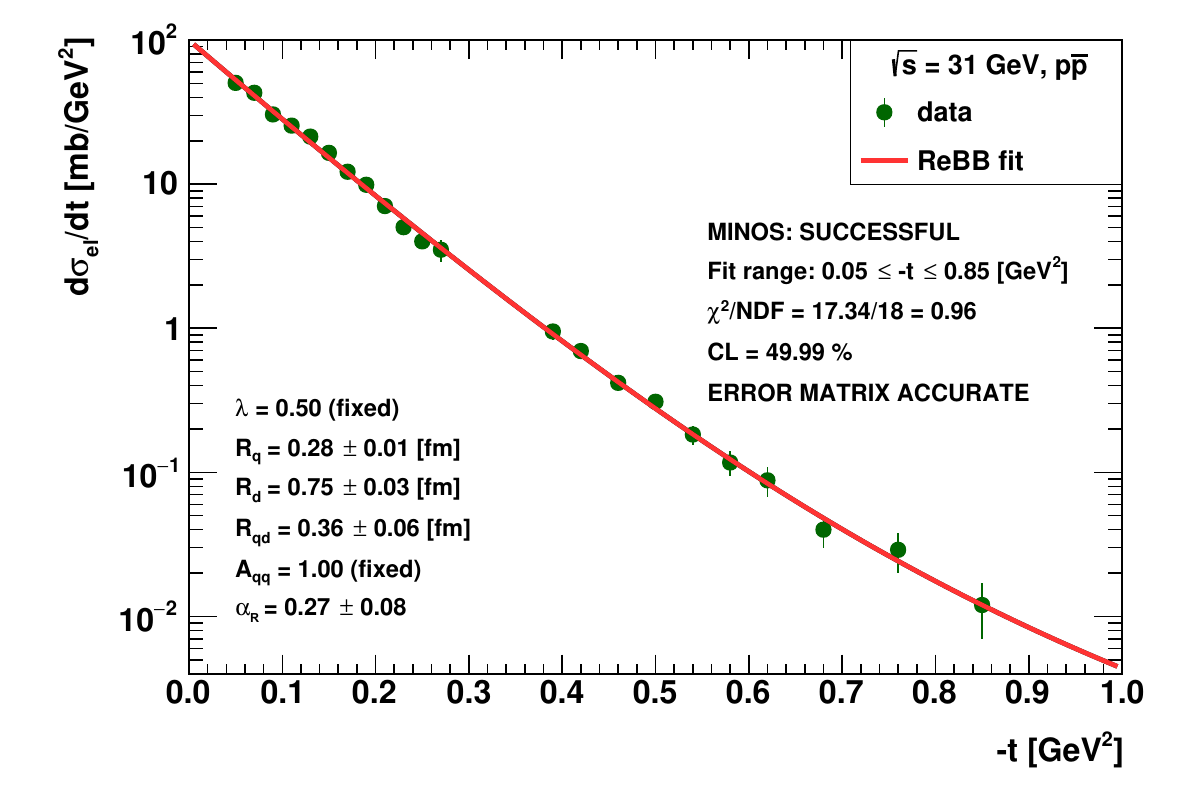}
\vspace{-0.35cm}
\caption{Fit of the ReBB model to $p\bar p$ differential cross section data at $\sqrt{s}=$ 31 GeV. The values of the fitted parameters and fit statistics are shown.}
\label{fig:rebb31}
\end{figure}

The result of the fit to the low-$|t|$ ISR 31 GeV data is shown in \cref{fig:rebb31}. The $CL$ of the description is 49.99\%. 

\begin{figure}[!hbt] 
\centering
\includegraphics[scale=0.65]{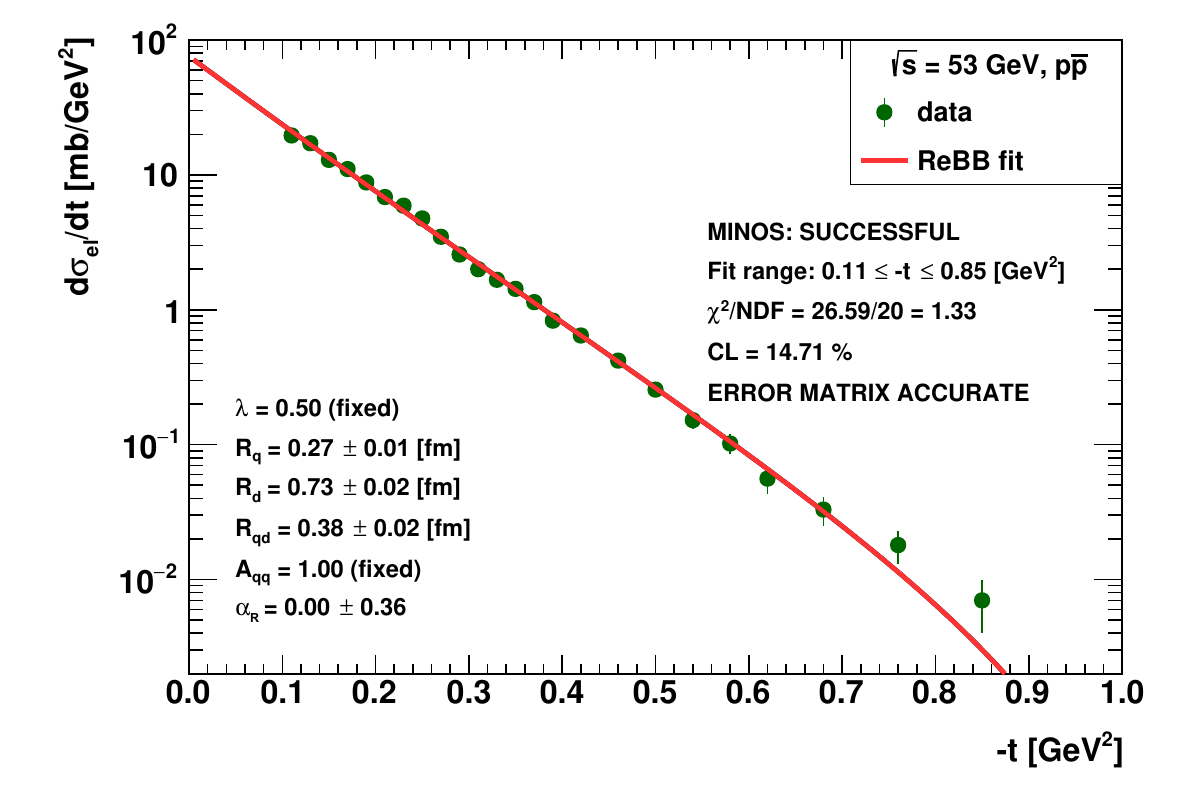}
\vspace{-0.35cm}
\caption{Fit of the ReBB model to low-$|t|$ $p\bar p$ differential cross section data at \mbox{$\sqrt{s}=$ 53 GeV.} The values of the fitted parameters and fit statistics are shown.}
\label{fig:rebb53low}
\end{figure}

\vspace{-0.5cm}

\begin{figure}[!hbt] 
\centering
\includegraphics[scale=0.65]{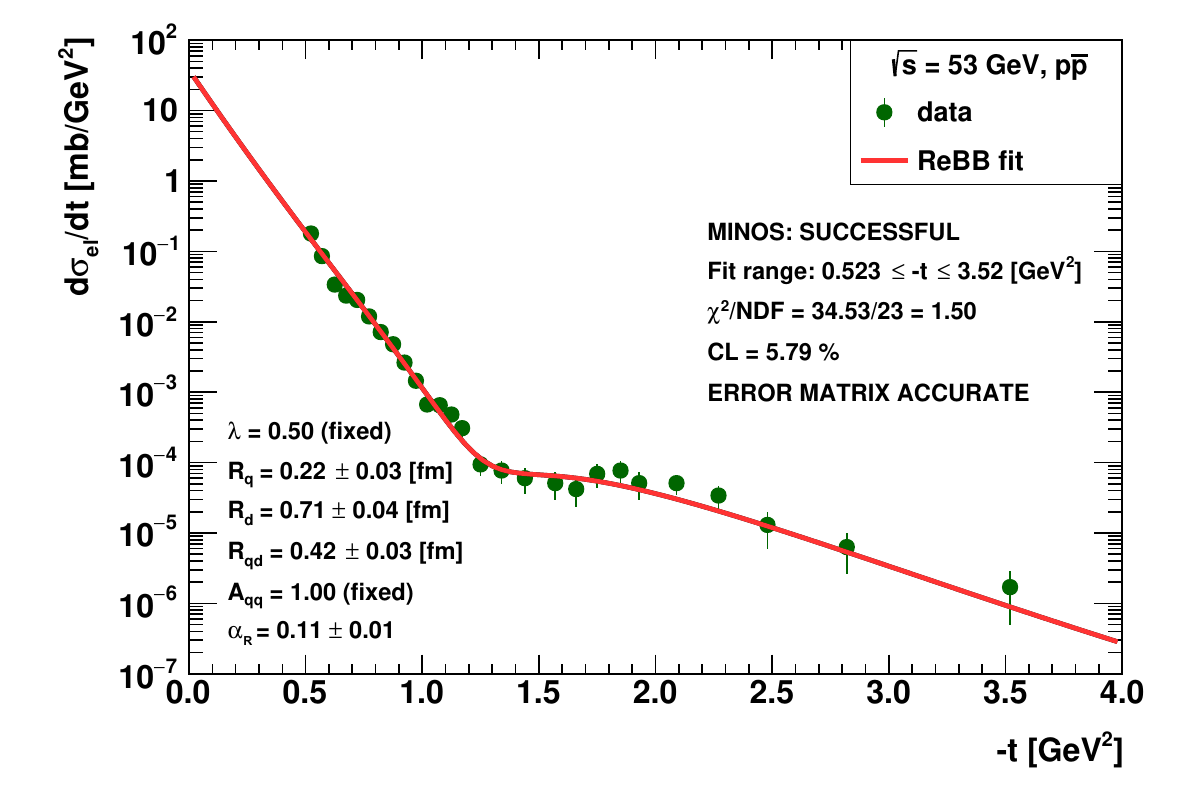}
\vspace{-0.35cm}
\caption{Fit of the ReBB model to high-$|t|$ $p\bar p$ differential cross section data at \mbox{$\sqrt{s}=$ 53 GeV.} The values of the fitted parameters and fit statistics are shown.}
\label{fig:rebb53h}
\end{figure}

The low-$|t|$ and high-$|t|$ ISR data at $\sqrt{s}=53$ GeV are published separately in Ref.~\cite{AMES-BOLOGNA-CERN-DORTMUND-HEIDELBERG-WARSAW:1984yby} and Ref.~\cite{Breakstone:1985pe}, respectively. The two datasets, possibly because of normalization differences in the two measurements, cannot be fitted simultaneously with \mbox{$CL\geq 0.1$\%} using the $\chi^2$ function given by \cref{eq:chi_trad0}. For this reason, I fitted these data sets separately. \cref{fig:rebb53low} shows the result of the ReBB fit to the low-$|t|$ data at $\sqrt{s}=53$ GeV while \cref{fig:rebb53h} displays the result of the fit to the high-$|t|$ data at $\sqrt{s}=53$ GeV. In the former case, the $CL$  of the fit is 14.71\%; in the latter case, it is 5.79\%. 

\begin{figure}[!hbt] 
\centering
\includegraphics[scale=0.65]{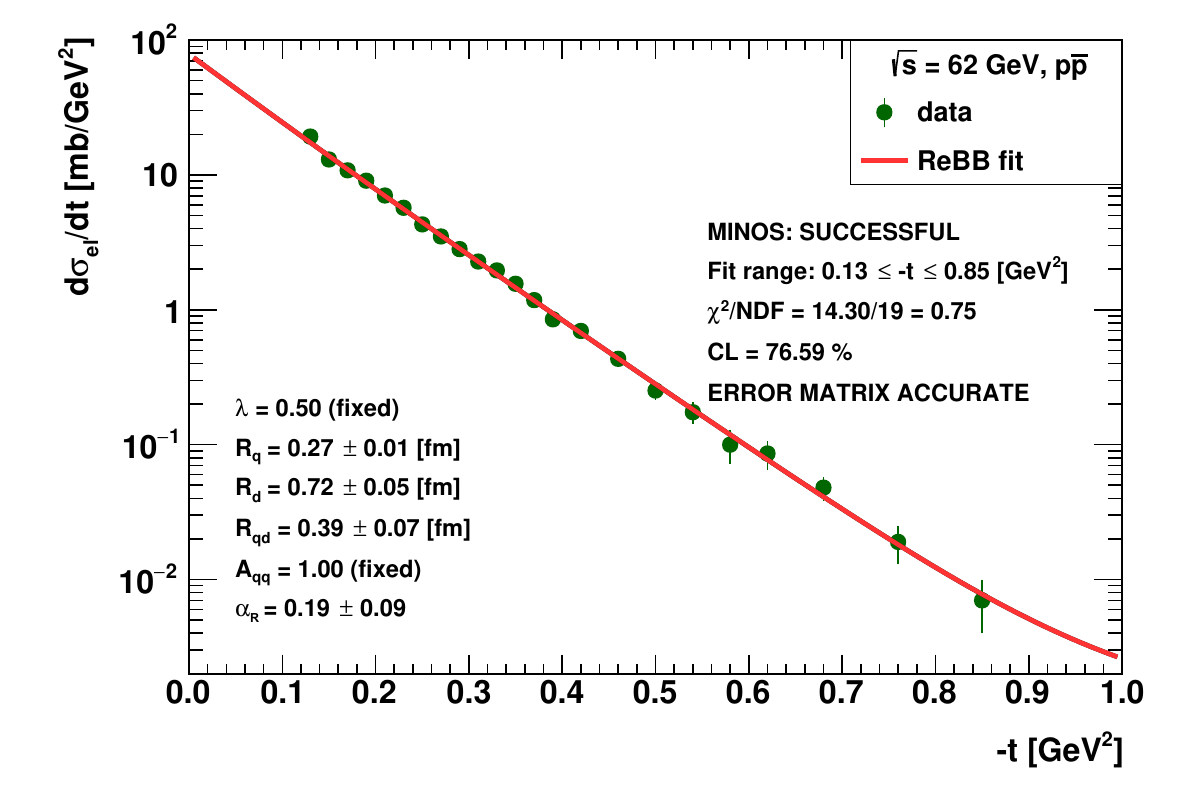}
\vspace{-0.35cm}
\caption{Fit of the ReBB model to $p\bar p$ differential cross section data at $\sqrt{s}=$ 62 GeV. The values of the fitted parameters and fit statistics are shown.}
\label{fig:rebb62}
\end{figure}

The result of the ReBB fit to the low-$|t|$ ISR 62 GeV data is shown in \cref{fig:rebb62}. The $CL$ is 76.59\%. 

\begin{figure}[!hbt] 
\centering
\includegraphics[scale=0.65]{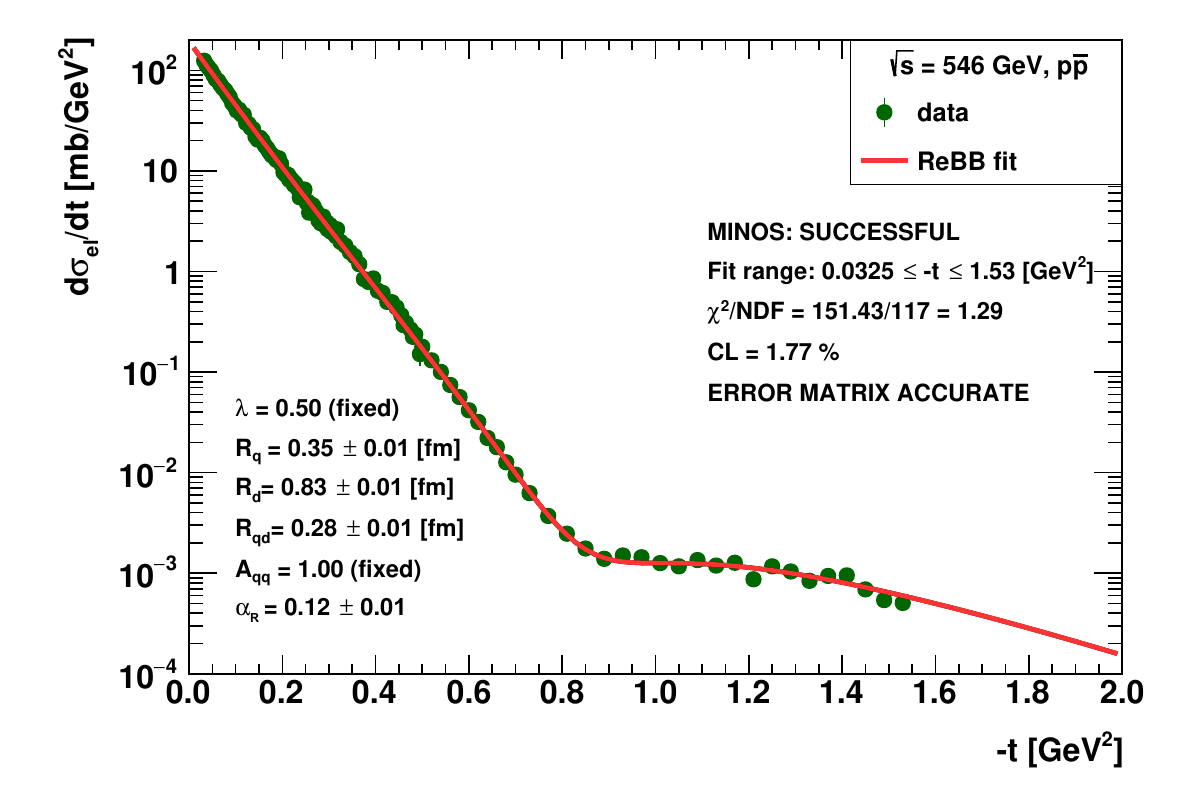}
\vspace{-0.35cm}
\caption{Fit of the ReBB model to the merged $\sqrt{s}=$~540~GeV and 546 GeV $p\bar p$ differential cross section data. The values of the fitted parameters and the fit statistics are shown.}
\label{fig:rebb546}
\end{figure}

The $\sqrt s$ = 540 GeV low-$|t|$ \cite{UA4:1983mlb}, the $\sqrt s$ = 546 GeV low-$|t|$ \cite{UA4:1984uui}, and the \mbox{$\sqrt s$ = 546 GeV} high-$|t|$ \cite{UA4:1985oqn} SPS data can be fitted simultaneously with $CL\geq0.1\%$ using the $\chi^2$ function given by \cref{eq:chi_trad0}. The result of the fit is shown in \cref{fig:rebb546}. The $CL$ is 1.77\% in the kinematic range of \mbox{$0.0325~{\rm GeV}^2 \leq |t|\leq 1.53~{\rm GeV}^2$.} By merging three datasets together, one can more precisely determine the values of the ReBB model parameters. The simultaneous fit to the SPS 540 GeV and 546 GeV datasets gives the values of the ReBB model parameters with the lowest uncertainty.

\begin{figure}[!hbt] 
\centering
\includegraphics[scale=0.65]{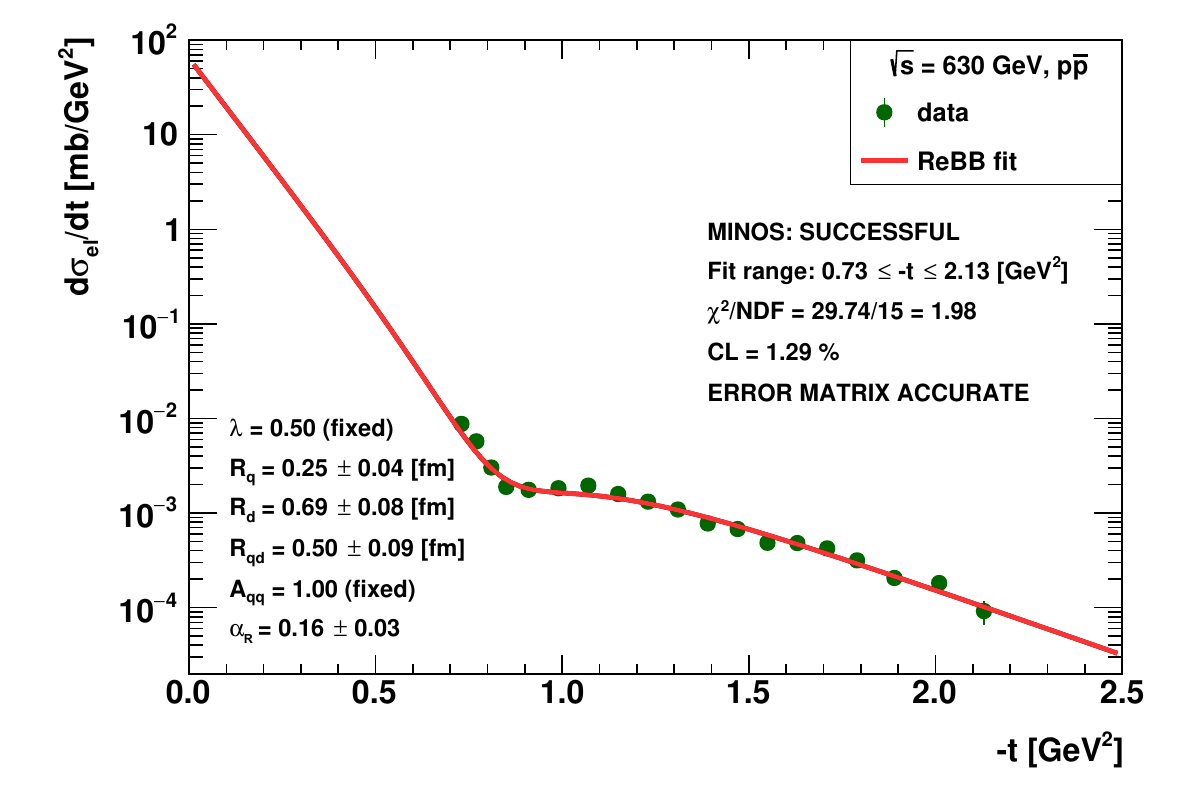}
\vspace{-0.35cm}
\caption{Fit of the ReBB model to $p\bar p$ differential cross section data at $\sqrt{s}=$ 630 GeV. The values of the fitted parameters and the fit statistics are shown.}
\label{fig:rebb630}
\end{figure}

\begin{figure}[!hbt] 
\centering
\includegraphics[scale=0.65]{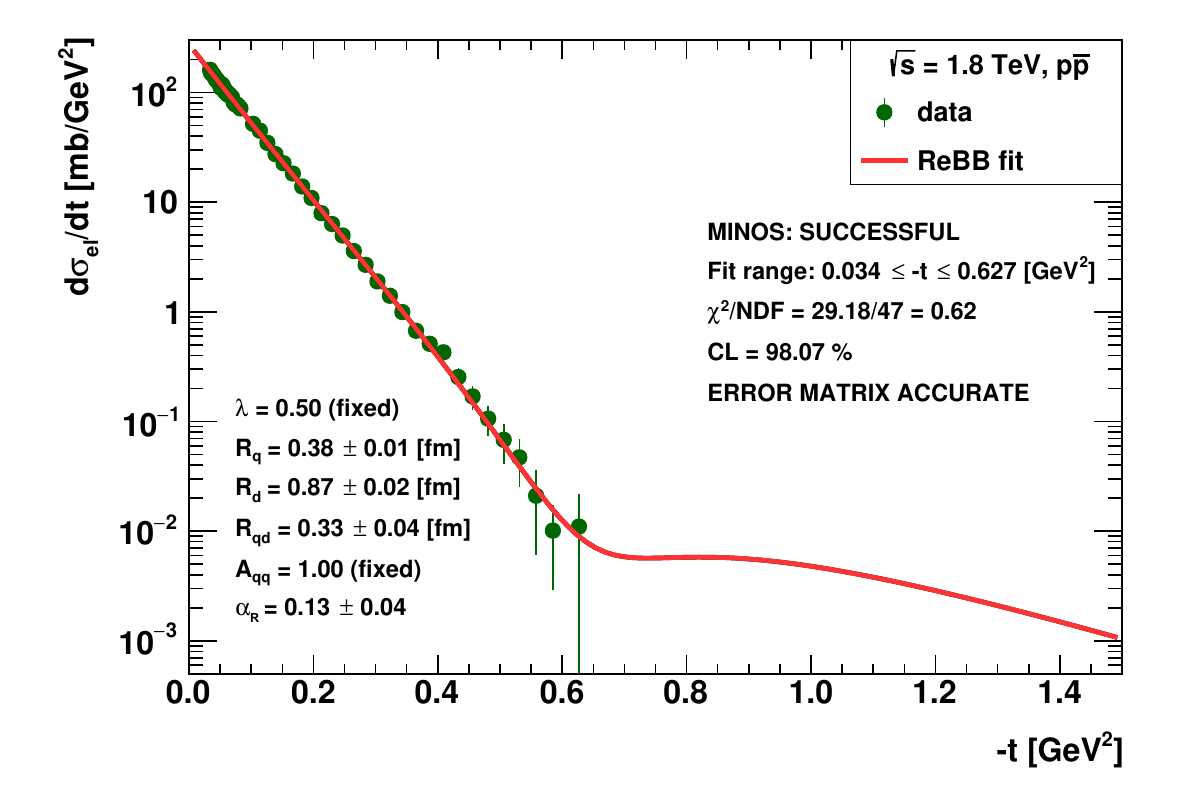}
\vspace{-0.35cm}
\caption{Fit of the ReBB model to $p\bar p$ differential cross section data at $\sqrt{s}=$ 1.8 TeV. The values of the fitted parameters and the fit statistics are shown.}
\label{fig:rebb180}
\end{figure}

\begin{figure}[!hbt] 
\centering
\includegraphics[scale=0.65]{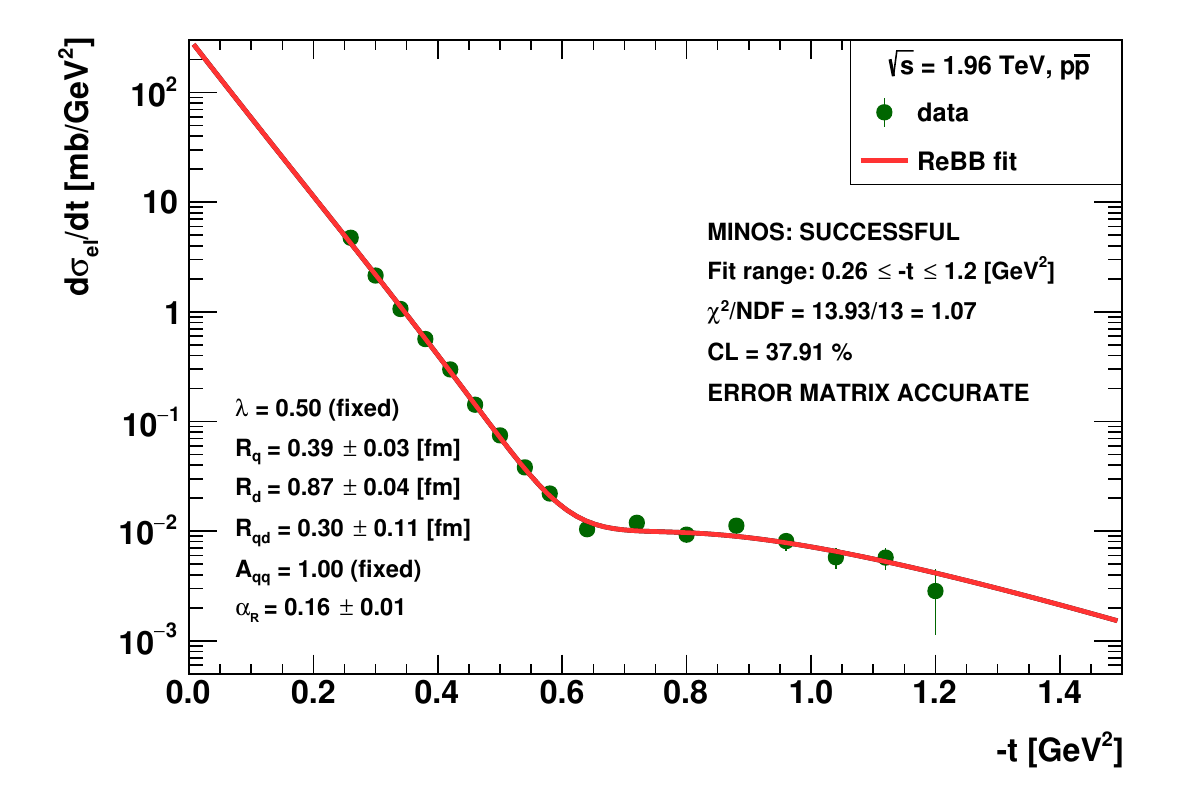}
\vspace{-0.35cm}
\caption{Fit of the ReBB model to $p\bar p$ differential cross section data at  $\sqrt{s}=$ 1.96 TeV. The values of the fitted parameters and the fit statistics are shown.}
\label{fig:rebb190}
\end{figure}

The result of the ReBB model fit of the SPS 630 GeV data in the kinematic range of $0.73~{\rm GeV}^2 \leq |t|\leq 2.13~{\rm GeV}^2$ is shown in \cref{fig:rebb630}. The $CL$ of the fit is 1.29\%. By fitting this high-$|t|$ dataset, one can extract the values of the ReBB model scale parameters with high uncertainty.  This shows that mainly the low-$|t|$ data constrain the values of the ReBB model scale parameters.

\cref{fig:rebb180} shows the result of the ReBB model fit of the Tevatron 1.8 TeV data in the kinematic range of $0.034~{\rm GeV}^2 \leq |t|\leq 0.627~{\rm GeV}^2$. The $CL$ of the fit is 98.07\%.

The Tevatron 1.96 TeV dataset is the highest energy $p\bar p$ elastic differential cross section dataset measured by the D0 Collaboration. The result of the ReBB model fit to this dataset in the squared four-momentum transfer range $0.26~{\rm GeV}^2 \leq |t|\leq 1.2~{\rm GeV}^2$ is shown in \cref{fig:rebb190}. The $CL$ of the fit is 37.91\%.

\section{Energy evolution in the $\rm p\bar{p}$ ReBB model}\label{sec:rebb_pbarp_en_evolution}

Knowing the values of the ReBB model parameters as determined from the $p\bar p$ differential cross section data at different energies, one can investigate the energy dependencies of these parameters. In Ref.~\cite{Nemes:2015iia}, the energy dependencies of the ReBB model parameters in the c.m. energy range of 23 GeV $\leq \sqrt{s}\leq 7$ TeV were determined by fitting a linear logarithmic functional form as given by \cref{eq:linlog} to the parameter values at different energies. It turns out that the same procedure works for elastic $p \bar p$ scattering. As detailed below, the energy dependencies of the parameters, $R_q$, $R_d$, $R_{qd}$, and $\alpha_R$, in the c.m. energy range \mbox{31 GeV $\leq \sqrt{s}\leq 1.96$ TeV,} are compatible with the linear-logarithmic functional form as given by \cref{eq:linlog}.

\begin{figure}[!hbt] 
\centering
\includegraphics[scale=0.65]{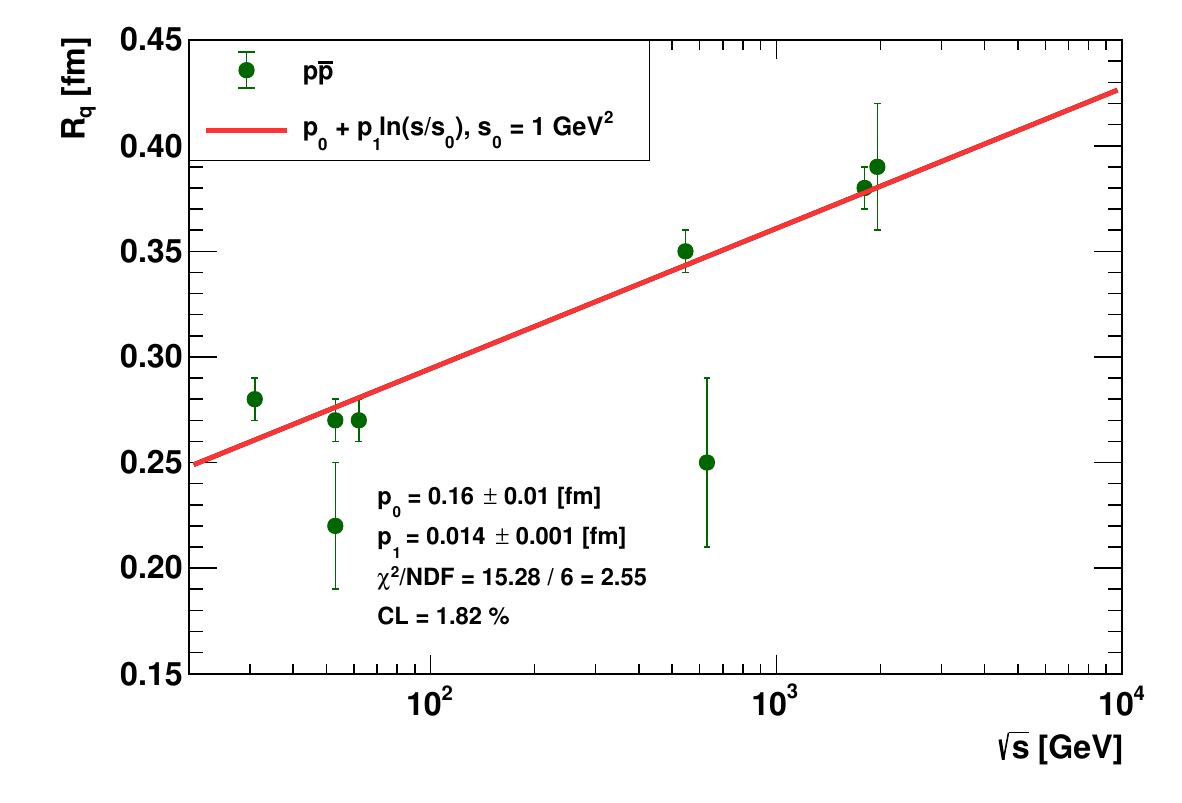}
\vspace{-0.35cm}
\caption{Energy dependence of the $R_{q}$ ReBB model parameter in the $p\bar p$ elastic scattering analysis. At $\sqrt{s}=53$ GeV, the bigger value is from the fit to the  low-$|t|$ dataset, while the smaller value is from the fit to the high-$|t|$ dataset.}
\label{fig:rabbpar_rq_pbarp}
\end{figure}

\begin{figure}[!hbt] 
\centering
\includegraphics[scale=0.65]{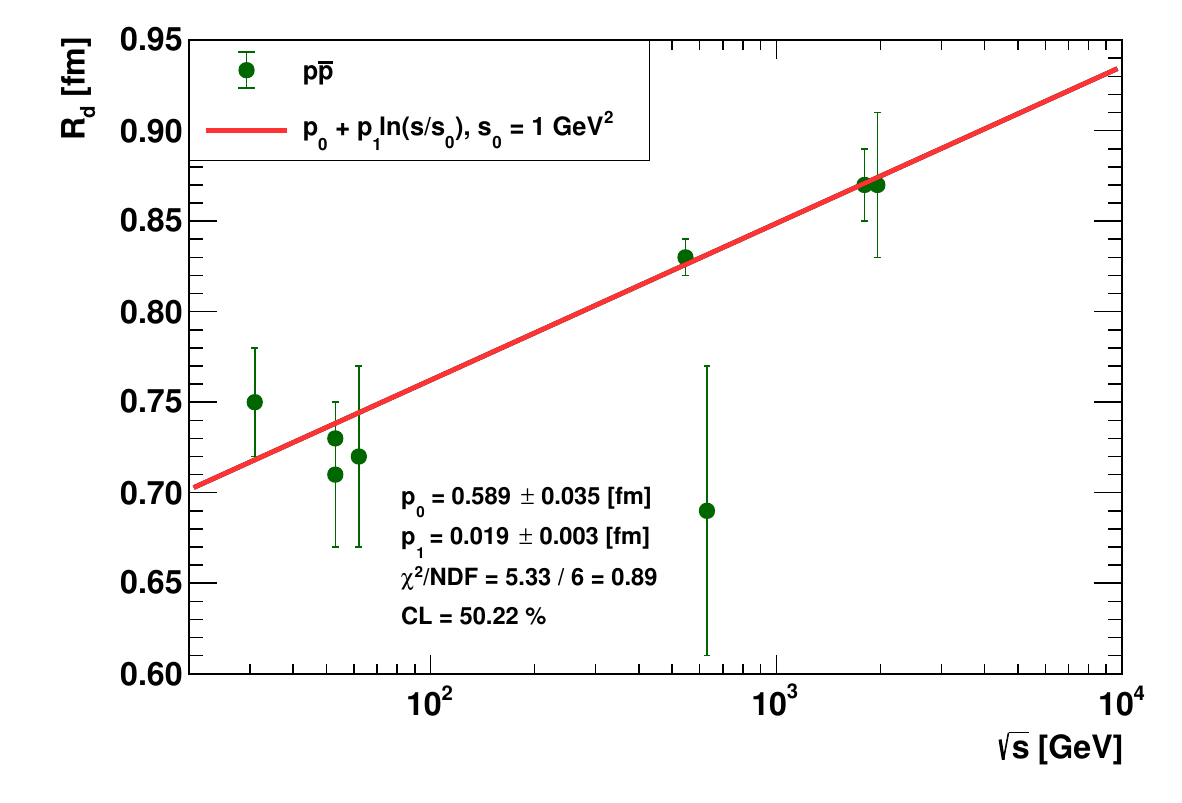}
\vspace{-0.35cm}
\caption{Energy dependence of the $R_{d}$ ReBB model parameter in the $p\bar p$ elastic scattering analysis. At $\sqrt{s}=53$ GeV, the bigger value is from the fit to the  low-$|t|$ dataset, while the smaller value is from the fit to the high-$|t|$ dataset.}
\label{fig:rabbpar_rd_pbarp}
\end{figure}

\begin{figure}[!hbt] 
\centering
\includegraphics[scale=0.65]{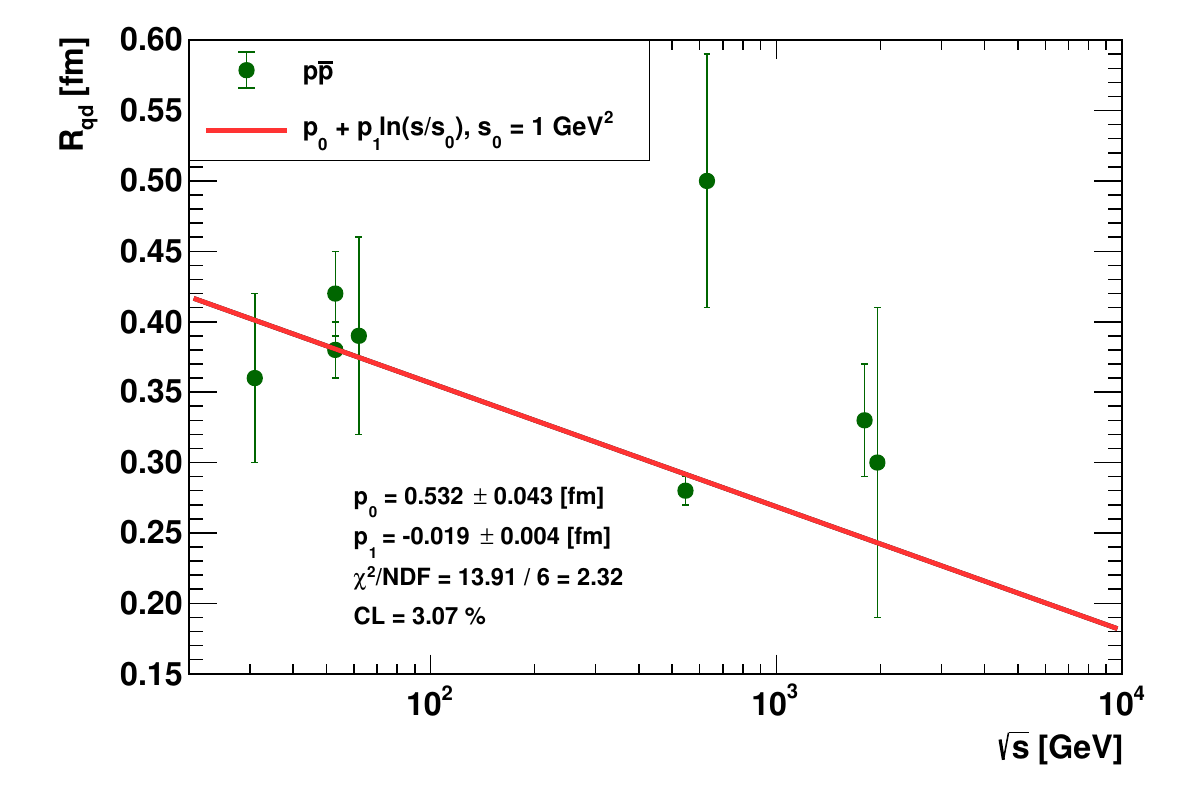}
\vspace{-0.35cm}
\caption{Energy dependence of the $R_{qd}$ ReBB model parameter in the $p\bar p$ elastic scattering analysis. At $\sqrt{s}=53$ GeV, the bigger value is from the fit to the  high-$|t|$ dataset, while the smaller value is from the fit to the low-$|t|$ dataset.}
\label{fig:rabbpar_rqd_pbarp}
\end{figure}

\begin{figure}[!hbt] 
\centering
\includegraphics[scale=0.65]{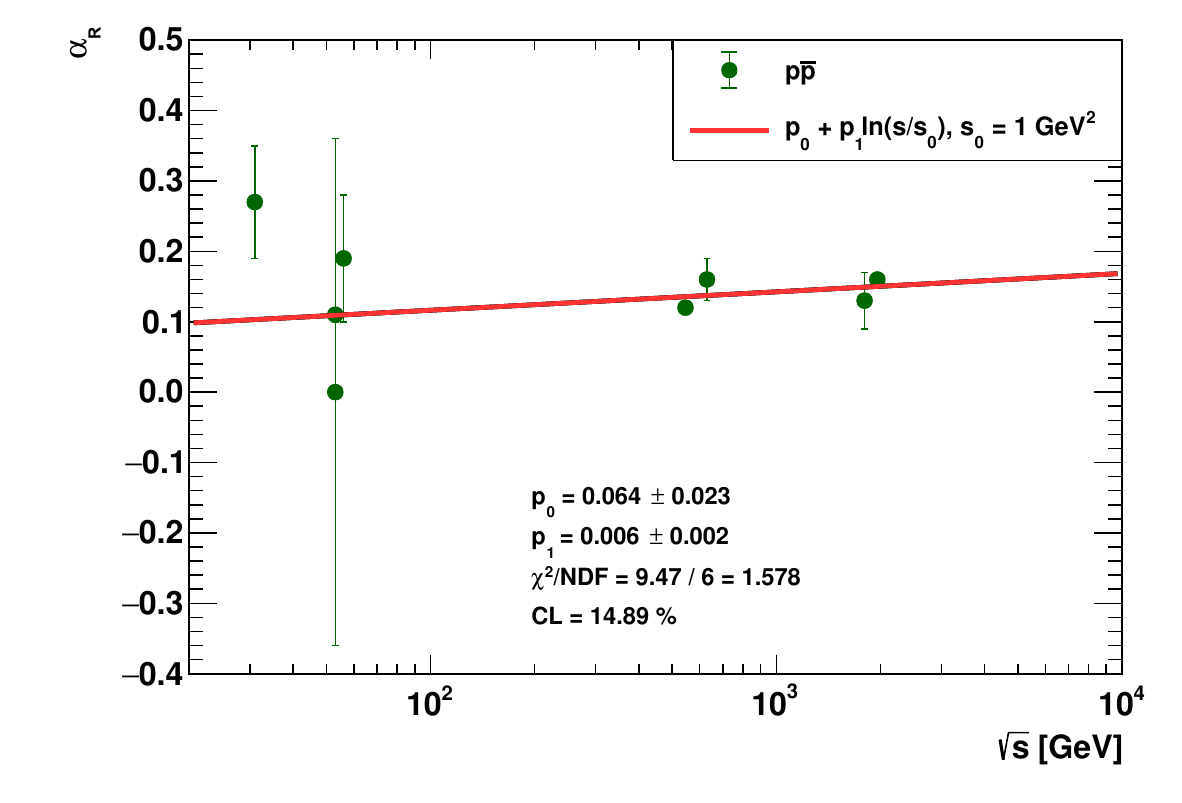}
\vspace{-0.35cm}
\caption{Energy dependence of the $\alpha_R$ ReBB model parameter in the $p\bar p$ elastic scattering analysis. At $\sqrt{s}=53$ GeV, the bigger value is from the fit to the  high-$|t|$ dataset, while the smaller value is from the fit to the low-$|t|$ dataset.}
\label{fig:rabbpar_alpha_pbarp}
\end{figure}

For $p\bar p$ scattering, the energy dependencies of the ReBB model parameters $R_q$, $R_d$, $R_{qd}$, and $\alpha_R$,  are shown on  \cref{fig:rabbpar_rq_pbarp}, \cref{fig:rabbpar_rd_pbarp}, \cref{fig:rabbpar_rqd_pbarp}, and \cref{fig:rabbpar_alpha_pbarp}, respectively. The parameters of the linear-logarithmic shape as given by \cref{eq:linlog} and the $CL$ values are summarised in \cref{tab:linlogparsCL}. In \cref{fig:rabbpar_rq_pbarp}, one can identify two obvious outlier points with high uncertainties. These points result from fits of higher-$|t|$ datasets at $\sqrt{s}=$ 53 GeV and 630~GeV. Similar behavior is seen in \cref{fig:rabbpar_rd_pbarp} and \cref{fig:rabbpar_rqd_pbarp} for the $R_d$ and $R_{qd}$ parameter values that are obtained from fits of higher-$|t|$ datasets at $\sqrt{s}=$ 53 GeV and 630~GeV. This is because the ReBB model scale parameters are mainly constrained by the lower-$|t|$ data, and the higher-$|t|$ data alone allows for a less precise determination of the values of the ReBB model scale parameters.

\begin{table}[!hbt]
        \centering
    \begin{tabular}{ccccc}
    \hline\hline\noalign{\smallskip}
Parameter      & $R_{q}$ [$\rm fm$]  & $R_{d}$ [$\rm fm$]  & $R_{qd}$ [$\rm fm$] & $\alpha_R$   \\ \hline
			$\chi^{2}/NDF$ 	& $15.28/6$& $5.33/6$ & $13.91/6$ & $9.47/6$	 \\	
			CL [\%]		& 1.82	& 50.22     & 3.07 & 14.89	 \\	\hline
			$p_{0}$ & $0.16\pm0.01$	  & $0.59\pm0.04$ & $0.53\pm0.04$ & $0.06\pm0.02$  \\ 
			$p_{1}$ & $0.014\pm0.001$ & $0.019\pm0.003$& $-0.019\pm0.004$ & $0.006\pm0.002$\\   \hline\hline
    \end{tabular}
    \caption{Parameter values that determine the linear-logarithmic energy dependencies of the ReBB model parameters according to \cref{eq:linlog} in the analysis of elastic $p\bar p$ scattering data utilizing the $\chi^2$ definition of \cref{eq:chi_trad0}. \label{tab:linlogparsCL}}
\end{table}


One may notice that an ever-growing $\alpha_R(s)$ as shown in \cref{fig:rabbpar_alpha_pbarp}, via \cref{eq:ReBB_b_ampl}, \cref{eq:relPWtoeik_2}, and \cref{eq:Ttot}, results in an ever-oscillating $\sigma_{\rm tot}$. However, this is actually not a problem since the ReBB model is found to be valid in a limited energy range below 13 TeV where the value of $\alpha_R$ is small, varying between 0.1 and 0.2 (see later \cref{chap:oddTD0} and \cref{chap:rebbdesc}). Also, the value of $\tilde\sigma_{\rm in}$ is small in the high-$b$ domain where $\sigma_{\rm tot}$ gets the main contribution dumping the oscillating effect following from the ever-growing $\alpha_R$. Though $p_1$ is not zero within errors, one can even fit the $\alpha_{R}$ parameter values also with a constant value of $\alpha_{R} = 0.132\pm0.006$ giving $CL=0.89\%$. 

Because of baryon-antibaryon symmetry, it is natural to expect that the values of the scale parameters of the ReBB model, $R_q$, $R_d$, and $R_{qd}$, at a given energy, have the same values in $pp$ and $p\bar p$ elastic scattering. Based on this expectation, the only ReBB model parameter that may be different in value in $pp$ and $p\bar p$ elastic scattering at a given energy is the opacity parameter, $\alpha_R$. As I will show below in \cref{chap:oddTD0} and later in a more refined analysis in \cref{chap:rebbdesc}, the experimental data on elastic  $pp$ and $p\bar p$ scattering are indeed compatible with such an expectation in the c.m. energy range of \mbox{0.546~TeV~$\leq \sqrt{s}\leq 7$~TeV.} Moreover, in \cref{chap:odderon}, I show that the ReBB model using the parameter values taken from the determined common $pp$ and $p\bar p$ energy dependence curves describe the experimental data with $CL>0.1\%$ in the kinematic range of \mbox{0.546~TeV~$\leq \sqrt{s}\leq 8$~TeV} and 0.38~GeV$^2\leq -t\leq 1.2$~GeV$^2$.

\vspace{0.5cm}
\textbf{Summary}
\vspace{0.2cm}

In this Chapter, I generalized the ReBB model of elastic $pp$ scattering to $p\bar p$: I fitted all the available elastic $p\bar p$ scattering data covering an energy range from ISR energies up to Tevatron energies. I found that the energy evolution of each of the ReBB model parameters in elastic $p\bar p$ scattering is compatible with linear logarithmic rise.



\chapter{
Model-dependent odderon effects at $\sqrt{s}=$ 1.96 TeV}\label{chap:oddTD0}

Participating in the joint project of the CERN LHC TOTEM and FNAL Tevatron D0 experimental collaborations on the odderon search, I studied the possibility of extrapolating the TOTEM measurements on $pp$ elastic differential cross section down to \mbox{$\sqrt{s}=1.96$~TeV} and I estimated the effects of the odderon exchange at $\sqrt{s}=1.96$ TeV. To do this, I used physical models. In \cref{sec:ReBB196}, I present the results of a ReBB model analysis. 
In \cref{sec:Regge196}, I present results obtained by utilizing a phenomenological model based on Regge theory.  In these analyses, I close the energy gap between the 1.96 TeV D0 elastic $p\bar p$ and the 2.76 TeV TOTEM elastic $pp$ scattering measurements. I find that the effect of the odderon exchange at $\sqrt{s}=1.96$ TeV manifests mainly in generating a prominent diffractive minimum in the $pp$  elastic differential cross section and filling in this minimum in the ${p\bar p}$ elastic differential cross section. My results presented in this Chapter are preliminary; however, these results served as a guide during the joint D0-TOTEM odderon analysis that finally led to an odderon signal observation with a statistical significance of at least 5.2$\sigma$. I present the final results of the ReBB model analysis of elastic $pp$ and $p\bar p$ scattering in \cref{chap:rebbdesc} and the final results of the ReBB model odderon analysis in \cref{chap:odderon}. 

This chapter is based on  Ref.~\cite{TOTEM:2020zzr} and twelve internal presentations I \mbox{gave \cite{talk1,talk2,talk3,talk4,talk5,talk6,talk7,talk8,talk9,talk10,talk11,talk12}} on my progress in this work in the TOTEM Experiment.



\newpage

\section{ReBB model results}\label{sec:ReBB196}

The joint TOTEM-D0 project aimed to close the energy gap between the 1.96 TeV D0 elastic $p\bar p$ scattering measurements and the 2.76 TeV TOTEM elastic $pp$ scattering measurements to analyze the effect of a possible odderon contribution. At the beginning of my work, I calculated ReBB model $pp$ elastic differential cross section curves based on the energy dependence trends obtained in Ref.~\cite{Nemes:2015iia} and compared them to new -- at that time preliminary -- $pp$ differential cross section data at \mbox{$\sqrt{s}$ = 2.76 TeV} and $13$ TeV. Since these interpolated and extrapolated curves described that data only at a qualitative level, the conclusion was that the energy dependence must be fine-tuned in order to use it to search for the signal of the odderon exchange. To achieve this goal, I fitted the ReBB model to all the $p\bar p$ elastic differential cross section data from ISR energies up to Tevatron energies (see \cref{chap:ReBBpbarp}) as well as to the new $pp$ differential cross section data measured at LHC. I used the obtained parameter values to fine-tune the energy dependence of the ReBB model. Each ReBB model scale parameter, as obtained at different energies by fitting $pp$ and $p\bar p$ scattering data, was compatible with a common energy dependence curve. However, the sanity tests failed: the curves calculated from the energy dependence trends did not represent all the experimental data in a statistically acceptable manner. To solve this problem, I reduced the analyzed energy range. I excluded the datasets measured at ISR energies and 
the $\sqrt s =$ 13 TeV TOTEM $pp$ differential cross section data since I found that the ReBB model cannot describe this very precise dataset in a statistically acceptable manner (see possible explanations in \cref{sec:rebb_desc_refind_TeV} of \cref{chap:rebbdesc}). 

In \cref{sec:fitdescription}, I discuss the details and results of the ReBB model fits to $pp$ and $p\bar p$ elastic differential cross section data.  In \cref{sec:rebb_TD_endep}, I study the energy dependencies of the ReBB model parameters. Finally, in \cref{sec:oddeffs_ReBB}, I present the results for the odderon effects within the ReBB model analysis of the data.   

\subsection{The fit procedure and its results}

\label{sec:fitdescription}

In the course of the minimization of the ReBB model, I considered statistical, systematic, and luminosity uncertainties  of the measured data using the following $\chi^{2}$ function:
\begin{equation}\label{eq:chi_Cj++}
\chi^2 =  \sum_{j=1}^{M} \sum_{i=1}^{n_j} \frac{(d_{ji} - N_j \, m_{ji}(\vec p\,))^2}{e_{ji}^2+s_{ji}^2} + 
\sum_{j=1}^M \frac{(N_j -1)^2}{\delta N_j^2} + \frac{(\sigma_{\rm tot} - \sqrt{N_1}\,m_{\sigma_{\rm tot}}(\vec p\,))^2}{\delta^2\sigma_{\rm tot}},
\end{equation}
where 
\begin{itemize}
    \item $M$ is the number of separately measured $d\sigma_{\rm el}/dt$ $t$ ranges at a given $\sqrt{s}$ energy; 
    \item $n_j$ is the number of $d\sigma_{\rm el}/dt$ datapoints in a separately measured subrange ($\sum_{j=1}^M n_j$ gives the number of fitted data points);
    \item $d_{ji}$ is the $i$th measured differential cross section data point in the $j$th subrange and $m_{ji}(\vec p\,)$ is the corresponding value calculated from the ReBB model, which depends on the fit parameters arranged in the vector $\vec{p}$;
    \item the values $e_{ji}$ and $s_{ji}$ are, correspondingly, the statistical and systematic uncertainty of the $i$th data point in the $j$th subrange;
    \item $\delta N_j$ is the luminosity uncertainty of the $j$th subrange and $N_j$ are additional normalization parameters to be minimized;
    \item $\sigma_{\rm tot}$ is the measured total cross section value, $\delta\sigma_{\rm tot}$ is its full error and $m_{\sigma_{\rm tot}}(\vec p\,)$ is its theoretical value calculated from the ReBB model (the proper normalization factor for the total cross section is $\sqrt{N_1}$, the square root of the normalization factor of the $|t|$-range with the lowest $|t|$ values).
\end{itemize}

The above $\chi^2$ definition is the same as \cref{eq:nemeschi} used in Ref.~\cite{Nemes:2015iia} but generalized for data with several separate $t$ subranges with fitting also the $\sigma_{\rm tot}$ value\footnote{In addition to the differential cross section data, data on \mbox{$t$-independent} quantities may be included in the fit to achieve a better agreement between the model results at the measured values of these \mbox{$t$-independent} quantities. It may be possible to determine the ReBB model parameters based on the $t$-independent quantities like $\sigma_{\rm tot}$, $\rho_0$, $B_0$, and compare the resulting differential cross section with experimental results; however, the $t$-distribution contains the greatest amount of information on the process, and it is more straightforward to fit the model to the differential cross section data too.}.  The $NDF$ is given by the number of fitted data points, including the measured value of $\sigma_{\rm tot}$, minus the number of fitted  ReBB model parameters. The effect of the free $N_i$ parameters of the $\chi^2$ definition of \cref{eq:chi_Cj++} cancels from the calculation of the $NDF$ value.

\begin{figure}[!hbt]
	\centering
	\includegraphics[width=0.8\linewidth]{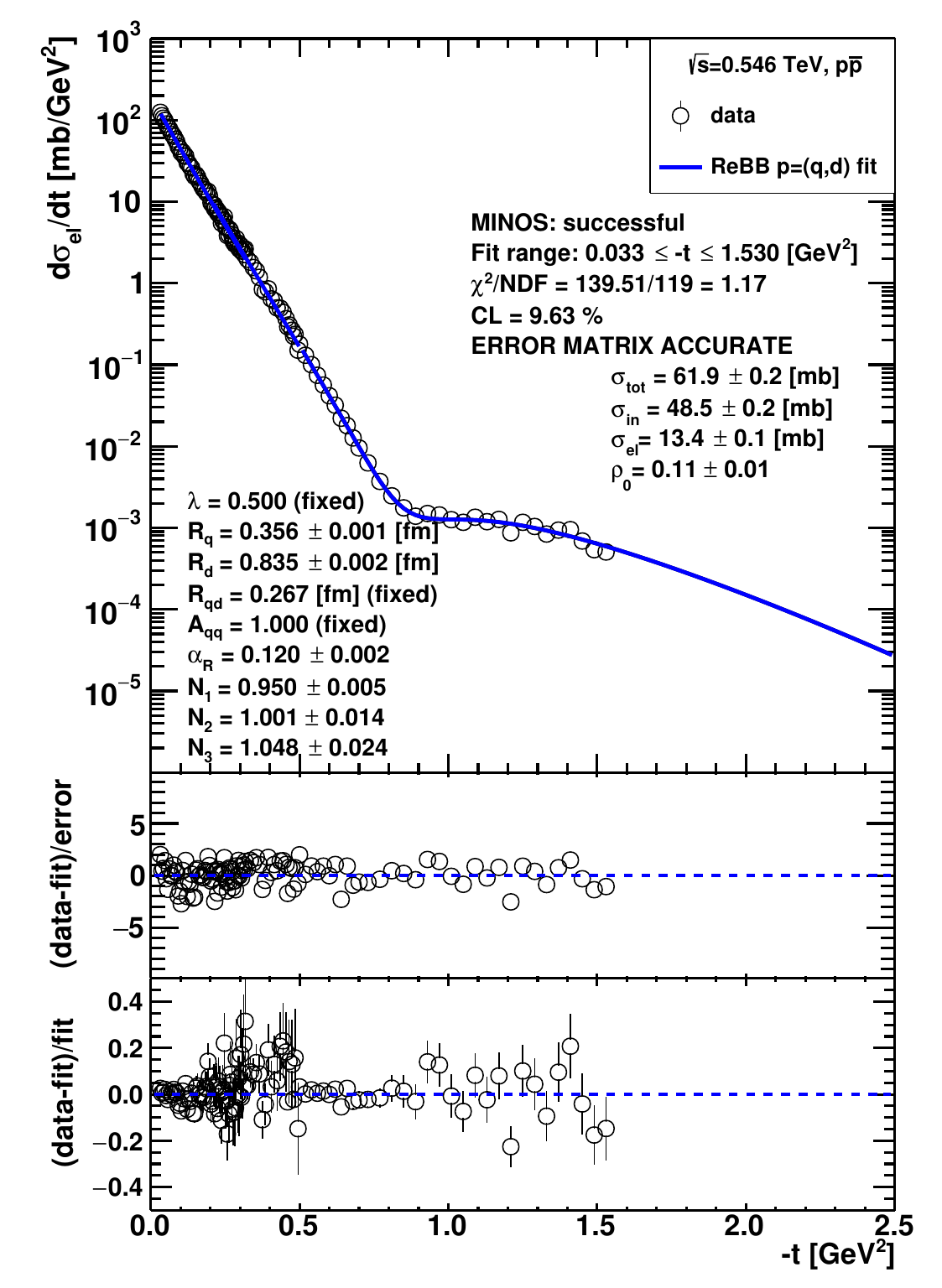}
	\caption{The fit of the ReBB model to the $p\bar p$ SPS UA4 $\sqrt{s}=0.546$~TeV data in the $|t|$ range of 0.033 GeV$^2$ $\leq|t|\leq1.53 $ GeV$^2$. The fit is performed using the $\chi^2$ definition of \cref{eq:chi_Cj++}. The ReBB model parameter values and the $N_i$ values are shown in the bottom left corner; the fixed parameters are indicated; the uncertainties of the fitted parameters are displayed. The curve is shifted by $N_i$ in each $|t|$ subrange of the data.}
	\label{fig:reBB_model_fit_0_546_TeV_fix}
\end{figure}

\begin{figure}[!hbt]
	\centering
	\includegraphics[width=0.8\linewidth]{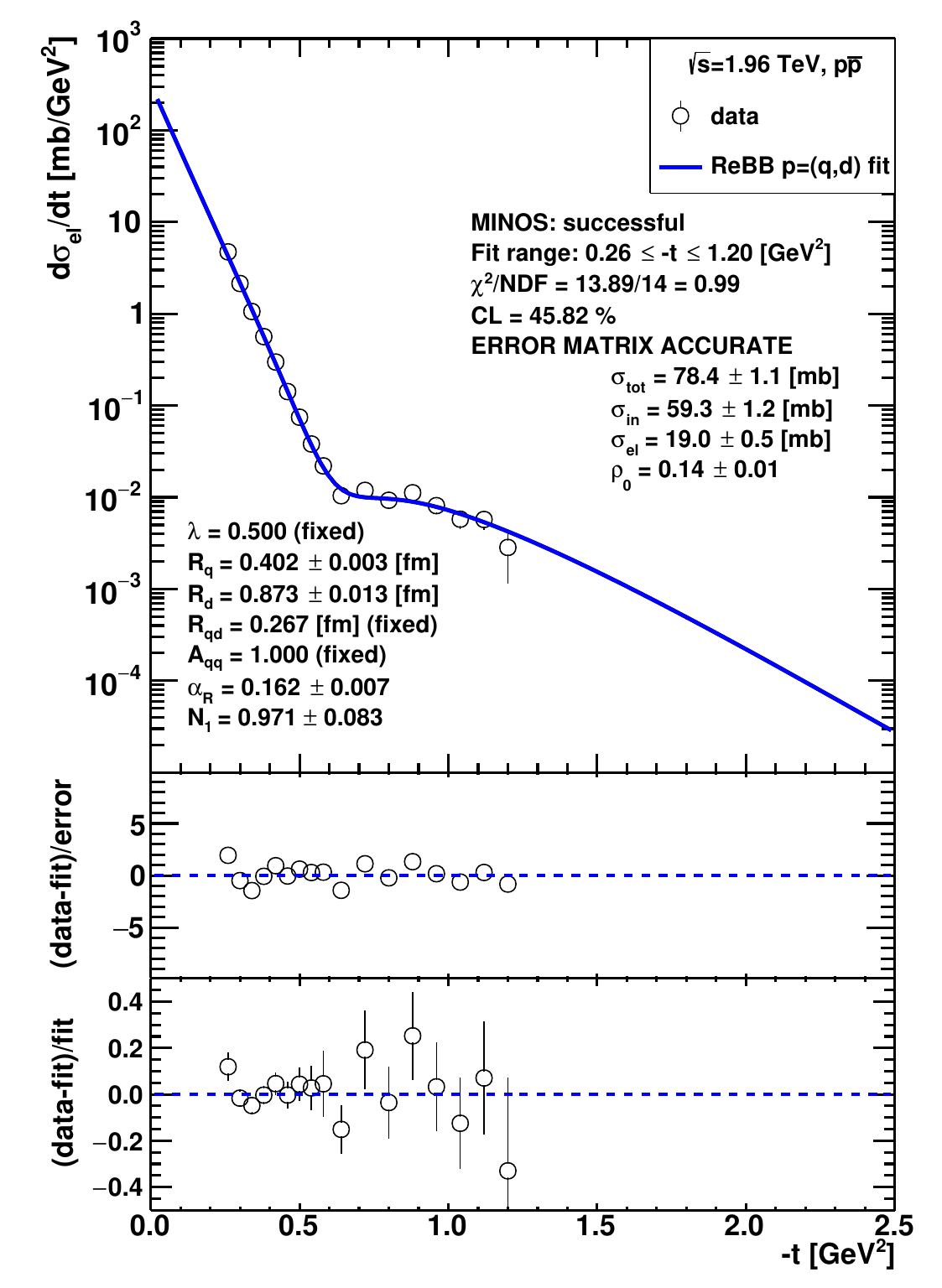}
	\caption{The fit of the ReBB model to the $p\bar p$ Tevatron D0 $\sqrt{s}=1.96$~TeV data in the $|t|$ range of 0.26 GeV$^2$ $\leq|t|\leq1.2 $ GeV$^2$. Otherwise, the same as \cref{fig:reBB_model_fit_0_546_TeV_fix}.}
	\label{fig:reBB_model_fit_1_96_TeV_fix}
\end{figure}

\begin{figure}[!hbt]
	\centering
	\includegraphics[width=0.8\linewidth]{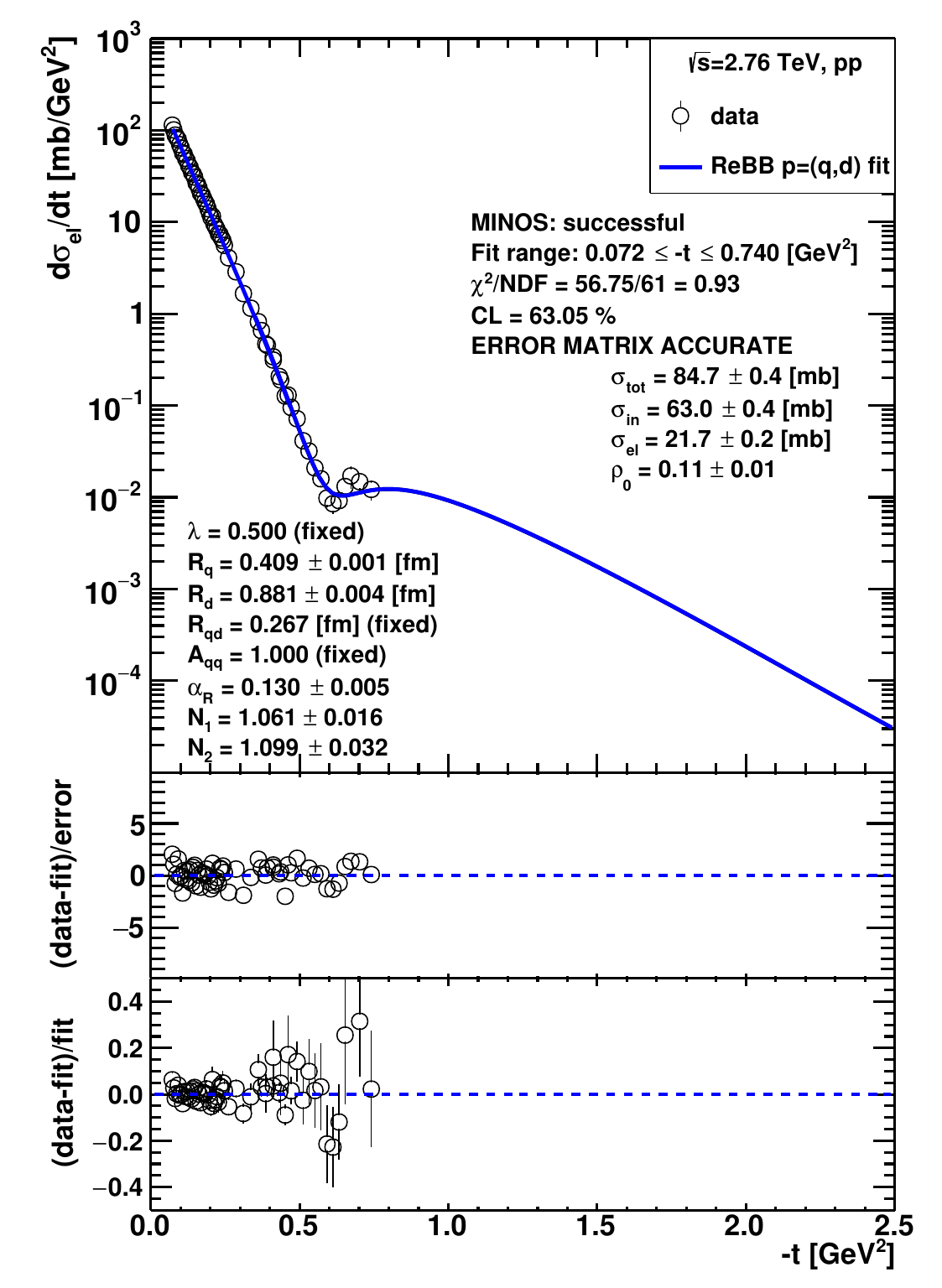}
	\caption{The fit of the ReBB model to the $pp$ LHC TOTEM $\sqrt{s}=2.76$~TeV data in the $|t|$ range of 0.072 GeV$^2$ $\leq|t|\leq0.74 $ GeV$^2$. Otherwise, the same as \cref{fig:reBB_model_fit_0_546_TeV_fix}.}
	\label{fig:reBB_model_fit_2_76_TeV_fix}
\end{figure}

\begin{figure}[!hbt]
	\centering
	\includegraphics[width=0.8\linewidth]{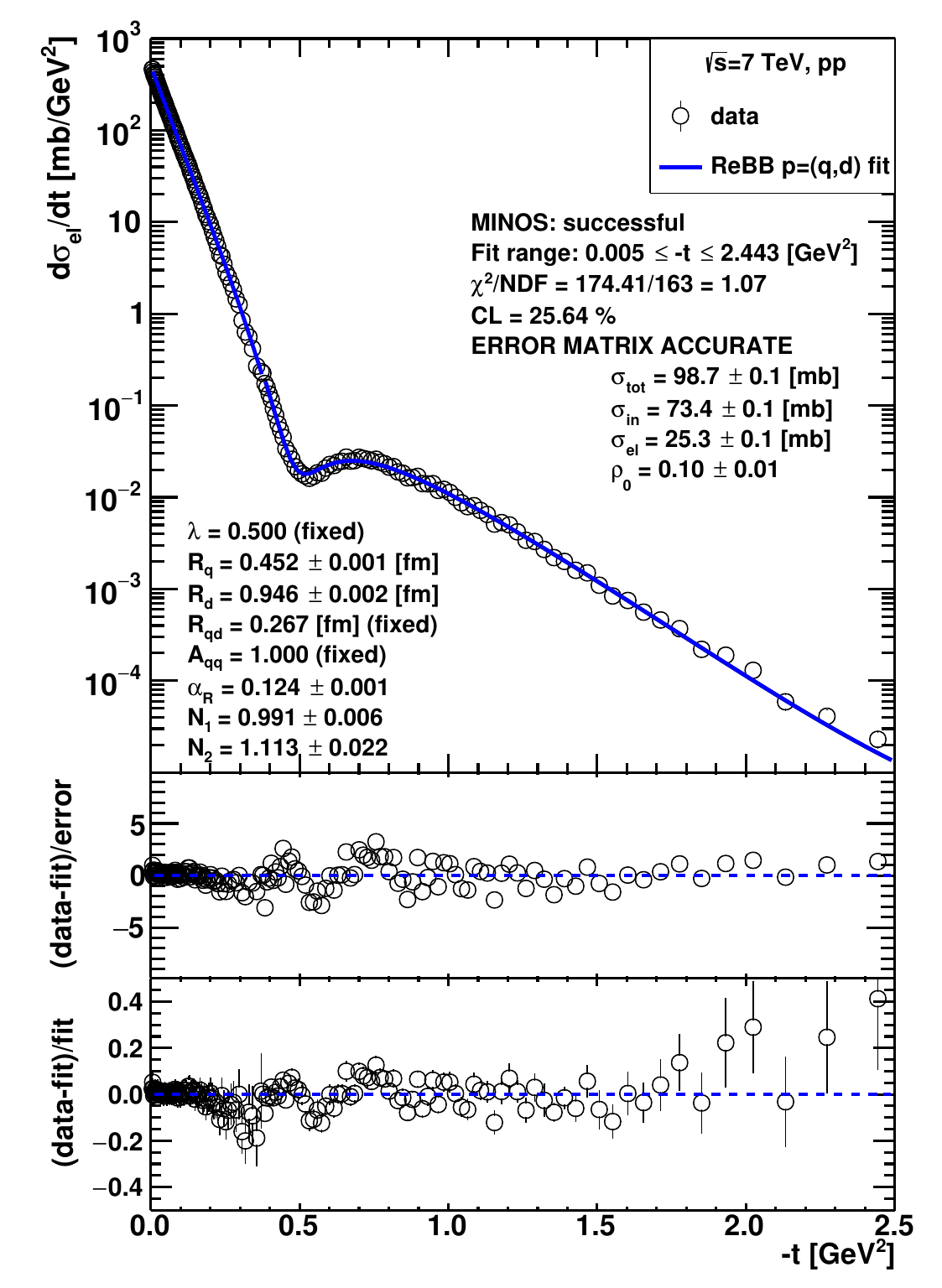}
	\caption{The fit of the ReBB model to the $pp$ LHC TOTEM $\sqrt{s}=7$~TeV data in the $|t|$ range of 0.005 GeV$^2$ $\leq|t|\leq2.45 $ GeV$^2$. Otherwise, the same as \cref{fig:reBB_model_fit_0_546_TeV_fix}.}
	\label{fig:reBB_model_fit_7_TeV_fix}
\end{figure}

In the first stage of the fitting procedure, in agreement with previous investigations in Ref. \cite{Nemes:2015iia}, the model parameters $A_{qq}=1$ and $\lambda=\frac{1}{2}$ were kept constant, which reduced the number of free parameters to four $R_{qd},\,R_{q},\,R_{d}$, and $\alpha_R$. In the first stage, I fitted two $pp$ differential cross section datasets at $\sqrt{s}$ = 2.76 TeV and 7 TeV and two $p\bar p$ differential cross section datasets at $\sqrt{s} =$ 0.546 TeV and 1.96 TeV. After performing these fits, I found that the value of the parameter $R_{qd}$ also can be kept fixed at 0.267 fm (see more details on this below). Thus, in the second stage of the fitting procedure, I kept fixed the value of $R_{qd}$. The results of these second stage fits are shown in \cref{fig:reBB_model_fit_0_546_TeV_fix}, \cref{fig:reBB_model_fit_1_96_TeV_fix}, \cref{fig:reBB_model_fit_2_76_TeV_fix} and \cref{fig:reBB_model_fit_7_TeV_fix}, while the values of the fitted parameters and their errors are summarized in \cref{table:fit_parameters}. The errors and the values are not rounded to two decimal places to maintain a higher precision in the analysis.

\begin{table}[!hbt]
        \centering
	{\begin{tabular}{ccccc} \hline\hline
		$\sqrt{s}$ [TeV] & 0.546 ($p\bar p$) & 1.96 ($p\bar p$)  & 2.76 ($pp$)	     & 7 ($pp$)  \\ \hline
		$|t|$ [GeV$^{2}$]&(0.033, 1.530)     &(0.260, 1.200)     & (0.072, 0.740)    &(0.005, 2.443)                             \\ 
		$\chi^{2}/NDF$	 &  139.51/119       &   13.89/14        & 56.75/61          & 174.41/163 \\ 
		CL [\%] 	     & 9.63              & 45.82	         & 63.05             & 25.64       \\ \hline
		$R_{q}$ [$fm$] 	 & 0.356 $\pm$ 0.001  & 0.402 $\pm$ 0.003 & 0.409 $\pm$ 0.001 & 0.452 $\pm$ 0.001   \\ 
		$R_{d}$ [$fm$] 	 & 0.835 $\pm$ 0.002 & 0.87 $\pm$ 0.01 & 0.881 $\pm$ 0.004 & 0.946 $\pm$ 0.002   \\ 
		$R_{qd}$ [$fm$]  & \multicolumn{4}{c}{0.267 (fixed)}  	\\
		$\alpha_R$ 	     & 0.120 $\pm$ 0.002 & 0.162 $\pm$ 0.007 & 0.130 $\pm$ 0.005 & 0.124 $\pm$ 0.001\\ \hline
		$N_1$ 	         & 0.950 $\pm$ 0.005 &0.97 $\pm$ 0.08  & 1.06 $\pm$ 0.02 & 0.991 $\pm$ 0.006\\ 
		$N_2$ 	         & 1.00 $\pm$ 0.01 & --                & 1.10 $\pm$ 0.03 & 1.11 $\pm$ 0.02   \\ 
		$N_3$ 	         & 1.05 $\pm$ 0.02 &--                 &    --   &--   \\ \hline\hline
	\end{tabular}}
 	\caption{The values of the fitted ReBB model parameters to $p\bar p$ and $pp$ data at different energies utilizing the $\chi^2$ definition of \cref{eq:chi_Cj++}. 
  \label{table:fit_parameters}}
\end{table}

I proceed with discussing each fit in a bit more detail. 

The SPS UA4 $p\bar p$~ differential cross section data at $\sqrt{s}$ = 540 GeV and 546 GeV are available in the squared four-momentum transfer range of \mbox{0.033 GeV$^2$ $\leq|t|\leq1.53 $ GeV$^2$} \cite{UA4:1983mlb,UA4:1984uui,UA4:1985oqn} which is subdivided into three subranges with different normalization uncertainties: \mbox{0.033 GeV$^2$ $\leq|t|\leq0.318 $ GeV$^2$} with $\delta N_1$ = 0.05, \mbox{0.215 GeV$^2$ $\leq|t|\leq0.495 $ GeV$^2$} with $\delta N_2$ = 0.03, and 0.46 GeV$^2$ $\leq|t|\leq1.53$ GeV$^2$ with $\delta N_3$ = 0.1 (the sub-ranges partly overlap). During the fit the \mbox{$t$-dependent} statistical errors, the normalization errors and the experimental value of the total cross section with its total uncertainty \mbox{($\sigma_{\rm tot}=61.26\pm0.93$ mb \cite{CDF:1993wpv})} were used according to \cref{eq:chi_Cj++}. For these data, the \mbox{$t$-dependent} systematic errors are not published. The obtained fit quality is satisfactory, $CL$ = 9.63\% (see \cref{fig:reBB_model_fit_0_546_TeV_fix}). The experimentally measured values of the total, elastic \mbox{($\sigma_{\rm el}=12.87\pm0.3$ mb \cite{ParticleDataGroup:2018ovx})} and inelastic \mbox{($\sigma_{\rm in}=48.39\pm1.01$~mb~\cite{ParticleDataGroup:2018ovx})} cross sections and the $\rho_0$ parameter \mbox{($\rho_0=0.135\pm0.015$ \cite{UA42:1993sta})} are reproduced by the ReBB model with a good accuracy. 

The $\sqrt{s}=630$ GeV and 1.8 TeV $p\bar p$ elastic scattering data are not used in the fitting stage of the analysis since the available $t$ range of these data is too narrow to determine the model parameters reliably: the analysis of the higher-$|t|$ 630 GeV data gives less precise values for the scale parameters and an off-trend value for the opacity parameter, while the analysis of the lower-$|t|$ 1.8~TeV data not covering the region of the shoulder structure do not allow to extract the value of $\alpha_R$ (the fitter gives $\alpha_R=0$ with a huge error bar). However, the $\sqrt{s}=630$ GeV and 1.8 TeV $p\bar p$ elastic scattering data is later used to cross-check the reliability of the determined energy evolution of the ReBB model.

The Tevatron $p\bar p$ differential cross section data at $\sqrt{s}=1.96$ TeV are available in the squared four-momentum transfer range of 0.26 GeV$^2$ $\leq|t|\leq1.20 $ GeV$^2$ \cite{D0:2012erd} with a normalization uncertainty $\delta N_1$ = 0.144. During the fit, the $t$-dependent statistical and systematical errors and the normalization error were used according to \cref{eq:chi_Cj++}. At this energy, the experimental value for the total cross section is not available. The obtained fit quality is satisfactory, $CL$ = 45.82\% (see \cref{fig:reBB_model_fit_1_96_TeV_fix}). The experimental values are not available for the forward observables at 1.96 TeV, however, according to the prediction of the COMPETE Collaboration \cite{Cudell:2002xe}, $\sigma_{\rm tot}=78.27\pm1.93$ mb and $\rho_0=0.145\pm0.006$. These values are reproduced by the ReBB model with good accuracy. 

The TOTEM $pp$ $\sqrt{s}=$ 2.76 TeV differential cross section dataset was measured in the squared four-momentum transfer range of 0.072 GeV$^2$ $\leq|t|\leq0.74 $ GeV$^2$ \cite{TOTEM:2018psk} which is subdivided into two subranges: 0.072 GeV$^2$ $\leq|t|\leq0.462 $ GeV$^2$ and \mbox{$0.372\leq|t|\leq0.74 $ GeV$^2$} each with a normalization uncertainty of $\delta N$ = 0.06 (the two sub-ranges partly overlap). During the fit in the first subrange, the $t$-dependent statistical and systematic errors, the normalization error, and the experimental value of the total cross section with its total uncertainty \mbox{($\sigma_{\rm tot}=84.7\pm3.3$ mb \cite{TOTEM:2017asr})} were used according to \cref{eq:chi_Cj++}. In the second higher-$|t|$ subrange, the $t$-dependent systematic errors were not considered to allow for the higher-$|t|$ subrange having a bigger weight during the fit\footnote{A refined treatment of systematic errors based on the $\chi^2$ formula given by \cref{eq:phenix_o} is applied in \cref{chap:rebbdesc}.} \mbox{(at $\sqrt{s}$ = 1.96 TeV,} the $t$-dependent systematic errors cannot be neglected as the statistical errors were not published separately \cite{D0:2012erd}). The obtained fit quality is satisfactory, $CL$ = 63.05\% (see \cref{fig:reBB_model_fit_2_76_TeV_fix}). The experimental values of the forward observables ($\sigma_{\rm in}=62.8\pm2.9$ mb, \mbox{$\sigma_{\rm el}=21.8\pm1.4$ mb \cite{TOTEM:2017asr,Nemes:2017gut}}) are reproduced by the ReBB model with a good accuracy. The experimental data is not available for parameter $\rho_0$, however, the calculated value is within the total error band of COMPETE \mbox{prediction \cite{Cudell:2002xe}}.

The TOTEM $pp$ $\sqrt{s}=$ 7 TeV differential cross section dataset was measured in the range of 0.005 GeV$^2$ $\leq|t|\leq2.443 $ GeV$^2$ \cite{TOTEM:2013lle} which can be subdivided into two subranges: 0.005 GeV$^2$ $\leq|t|\leq0.371$ GeV$^2$ and 0.377 GeV$^2$ $\leq|t|\leq2.443 $ GeV$^2$ each with a normalization uncertainty of $\delta N$ = 0.042. During the fit, similarly to the case of the $\sqrt{s}=$ 2.76 TeV data, in the first subrange the $t$-dependent statistical and systematic errors, the normalization errors and the experimental value of the total cross section with its total uncertainty ($\sigma_{\rm tot}=98.0\pm2.5$ mb \cite{TOTEM:2013vij}) were used according to \cref{eq:chi_Cj++}, while in the second, higher-$|t|$ subrange the $t$-dependent systematic errors were not considered. The obtained fit quality is satisfactory, $CL$ = 25.64\% (see \cref{fig:reBB_model_fit_7_TeV_fix}). The total, elastic \mbox{($\sigma_{\rm el}=25.1\pm1.1$ mb \cite{TOTEM:2013vij})} and inelastic ($\sigma_{\rm in}=72.9\pm1.5$ mb \cite{TOTEM:2013vij}) cross sections and the $\rho_0$ parameter \mbox{($\rho_0=0.145\pm0.091$ \cite{TOTEM:2013vij})} are reproduced by the ReBB model with a good accuracy. 

In conclusion, all of the above fits are physically and statistically acceptable. 

\subsection{Energy dependence of the parameters}\label{sec:rebb_TD_endep}

\begin{figure}[!hbt]
	\centering
		\includegraphics[width=0.8\linewidth]{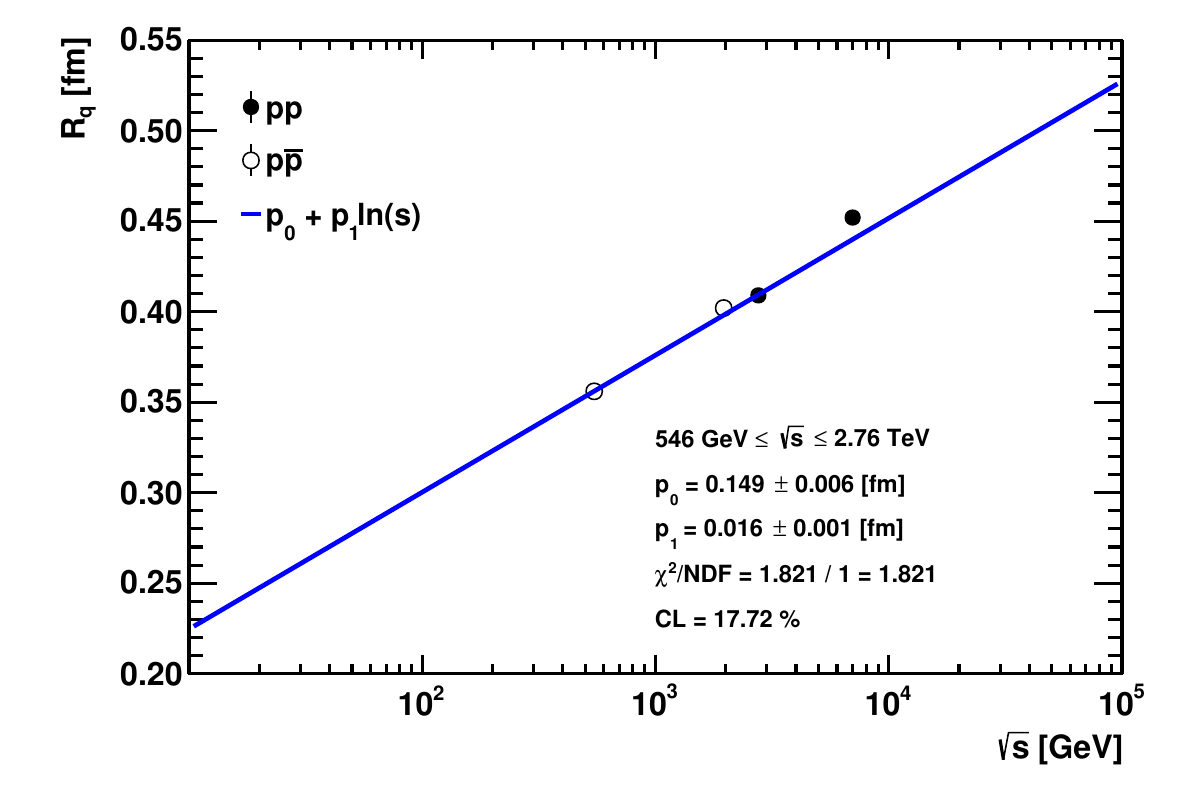}
\caption{Energy dependence of the $R_q$ ReBB model parameter in the TeV scale in the analysis of elastic $pp$ and $p\bar p$ scattering data utilizing the $\chi^2$ definition of \cref{eq:chi_Cj++}.}
	\label{fig:Rqpar196}
\end{figure}

\begin{figure}[!hbt]
	\centering
		\includegraphics[width=0.8\linewidth]{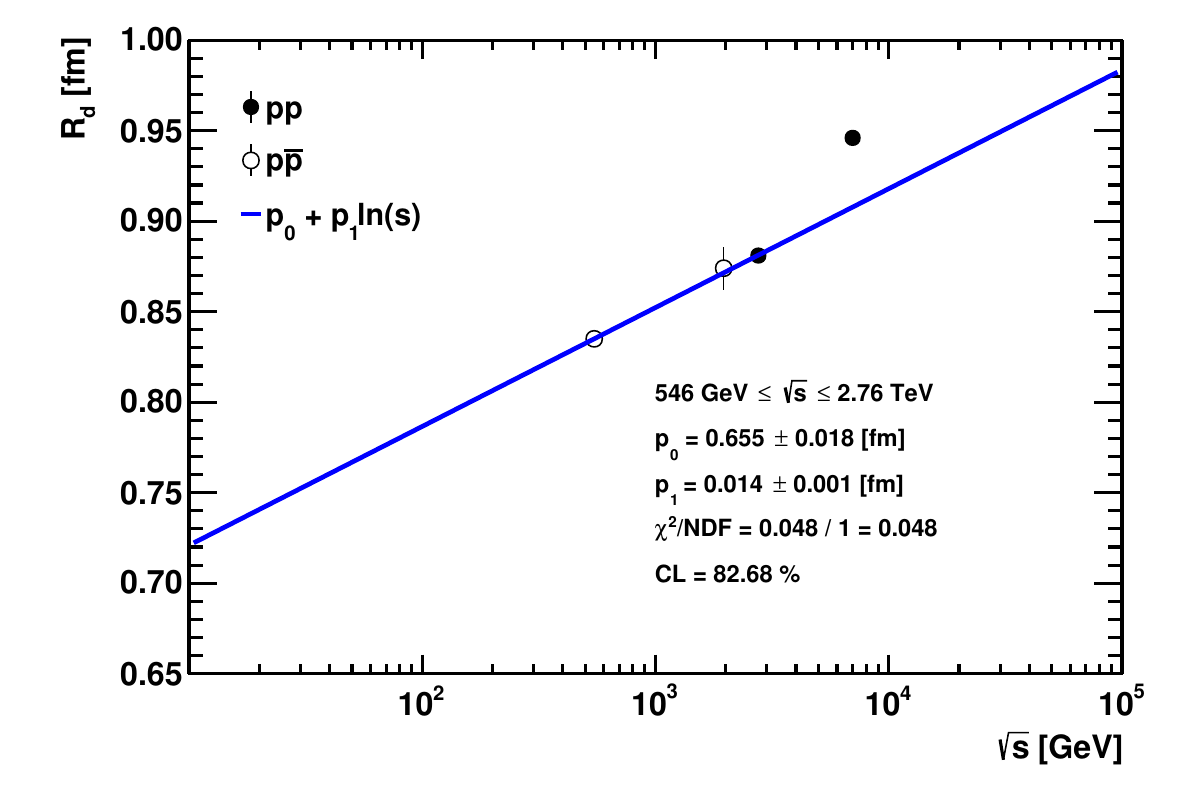}
\caption{Energy dependence of the $R_d$ ReBB model parameter in the TeV scale in the analysis of elastic $pp$ and $p\bar p$ scattering data utilizing the $\chi^2$ definition of \cref{eq:chi_Cj++}.}
	\label{fig:Rdpar196}
\end{figure}

\begin{figure}[!hbt]
	\centering
		\includegraphics[width=0.8\linewidth]{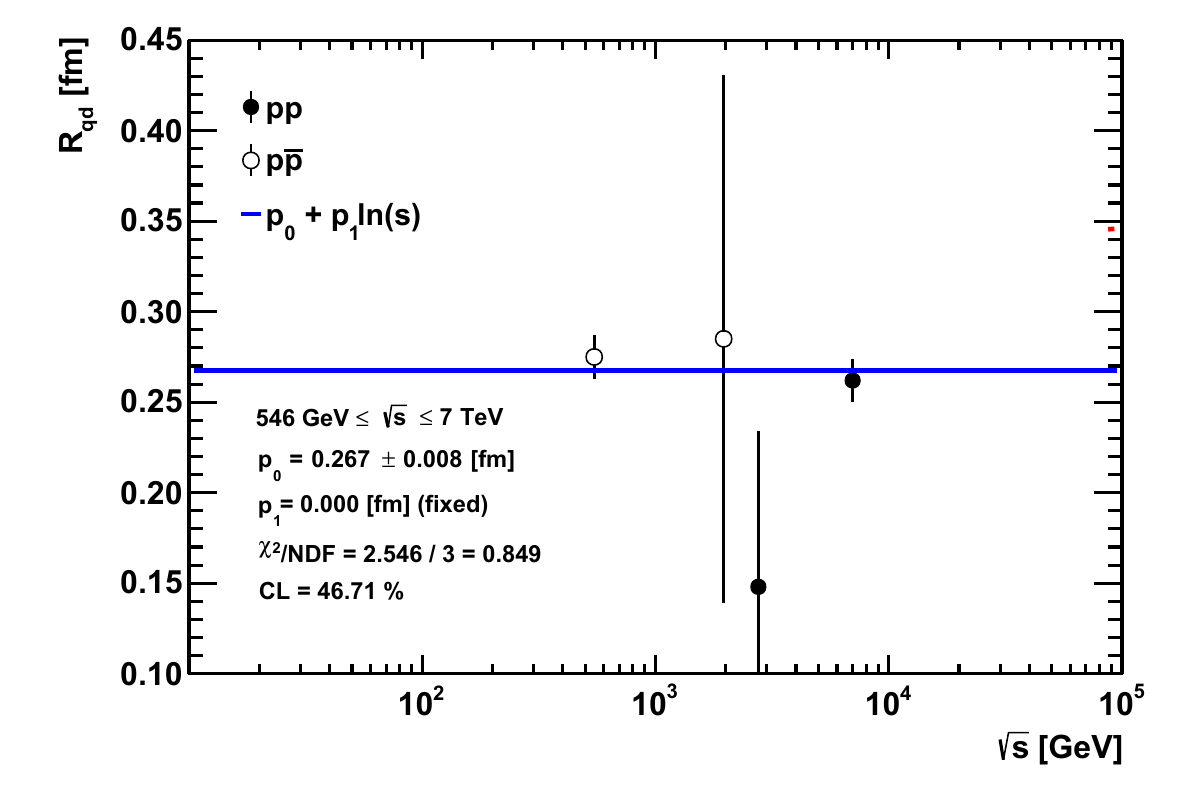}
\caption{Energy dependence of the $R_{qd}$ ReBB model parameter in the TeV scale in the analysis of elastic $pp$ and $p\bar p$ scattering data utilizing the $\chi^2$ definition of \cref{eq:chi_Cj++}.}
	\label{fig:Rqdpar196}
\end{figure}

\begin{figure}[!hbt] 
	\centering
		\includegraphics[width=0.8\linewidth]{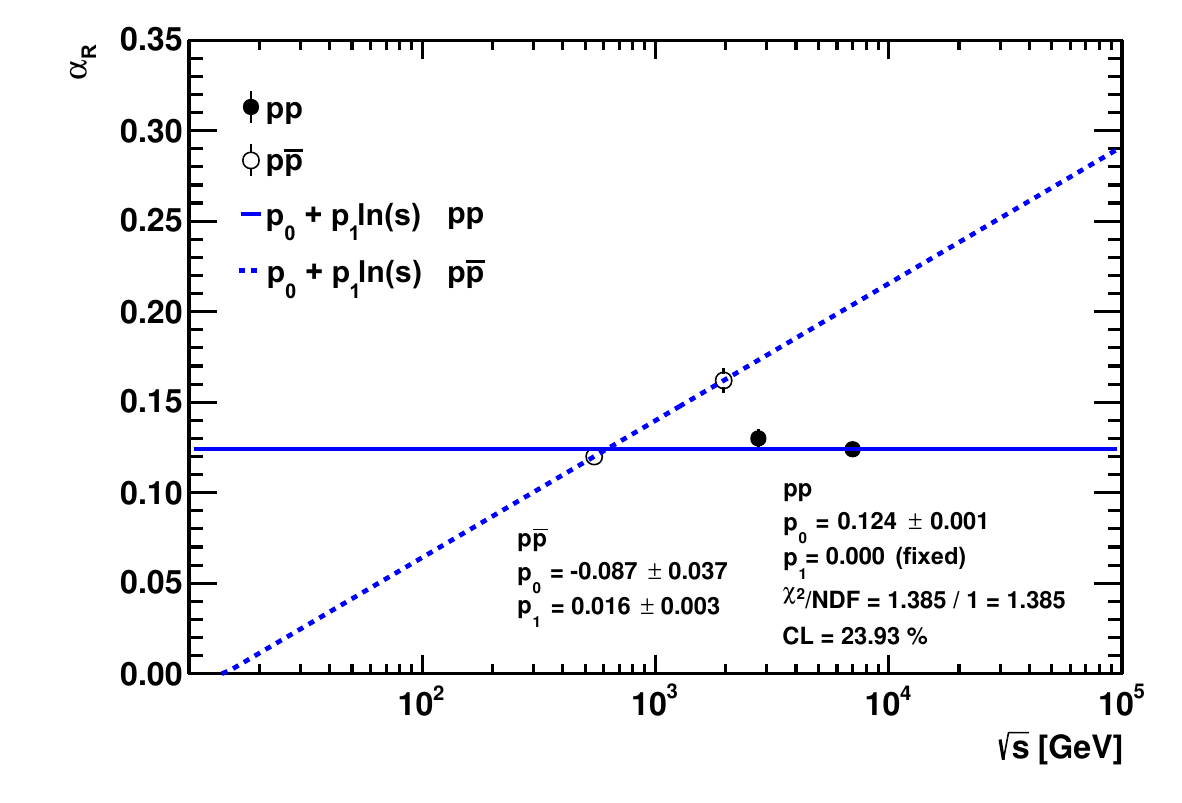}
\caption{Energy dependence of the $\alpha_R$ ReBB model parameter in the TeV scale in the analysis of elastic $pp$ and $p\bar p$ scattering data utilizing the $\chi^2$ definition of \cref{eq:chi_Cj++} (the parameter errors are estimated by the MINOS algorithm of Minuit as in all of the other cases).}
	\label{fig:alphapar196}
\end{figure}

In order to determine the energy dependencies of the model parameters, the fit results of \cref{table:fit_parameters} and the parametrization, used originally in Ref.~\cite{Nemes:2012cp}, and given by \cref{eq:linlog}, are applied for each parameter, $R_{q}$, $R_{d}$, $R_{qd}$, and $\alpha_R$. The results are shown in \cref{fig:Rqpar196}, \cref{fig:Rdpar196}, \cref{fig:Rqdpar196} and \cref{fig:alphapar196}.
 The fit parameters that determine the energy dependence of each parameter via the parametrization \cref{eq:linlog} are collected in \cref{tab:excitation_pars}. The compatibility of the energy trends with linear-logarithmic functional forms or with a constant value is quantified by the $CL$ values given in the corresponding figures and in \cref{tab:excitation_pars}.

\begin{table}[!hbt]
     \begin{subtable}[h]{0.99\textwidth}
        \centering
    \begin{tabular}{cccc}
    \hline\hline\noalign{\smallskip}
Parameter      & $R_{q}$ [$\rm fm$]  & $R_{d}$ [$\rm fm$]  & $R_{qd}$ [$\rm fm$]    \\ \hline
			$\chi^{2}/NDF$ 	& $1.821/1$& $0.048/1$ & $2.546/3$ 	 \\	
			CL [\%]		& 17.72	& 82.68     & 46.71 	 \\	\hline
			$p_{0}$ & $0.149\pm0.006$	  & $0.66\pm0.02$ & $0.267\pm0.008$  \\ 
			$p_{1}$ & $0.016\pm0.001$ & $0.014\pm0.001$& $0.000~(fixed)$ 	\\   \hline\hline
    \end{tabular}
           \caption{}
       \label{tab:sub1b}
    \end{subtable}
    \vspace{0.5cm}
    \vfill
    \begin{subtable}[h]{0.99\textwidth}
        \centering
    \begin{tabular}{ccc}
    \hline\hline\noalign{\smallskip}
Parameter       & $\alpha_R$ ($p\bar p$)      &$\alpha_R$ ($p p$)  \\ \hline
			$\chi^{2}/NDF$ 	&  --&$1.385/1$ \\	
			CL [\%]		&  --&23.93 \\	\hline
			$p_{0}$ &  $-0.09\pm0.04$&$0.124\pm0.001$ \\ 
			$p_{1}$ &  $0.016\pm0.003$&$0.000~({\rm fixed})$ 	\\   \hline\hline
    \end{tabular}
           \caption{}
       \label{tab:sub2b}
    \end{subtable}
    \caption{Parameter values that determine the linear-logarithmic energy dependencies of the ReBB model parameters according to \cref{eq:linlog} in the analysis of elastic $pp$ and $p\bar p$ scattering data utilizing the $\chi^2$ definition of \cref{eq:chi_Cj++}. 
    \label{tab:excitation_pars}}
\end{table}

Chronologically, the energy dependence of the parameter $R_{qd}$ was determined first using the first stage fits, where the parameter $R_{qd}$ was free. Surprisingly, as we can see the result shown in \cref{fig:Rqdpar196}, this parameter is compatible with a constant \textit{i.e.} does not depend on the energy with a confidence level of $CL$ = 46.71\%.

In the case of parameters $R_q$, $R_d$, and $\alpha_R$, the energy dependence is obtained from the second stage fits, where the parameter $R_{qd}$ was fixed. 

The energy dependencies of the parameters $R_q$ and $R_d$ are shown in \cref{fig:Rqpar196} and \cref{fig:Rdpar196}, respectively. The curves show the trends obtained from fitting only the \mbox{546 GeV,} \mbox{1.96 TeV,} and 2.76 TeV points since the 7 TeV points do not lie on a linear trend with the lower energy points. The slope parameter $B_0$ is related to the ReBB model scale parameters and the TOTEM Collaboration found in Ref.~\cite{TOTEM:2017asr} that there is a jump in the energy dependence of $B_0$ in the energy interval of \mbox{3~TeV $\lesssim \sqrt s \lesssim$ 4 TeV}. This explains why the \mbox{$\sqrt{s}=$ 7 TeV} $R_q$ and $R_d$ points do not lie on a linear trend with the lower energy points. In the c.m. energy range of 546~GeV $\leq \sqrt s \leq$ 2.76 TeV, the $R_q$ and $R_d$ parameter values at different energies are compatible with linearly-logarithmic energy dependence trends with $CL$ = 17.72\% for the $R_q$ parameter and with $CL$ = 82.86\% for the $R_d$ parameter.

The energy dependence of the parameter $\alpha_R$ is shown in \cref{fig:alphapar196}. The value of this parameter is highly correlated with the depth of the diffraction minimum. Since the diffractive minimum apparently fills in in $p\bar p$ elastic scattering, I fitted separately the parameter $\alpha_R$ obtained for $pp$ and $p\bar p$ scattering. For $p\bar p$ scattering, I obtained a trend growing with energy, while for $pp$, I got a constant, energy-independent trend with a confidence level of \mbox{CL = 23.93\%.}

I tested the determined energy dependencies of the ReBB model parameters by comparing the resulting $pp$ and $p\bar p$ differential cross section curves to the existing measurements in the c.m. energy range of  546~GeV $< \sqrt s <$ 2.76 TeV. In the case of the $p\bar p$ data at $\sqrt s$ = 630 GeV, the first two and the last two points of the $|t|$-acceptance must be dropped to have a description with $CL$ $\geq$ 0.1\%. All the other $p\bar p$ ($\sqrt s$ = 546 TeV, \mbox{1.8 TeV,} 1.96 TeV) and $pp$ ($\sqrt s$ =2.76 TeV) differential cross section datasets are reproduced with \mbox{$CL$ $\geq$ 0.1\%} in the full $|t|$ range of the measurement.

\vspace{-2mm}
\subsection{Odderon effects at 1.96 TeV}\label{sec:oddeffs_ReBB}

\begin{figure}[!t]
	\centering
	\includegraphics[width=0.8\linewidth]{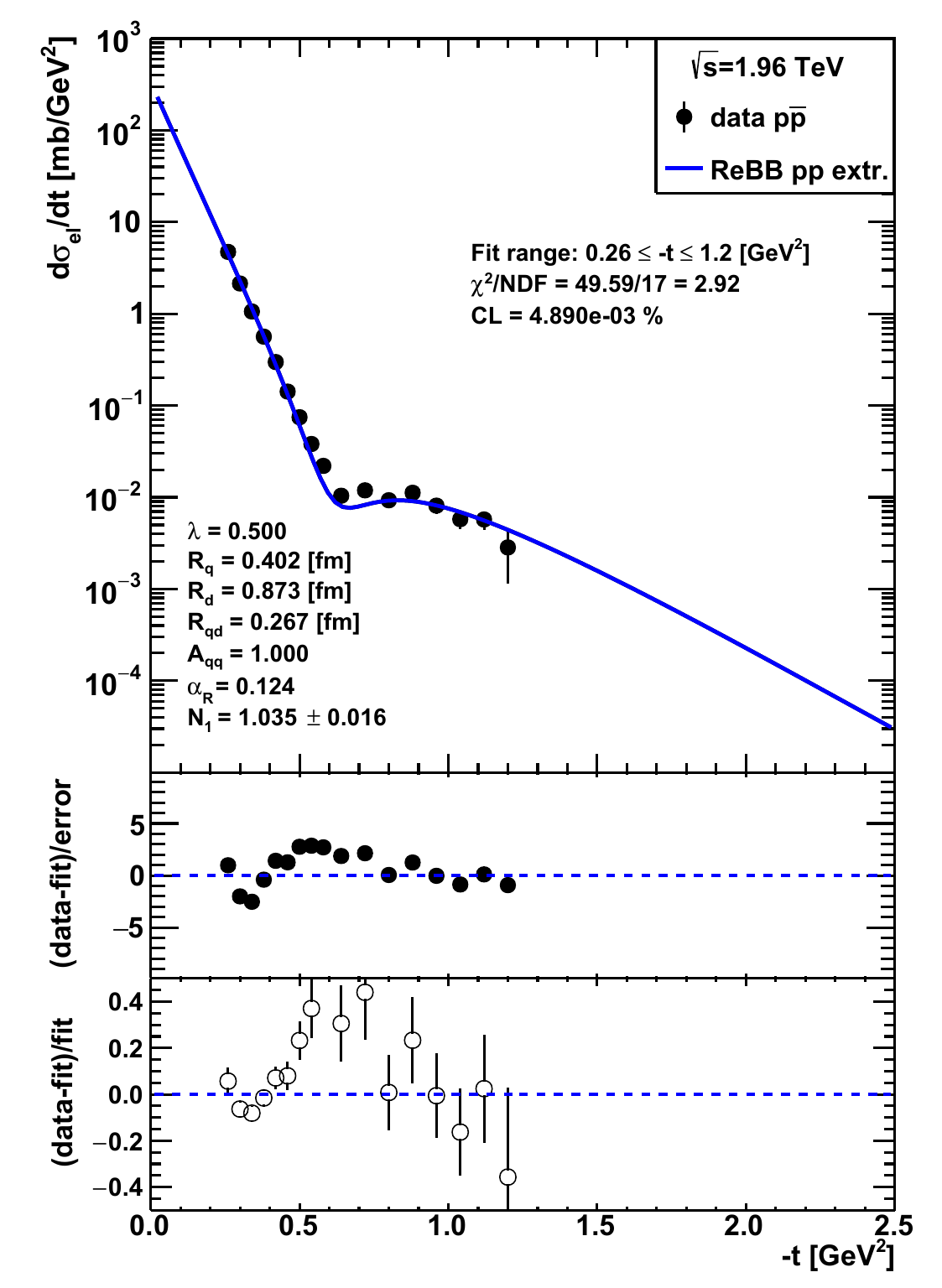}
	\caption{The ReBB model $pp$ elastic differential cross section at $\sqrt{s}=1.96$~TeV compared to the $p\bar p$ elastic differential cross section data measured at the same energy.}
	\label{fig:reBB_model_extr_1_96_TeVb}
\end{figure}

The above results show that the energy dependence of each ReBB model scale parameter, $R_{q}$, $R_{d}$, and $R_{qd}$, is consistent with the same energy evolution in $pp$ and $p\bar p$ elastic scattering, only the energy dependence of the opacity parameter $\alpha_R$ differs in $pp$ and $p\bar p$ elastic scattering. Based on these results, I assumed that the values of the ReBB model scale parameters are the same in $pp$ and $p\bar p$ scattering, and I calculated the ReBB $pp$ differential cross section by taking the values of the scale parameters $R_{q}$, $R_{d}$, and $R_{qd}$ from the fit of the $\sqrt{s}$ = 1.96 TeV $p\bar p$ data shown in \cref{fig:reBB_model_fit_1_96_TeV_fix} while the value of the $\alpha_R$ parameter from the trend seen in \cref{fig:alphapar196}. The result is shown in \cref{fig:reBB_model_extr_1_96_TeVb}. 
Looking at the middle and bottom panels of \cref{fig:reBB_model_extr_1_96_TeVb}, one can observe a substantial difference between $pp$ and $p\bar p$ differential cross sections in the region of the diffractive minimum. The minimum is filled in in case of $p\bar p$ scattering. The quantitative proof for the difference is the value of the $CL$. 
I compared the calculated $pp$ ReBB model curve with the $\sqrt{s}$ = 1.96 TeV $p\bar p$ data by utilizing the $\chi^2$ definition of \cref{eq:chi_Cj++} where, in this case, the third term, which concerns the total cross section, was neglected. The $CL$ of the compatibility of the $pp$ ReBB model curve with the $p\bar p$ data is 4.619$\times$10$^{-3}$\%. This difference can be attributed only to the effect of a $C$-odd exchange. Since in the TeV c.m. energy region, the secondary Reggeon exchanges are negligibly small, the only candidate can be the odderon exchange (see \cref{sec:oddintro} and later also \cref{chap:odderon} for a detailed discussion of the odderon within the framework of the ReBB model). The $CL$ value of 4.619$\times$10$^{-3}$\% corresponds to an odderon signal observation with a statistical significance of 4.06$\sigma$.

\begin{figure}[!hbt]
	\centering
	\includegraphics[width=0.8\linewidth]{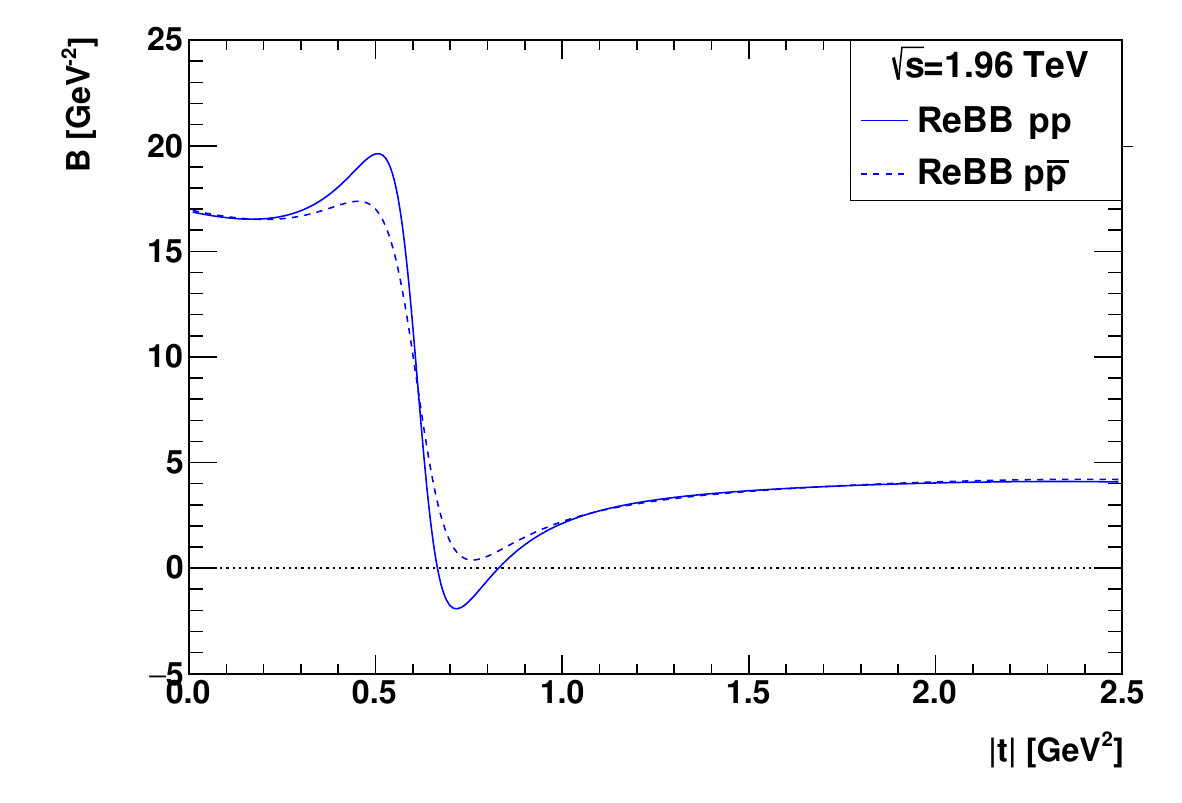}
 \vspace{-0.5cm}
	\caption{The $pp$ and $p \bar p$  $t$-dependent elastic slope as calculated from the ReBB model at $\sqrt{s}=1.96$~TeV.}
	\label{fig:reBB_model_slope_1_96_TeV}
\end{figure}

\cref{fig:reBB_model_slope_1_96_TeV} shows the $pp$ and $p\bar p$ $t$-dependent elastic slope as calculated from the ReBB model. One can see that the $t$-dependent slope of the differential cross section of elastic $pp$ scattering crosses zero while that of elastic $p\bar p$ scattering does not cross zero. This effect of the odderon was first discussed in Ref.~\cite{Csorgo:2018uyp}. However, in Ref.~\cite{Csorgo:2018uyp}, the authors did not calculate the $pp$ elastic differential cross section and its $t$-dependent slope at exactly the same energies in the TeV c.m. energy range. My calculation at $\sqrt{s}=$ 1.96 TeV explicitly shows that this is an odderon effect in the $t$-dependent elastic slope and is not the effect of the energy difference.

\section{Regge model results}\label{sec:Regge196}

In this section, I apply a $t$-channel phenomenological model to close the energy gap between the $p\bar p$ \mbox{$\sqrt s$ = 1.96 TeV} and $pp$ $\sqrt s$ = 2.76 TeV elastic differential cross section measurements. Here, the inputs are the pomeron and odderon exchange amplitudes from which the $pp$ and $p\bar p$ scattering amplitudes are constructed. In contrast, the ReBB model is an $s$-channel model where the inputs are the $pp$ and $p\bar p$ amplitudes from which the pomeron and odderon amplitude can be recovered. In \cref{sec:reggemodel}, I introduce the model. Then in \cref{sec:reggeOddeffects}, I show the results of the fits to the $p\bar p$ $\sqrt s$ = 1.96 TeV and $pp$ $\sqrt s$ = 2.76 TeV elastic differential cross section data and discuss the odderon effects.

\subsection{A dipole Regge exchange model}\label{sec:reggemodel}

The model I apply in this section is a dipole (DP) Regge exchange model. The dipole pomeron amplitude was developed in the 1970s by László Jenkovszky and his coauthors \cite{Jenkovszky:1974yv,Jenkovszky:1976sf}. The first version of a dipole odderon amplitude was introduced in the \mbox{1980s \cite{Jenkovszky:1987gv}.} The model that included dipole pomeron and odderon amplitude components was used to analyze $pp$ and $p\bar p$ elastic scattering data below LHC energies in a number of \mbox{papers \cite{Jenkovszky:1987gv,Desgrolard:1992ri,Jenkovszky:2011hu}.}
Later this dipole model was used also to analyze LHC measurements in a preprint \cite{Lengyel:2012sw}, and by myself and my coauthors in published papers \cite{Broniowski:2018xbg, Szanyi:2018pew, Szanyi:2019kkn}. This rather simple model with dipole pomeron and odderon components is sometimes referred to as the Jenkovszky model.


In the dipole Regge exchange model, the high-energy behavior of the elastic scattering amplitude, 
\begin{equation}\label{eq:amplA}
    A_{\rm el}(s,t)=\frac{\mathcal{M}_{\rm el}(s,t)}{4\pi},
\end{equation}
is determined by an isolated $\ell$-plane second-order pole (dipole) of the partial wave amplitude \cite{Jenkovszky:1986vh},  
\begin{equation}\label{eq:DPpartial}
    f_\ell(t)\equiv f(\ell,t)=\frac{d}{d\alpha(t)}\left[\frac{\beta(\ell)}{\ell-\alpha(t)}\right]=\frac{\beta(\ell)}{\left[\ell-\alpha(t)\right]^2},
\end{equation}
where the residue $\beta(\ell)$ is $t$-independent and non-singular at $\ell=\alpha(t)$; $\alpha(t)$ is the Regge trajectory (see \cref{sec:Regge}). After performing a Sommerfeld-Watson transformation, the resulting pomeron scattering amplitude is \cite{Jenkovszky:1976sf,Jenkovszky:1986vh}
\begin{eqnarray}
A(s,t)={d\over{d\alpha}}\Bigl[{\rm
e}^{-i\pi\alpha/2}G(\alpha)\Bigl(s/s_0\Bigr)^{\alpha}\Bigr]\\ \nonumber
={\rm e}^{-i\pi\alpha/2}\Bigl(s/s_0\Bigr)^{\alpha(t)}\Bigl[G'(\alpha)+\Bigl(L-i\pi
/2\Bigr)G(\alpha)\Bigr],
\end{eqnarray}
where $L\equiv\ln\frac{s}{s_0}$ and $\alpha\equiv\alpha(t)$. The parameterization of  $G'(\alpha)$  is not fixed, and generally chosen as 
\begin{equation}
    G'(\alpha)=-ae^{b[\alpha-1]}
\end{equation}
motivated by the shape of the hadronic differential cross sections. Then $G(\alpha)$ is recovered by integration up to an integration constant:
\begin{equation}
    G(\alpha)=-a\left(e^{b[\alpha-1]}/b -\gamma\right).
\end{equation}
Introducing the parameter $\epsilon=b\gamma$, the pomeron component of the scattering amplitude reads \cite{Jenkovszky:1986vh,Wall:1988pa}
\begin{eqnarray}\label{eq:GP}
A_P(s,t)=i{a_P\ s\over{b_P\ s_{0P}}}\left[r_{1P}^2(s){\rm e}^{r_{1P}^2(s)[\alpha_P-1]}-\varepsilon_P r_{2P}^2(s){\rm e}^{r_{2P}^2(s)[\alpha_P-1]}\right],
\end{eqnarray} 
where 
\begin{equation}
r_{1P}^2(s)=b_P+L_P-i\pi/2,    
\end{equation}
\begin{equation}
r_{2P}^2(s)=L_P-i\pi/2,  
\end{equation}
and
\begin{equation}
    L_P\equiv L_P(s)=\ln{\left(\frac{s}{s_{0P}}\right)}.
\end{equation}



The shape of the odderon amplitude is chosen to be the same as that of the pomeron amplitude through the following relation \cite{Lengyel:2012sw, Broniowski:2018xbg, Szanyi:2018pew, Szanyi:2019kkn}: 
\begin{equation}\label{eq:GO}
    A_O(s,t)=\frac{1}{i}A_{P\rightarrow O}(s,t).
\end{equation}
The parameters now have ``$O$'' index instead of a ``$P$'' index. Thus explicitly we have
\begin{eqnarray}\label{eq:GOexpl}
A_O(s,t)={a_O\ s\over{b_O\ s_{0O}}}\left[r_{1O}^2(s){\rm e}^{r_{1O}^2(s)[\alpha_O-1]}-\varepsilon_O r_{2O}^2(s){\rm e}^{r_{2O}^2(s)[\alpha_O-1]}\right],
\end{eqnarray} 
where
\begin{equation}
r_{1O}^2(s)=b_O+L_O-i\pi/2,    
\end{equation}
\begin{equation}
r_{2O}^2(s)=L_O-i\pi/2,    
\end{equation}
\begin{equation}
    L_O\equiv L_O(s)=\ln{\left(\frac{s}{s_{0O}}\right)}.
\end{equation}

The pomeron and odderon trajectories are chosen to have linear shapes \cite{Broniowski:2018xbg,Szanyi:2018pew}:
\begin{equation}\label{Ptray}
\alpha_P\equiv \alpha_P(t)= 1+\delta_P+\alpha'_{P}t,
\end{equation}
\begin{equation}\label{Eq:Otray}
\alpha_O\equiv \alpha_O(t)=1+\delta_O+\alpha'_{O}t.
\end{equation}

The $pp$ and $p\bar p$ elastic scattering amplitudes read: 
\begin{equation}\label{eq:amplApp}
A^{pp}_{\rm el}(s,t)  =  A_P(s,t) - A_O(s,t).
\end{equation}
\begin{equation}\label{eq:amplApbarp}
A^{p\bar{p}}_{\rm el}(s,t)  =  A_P(s,t) + A_O(s,t).
\end{equation}
Because of the relation given by \cref{eq:amplA}, the elastic differential cross section and the total cross section, respectively, read:
\begin{equation}
\frac{{\rm d}\sigma_{\rm el}}{{\rm d} t}(s,t)=\frac{\pi}{s^2}\left|A_{\rm el}\left(s,t\right)\right|^2\, ,
\label{eq:differential_cross_sectionA}
\end{equation}
and
\begin{equation}
\sigma_{tot}(s)=\frac{4\pi}{s}\, {\rm Im} \, A_{\rm el}(s,t=0),
\label{eq:total_cross_section}
\end{equation}
where $A_{\rm el}\left(s,t\right)\equiv A^{pp}_{\rm el}(s,t)$ or $A_{\rm el}\left(s,t\right)\equiv A^{p\bar p}_{\rm el}(s,t)$ depending on whether we want to calculate the $pp$ or the $p\bar p$ observables\footnote{To get the differential cross section in units of mb/GeV$^2$ and the total cross section in units of mb, the conversion \mbox{$\hbar^2c^2$/GeV$^2$~=~0.3894~mb} is applied. Since in natural units $\hbar = c = 1$, \mbox{1 GeV$^{-2} = 0.3894$ mb.}}.

This model contains 12 free parameters. These are:  $\delta_P$, $\alpha_P'$ [GeV$^{-2}$], $a_P$, $b_P$, $\epsilon_P$, \mbox{$s_{0P}$ [GeV$^{2}$]}, $\delta_O$, $\alpha_O'$ [GeV$^{-2}$], $a_O$, $b_O$, $\epsilon_O$, and $s_{0O}$ [GeV$^{2}$].

\subsection{Fits to the data and odderon effects}\label{sec:reggeOddeffects}

\begin{figure}[!hbt]
	\centering
\includegraphics[width=0.8\linewidth]{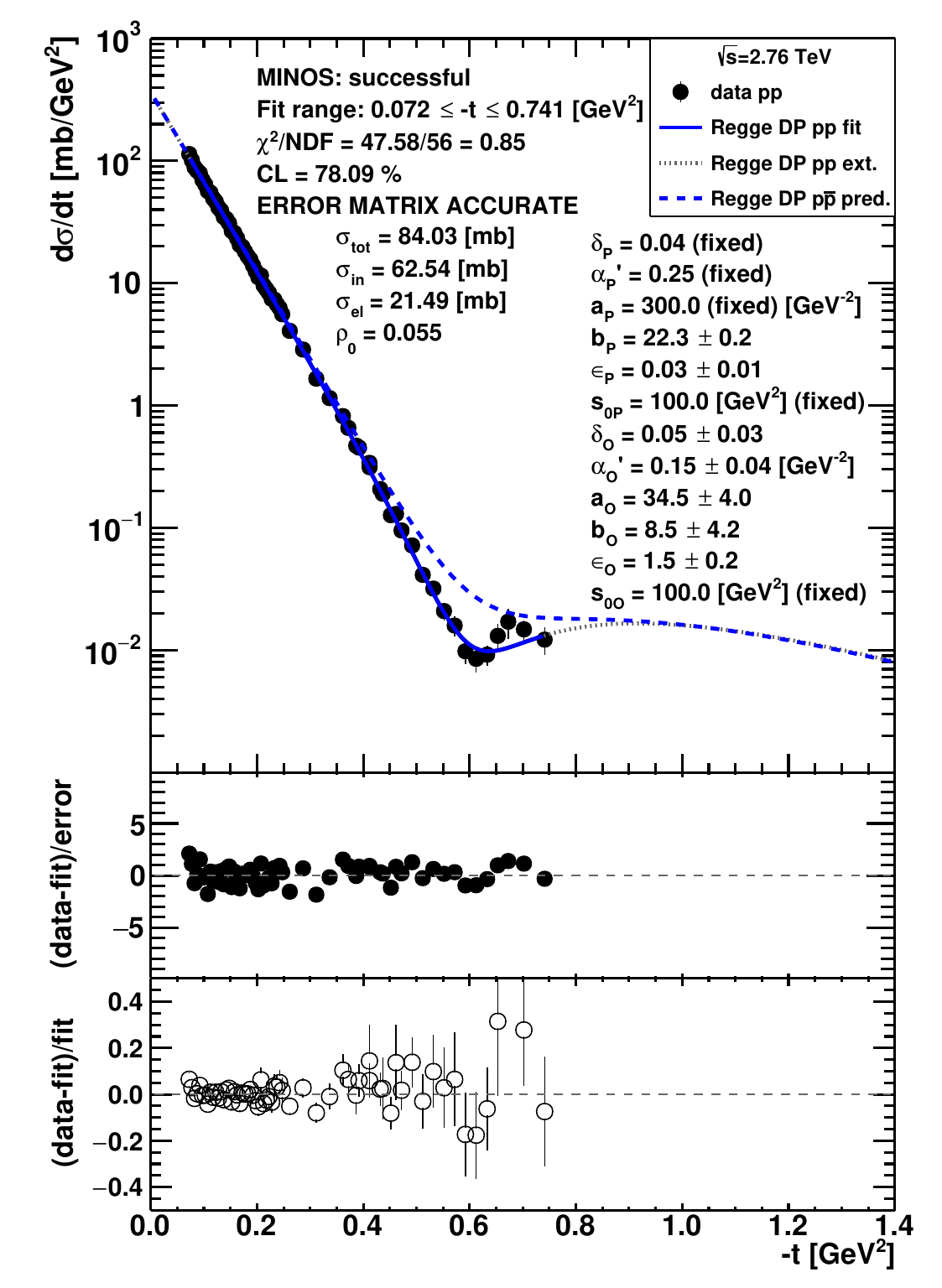}
\vspace{-2mm}
	\caption{Fit of the dipole Regge exchange model to the LHC TOTEM $\sqrt{s}$ = 2.76 TeV $pp$ differential cross section data \cite{TOTEM:2018psk} and prediction to the $p\bar p$ differential cross section at the same energy. }
	\label{fig:ReggeDP_fit_2_76_TeV}
\end{figure}

\begin{figure}[!hbt]
	\centering
\includegraphics[width=0.8\linewidth]{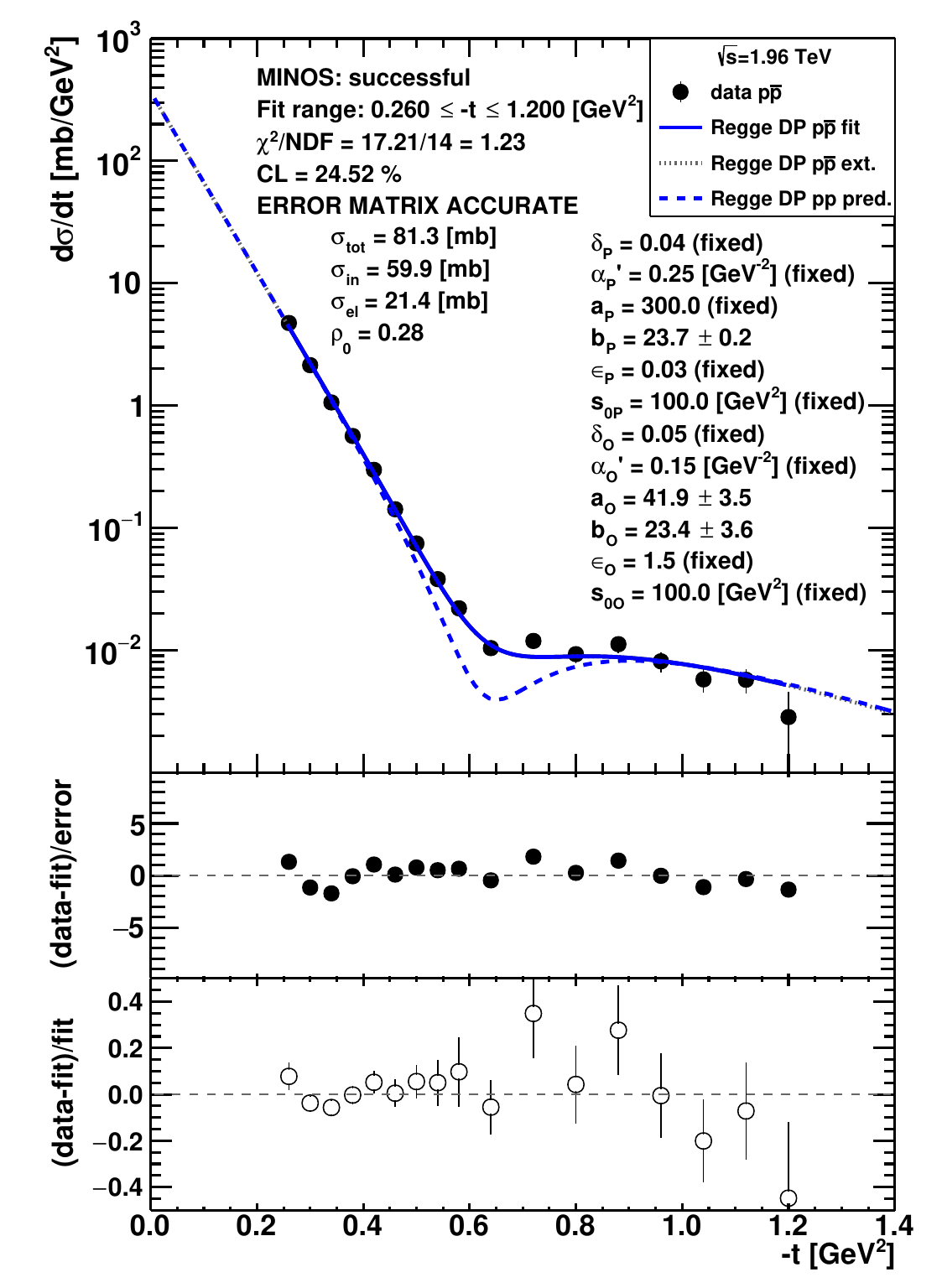}
	\caption{Fit of the dipole Regge exchange model to the Tevatron D0 \mbox{$\sqrt{s}$ = 1.96 TeV} $p\bar p$ differential cross section data \cite{D0:2012erd} and prediction to the $pp$ differential cross section at the same energy.}
	\label{fig:ReggeDP_fit_1_96_TeV}
\end{figure}

The results of the fits to the $pp$ $\sqrt s$ = 2.76 TeV and $p\bar p$ $\sqrt s$ = 1.96 TeV elastic differential cross section data are shown in \cref{fig:ReggeDP_fit_2_76_TeV} and \cref{fig:ReggeDP_fit_1_96_TeV}, respectively.

First, I performed the fit of the more restrictive $pp$ $\sqrt s$ = 2.76 TeV dataset. To fit this dataset, 12 parameters are too much: the fitter algorithm was not able to minimize the $\chi^2$ function. To solve this problem, I fixed some of the parameters at their typical value (the results are not very sensitive to changing these values by about 20\%; bigger changes deteriorate the model-data agreement). The value of the $\delta_P$ in the dipole model is about half of the value found in a simple pole model analysis of the data \cite{Szanyi:2018pew}. In the simple pole model analysis $\delta_P\approx 0.08$ (see \cref{sec:Regge}), thus I fixed the value of $\delta_P$ at $0.04$. The value of $\alpha_P'$ is fixed at the same value of 0.25 GeV$^{-2}$ as obtained in the simple pole model analysis \cite{donnachie_dosch_landshoff_nachtmann_2002}. The typical value of $a_P$ is a few hundred when $s_{0P}$ is fixed at the value of 100 GeV$^2$ \cite{Jenkovszky:2011hu,Lengyel:2012sw,Szanyi:2018pew}. I fixed the value of $s_{0P}$ at 100 GeV$^2$ and the value of $a_P$ at 300. I fixed also the value of $s_{0O}$ at  100 GeV$^2$ as it was done in Refs.~\cite{Jenkovszky:2011hu,Lengyel:2012sw,Szanyi:2018pew}. By this, I reduced the number of free parameters by about half. I left all the other parameters free, resulting in a 7-parameter fit. I added systematic and statistical uncertainties in quadrature and used the traditional $\chi^2$ definition of \cref{eq:chi_trad0}. The $CL$ value of the fit is 78.09\%, and the values of $\sigma_{\rm tot}$, $\sigma_{\rm el}$, and $\sigma_{\rm in}$ are reproduced with good accuracy (see \cref{fig:ReggeDP_fit_2_76_TeV}). The value of $\rho_0$ as calculated from the fitted dipole Regge exchange model is outside of the total error band of the prediction of the COMPETE Collaboration \cite{Cudell:2002xe}. 

\vspace{-2mm}

After fitting the model with the amplitude of \cref{eq:amplApp}, one obtains \cref{eq:amplApbarp} by changing the sign of the odderon component. This makes it possible to calculate the $p\bar p$ differential cross section by fitting only the $pp$ data. The result of this calculation is shown by the dashed blue curve in \cref{fig:ReggeDP_fit_2_76_TeV}. We can immediately conclude that, in this model, the effect of the odderon manifests in filling in the diffractive minimum in the $t$-distribution of $p\bar p$ scattering.

During the fit of the Tevatron D0 $p\bar p$ differential cross section data I fixed three additional fit parameters $\epsilon_P$, $\delta_O$, $\alpha_O'$, and $\epsilon_O$ at the values as obtained in the fit to the $pp$ \mbox{$\sqrt s$ = 2.76 TeV} dataset. This fit to the Tevatron D0 $p\bar p$ differential cross section data with three parameters resulted in a description with $CL$ = 24.58\% and the values of $\sigma_{\rm tot}$, $\sigma_{\rm el}$, and $\sigma_{\rm in}$ are reasonable (see \cref{fig:ReggeDP_fit_1_96_TeV}). The value of $\rho_0$ calculated from the fitted dipole Regge exchange model is twice as large as predicted by the COMPETE Collaboration \cite{Cudell:2002xe}. \cref{fig:ReggeDP_fit_1_96_TeV} also shows the result of the calculated $pp$ $\sqrt{s}$ = 1.96 TeV elastic differential cross section. The effect of the odderon manifests in producing a prominent diffractive minimum in the $t$-distribution of elastic $pp$ collisions at $\sqrt{s}$ = 1.96 TeV. 

Looking at the parameter values as displayed on \cref{fig:ReggeDP_fit_2_76_TeV} and \cref{fig:ReggeDP_fit_1_96_TeV}, we see that the value of $b_P$ and $b_O$ at 1.96 TeV and 2.76 TeV does not coincide even within their uncertainties. However, since there is a built-in energy dependence in the Regge model, the values of these, in principle, energy-independent parameters should coincide. 

\begin{figure} [hbt!]
	\centering
	\subfloat[\label{fig:1}]{%
		\includegraphics[width=0.8\linewidth]{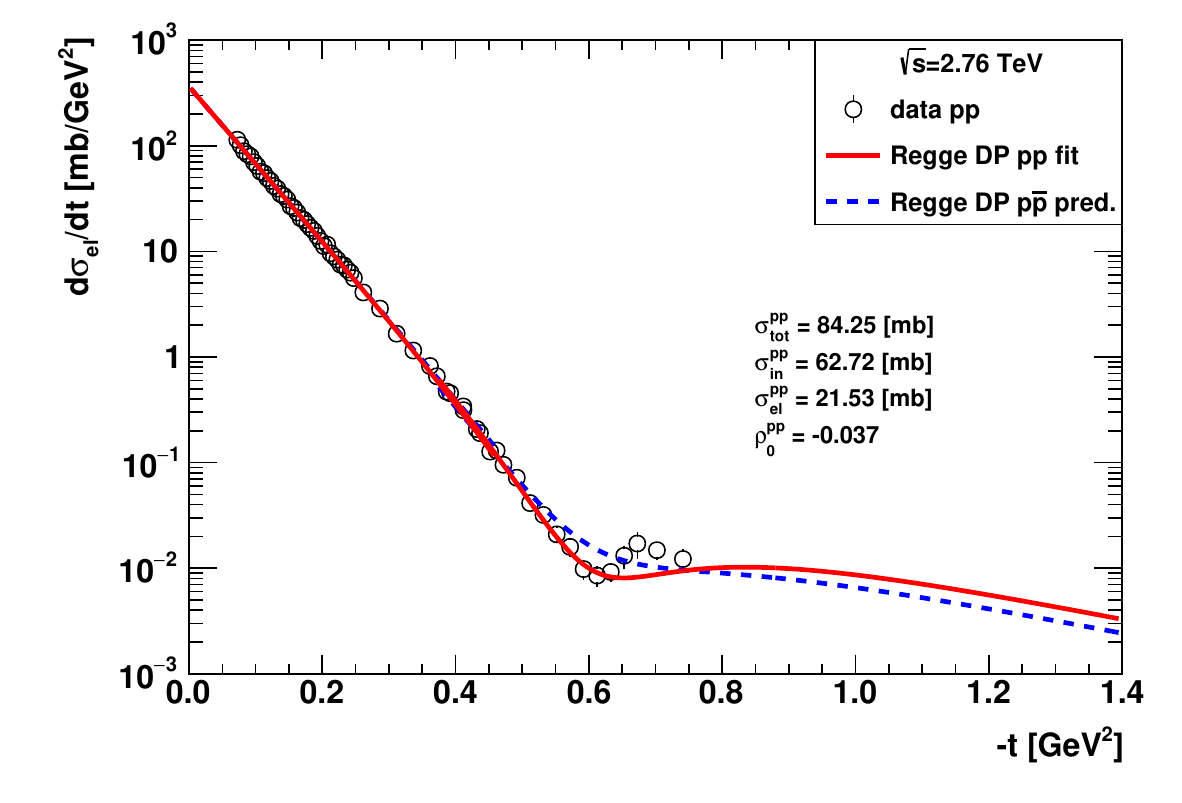}%
	}\hfill
	\subfloat[\label{fig:2}]{%
		\includegraphics[width=0.8\linewidth]{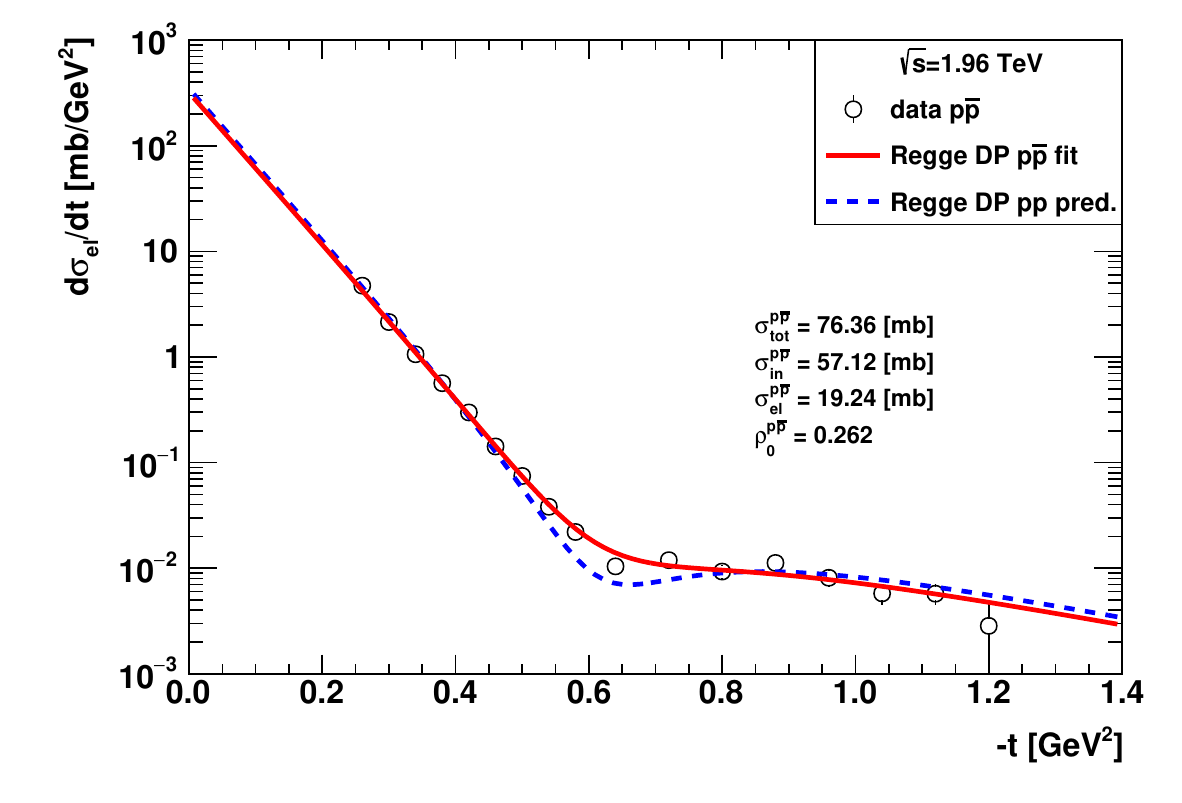}%
	}
	\caption{Simultaneous fit of the dipole Regge model to the (a) Tevatron D0 \mbox{$\sqrt{s}$ = 1.96 TeV} $p\bar p$ \cite{D0:2012erd} and (b) the LHC TOTEM $\sqrt{s}$ = 2.76 TeV $pp$ \cite{TOTEM:2018psk} elastic differential cross section data with predictions to the $\sqrt{s}$ = 1.96 TeV $pp$ and the \mbox{$\sqrt{s}$ = 2.76 TeV} $p\bar p$ differential cross sections. The curves are shifted by $N_i$ values of the $\chi^2$ definition of \cref{eq:chi_Cj++} in each $|t|$ subrange of the data.}
	\label{fig:ReggeDp_fitextr}
 \vspace{-2mm}
\end{figure}

I tested whether the built-in energy dependence of the model works or not if I simultaneously optimize the values of the fit parameters of the Regge model to the $pp$ \mbox{$\sqrt s$ = 2.76 TeV} and $p\bar p$ $\sqrt s$ = 1.96 TeV elastic differential cross section data. It turned out that by fixing $\delta_P$, $\alpha_P'$, $a_P$, $s_{0P}$ and $s_{0O}$ at the values as before and fixing $a_O$ at the value as obtained in the fit of the  2.76 TeV $pp$ data, the simultaneous fit is successful with an accurate error matrix and $CL$ value of 43.2\%. I used the $\chi^2$ definition of \cref{eq:chi_Cj++} without the term that concerns the total cross section. The parameter values are shown in \cref{table:Regge_fit_parameters} while the results are graphically displayed on \cref{fig:ReggeDp_fitextr}. The calculated values of $\sigma_{\rm tot}$, $\sigma_{\rm el}$, and $\sigma_{\rm in}$ are quite reasonable, the experimental values at $\sqrt s$ = 2.76 TeV are reproduced with good accuracy (see \cref{fig:ReggeDp_fitextr}). However, the calculated $\rho_0$ values are completely outside of the trend as determined by the $\rho_0$ measurements at lower and higher energies \cite{Cudell:2002xe,TOTEM:2017sdy}. It was demonstrated in previous \mbox{studies \cite{Jenkovszky:1987gv,Desgrolard:1992ri,Jenkovszky:2011hu,Lengyel:2012sw,Broniowski:2018xbg,Szanyi:2018pew,Szanyi:2019kkn}} that the dipole Regge exchange model describes $\rho_0$ data reasonably well up to LHC energies. However, when the model describes $\rho_0$ data reasonably well, the description in the minimum-maximum region of the differential cross section becomes less precise. This deficiency of the model should be fixed in the future. Though the model has deficiencies, it describes the data on elastic $pp$ and $p\bar p$ differential cross sections at $\sqrt{s}=2.76$ TeV and 1.96 TeV, and it is interesting to see what we can conclude about the odderon effects that may be observed in the $t$-distribution of elastic scattering based on this Regge model.

\begin{table}[!t]
        \centering
	{\begin{tabular}{cccc} \hline\hline
 \multicolumn{2}{c}{odderon}&\multicolumn{2}{c}{pomeron}\\\hline
$\delta_P$  & 0.04 (fixed)&$\alpha_{0O}$  & 0.04 $\pm$ 0.02\\
$\alpha_P'$  & 0.25 (fixed)&$\alpha_{1O}$  & 0.13 $\pm$ 0.05\\
$a_P$       & 300 (fixed)&$a_O$       & 34.5 (fixed)\\
$b_P$      & 22.5 $\pm$ 0.5&$b_O$       & 33 $\pm$ 18\\
$\epsilon_P$ & 0.06 $\pm$ 0.03&$\epsilon_O$& 2.0 $\pm$ 0.5\\
$s_{0P}$      & 100 (fixed)&$s_{0O}$     & 100 (fixed)\\ \hline\hline
\multicolumn{2}{c}{for fit of data}&\multicolumn{2}{c}{for prediction vs. data}\\\hline
$N_1$      & 0.99 $\pm$ 0.04&\multicolumn{2}{c}{0.976 $\pm$ 0.005}\\
$N_2$      & 1.07 $\pm$ 0.05&\multicolumn{2}{c}{0.84 $\pm$ 0.02}\\
$N_3$      & 0.94 $\pm$ 0.03&\multicolumn{2}{c}{1.04 $\pm$ 0.02}\\	\hline
$\chi^2$   &75.41&\multicolumn{2}{c}{179.02}\\
NDF        &74&\multicolumn{2}{c}{80}\\
CL         &43.2\%&\multicolumn{2}{c}{1.5$\times10^{-7}$\%}\\\hline
\multicolumn{4}{c}{odderon significance: 6.04$\sigma$}\\ \hline\hline
	\end{tabular}}
 	\caption{The values of the dipole Regge exchange model parameters simultaneously fitted to the Tevatron D0 \mbox{$\sqrt{s}$ = 1.96 TeV} $p\bar p$ \cite{D0:2012erd} and the LHC TOTEM \mbox{$\sqrt{s}$ = 2.76 TeV} $pp$ \cite{TOTEM:2018psk} elastic differential cross section data utilizing the $\chi^2$ definition of \cref{eq:chi_Cj++}. The values of the $N_i$ parameters of the $\chi^2$ definition are given also for the cases when the Regge model $pp$ ($p\bar p$) prediction is compared to the $p\bar p$ ($pp$) data.  \label{table:Regge_fit_parameters}}
\end{table}

Let me now discuss the odderon effects observed. \cref{fig:ReggeDp_fitextr} displays also the calculated $p\bar p$ elastic differential cross section at $\sqrt s$ = 2.76 TeV and the calculated $pp$ elastic differential cross section at $\sqrt s$ = 1.96 TeV. These calculated curves are compared to the experimental data utilizing the $\chi^2$ definition of \cref{eq:chi_Cj++}. The calculated curves are shifted by $N_i$ values of the $\chi^2$ definition of \cref{eq:chi_Cj++} in each $|t|$ subrange of the data. The values of $N_i$ for comparing the Regge models with the data are given in \cref{table:Regge_fit_parameters}. The hypothesis that the $pp$ and $p\bar p$ differential cross sections at a given energy can be described by the same curve has a $CL$ value of 1.5$\times10^{-7}$\% corresponding to an odderon signal with a statistical significance of 6.04$\sigma$. Thus, closing the energy gap between the $p\bar p$ \mbox{$\sqrt s$ = 1.96 TeV} and $pp$ $\sqrt s$ = 2.76 TeV elastic differential cross section measurements using the dipole Regge exchange model leads to a discovery level odderon signal observation.  However, one has to keep in mind that this model has some deficiencies, and this should be considered as a preliminary/quantitative result.

The result for the statistical significance within the Regge model study may vary a bit when the $\chi^2$ definition of \cref{eq:phenix_o} is used. The evaluation of how the results change utilizing \cref{eq:phenix_o} is a subject of a future study.

\vspace{0.5cm}
\textbf{Summary}
\vspace{0.2cm}


In this Chapter, Using physical models, the ReBB model and a dipole Regge exchange model, I showed that the TOTEM measured $pp$ elastic differential cross section can be extrapolated to $\sqrt{s}=1.96$ TeV: I closed the energy gap between the D0 $p\bar p$ \mbox{$\sqrt s$ = 1.96 TeV} and TOTEM $pp$ \mbox{$\sqrt s$ = 2.76 TeV} elastic differential cross section measurements. In my model-dependent study, I estimated the effects of the odderon exchange at $\sqrt{s}=1.96$ TeV. Both models I used predict that the effect of the odderon exchange manifests mainly in \mbox{(i) generating} a prominent diffractive minimum in the $t$-distribution of elastic $pp$ scattering and (ii) filling in this minimum in the $t$-distribution of elastic $p\bar p$ scattering.

The joint team of the CERN LHC TOTEM and the FNAL Tevatron D0 collaborations, with my participation, finally published results on odderon effects at $\sqrt{s}$ = 1.96 TeV obtained by extrapolations based on fits with functional forms without physical content behind \cite{TOTEM:2020zzr}. My results presented in this Chapter served as a guide during the joint D0-TOTEM analysis that finally observed the signal of the odderon exchange with a statistical significance of at least 5.2$\sigma$ by the comparison of D0 measured $p\bar p$ $d\sigma_{\rm el}/dt$ with D0-TOTEM extrapolated $pp$ $d\sigma_{\rm el}/dt$ at $\sqrt{s} = 1.96$ TeV together with the comparison of TOTEM measured $\rho_{0}$ parameter and total cross section at $\sqrt{s} = 13$ TeV with a set of models not containing odderon contribution.

\newpage
\thispagestyle{empty}

\chapter{Refined ReBB model analysis of elastic $\rm pp$ and $\rm p\bar p$ scattering}\label{chap:rebbdesc}


The ReBB model describes both $pp$
and $p\bar p$ elastic differential cross section data as discussed in \cref{chap:ReBBpbarp} and \cref{chap:oddTD0}. In this Chapter, I present a refined and final analysis of both $pp$
and $p\bar p$ elastic differential cross section data in the TeV c.m. energy domain. In \cref{sec:refindchi}, I present a refined form of the $\chi^2$ function that properly handles both statistical and systematic errors based on the method developed by the PHENIX Collaboration and summarised in \cref{sec:fittingmethod} of this dissertation. Then, in \cref{sec:rebb_desc_refind_TeV}, using this refined $\chi^2$ function, I fit the ReBB model to the measured elastic $pp$ and $p\bar p$ differential cross section data in the squared four-momentum transfer range of 0.37~GeV$^2$~$\leq -t\leq1.2$~GeV$^2$ at $\sqrt{s}=$ 546 GeV, 1.96 TeV, 2.76 TeV and 7 TeV. I find that the ReBB model gives a quantitative description, \textit{i.e.}, a description with a confidence level greater than 0.1\% to all datasets. In \cref{sec:rebb_endep_TeV}, I show that the geometrical parameters of the model, $R_q$, $R_d$, and $R_{qd}$, as determined from elastic $pp$ and $p\bar p$ scattering data lie on the same linear-logarithmic energy dependence curves. I find that the opacity parameter, $\alpha_R$, is the only ReBB model parameter whose energy evolution is not compatible with the same curve in elastic $pp$ and $p\bar p$ processes. In the c.m. energy range of 0.546~TeV~$\leq\sqrt{s}\leq 7$~TeV, for $pp$ scattering, $\alpha_R$ is even compatible with a constant value, while for $p\bar p$ scattering, $\alpha_R$ rises with increasing $\sqrt{s}$ according to a linear-logarithmic functional form.


This chapter is based on Ref.~\cite{Csorgo:2020wmw}.

\vfill

\section{A refined $\chi^2$ function}\label{sec:refindchi}

The main goal of the study presented in Ref.~\cite{Csorgo:2020wmw} was to determine the statistical significance of a possible odderon signal based on the ReBB model analysis of elastic $pp$ and $p\bar p$ elastic differential cross section data. Such an investigation required a precise comparison of $pp$ and $p\bar p$ elastic differential cross sections at the same c.m. energy in the TeV energy scale in a common $-t$ range that includes the minimum-maximum structure of $pp$ scattering. To perform the analysis, I used a more advanced version of the $\chi^2$ function, which is essentially the same as \cref{eq:phenix_o} introduced in Ref.~\cite{PHENIX:2008ove} but generalized for data that consists of sub-datasets corresponding to several, separately measured kinematic ranges and have both vertical and horizontal errors. This allowed for a more precise analysis of the data. During the fitting procedure, I minimized the function 
\begin{align}\label{eq:chi2_refind}
 \chi ^{2}=&\sum _{i=1}^{M} \left(\sum _{j=1}^{n_{i}}\chi_{ij}^2+ \epsilon _{bi}^{2}+ \epsilon _{ci}^{2}\right) + \left( \frac{d_{ \sigma _{tot}}-m_{ \sigma _{tot}}(\vec p\,)}{ \delta  \sigma _{tot}} \right) ^{2}+ \left( \frac{d_{ \rho_{0}}-m_{ \rho _{0}}(\vec p\,)}{ \delta  \rho _{0}} \right) ^{2},
\end{align}
where 
\begin{equation}
\chi_{ij} = \frac{  d_{ij}+ \epsilon _{bi} \widetilde\sigma _{bij}+ \epsilon _{ci}d_{ij} \sigma _{ci}-m_{ij} (\vec p\,)}{\widetilde{ \sigma }_{ij}},
\end{equation}
\begin{equation}
    \widetilde{ \sigma }_{ij}^{2}= \widetilde\sigma _{aij} \left( \frac{d_{ij}+ \epsilon _{bi} \widetilde\sigma _{bij}+ \epsilon _{ci}d_{ij} \sigma _{ci}}{d_{ij}} \right),
\end{equation}
\begin{equation}\label{eq:vherr}
    \widetilde\sigma_{kij} =\sqrt{\sigma _{kij}^2+ (d^{\prime}_{ij} \delta_{k}t_{ij})^2}, \ \ k\in\{a,b\},
\end{equation}
     \begin{equation}\label{eq:deriv}
     d^{\prime}_{ij}=\frac{d_{(i+1)j}-d_{ij}}{t_{(i+1)j}-t_{ij}},
     \end{equation}
and furthermore: 
\begin{itemize}
\item $M$ is the number of $d\sigma_{\rm el}/dt$ sub-datasets measured at a given c.m. energy but in different $t$ intervals;
\item $n_i$ is the number of $d\sigma_{\rm el}/dt$ datapoints in the $i$-th sub-dataset (consequently, the number of $d\sigma_{\rm el}/dt$ data points in the fit is $\sum_{i=1}^M n_i$);
\item$d_{ij}$ are the $j$th measured $d\sigma_{\rm el}/dt$ data point in the  $i$th sub-dataset  and $m_{ij}(\vec p\,)$ is the corresponding value calculated from the ReBB model which depends on fit parameters arranged in the vector $\vec{p}$; 
\item $\sigma_{kij}$ are the point-to-point varying type $k\in\{a,b\}$ (vertical) errors of the $j$th measured data point in the $i$th sub-dataset (for the discussion of the various types of errors ($a$, $b$, and $c$), see \cref{sec:fittingmethod}); 
\item $\sigma_{ci}$ is the point independent, overall relative error for the $i$th sub-dataset;  
\item $\delta_{kij}$ is the point-to-point varying type $k\in\{a,b\}$ (horizontal) error of the $t$ value corresponding to the $j$th measured  data point in the sub-dataset $i$ and denoted as $t_{ij}$; 
\item $d'_{ij}$ is the numerical derivative\footnote{\cref{eq:vherr} is based on the effective variance method utilized by the CERN ROOT data analysis framework \cite{Brun:1997pa}. This method assumes that there is no correlation between the vertical and horizontal coordinate.} of $d\sigma_{\rm el}/dt$ at the point $t_{ij}$; 
\item
$\epsilon_{bi}$ and $\epsilon_{ci}$ are the fractions of type $b$ and type $c$ errors, respectively, in the $i$th sub-dataset; the values of these parameters are optimized during the fitting procedure; 
\item $d_{\sigma_{\rm tot}}$  is the value of the measured $\sigma_{\rm tot}$, $m_{\sigma_{\rm tot}}(\vec p\,)$  is the corresponding value calculated from the fit model,  and $\delta\sigma_{\rm tot}$ is the total error (statistical and systematic errors added quadratically) of the measured $\sigma_{\rm tot}$ at the c.m. energy of the $d\sigma_{\rm el}/dt$ data;     

\item $d_{\rho_{\rm 0}}$ is the value of the measured $\rho_{\rm 0}$,  $m_{\rho_{\rm 0}}(\vec p\,)$ is the corresponding value calculated from the fit model, and $\delta\rho_{\rm 0}$ is the total error (statistical and systematic errors added quadratically) of the measured $\rho_{\rm 0}$ at the c.m. energy of the $d\sigma_{\rm el}/dt$ data. 
\end{itemize}
The presence of the last two terms in \cref{eq:chi2_refind} allowed to adjust the values of the ReBB model parameters not only to the measured $d\sigma_{\rm el}/dt$  but also to the measured $\sigma_{\rm tot}$ and $\rho_{\rm 0}$ values. The $NDF$ is calculated by the formula 
\begin{equation}\label{eq:NDF_refchi}
    {\rm NDF} = N_{d,\,d\sigma_{\rm el}/dt} + N_{d,\,\sigma_{\rm tot}} + N_{d,\,\rho_{\rm 0}} - N_{p,\,ReBB},
\end{equation}
where $N_{d,\,d\sigma_{\rm el}/dt}$ is the number of differential cross section data points, $N_{d,\,\sigma_{\rm tot}}$ is the number of total cross section data points and $N_{d,\,\rho_{\rm 0}}$ is the number of $\rho_{\rm 0}$ data points included in the fitting procedure; $N_{p,\,ReBB}$ is the number of the free ReBB model parameters. The effect of the $\epsilon$ parameters cancels from the final $NDF$ \cite{PHENIX:2008ove} (see also \cref{sec:fittingmethod}). In Ref.~\cite{Csorgo:2020wmw}, the NDF (i) was decreased by the number of the $\epsilon$ parameters and (ii) was not increased by the number of total cross section $\rho_{\rm 0}$ data points. Point (i) was indicated and corrected in a subsequent paper \cite{Szanyi:2022ezh}. This dissertation contains the corrected results, including the corrections mentioned in point (ii). These corrections do not affect the final conclusions drawn in Ref.~\cite{Csorgo:2020wmw}.

The $\sqrt{s} = 7 $ TeV TOTEM  data on $pp$ $d\sigma_{\rm el}/dt$ have asymmetric type $b$ (vertical) errors~\cite{TOTEM:2013lle}. To handle this situation, I chose the upward type $b$ errors if $\epsilon_b$ had a positive value, and I chose the downward type $b$ errors if $\epsilon_b$ had a negative value.
The $\chi^2$ fitting method assumes symmetric type A errors, and this condition is satisfied for the \mbox{$\sqrt{s} = 7$ TeV data.} This allows the use of the $\chi^2$ fitting method even if the type $b$ errors are asymmetric.

To inspect the fit results obtained using the $\chi^2$ given by \cref{eq:chi2_refind}, I plot the points
\begin{equation}\label{eq:chiij}
    \chi_i \equiv \left\{\chi_{ij}\right\}
\end{equation}
and
\begin{equation}\label{eq:d-f}
\frac{  d_{i}+ \epsilon _{b} \sigma _{bi}+ \epsilon _{c}d_{i} \sigma _{c}-m_{i} }{m_{i} } \equiv \left\{\frac{  d_{ij}+ \epsilon _{bi} \widetilde\sigma _{bij}+ \epsilon _{ci}d_{ij} \sigma _{ci}-m_{ij} (\vec p\,)}{m_{ij} (\vec p\,)} \right\}
\end{equation}
using the optimized parameter values. The notations used in the right-hand sides of \cref{eq:chiij} and \cref{eq:d-f} coincide with the notations used in \cref{eq:chi2_refind} and already explained above. The left-hand sides of \cref{eq:chiij} and \cref{eq:d-f} are just notations for sets of values. The values given by \cref{eq:chiij} are actually the (data $-$ fit)/error values, and the values given by \cref{eq:d-f} are basically the (data $-$ fit)/fit values.  

\vspace{-5mm}
\section{Refined fits to $\rm pp$ and $\rm p\bar p$ $d\sigma_{\rm el}/dt$ data at TeV energies}\label{sec:rebb_desc_refind_TeV}
\vspace{-3mm}


In this section, I present the results of the ReBB model fits to the available elastic $pp$ and $p\bar p$ differential cross section data in the squared four-momentum transfer range of \mbox{$0.37~{\rm GeV^2}\leq -t \leq 1.2~{\rm GeV^2}$} at $\sqrt{s}$ = 540 GeV, 546~GeV, 1.96 TeV, 2.76 TeV, 7 TeV, and 13 TeV utilizing the $\chi^2$ function as defined by \cref{eq:chi2_refind}.  The choice of the $t$-range is motivated by (i) the expected signal region of the odderon exchange based on the preliminary study presented in \cref{chap:oddTD0};
(ii) the lower edge of the acceptance of the \mbox{high-$|t|$} $pp$ elastic differential cross section measurement at 7 TeV ($|t|\simeq$ 0.37 GeV$^2$);
\mbox{(iii) the} higher edge of the acceptance of the $p\bar p$ elastic differential cross section measurement at 1.96 TeV ($|t|=$ 1.2 GeV$^2$).

\begin{figure}[!hbt]
	\centering
\includegraphics[width=0.8\linewidth]{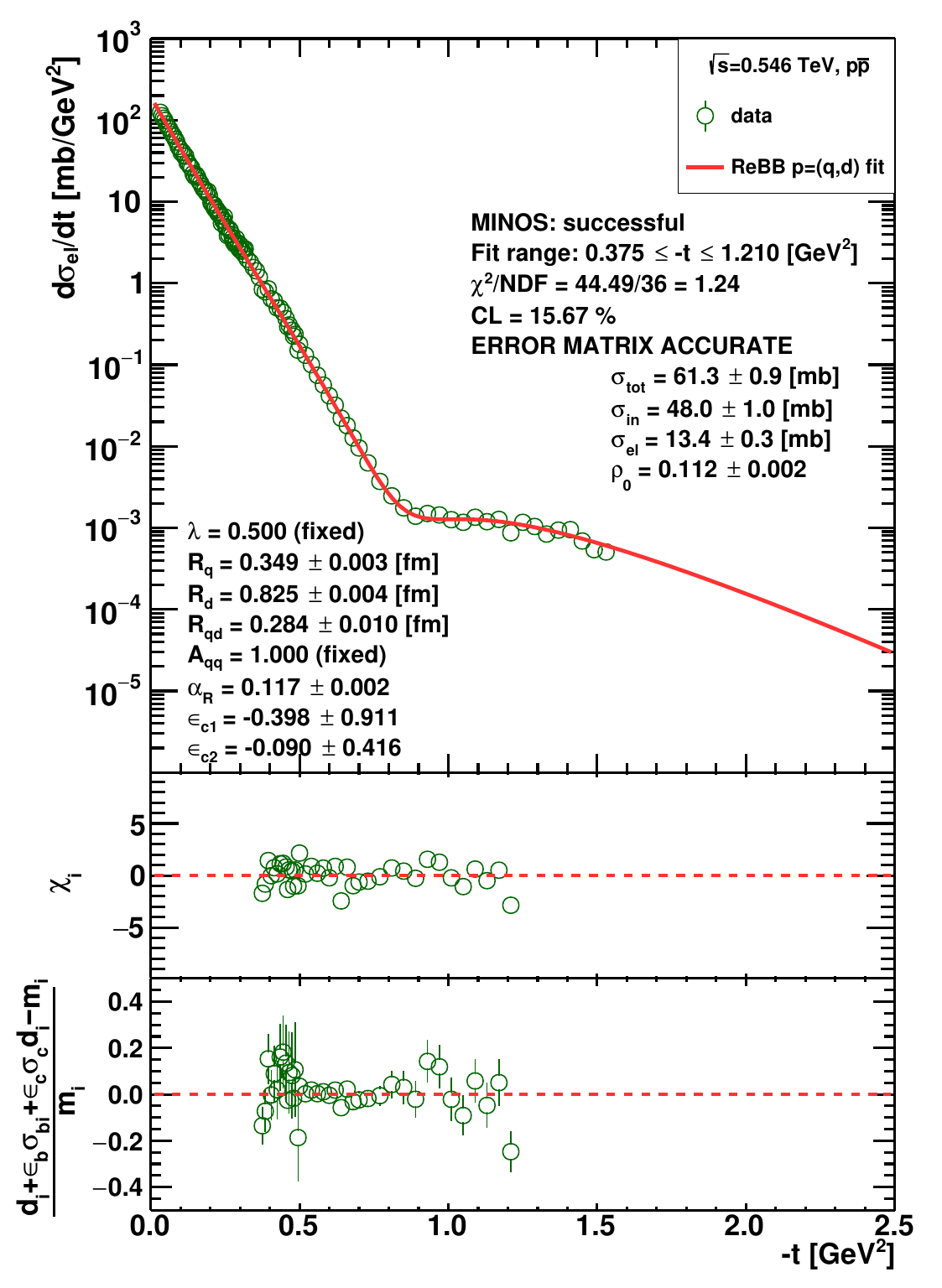}
\vspace{-4mm}
 \caption{ReBB model fit to the $p\bar p$ merged SPS UA4 $\sqrt{s}=0.54$~TeV \mbox{low-$|t|$ \cite{UA4:1983mlb,UA4:1984uui}} and $\sqrt{s}=0.546$~TeV high-$|t|$~\cite{UA4:1985oqn} differential cross section data in the squared four-momentum transfer range of 0.375 GeV$^2$ $\leq -t \leq 1.210$ GeV$^2$ utilizing the $\chi^2$ definition as given by \cref{eq:chi2_refind}. Only type $a$ vertical errors of the data points are shown. The ReBB model parameter values and the $\epsilon$ values are shown in the bottom left corner; the fixed parameters are indicated; the uncertainties of the fitted parameters are displayed. 
 The fit quality parameters and the values of $\sigma_{\rm tot}$ and $\rho_0$ calculated from the fitted ReBB model are displayed in the top right part of the plot. The middle and bottom panels show the values given by \cref{eq:chiij} and \cref{eq:d-f}, respectively. } 
	\label{fig:reBB_model_fit_0_546_TeV}
\end{figure}

The result of the fit to the merged SPS UA4 lower-$|t|$ 540 GeV \cite{UA4:1983mlb,UA4:1984uui} and \mbox{higher-$|t|$} 546~GeV \cite{UA4:1985oqn} elastic $p\bar p$ differential cross section data is shown in \cref{fig:reBB_model_fit_0_546_TeV}. In the fitted range, the data is composed of two separately measured sub-ranges with different normalization uncertainties (type $c$ errors): for 0.375~GeV$^2$ $\leq|t|\leq0.495 $ GeV$^2$, $\sigma_{c}$ = 0.03, and, for 0.46 GeV$^2$ $\leq|t|\leq1.2$ GeV$^2$, $\sigma_{c}$ = 0.1. In addition to the type $c$ error, only type $a$ vertical errors ($\sigma_{ai}$) are published and consequently used in the fit. The measured value of $\sigma_{\rm tot}$ at $\sqrt{s}=546$ GeV is $61.26\pm0.93$ mb \cite{CDF:1993wpv}. This value was used during the fitting procedure. The $CL$ of the fit is 15.67\%. The experimentally measured values of the total, elastic ($\sigma_{\rm el}=12.87\pm0.3$ mb \cite{ParticleDataGroup:2018ovx}) and inelastic ($\sigma_{\rm in}=48.39\pm1.01$ mb \cite{ParticleDataGroup:2018ovx}) cross sections reproduced by the ReBB model with good accuracy. The value of $\rho_0$ was measured at $\sqrt{s} = 541$ GeV to be $0.135\pm0.015$ \cite{UA42:1993sta}. The calculated value is very close to the measured value. The uncertainties of these cross sections and $\rho_0$ are estimated by shifting up and down the ReBB model parameters by their error values and choosing the maximal upward and downward deviations compared to the calculated central values of the observables.  In Ref.~\cite{Csorgo:2020wmw}, a slightly different method was used to estimate the uncertainties of the calculated cross sections and $\rho_0$: all upward and downward deviations were averaged.

\begin{figure}[hbt!]
	\centering
\includegraphics[width=0.8\linewidth]{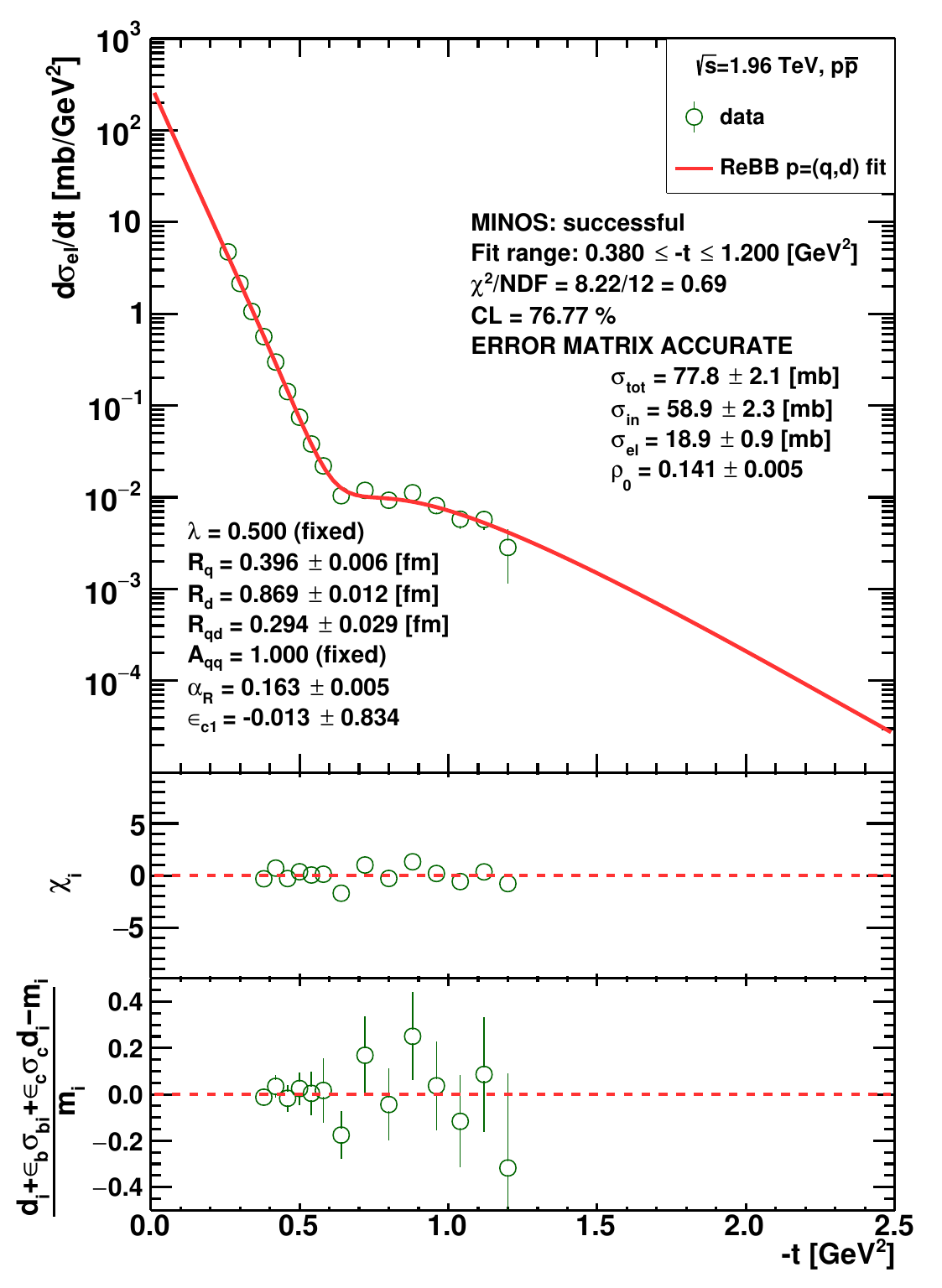}
	\caption{ReBB model fit to the $p\bar p$ Tevatron D0 $\sqrt{s}=1.96$~TeV data \cite{D0:2012erd} in the range of $0.37$ GeV$^2\leq-t\leq1.2 $ GeV$^2$ utilizing the $\chi^2$ definition as given by \cref{eq:chi2_refind}. Otherwise, same as Fig.~\ref{fig:reBB_model_fit_0_546_TeV}.
	}
	\label{fig:reBB_model_fit_1_96_TeV}
 \vspace{-5mm}
\end{figure}

The result of the fit to the Tevatron D0 1.96 TeV elastic $p\bar p$ differential cross section data \cite{D0:2012erd} is shown in \cref{fig:reBB_model_fit_1_96_TeV}. This dataset has a type $c$ error of $\sigma_{c}$ = 0.144. Only quadratically added statistical and systematic uncertainties are published. These merged errors are considered to be type $a$ errors ($\sigma_{ai}$) during the fit.  $\sigma_{\rm tot}$ and $\rho_0$ values are not measured at  $\sqrt{s} = 1.96$ TeV. In the fitting procedure, I utilized the values as predicted by the COMPETE Collaboration in Ref.~\cite{Cudell:2002xe}: $\sigma_{\rm tot}=78.27\pm1.93$ mb and $\rho_0=0.145\pm0.006$. The $CL$ of the fit is 76.77\%, and the experimental values of $\sigma_{\rm tot}$ and $\rho_0$ are reproduced by the ReBB model with good accuracy.

\begin{figure}[hbt!]
	\centering
\includegraphics[width=0.8\linewidth]{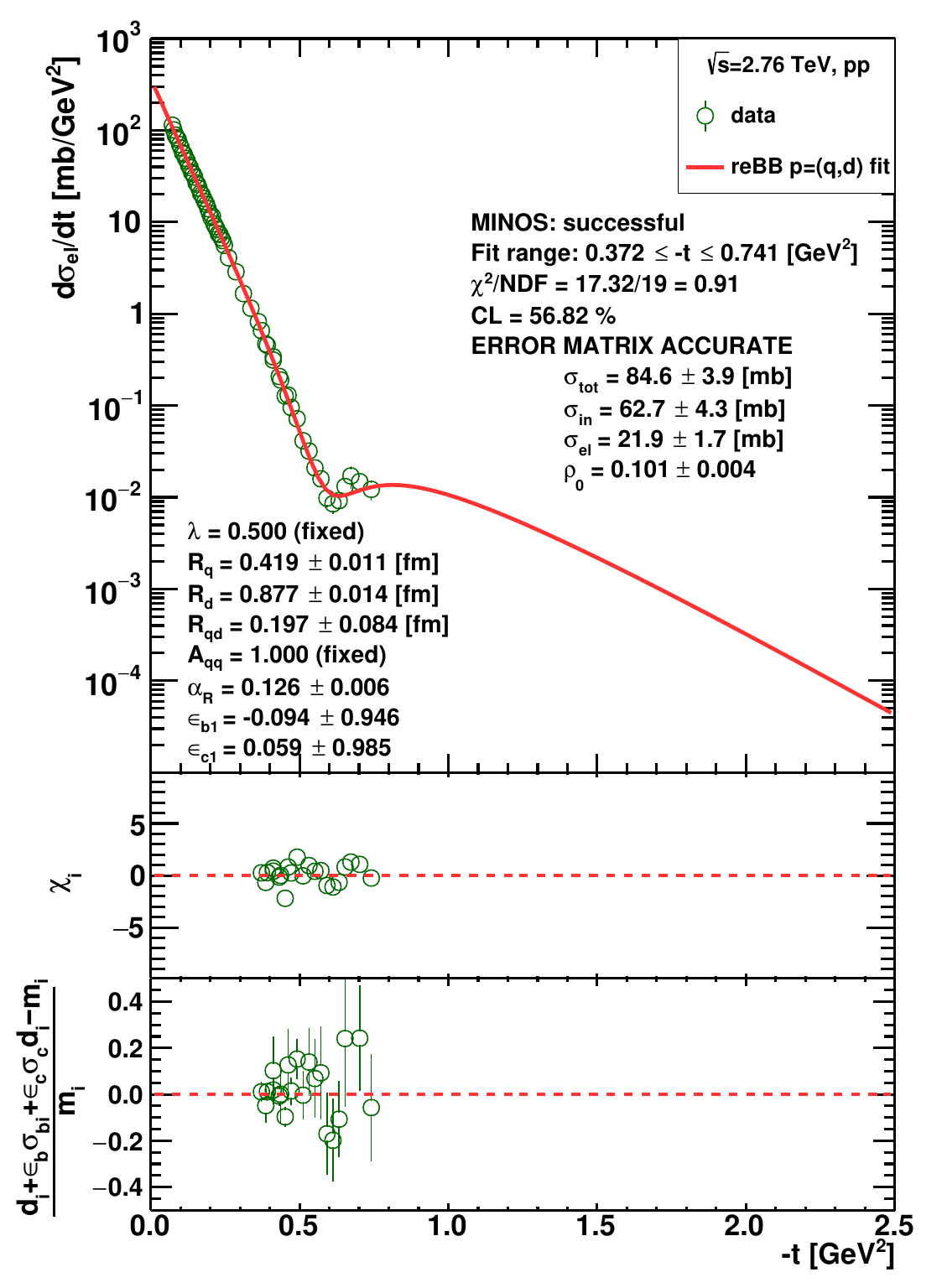}
	\caption{ReBB model fit to the $pp$ LHC TOTEM $\sqrt{s}=2.76$~TeV data \cite{TOTEM:2018psk} in the range of $0.372$ GeV$^2\leq-t\leq0.741 $ GeV$^2$  utilizing the $\chi^2$ definition as given by \cref{eq:chi2_refind}. 
	Otherwise, same as Fig.~\ref{fig:reBB_model_fit_0_546_TeV}.
}
	\label{fig:reBB_model_fit_2_76_TeV}
  \vspace{-5mm}
\end{figure}

The result of the fit to the LHC TOTEM 2.76 TeV elastic $p p$ differential cross section \mbox{data \cite{TOTEM:2018psk}} is shown in \cref{fig:reBB_model_fit_2_76_TeV}. In the fitted range, the data are measured in two sub-ranges of~$|t|$: $0.375~{\rm GeV}^2\leq|t|\leq0.462 $ GeV$^2$ and $0.372~{\rm GeV}^2\leq|t|\leq0.74 $ GeV$^2$ (the second sub-range contains the first one).  The sub-ranges have the same type $c$ error of $\sigma_c=0.06$. Statistical (type $a$, $\sigma_{ai}$) and systematic (type $b$, $\sigma_{bi}$) uncertainties are published and used during the fit. The measured value of $pp$ $\sigma_{\rm tot}$ at $\sqrt{s}=2.76$ TeV is $84.7\pm3.3$ mb \cite{TOTEM:2017asr}. The value of $\rho_0$ is not measured at $\sqrt{s}=$ 2.76 TeV. The $CL$ of the fit is 56.82\%. The value of $\rho_0$ as calculated from the fitted ReBB model is within the total error band of the prediction of the COMPETE Collaboration \cite{Cudell:2002xe}. The experimental values of the total, elastic, and inelastic cross sections ($\sigma_{\rm in}=62.8\pm2.9$ mb, $\sigma_{\rm el}=21.8\pm1.4$ mb \cite{TOTEM:2017asr,Nemes:2017gut}) are reproduced by the ReBB model with good accuracy. 

\begin{figure}[hbt!]
	\centering
\includegraphics[width=0.8\linewidth]{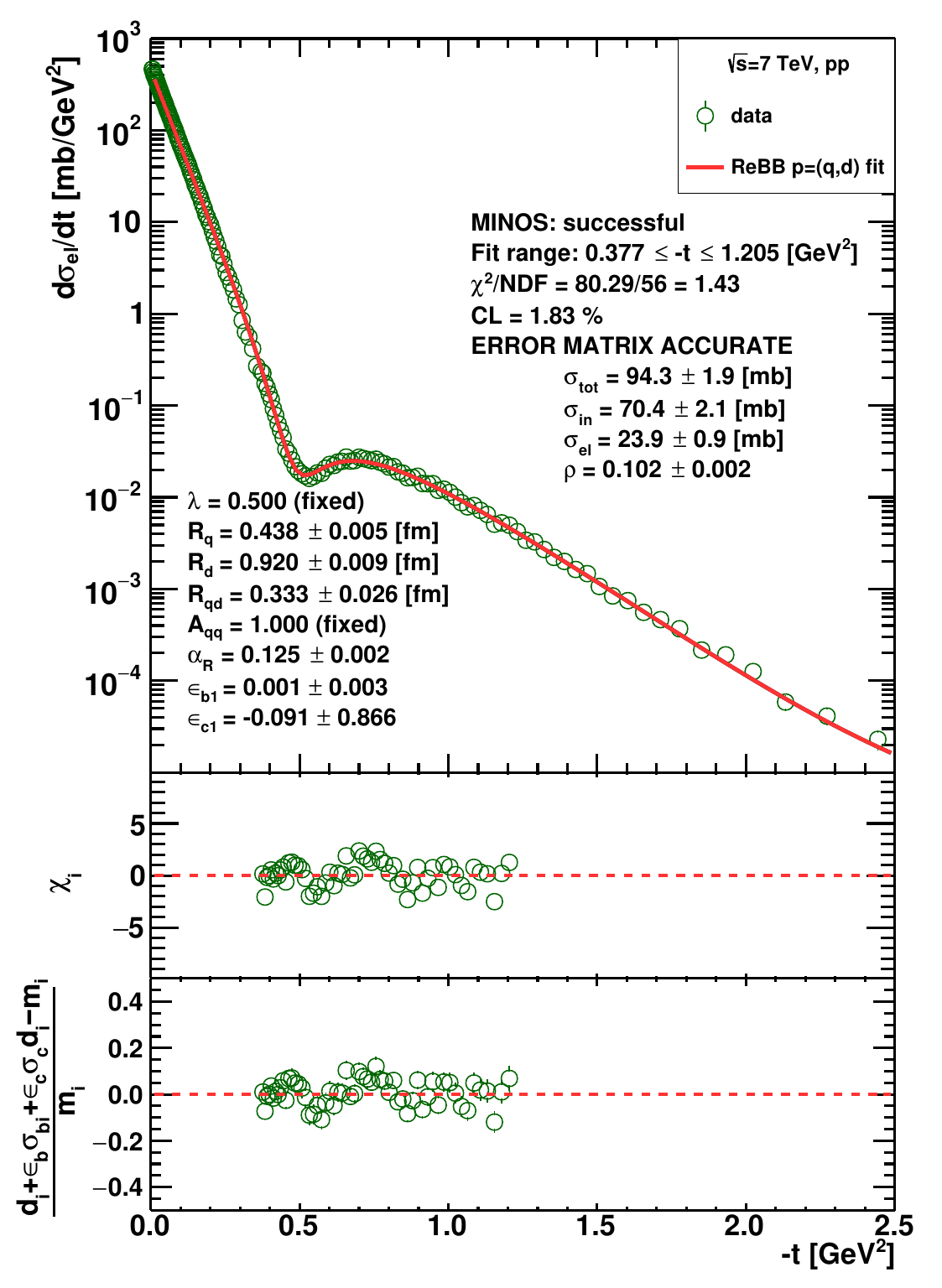}
	\caption{ReBB model fit to the $pp$ LHC TOTEM $\sqrt{s}=7$~TeV data \cite{TOTEM:2013lle} in the range of $0.377$ GeV$^2\leq-t\leq1.205 $ GeV$^2$ utilizing the $\chi^2$ definition as given by \cref{eq:chi2_refind}.
	Otherwise, same as Fig.~\ref{fig:reBB_model_fit_0_546_TeV}.
 \vspace{-0.5 cm}
}
	\label{fig:reBB_model_fit_7_TeV}
\end{figure}

The result of the fit to the TOTEM 7 TeV elastic $p p$ differential cross section \mbox{data \cite{TOTEM:2013lle}} is shown in \cref{fig:reBB_model_fit_7_TeV}. In the fitted range of 0.377 GeV$^2$ $\leq|t|\leq1.205 $ GeV$^2$ the data has a type $c$ error of $\sigma_c$ = 0.042. Vertical and horizontal statistical (type $a$, $\sigma_{ai}$ and $\delta_a t_{i}$) and systematic (type $b$, $\sigma_{bi}$ and $\delta_b t_{i}$ \cite{TOTEM:2011vxg}) uncertainties are published and used during the fit. The measured value of $pp$ $\sigma_{\rm tot}$ at $\sqrt{s}=7$ TeV is $98.0\pm2.5$ mb \cite{TOTEM:2013vij}, the measured value of $\rho_0$ is $\rho_0=0.145\pm0.091$ \cite{TOTEM:2013dtg}. The $CL$ of the fit is 1.83\% and the experimental values of the total, elastic ($\sigma_{\rm el}=25.1\pm1.1$ mb \cite{TOTEM:2013vij}) and inelastic ( \mbox{$\sigma_{\rm in}=72.9\pm1.5$ mb \cite{TOTEM:2013vij}}) cross sections and the $\rho_0$ parameter are reproduced by the ReBB model with good accuracy. 

\begin{figure}[hbt!]
	\centering
	\includegraphics[width=0.8\linewidth]{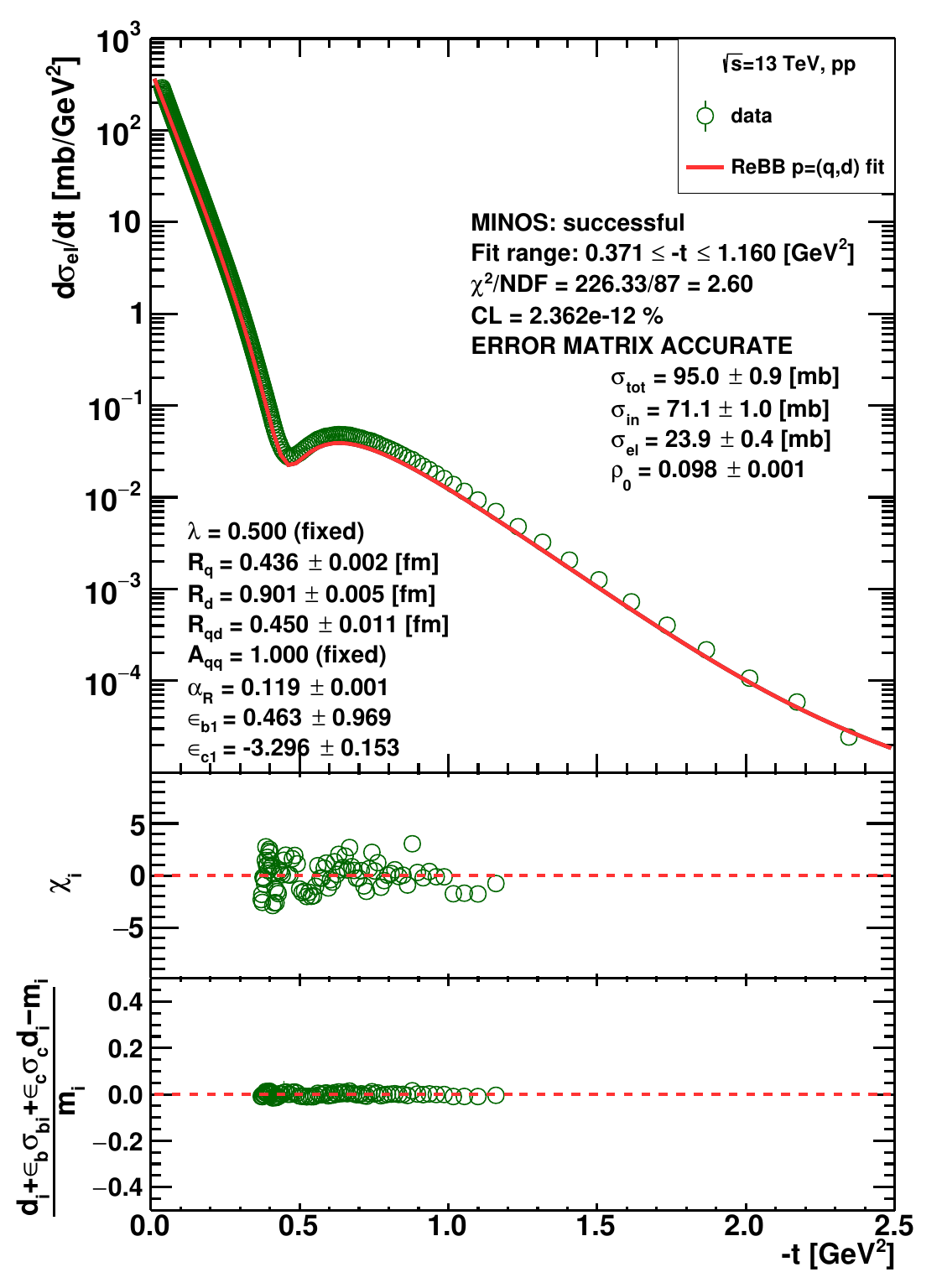}
	\caption{ReBB model fit to the $pp$ LHC TOTEM $\sqrt{s}=13$~TeV data \cite{TOTEM:2018hki} in the range of $0.371$ GeV$^2\leq-t\leq1.16$ GeV$^2$ utilizing the $\chi^2$ definition as given by \cref{eq:chi2_refind}.
		Otherwise, same as Fig.~\ref{fig:reBB_model_fit_0_546_TeV}.
	}
	\label{fig:reBB_model_fit_13_TeV}
\end{figure}

The result of the fit to the TOTEM 13 TeV elastic $p p$ differential cross section  \mbox{data \cite{TOTEM:2018hki}} is shown in \cref{fig:reBB_model_fit_13_TeV}. This dataset has a type $c$ error of $\sigma_c$ = 0.055. Statistical (type $a$, $\sigma_{ai}$) and systematic (type $b$, $\sigma_{bi}$) uncertainties are published and used during the fit. The measured value of $pp$ $\sigma_{\rm tot}$ at $\sqrt{s}=13$ TeV is $110.5\pm2.4$ mb \cite{TOTEM:2017sdy}; the measured value of $\rho_0$ is $0.09\pm0.01$ or $0.1\pm0.01$, depending on different physics assumptions and mathematical modeling \cite{TOTEM:2017sdy}. The $CL$ of the fit is 3.17$\times10^{-11}$\% $\ll0.1$\% (releasing the parameters $A_{qq}$ and $\lambda$ only a slight improvement can be achieved). This $CL$ value indicates that the ReBB model can not describe the most precise data on $pp$ elastic scattering measured at  \mbox{$\sqrt{s}=13$ TeV}. A $CL$ $\geq$ 0.1\% description of this dataset was obtained in Ref.~\cite{Csorgo:2018uyp} by utilizing the model-independent Lévy series technique.

A possible explanation of the statistically insufficient ReBB model description of the $\sqrt{s}=13$ TeV data is the so-called hollowness effect \textit{i.e.} a minimum in $\tilde\sigma_{\rm in}(s,b)$ observed at low values of $b$. A hollowness effect was observed in Ref.~\cite{Csorgo:2019egs} at $\sqrt{s}=13$ TeV with a statistical significance higher than $5\sigma$. Changes in the $s$-dependence of $B_0$ and $\sigma_{\rm el}/\sigma_{\rm tot}$ are also observed experimentally, suggesting that a new domain of QCD may emerge at higher energies, including 13 TeV \cite{Csorgo:2019fbf}. A generalization of the ReBB model that may account for some of these new features is presented in \cref{chap:levy}.

The ReBB model parameter values obtained in this section from statistically acceptable, $CL$ $\geq$ 0.1\% fits to the data at $\sqrt s$ = 0.546 TeV, 1.96 TeV, 2.76 TeV, and 7 TeV in the squared four-momentum range of $0.38~{\rm GeV^2}\leq -t \leq 1.2~{\rm GeV^2}$ are used to determine the energy dependencies of the ReBB model parameters in \cref{sec:rebb_endep_TeV}.

I did not present fits to $p\bar p$ differential cross section data at $\sqrt s$ = 630 GeV and 1.8 TeV. At \mbox{$\sqrt s = 630~{\rm GeV}$}, the data is only measured at higher-$|t|$ values, while at \mbox{$\sqrt s$ = 1.8 TeV,} the data is only measured at lower-$|t|$ values. For this reason, in both cases, the parameters obtained from these fits are unreliable for determining the parameters' energy dependence trends. However, as I will demonstrate in \cref{sec:rebb_test}, the ReBB model calibrated to data not including the $\sqrt s$ = 630 GeV and 1.8 TeV measurements describes the available $p\bar p$ differential cross section data at $\sqrt s$ = 630 GeV and 1.8 TeV in the squared four-momentum range of $0.38~{\rm GeV^2}\leq -t \leq 1.2~{\rm GeV^2}$ with a $CL$ $\geq$ 0.1\% when utilizing the $\chi^2$ definition as given by \cref{eq:chi2_refind}.

~
~

I did not present the ReBB model fit to $pp$ differential cross section data at \mbox{$\sqrt s$ = 8 TeV} either. The 8 TeV data at higher-$|t|$ values in the region of the diffractive minimum-maximum structure were published in Ref.~\cite{TOTEM:2021imi} only after the ReBB model study described in Ref.~\cite{Csorgo:2020wmw} was completed. However, as I will demonstrate in \cref{chap:odderon}, the ReBB model calibrated to data not including the   $\sqrt s$ = 8 TeV measurements describes the $pp$ differential cross section data at $\sqrt s$ = 8 TeV in the squared four-momentum transfer range of $0.38~{\rm GeV^2}\leq -t \leq 1.2~{\rm GeV^2}$ with a $CL$ $\geq$ 0.1\% when utilizing the $\chi^2$ definition as given by \cref{eq:chi2_refind}.


\section{Energy dependence in the ReBB model}\label{sec:rebb_endep_TeV}

In this section, I determine the energy dependencies of the ReBB model parameters, $R_q$, $R_d$, $R_{qd}$, and $\alpha_R$, utilizing the values and uncertainties of the parameters as obtained in the previous section from statistically acceptable, $CL$ $\geq$ 0.1\% fits of $pp$ and $p\bar p$ elastic scattering data at $\sqrt s$ = 0.546 TeV, 1.96 TeV, 2.76 TeV, and 7 TeV in the squared four-momentum transfer range of $0.38~{\rm GeV^2}\leq -t \leq 1.2~{\rm GeV^2}$.  Thus, four datasets at different energies are analyzed, and, at each energy, the values of four different physical parameters are determined. The resulting sixteen parameter values and their uncertainties, utilized in determining the energy dependencies of the ReBB model parameters, are summarized in \cref{tab:rebb_fit_parameters}.

\begin{table}[!htb]
\begin{center}
	{\begin{tabular}{ccccc} \hline\hline\noalign{\smallskip}
	$\sqrt{s}$ [TeV]  &0.546 ($p\bar p$)  &1.96 ($p\bar p$)   &2.76 ($pp$)	      &7.0 ($pp$)          \\  
	\noalign{\smallskip}\hline\noalign{\smallskip}
	$|t|$ [GeV$^{2}$] &(0.375, 1.210)     &(0.380, 1.200)     &(0.372, 0.741)     &(0.377, 1.205)    \\ 
	$\chi^{2}/NDF$	  &44.49/36           &8.22/12             &17.32/19           & 80.29/56         \\ 
	CL [\%] 	      &15.67              &76.77	          &56.82              & 1.83            \\ 
	$R_{q}$ [fm] 	  &0.349 $\pm$ 0.003  &0.396 $\pm$ 0.006  &0.42 $\pm$ 0.01  & 0.438 $\pm$ 0.005 ($\pm$ 0.001)\\ 
	$R_{d}$ [fm] 	  &0.825 $\pm$ 0.004  &0.87 $\pm$ 0.01  &0.88 $\pm$ 0.01  & 0.920 $\pm$ 0.009 ($\pm$ 0.002)\\ 
	$R_{qd}$ [fm]   &0.28 $\pm$ 0.01  &0.29 $\pm$ 0.03  &0.20 $\pm$ 0.08    & 0.33 $\pm$ 0.03 ($\pm$ 0.002)\\ 
	$\alpha_R$ 	      &0.117 $\pm$ 0.002  &0.163 $\pm$ 0.005  &0.126 $\pm$ 0.006  & 0.125 $\pm$ 0.002 ($\pm$ 0.001)\\ \hline
	$\epsilon_{b1}$   &--                 &--                 &-0.09 $\pm$ 0.95 &0.001 $\pm$ 0.003\\ 
	$\epsilon_{c1}$   &-0.4 $\pm$ 0.9 &-0.01 $\pm$ 0.83 & 0.06 $\pm$ 0.99 &-0.09 $\pm$ 0.87\\ 
	$\epsilon_{c2}$   &-0.1 $\pm$ 0.4 &--                 &--                 &--                \\ \hline \hline
	\end{tabular}}
	\end{center}
 \vspace{-0.5cm}
\caption{ReBB model and $\chi^2$ function parameter values with uncertainties as obtained by fitting the  $pp$ and $p\bar p$ elastic scattering data at $\sqrt s$ = 0.546 TeV, 1.96 TeV, 2.76 TeV, and 7 TeV in the squared four-momentum range of $0.38~{\rm GeV^2}\lesssim-t \lesssim 1.2~{\rm GeV^2}$ by utilizing the $\chi^2$ definition as given by \cref{eq:chi2_refind}.
For 7 TeV, the parameter error values shown in parenthesis do not include the contribution from the parameter correlations, i.e., these error values are less than the MINOS errors. 
}\vspace{-0.4cm}\label{tab:rebb_fit_parameters}
\end{table}

It is natural to expect that the scale parameters $R_q$, $R_d$, and $R_{qd}$ have the same values within errors in $pp$ and $p\bar p$ scattering. Consequently, the energy dependencies of the ReBB scale parameters within errors are expected to be the same in $pp$ and $p\bar p$ scattering. Thus, I fitted each ReBB model scale parameter known at four different energies with a linear-logarithmic shape as defined by \cref{eq:linlog}.  As I detail below, it turns out that indeed (i)~all ReBB model scale parameters in the c.m. energy range of \mbox{0.546 TeV $\leq \sqrt s \leq$ 7 TeV} are compatible with a linear-logarithmic energy dependence with confidence levels higher than 32\% and (ii)~these energy dependencies are the same within errors in $pp$ and $p\bar p$ scattering.

\begin{figure}[b!]
	\centering
	\includegraphics[width=0.8\linewidth]{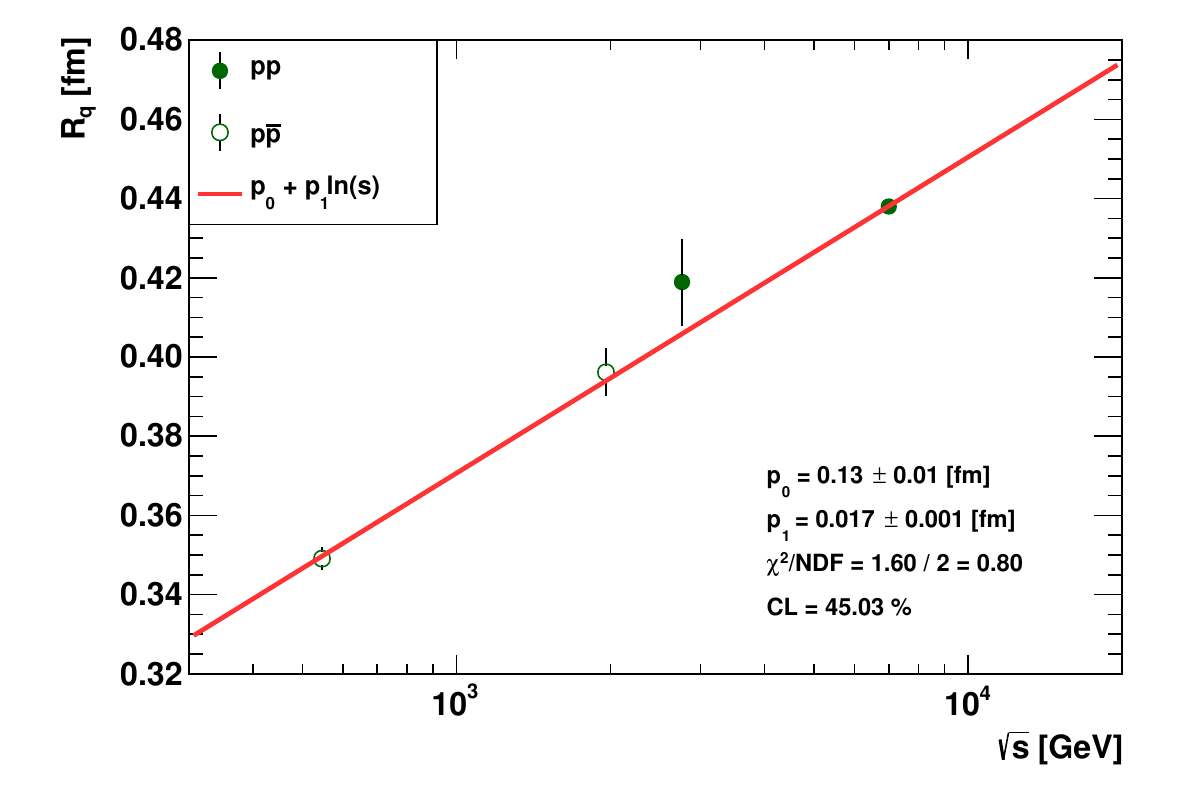}
 \vspace{-0.35cm}
	\caption{Energy dependence of the  ReBB model parameter $R_{q}$ in the TeV scale in the analysis of elastic $pp$ and $p\bar p$ scattering data utilizing the $\chi^2$ definition of \cref{eq:chi2_refind}.}
    \label{fig:reBB_model_log_lin_Rq}
\end{figure}

\begin{figure}[b!]
	\centering
	\includegraphics[width=0.8\linewidth]{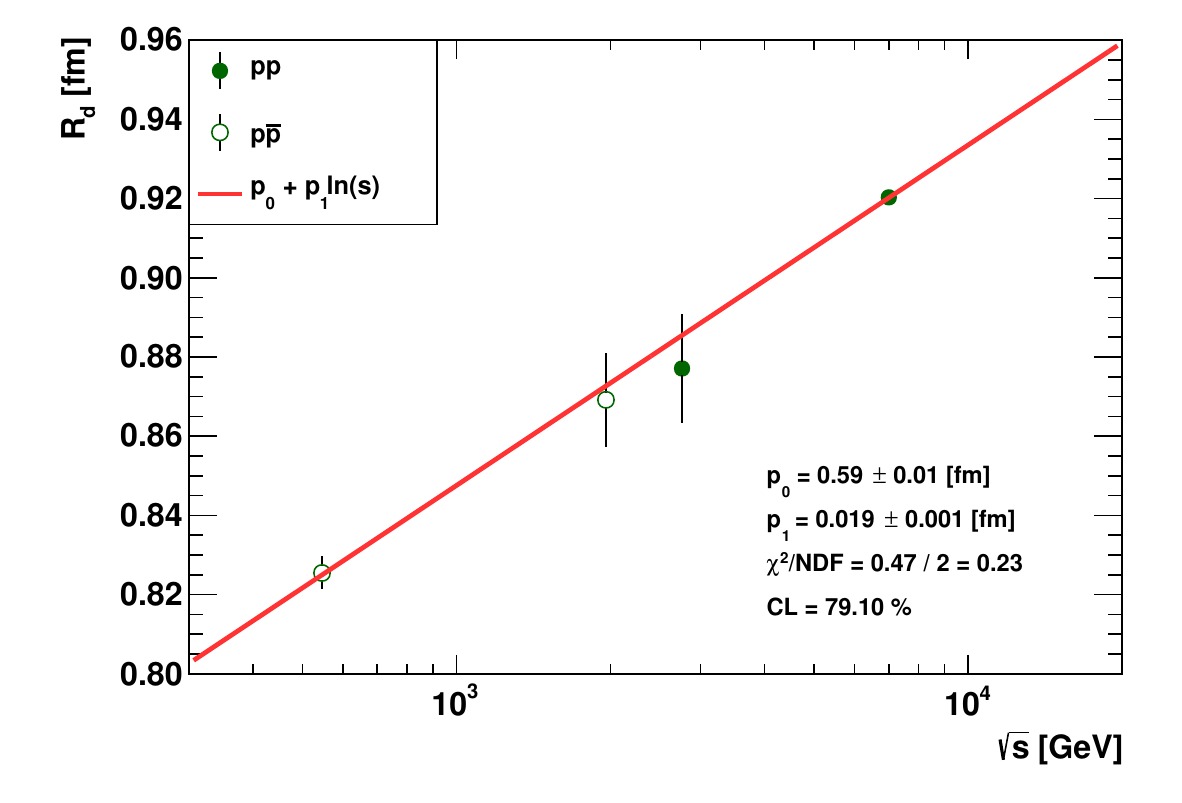}
 \vspace{-0.35cm}
	\caption{Energy dependence of the  ReBB model parameter $R_{d}$ in the TeV scale in the analysis of elastic $pp$ and $p\bar p$ scattering data utilizing the $\chi^2$ definition of \cref{eq:chi2_refind}.}
	\label{fig:reBB_model_log_lin_Rd}
\end{figure}

\begin{figure}[hbt!]
	\centering
	\includegraphics[width=0.8\linewidth]{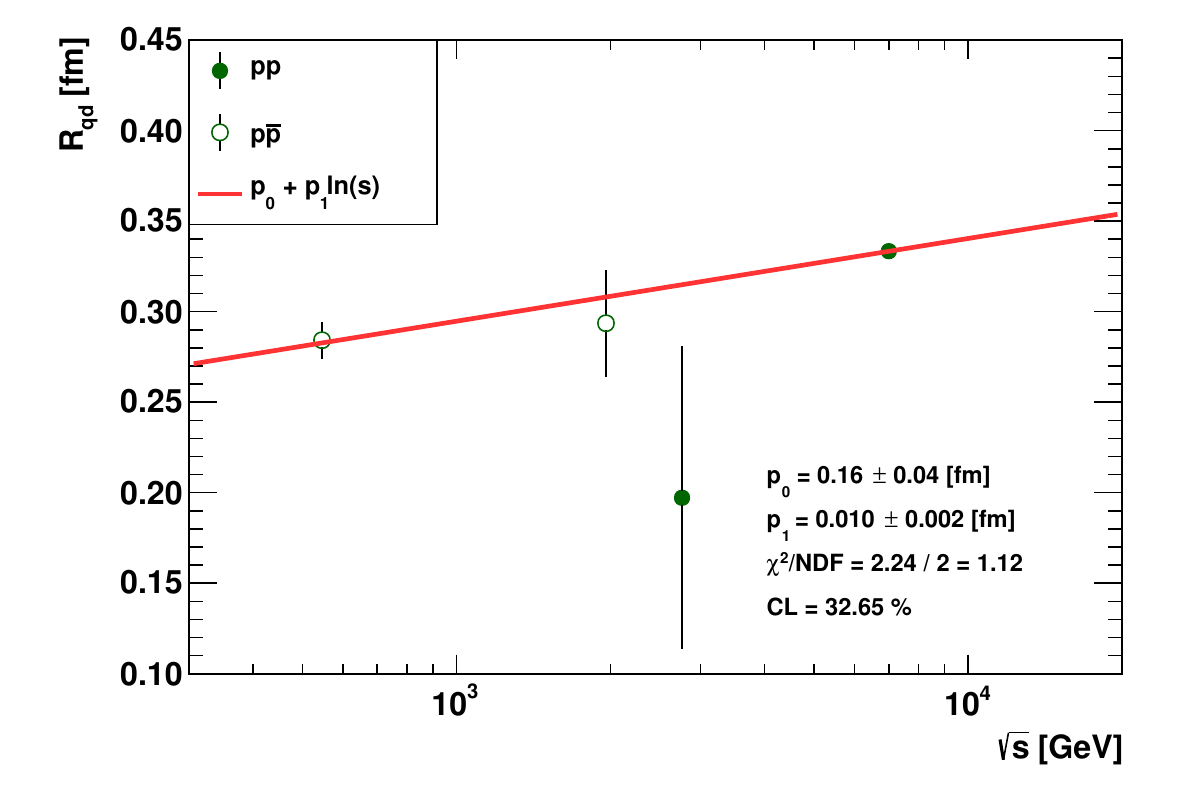}
 \vspace{-0.35cm}
	\caption{Energy dependence of the  ReBB model parameter $R_{qd}$ in the TeV scale in the analysis of elastic $pp$ and $p\bar p$ scattering data utilizing the $\chi^2$ definition of \cref{eq:chi2_refind}.}
	\label{fig:reBB_model_log_lin_Rqd}
\end{figure}

The energy dependence of each of the ReBB model scale parameters ($R_{q}$, $R_{d}$, and $R_{qd}$) in the TeV energy range is displayed in \cref{fig:reBB_model_log_lin_Rq}, \cref{fig:reBB_model_log_lin_Rd}, and \cref{fig:reBB_model_log_lin_Rqd}, respectively. The energy dependence of each scale parameter in $pp$ and $p\bar p$ scattering is compatible with a linear logarithmic shape as given by \cref{eq:linlog}. The $CL$ of the compatibility for the three different scale parameters ranges between 32.65\% and 79.10\%. The parameter values in \cref{eq:linlog} and the $CL$ value for each of the ReBB model scale parameters are summarised in \cref{tab:rebb_par_en_deps}.

\begin{figure}[hbt!]
	\centering
	\includegraphics[width=0.8\linewidth]{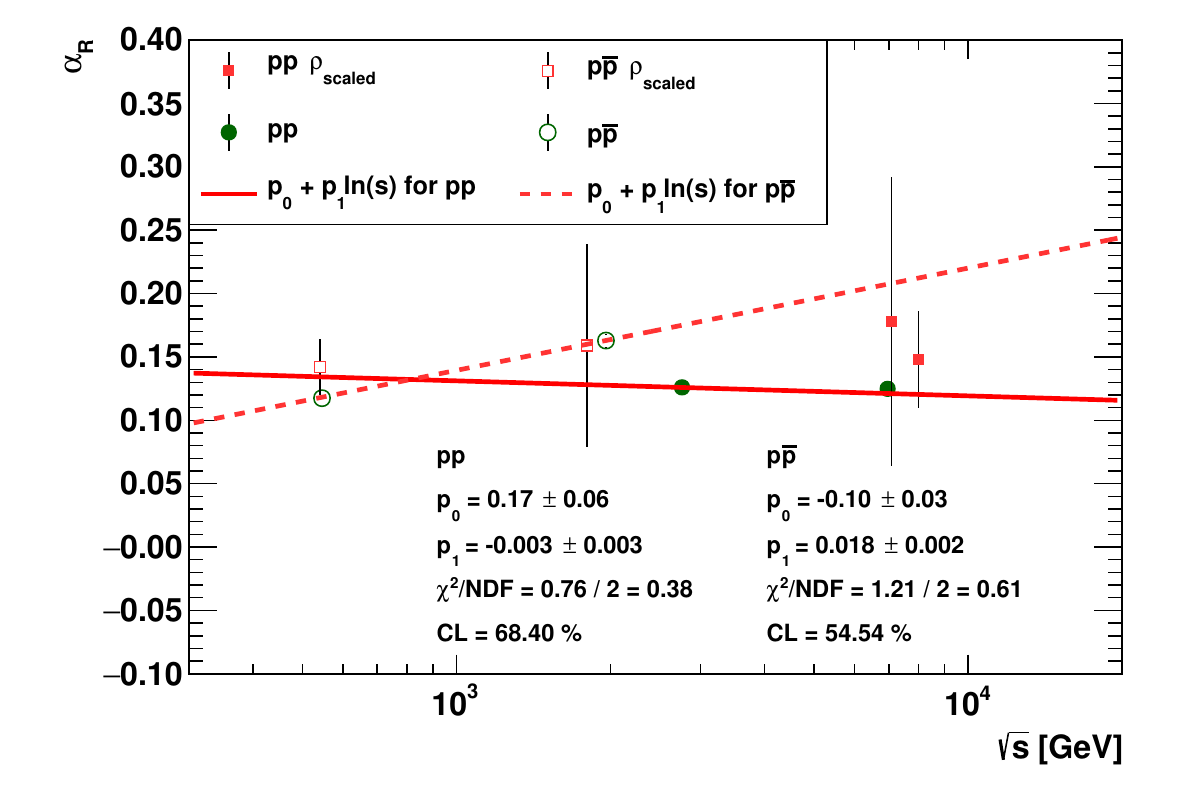}
  \vspace{-0.35cm}
	\caption{Energy dependence of the  ReBB model parameters $\alpha_R^{pp}$ and $\alpha_R^{p\bar p}$ in the TeV scale in the analysis of elastic $pp$ and $p\bar p$ scattering data utilizing the $\chi^2$ definition of \cref{eq:chi2_refind}. Data points at 7 TeV are slightly shifted to be separated for better visibility.}
	\label{fig:reBB_model_log_lin_alpha}
\end{figure}

In \cref{chap:oddTD0}, the scale parameter values at 7 TeV were not included when determining the energy evolution of the scale parameters. The scale parameter values at 7 TeV obtained by fitting the data down to $|t|=0.005$ GeV$^2$ were not compatible with lower energy data within a linear logarithmic energy dependence. This may be related to the jump in the energy dependence of $B_0$ in the energy interval of \mbox{3~TeV $\lesssim \sqrt s \lesssim$ 4 TeV} \cite{TOTEM:2017asr}. Here only the higher-$|t|$ $\sqrt{s}=$ 7 TeV data is included in the fit, which results in scale parameter values compatible with the lower energies within a linear trend. This may signal an incompatibility between lower-$|t|$ and higher-$|t|$ TOTEM data observed before also in the ReBB model analysis of Ref.~\cite{Nemes:2015iia}. Very interestingly, as it is presented in \cref{sec:rebb_test}, the ReBB model calibrated to the SPS UA4 $p\bar p$, Tevatron D0 $p\bar p$, and LHC TOTEM $pp$ $d\sigma_{el}/dt$ data in the higher-$|t|$ range
perfectly describes the LHC ATLAS $pp$ $\sigma_{tot}$ data being slightly below the LHC TOTEM $pp$ $\sigma_{tot}$ data. This shows that the ReBB model calibrated to the higher-$|t|$ data predicts a more moderate rise in energy nicely describing LHC ATLAS $pp$ $\sigma_{tot}$ data.  Further studies on the discrepancy between the lower-$|t|$ TOTEM and ATLAS data \cite{ATLAS:2022mgx} are needed in the future, but this issue with the data in the lower-$|t|$ kinematic domain does not affect the study of the odderon contribution at the higher-$|t|$ kinematic domain.

The energy dependence of the opacity parameter, $\alpha_R$,  in the TeV energy range for $pp$ and $p\bar p$ scattering is shown in \cref{fig:reBB_model_log_lin_alpha}. One can see that there is an $\alpha_R^{pp}$ and $\alpha_R^{p\bar p}$ opacity parameter, $i.e.$, the value of the opacity parameter is in general different in $pp$ and $p \bar p$ scattering. By fitting the $pp$ and $p\bar p$ elastic differential cross section data at four different c.m. energies, I determined two $\alpha_R^{pp}$ parameter values and two $\alpha_R^{p\bar p}$ parameter values. Since any two points fit into a linear curve, we need more information to more reliably determine the energy dependencies of the $pp$ and $p\bar p$ opacity parameters. This information can be obtained by rescaling the measured values of $\rho_0$ parameters. 

A simple analytical calculation allows us to demonstrate the relation between $\alpha_R$ and the $\rho_0$ in the ReBB model as follows. 
As we saw from the above fits summarized in \cref{tab:rebb_fit_parameters}, $\alpha_R \lesssim 0.2$ in all cases. The leading order term of the Taylor expansion of \cref{eq:ReBB_b_ampl}  in $\alpha_R$ gives the following approximations for the real and imaginary parts of the scattering amplitude: 
\begin{equation}\label{ReTelsb}
\text{Re} \, \widetilde T_{\rm el}(s,b) \simeq \alpha_R\, \tilde\sigma_{\rm in}(s,b)\, \sqrt{1-\tilde\sigma_{\rm in}(s,b)}
\end{equation}
and
\begin{equation}\label{ImTelsb}
\text{Im} \, \widetilde T_{\rm el}(s,b) \simeq 1-\sqrt{1-\tilde\sigma_{\rm in}(s,b)}.
\end{equation}

Approximating the imaginary part of the scattering amplitude with a Gaussian form, 
\begin{equation}\label{ImTelsbGauss}
\text{Im} \, \widetilde T_{\rm el}(s,b) \simeq \kappa(s) \, \exp\left(-\frac{b^2}{2R^2(s)}\right), 
\end{equation}
\cref{ImTelsb} results
\begin{equation}
\tilde\sigma_{\rm in}(s,b)\simeq 2 \kappa(s) \exp\left(-\frac{b^2}{2R^2(s)}\right) - \kappa(s)^2 \exp\left(-\frac{ b^2}{R^2(s)}\right) .
\end{equation}
Then using \cref{eq:relPWtoeik_2}, \cref{ReTelsb} and \cref{ImTelsb} gives
\begin{equation}\label{ReTelst}
\text{Re} \, T_{\rm el}(s,t) \simeq 2\pi R^2(s) \kappa(s)\alpha_R(s)\left(2-\frac{3}{2}\kappa(s)\, e^{\frac{tR^2(s)}{4}}+\frac{1}{3}\kappa^2(s)e^{\frac{tR^2(s)}{3}}\right) \, e^{\frac{tR^2(s)}{2}},
\end{equation}
and
\begin{equation}\label{ImTelst}
\text{Im} \, T_{\rm el}(s,t) \simeq 2\pi R^2(s) \kappa(s) \, e^{\frac{tR^2(s)}{2}}.
\end{equation}
Finally,
\begin{equation}
\rho_0(s)\simeq\alpha_R(s)\,\left(2-\frac{3}{2}\kappa(s)+\frac{1}{3}\kappa^2(s)\right) .
\label{eq:rho-vs-alpha-appendix}
\end{equation}

\begin{figure}[hbt!]
	\centering
	\includegraphics[width=0.8\linewidth]{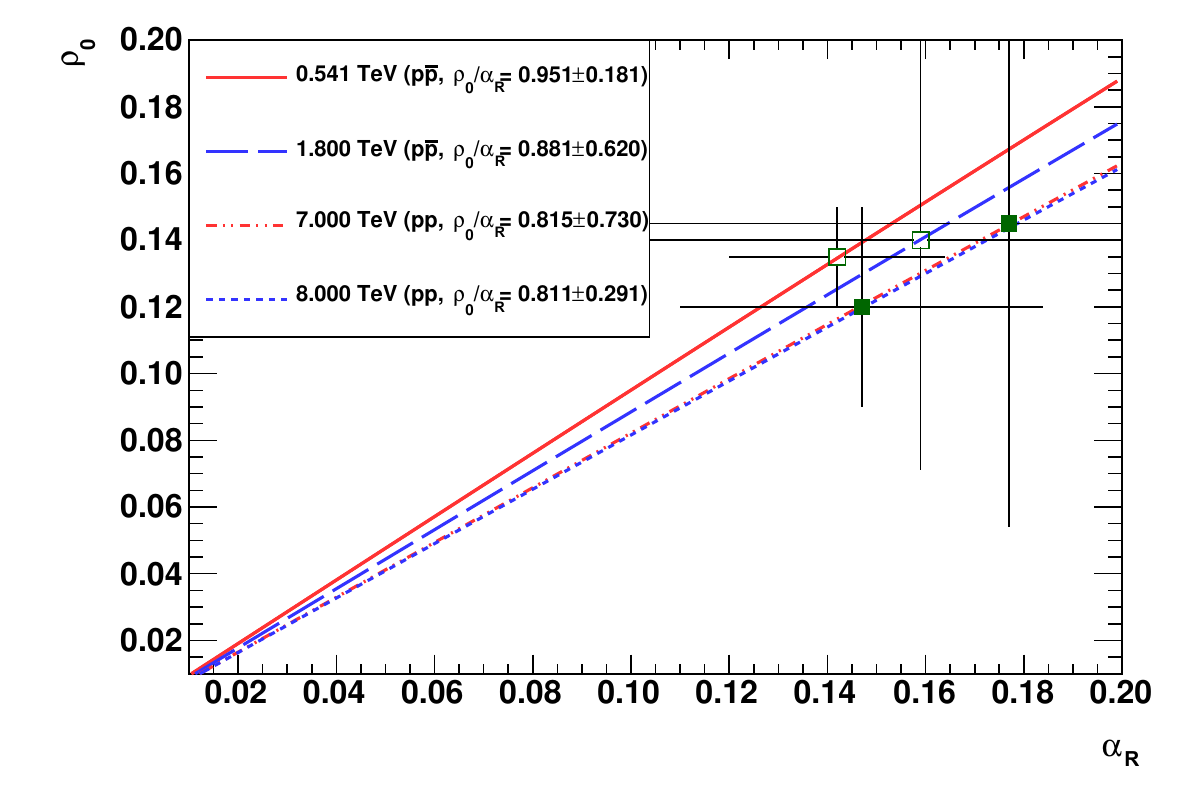}
  \vspace{-0.35cm}
	\caption{The $\alpha_R$ dependence of $\rho_0$  in the TeV energy range. The data points are generated numerically by using the trends of the ReBB model parameters, $R_q$, $R_d$, and $R_{qd}$, shown in \cref{fig:reBB_model_log_lin_Rq}, \cref{fig:reBB_model_log_lin_Rd}, and \cref{fig:reBB_model_log_lin_Rqd}, respectively, as well as the experimentally measured $\rho_0$ values. The curves are again generated numerically using the mentioned ReBB model scale parameter energy dependence trends.}
	\label{fig:rho0-vs-alpha-LHC}
\end{figure}

\cref{eq:rho-vs-alpha-appendix} gives the relation between $\rho_0$ and $\alpha_R$ when the imaginary part of the scattering amplitude in the impact parameter representation is approximated by a Gaussian form of \cref{ImTelsbGauss}.  Alternatively, one can calculate the $\alpha_R$ dependence of $\rho_0$ numerically by using the trends of the ReBB model parameters, $R_q$, $R_d$, $R_{qd}$, shown in \cref{fig:reBB_model_log_lin_Rq}, \cref{fig:reBB_model_log_lin_Rd}, and \cref{fig:reBB_model_log_lin_Rqd} from the full ReBB model without any assumptions and/or approximations. This result is shown on \cref{fig:rho0-vs-alpha-LHC} together with data points generated numerically using the mentioned ReBB model scale parameter energy dependence trends and the experimentally measured $\rho_0$ values listed in \cref{tab:rho0toalpha_R}.  

The one-to-one correspondence between $\alpha_R$ and the $\rho_0$ in the ReBB model at a given energy allows us to rescale the experimentally measured $\rho_0$ parameter to a corresponding $\alpha_R$ value. The rescaled values are listed in \cref{tab:rho0toalpha_R}, while the scaling factors rounded to three decimal places, are shown in the legends of \cref{fig:rho0-vs-alpha-LHC}.

Note that the opacity parameter $\alpha_R$ regulates the value of the differential cross section in the diffractive minimum. Since $\alpha_R\sim\rho_0$, there is a deep connection between the $t=0$ and the $t\neq0$ domain within the ReBB model.

\begin{table}[h]\small
\vspace{0.5cm}
	\centering
	\begin{tabular}{ccc} \hline\hline
		$\sqrt{s}$ [TeV]   & $\rho_0$      & $\alpha_R$	  \\ \hline
		0.541 ($p\bar p$)  &0.135$\pm$0.015   & 0.142$\pm$0.022  \\
		1.800 ($p\bar p$)  &0.140$\pm$0.069   & 0.159$\pm$0.080   \\ 
		7.000 	  ($pp$)   & 0.145$\pm$0.091  & 0.177$\pm$0.114		  \\  
            8.000 	  ($pp$)   & 0.120$\pm$0.030  & 0.147$\pm$0.037  \\  \hline\hline
	\end{tabular}
	\caption{Values of the measured $\rho_0$ at $\sqrt{s}=$ 541 GeV ($p\bar p$) \cite{UA42:1993sta}, 1.8 TeV ($p\bar p$) \cite{E710:1991bcl}, \mbox{7 TeV ($pp$) \cite{TOTEM:2013vij}} and 8 TeV ($pp$) \cite{TOTEM:2016lxj} with the corresponding rescaling to $\alpha_R$ parameter. The scaling factors rounded to three decimal places are shown in the legends of \cref{fig:rho0-vs-alpha-LHC}.}
	\label{tab:rho0toalpha_R}
\end{table} 

\begin{figure}[hbt!]
	\centering
   \vspace{-0.15cm}
	\includegraphics[width=0.8\linewidth]{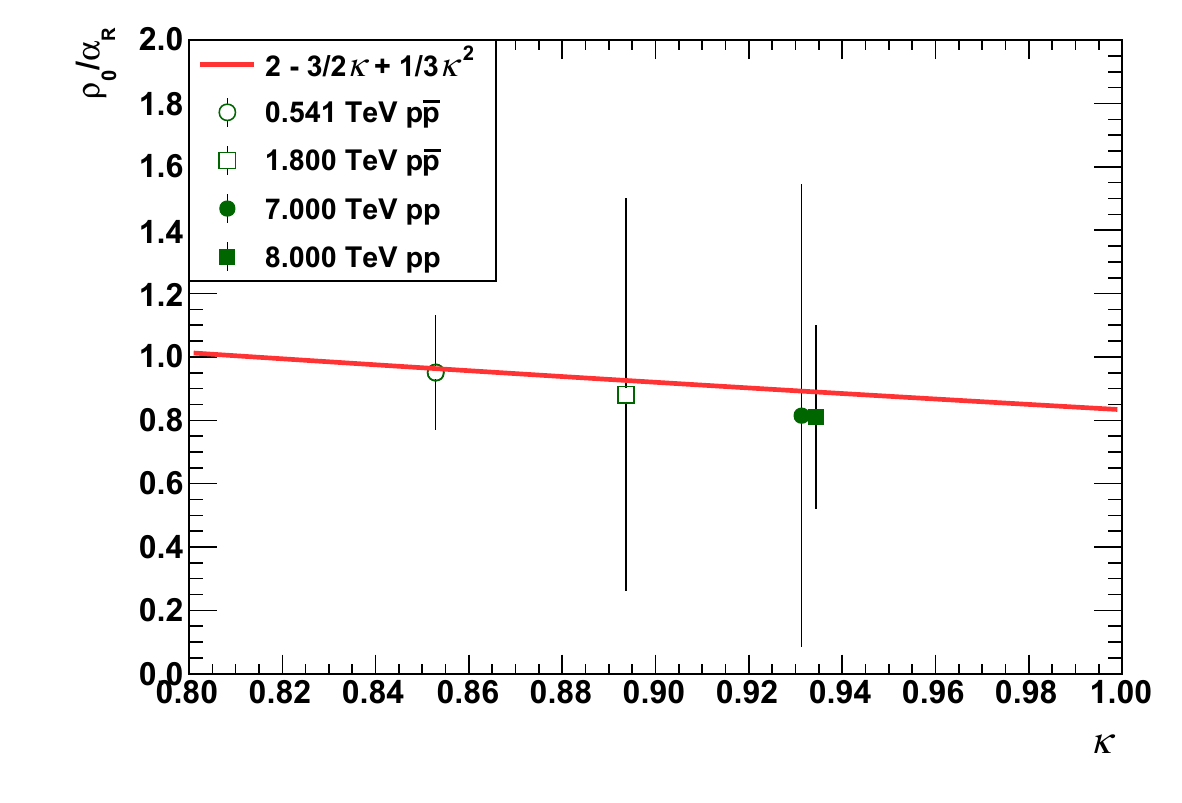}
  \vspace{-0.55cm}
	\caption{The $\kappa(s)=\tilde\sigma_{\rm in}(s,b=0)$ dependence of $\rho_0/\alpha_R$ in the TeV energy range. The data points are generated numerically by using the trends of the ReBB model parameters, $R_q$, $R_d$, $R_{qd}$, shown in \cref{fig:reBB_model_log_lin_Rq}, \cref{fig:reBB_model_log_lin_Rd}, and \cref{fig:reBB_model_log_lin_Rqd}, as well as the experimentally measured $\rho_0$ values. The red curve represents the result of the analytical result of  \cref{eq:rho-vs-alpha-appendix} that shows a good agreement with the numerical calculations.}
	\label{fig:rho0-per-alpha-vs-P0-LHC}
\end{figure}

\cref{fig:rho0-per-alpha-vs-P0-LHC} shows the comparison of the analytical approximation of \cref{eq:rho-vs-alpha-appendix} to the $\rho_0/\alpha_R$ data points generated numerically by using the trends of the ReBB model parameters, $R_q$, $R_d$, $R_{qd}$, shown in \cref{fig:reBB_model_log_lin_Rq}, \cref{fig:reBB_model_log_lin_Rd}, and \cref{fig:reBB_model_log_lin_Rqd}, as well as the experimentally measured $\rho_0$ values. One can conclude that the analytical approximation of \cref{eq:rho-vs-alpha-appendix} works remarkably well. 

Let us now discuss in more detail the energy dependencies of $\alpha_R^{pp}$ and $\alpha_R^{p\bar p}$. 

The energy dependence of $\alpha_R^{pp}$ is compatible with a linear-logarithmic shape as given by \cref{eq:linlog} with parameter values $p_0$ = 0.17 $\pm$ 0.06  and $p_1$ = $-$0.003 $\pm$ 0.003. The $CL$ of the compatibility is 68.40\% (see \cref{fig:reBB_model_log_lin_alpha} and \cref{tab:rebb_par_en_deps}). We can see that the value of $p_1$ is zero within its uncertainty. This suggests that in the energy range \mbox{0.546~TeV~$\leq\sqrt{s}\leq 7$~TeV,} the $\alpha_R^{pp}$ is practically constant.  

The energy dependence of $\alpha_R^{p\bar p}$ is again compatible with a linear-logarithmic shape as given by \cref{eq:linlog} with parameter values $p_0$ = $-$0.10 $\pm$ 0.03  and $p_1$ = 0.018 $\pm$ 0.002. 
The $CL$ of the compatibility is 54.54\% (see \cref{fig:reBB_model_log_lin_alpha} and \cref{tab:rebb_par_en_deps}). These values are significantly different from those obtained for $\alpha_R^{pp}$, consequently, the energy dependence of $\alpha_R^{pp}$ is significantly different from that of $\alpha_R^{p\bar p}$. This difference is clearly the effect of the odderon exchange since all the other ReBB model parameter values at a given energy are the same in $pp$ and $p\bar p$ scattering. A common fit to the $\alpha_R^{pp}$ and $\alpha_R^{p\bar p}$ values in the energy range of \mbox{0.546~TeV~$\leq\sqrt{s}\leq 8$~TeV} by a linear-logarithmic shape of \cref{eq:linlog} is characterized by a confidence level value of $CL=1.62\times10^{-11}$\% corresponding to a $t$-channel odderon exchange signal with a statistical significance higher than $7\sigma$. We can also see in \cref{fig:reBB_model_log_lin_alpha} that, at $\sqrt{s}\approx 800$ GeV, $\alpha_R^{pp}=\alpha_R^{p\bar p}$ predicting a vanishing odderon contribution at $\sqrt{s}\approx 800$ GeV.


The $\alpha_R$ were not used in the numerics used to rescale $\rho_0$. Only the scale parameters, $R_q$, $R_d$, $R_{qd}$, were used in the numerics to rescale $\rho_0$. Thus, the rescaled points can be considered as independent from the $\alpha_R$ values obtained by fitting the differential cross section data. Since the rescaled values have much higher uncertainties than the $\alpha_R$ values obtained by fitting the differential cross section data, the energy dependence trends of $\alpha_R$ in $pp$ and $p\bar p$ processes are dominated by the $\alpha_R$ values obtained by fitting the differential cross section data.

Looking~ at~ the~ above~ results,~ one~ may~ conclude~ that~ the~ ReBB~ model describes elastic $pp$ and $p\bar p$ scattering data in the kinematic range of \mbox{0.546~TeV~$\leq\sqrt{s}\leq 7$~TeV} and \mbox{0.38~GeV$^2$~$\leq -t\leq1.2$~GeV$^2$} based on $5\times2=10$ parameters. The values of these 10 parameters are summarized in \cref{tab:rebb_par_en_deps}. In \cref{chap:odderon}, these results are validated and utilized to perform a model-dependent odderon search.

\begin{table}[!hbt]
     \begin{subtable}[h]{0.99\textwidth}
        \centering
    \begin{tabular}{cccc}
    \hline\hline\noalign{\smallskip}
Parameter      & $R_{q}$ [$\rm fm$]  & $R_{d}$ [$\rm fm$]  & $R_{qd}$ [$\rm fm$]    \\ \hline
	$\chi^{2}/NDF$ & $1.596/2$       & $0.469/2$       & $2.239/2$ 	      \\	
	CL [\%]		   & 45.03	         & 79.10           & 32.65 	               \\	\hline
	$p_{0}$         & $0.13\pm0.01$ & $0.59\pm0.02$ & $0.16\pm0.04$  \\ 
	$p_{1}$         & $0.017\pm0.001$ & $0.019\pm0.001$ & $0.010\pm0.002$ 	\\\hline\hline
    \end{tabular}
           \caption{}
       \label{tab:sub1}
    \end{subtable}
    \vfill
    \vspace{0.7cm}
    \begin{subtable}[h]{0.99\textwidth}
        \centering
    \begin{tabular}{ccc}
    \hline\hline\noalign{\smallskip}
Parameter       & $\alpha_R$ ($pp$)      &$\alpha_R$ ($p\bar p$)  \\ \hline
	$\chi^{2}/NDF$	      & $0.760/2$             &$1.212/2$        \\	
	CL [\%]		  	          & 68.40                 &54.54            \\	\hline
	$p_{0}$          & $0.17\pm0.06$      &$-0.10\pm0.03$  \\ 
	$p_{1}$         & $-0.003\pm0.003$     &$0.018\pm0.002$ 	\\ \hline\hline
    \end{tabular}
           \caption{}
       \label{tab:sub2}
    \end{subtable}
    \caption{Parameter values that determine the linear-logarithmic energy dependencies of the ReBB model parameters according to \cref{eq:linlog} in the analysis of elastic $pp$ and $p\bar p$ scattering data utilizing the $\chi^2$ definition of \cref{eq:chi2_refind}. 
    \label{tab:rebb_par_en_deps}}
\end{table}

\textbf{Summary}
\vspace{0.2cm}

In this Chapter, we saw that the ReBB model describes elastic $pp$ and $p\bar p$ scattering in a statistically acceptable manner utilizing the refined $\chi^2$ formula of \cref{eq:chi2_refind}. The analysis of the data in the kinematic range of 0.546~TeV~$\leq\sqrt{s}\leq 7$~TeV and \mbox{0.38~GeV$^2$~$\leq -t\leq1.2$~GeV$^2$}  shows that: (i) the values of the ReBB model parameters, \mbox{$\lambda=1/2$} and $A_{qq}=1$, are compatible with the analyzed data independently from the energy; (ii) the values of the ReBB model scale parameters, $R_q$, $R_d$, and $R_{qd}$, rise with increasing $s$, and their energy dependencies are compatible with the same linear-logarithmic functional forms in elastic $pp$ and $p\bar p$ scattering;  (iii) the parameter $\alpha_R^{pp}$ is compatible with a constant value in the analyzed energy range while the value of the parameter $\alpha_R^{p\bar p}$ rises with increasing $s$ and its energy dependence is again compatible with a linear-logarithmic functional form. The rise of the ReBB model scale parameters with increasing $s$ implies that the constituent quark sizes and their distance in a(n) (anti)proton are rising with increasing $s$. We can conclude that, in the ReBB model, all the differences between elastic $pp$ and  $p\bar p$ scattering is encoded in the value of the opacity parameter $\alpha_R$ given that this is the only ReBB model parameter whose energy dependence is incompatible with the same linear curve in elastic $pp$ and  $p\bar p$ scattering. 

Finally, I emphasize that: (i) in \cref{chap:ReBBpbarp}, I demonstrated using the traditional $\chi^2$ formula \cref{eq:chi_trad0} that the ReBB model gives a reasonable description to all the available $p\bar p$ elastic scattering data; (ii) in \cref{chap:oddTD0}, I presented preliminary results on a joint analysis of elastic $pp$ and $p\bar p$ scattering data in the TeV energy domain using the ReBB model and the $\chi^2$ formula \cref{eq:chi_Cj++} that treats not only the statistical errors of the measured data but also the systematic errors; (iii) in this Chapter, I presented the final results on the joint ReBB model analysis of elastic $pp$ and $p\bar p$ scattering data in the TeV energy domain using the $\chi^2$  formula \cref{eq:chi2_refind} derived by the PHENIX Collaboration that again treats not only the (vertical and horizontal) statistical errors of the measured data but also the (vertical and horizontal) systematic errors in a way equivalent to the use of the covariance matrix of errors.


\clearpage
\chapter{Model-dependent odderon observation}\label{chap:odderon}

Any statistically significant difference between $pp$ and $p\bar p$ elastic scattering in the TeV c.m. energy region in the same kinematic ($s$, $t$) domain is a sign of odderon exchange. 
In this chapter, using the results of the refind ReBB model analysis of elastic $pp$ and $p\bar p$ scattering data, presented in \cref{chap:rebbdesc}, I compare the $pp$ and $p\bar p$ elastic differential cross sections at exactly the same energies in a common squared four-momentum transfer domain in the TeV c.m. energy range. By comparing the ReBB model extrapolations to experimental $pp$ and $p\bar p$ elastic scattering data, I find model-dependent, discovery-level odderon exchange signals. 

In \cref{sec:rebb_pomodd}, I detail the pomeron and odderon within the framework of the ReBB model. In  \cref{sec:rebb_test}, I show that the ReBB model with the energy dependence as determined in \cref{sec:rebb_endep_TeV} of \cref{chap:rebbdesc} describes all the experimentally measured $pp$ and $p\bar p$ elastic differential cross section datasets in the kinematic range of \mbox{0.546~TeV~$\leq\sqrt{s}\leq 8$~TeV} and \mbox{0.38~GeV$^2$~$\leq -t\leq1.2$~GeV$^2$} in a statistically acceptable manner, $i.e.$, with \mbox{$CL$ $>$ 0.1\%.} By using this validated ReBB model description, in \cref{sec:rebbextraps}, I find that the $pp$ and $p\bar p$ elastic differential cross sections differ with a statistical significance of at least 6.3$\sigma$ when significances obtained at 1.96 TeV and 2.76 TeV are combined. Then I find, by combining the significances obtained at all the four analyzed energies, $i.e.$, 1.96~TeV, 2.76~TeV, 7 TeV and 8 TeV, that the statistical significance of the difference between $pp$ and $p\bar p$ elastic differential cross sections, \textit{i.e.}, the statistical significance of the $t$-channel odderon exchange signal is higher than $30\sigma$. While in \cref{chap:oddTD0}, I presented preliminary results on the ReBB model odderon analysis, in the present Chapter, I discuss the final results of the odderon observation within the ReBB model analysis of elastic $pp$ and $p\bar p$ scattering data in the TeV c.m. energy domain.

This chapter is based on Refs.~\cite{Szanyi:2022ezh,Csorgo:2020wmw,Szanyi:2022qgx}.

\newpage


\section{Pomeron and odderon in the ReBB model}\label{sec:rebb_pomodd}

Associating the crossing-even part of the elastic $pp$ and $p\bar p$  scattering amplitudes with the pomeron exchange, and the crossing-odd part with the odderon exchange, I present in this section the explicit analytic formulas for the pomeron and odderon amplitudes in the impact parameter representation, within the framework of the ReBB model. These associations are valid in the TeV c.m. energy domain where the reggeon contributions are negligibly small (see \cref{sec:Regge} and \cref{sec:oddintro}).

\cref{eq:M_ampl_pp} and \cref{eq:M_ampl_pbarp} gives the decomposition of the $pp$ and $p\bar p$ elastic scattering amplitudes into crossing-even and crossing-odd parts. Using \cref{eq:MtoT} we can rewrite \cref{eq:M_ampl_pp} and \cref{eq:M_ampl_pbarp} as:
\begin{equation}\label{eq:pp_ampl}
    T_{\rm el}^{pp}  =  T_{\rm el}^+ - T_{\rm el}^- ,
\end{equation}
\vspace{-0.7cm}
\begin{equation}\label{eq:pbarp_ampl}
    T_{\rm el}^{p\bar p}  =  T_{\rm el}^+ + T_{\rm el}^- .
\end{equation}
Then using \cref{eq:pp_ampl} and \cref{eq:pbarp_ampl} we can rewrite \cref{eq:ampl_M_even} and \cref{eq:ampl_M_odd} as
\begin{equation}\label{eq:even-amplitude}
    T_{\rm el}^{+}  =  \frac{1}{2} \left(T_{\rm el}^{p\bar p} + T_{\rm el}^{p p}\right) ,
\end{equation}
\vspace{-0.7cm}
\begin{equation}\label{eq:odd-amplitude}
    T_{\rm el}^{-}  =  \frac{1}{2}  \left(T_{\rm el}^{p\bar p} - T_{\rm el}^{pp} \right).
\end{equation}
In the $\sqrt s \geq 1$ TeV domain, we have:
\begin{equation}\label{eq:pomeron-amplitude}
    T_{\rm el}^{\mathbb P}\equiv T_{\rm el}^{+}
\end{equation}
and
\begin{equation}\label{eq:odderon-amplitude}
    T_{\rm el}^{\mathbb O}\equiv T_{\rm el}^{-}.
\end{equation}

It is obvious from the above formulas that the knowledge of the elastic $pp$ and $p\bar p$ scattering amplitudes straightforwardly yields the $s$ and $t$ dependence of the pomeron and odderon exchange amplitudes. Based on the eikonal form of the elastic scattering amplitude, \cref{eq:impact_ampl_eik_sol}, we can write model-independently  the elastic $pp$ and $p\bar p$  scattering amplitudes in the impact parameter representation as
\begin{eqnarray}
    \widetilde T_{\rm el}^{pp}(s,b)  & = & i \, \left(1 - e^{-\Omega_{pp}(s,b)} \right) \,, \\  
    \widetilde T_{\rm el}^{p\bar p}(s,b)  & = & i \, \left(1 - e^{-\Omega_{p\bar p}(s,b)} \right) \,,
\end{eqnarray}
where $\Omega_{pp}(s,b)$ is the opacity function for elastic $pp$ scattering and $\Omega_{p\bar p}(s,b)$ is that for elastic $p\bar p$ scattering.
Then \cref{eq:pomeron-amplitude} and \cref{eq:odderon-amplitude} with  \cref{eq:even-amplitude} and \cref{eq:odd-amplitude} results 
\begin{equation}\label{eq:pomeron-amp-op}
    \widetilde T_{\rm el}^{\mathbb P}(s,b)   = i \, \left(1 - \frac{1}{2} \left(e^{-\Omega_{pp}(s,b)} + e^{-\Omega_{p\bar p}(s,b) }\right)\right) \,,
\end{equation}
\vspace{-0.6cm}
\begin{equation}\label{eq:odderon-amp-op}
    \widetilde T_{\rm el}^{\mathbb O}(s,b)  = \frac{i}{2}
                            \left(e^{-\Omega_{pp}(s,b)} - e^{-\Omega_{p\bar p}(s,b)}\right) \,.
\end{equation}

It was concluded in \cref{chap:rebbdesc} that the energy dependencies of the ReBB model parameters, except for the opacity parameter $\alpha_{R}$, are compatible with the same linear-logarithmic curves in $pp$ and $p\bar p$ scattering. Since the opacity parameter shows up only in the imaginary part of the opacity function as given by \cref{eq:im_omega}, one can write
\begin{equation}\label{eq:im_omega_pp}
     {\rm Im} \Omega^{pp}(s,b)  =  - \alpha^{pp}_R(s) \tilde\sigma_{\rm in}(s,b),
\end{equation}
\begin{equation}\label{eq:im_omega_pbarp}
    {\rm Im} \Omega^{p\bar p}(s,b)  =  - \alpha^{p\bar p}_R(s) \tilde\sigma_{\rm in}(s,b).
\end{equation}
Then substituting \cref{eq:im_omega_pp}, \cref{eq:im_omega_pbarp}, and \cref{eq:re_omega} into  \cref{eq:impact_ampl_eik_sol} we obtain
\begin{equation}\label{eq:rebbppb}
    \widetilde T_{\rm el}^{pp}(s,b) = i\left(1-e^{i\, \alpha^{pp}_R(s) \, \tilde\sigma_{\rm in}(s,b)} \, 
    \sqrt{1-\tilde\sigma_{\rm in}(s,b)}\right) ,
\end{equation}
\begin{equation}\label{eq:rebbpbarpb}
    \widetilde T_{\rm el}^{p\bar p}(s,b) = i\left(1-e^{i\, \alpha^{p\bar p}_R(s) \, \tilde\sigma_{\rm in}(s,b)} \, 
    \sqrt{1-\tilde\sigma_{\rm in}(s,b)}\right).
\end{equation}
\cref{eq:rebbppb} and \cref{eq:rebbpbarpb} with \cref{eq:pomeron-amplitude},  \cref{eq:odderon-amplitude}, \cref{eq:even-amplitude}, and \cref{eq:odd-amplitude} yields the explicit form of the real and imaginary part of the ReBB model pomeron amplitude,
\begin{equation}\label{eq:Re-Pomeron}
    {\rm Re}\,  \widetilde T_{\rm el}^{\mathbb P}(s,b)   = 
        \sqrt{1-\tilde\sigma_{\rm in} } \, 
        \sin\left( \frac{\alpha^{pp}_R + \alpha^{p\bar p}_R}{2}\tilde\sigma_{\rm in}  \right) 
        \cos\left( \frac{\alpha^{p\bar p}_R - \alpha^{pp}_R}{2}\tilde\sigma_{\rm in}  \right) \,,
\end{equation}
\begin{equation}\label{eq:Im-Pomeron}
    {\rm Im }\, \widetilde T_{\rm el}^{\mathbb P}(s,b)   = 
        1 - \sqrt{1-\tilde\sigma_{\rm in} } \,
        \cos\left( 
        \frac{\alpha^{pp}_R + \alpha^{p\bar p}_R}{2} \tilde\sigma_{\rm in} \right)
            \cos\left( 
            \frac{\alpha^{p\bar p}_R - \alpha^{pp}_R}{2} \tilde\sigma_{\rm in}  \right) \,,
\end{equation}
and those of the ReBB model odderon amplitude,
\begin{equation}\label{eq:Re-Odderon}
    {\rm Re }\, \widetilde T_{\rm el}^{\mathbb O}(s,b)   =  \sqrt{1-\tilde\sigma_{\rm in} } 
        \,
        \sin\left( \frac{\alpha^{p\bar p}_R - \alpha^{pp}_R}{2}\tilde\sigma_{\rm in}  \right)
            \cos\left( \frac{\alpha^{p\bar p}_R + \alpha^{pp}_R}{2}\tilde\sigma_{\rm in}  \right),
\end{equation}
\begin{equation}\label{eq:Im-Odderon}
    {\rm Im}\, \widetilde T_{\rm el}^{\mathbb O}(s,b)   =  \sqrt{1-\tilde\sigma_{\rm in} }   \,
        \sin\left( \frac{\alpha^{p\bar p}_R - \alpha^{pp}_R}{2}\tilde\sigma_{\rm in}  \right) 
        \sin\left( \frac{\alpha^{pp}_R + \alpha^{p\bar p}_R}{2}\tilde\sigma_{\rm in}  \right),
\end{equation}
where I used the shorthand notations $\tilde\sigma_{\rm in}\equiv\tilde\sigma_{\rm in}(s,b)$, $\alpha^{pp}_R\equiv \alpha^{pp}_R(s)$ and $\alpha^{p\bar p}_R\equiv\alpha^{p\bar p}_R(s)$.  Alternatively these formulas can be obtained by substituting \cref{eq:im_omega_pp}, \cref{eq:im_omega_pbarp}, and \cref{eq:re_omega} into \cref{eq:pomeron-amp-op} and \cref{eq:odderon-amp-op}. 

What can we say about the pomeron and odderon amplitudes within the ReBB model? In the TeV c.m. energy domain both $\alpha^{pp}_R$ and $\alpha^{p\bar p}_R$ are smaller than 0.2. Also the difference between $\alpha^{pp}_R$ and $\alpha^{p\bar p}_R$ is small, smaller than 0.1. Unitarity sates that $\tilde\sigma_{\rm in}(s,b)\leq1$. Then it follows from \cref{eq:Re-Pomeron} and \cref{eq:Im-Pomeron} that the pomeron is predominantly imaginary with a small real part proportional to $\sin\left( \left(\alpha^{pp}_R + \alpha^{p\bar p}_R\right)\tilde\sigma_{\rm in} /2 \right)$. Looking at \cref{eq:Re-Odderon} and \cref{eq:Im-Odderon}, we see that the odderon amplitude is predominantly real with a small imaginary part proportional to $\sin\left( \left(\alpha^{pp}_R + \alpha^{p\bar p}_R\right)\tilde\sigma_{\rm in} /2 \right)$. If $\alpha^{pp}_R$ = $\alpha^{p\bar p}_R$, the odderon amplitude vanishes since both the real and imaginary parts of the odderon amplitude are proportional to $\sin\left( \left(\alpha^{pp}_R - \alpha^{p\bar p}_R\right)\tilde\sigma_{\rm in} /2 \right)$.

\begin{figure}[!b]
	\centering
	\subfloat[
	\label{fig:par_Rq_lin}]{%
\includegraphics[width=0.5\linewidth]{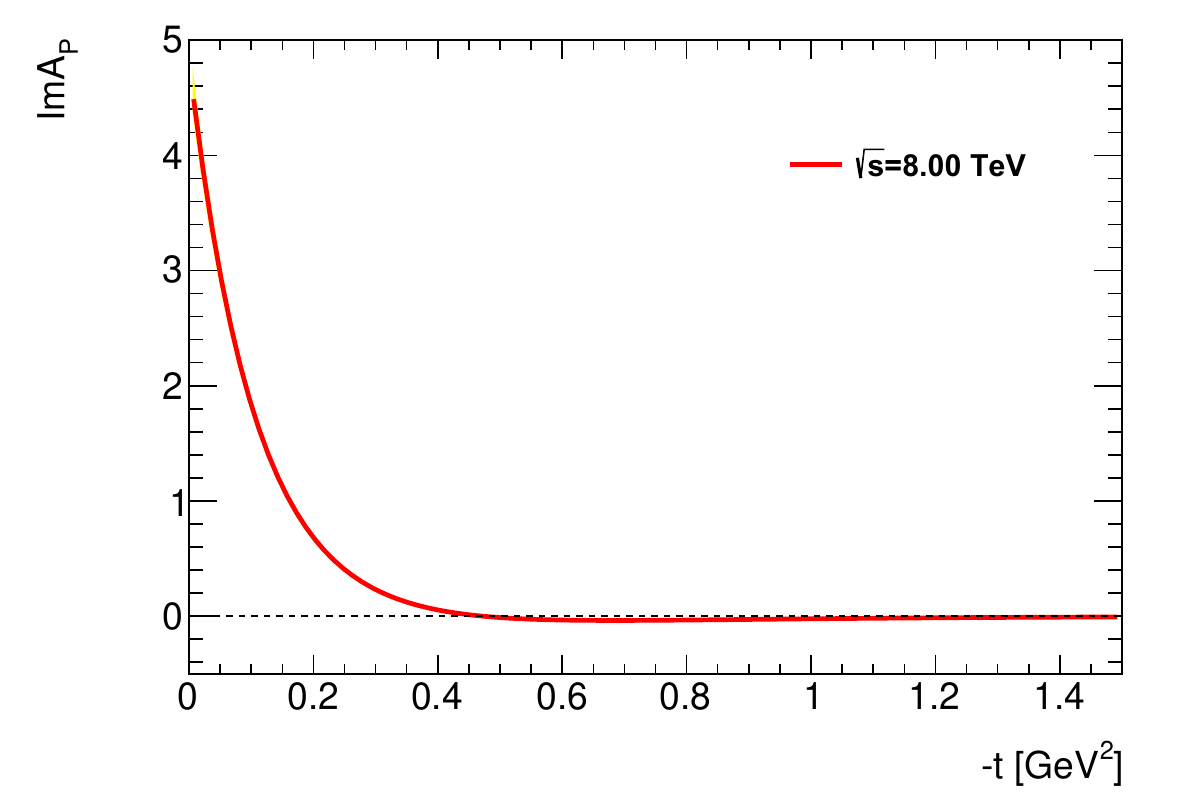}%
	}\hfill
	\subfloat[
        \label{fig:par_Rd_lin}]{%
\includegraphics[width=0.5\linewidth]{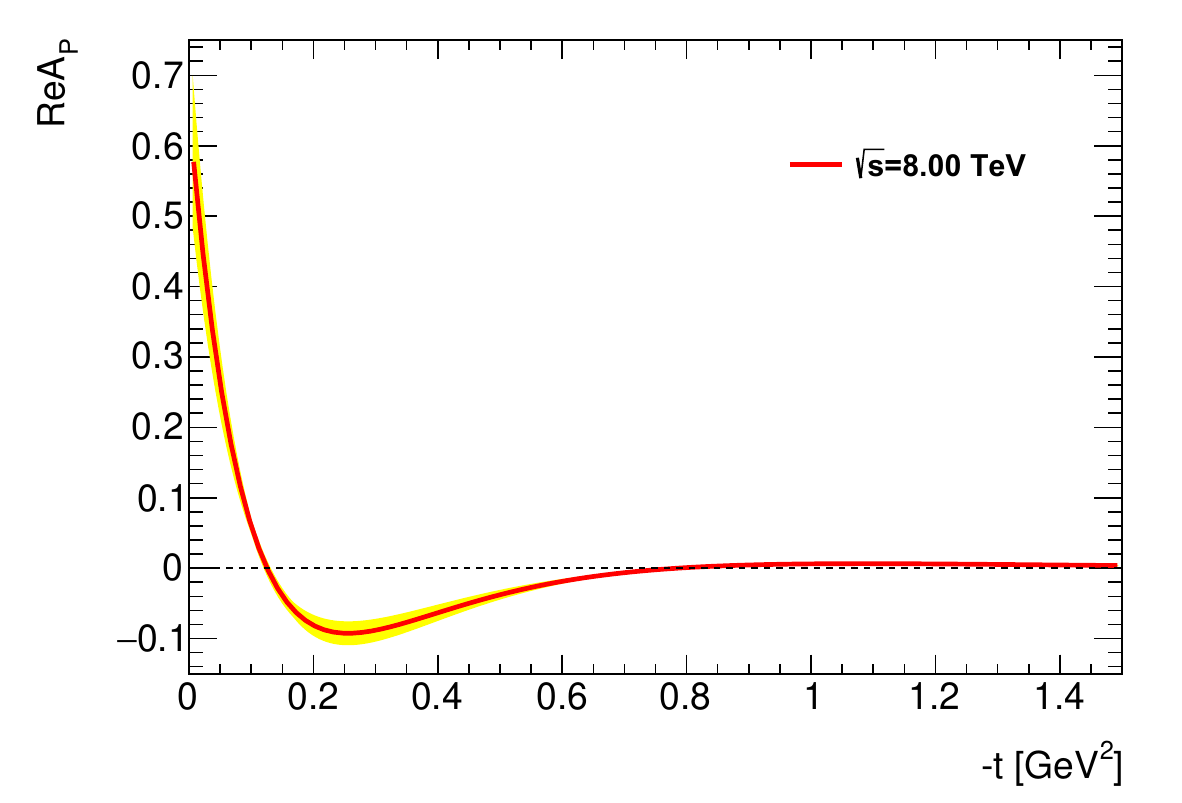}%
	}\hfill
	\subfloat[
          \label{fig:par_Rqd_lin}]{%
\includegraphics[width=0.5\linewidth]{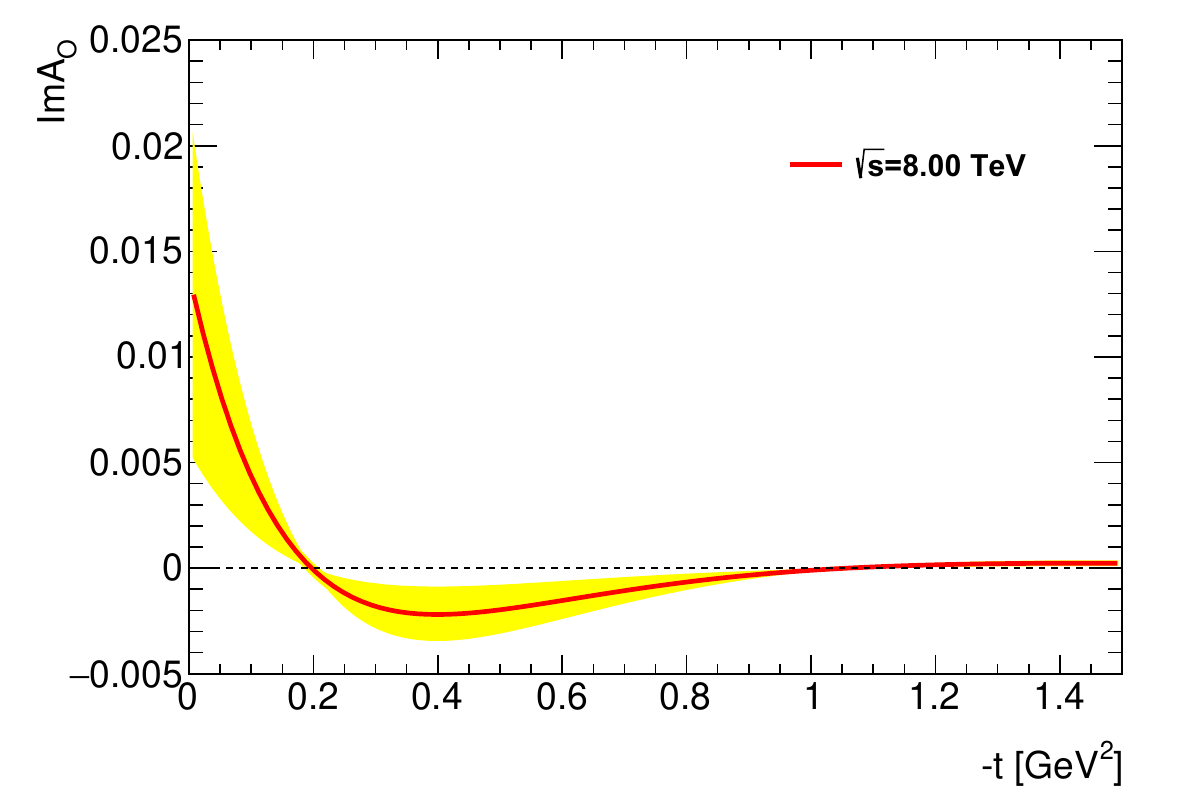}%
	}\hfill
	\subfloat[
           \label{fig:par_alpha_lin}]{%
\includegraphics[width=0.5\linewidth]{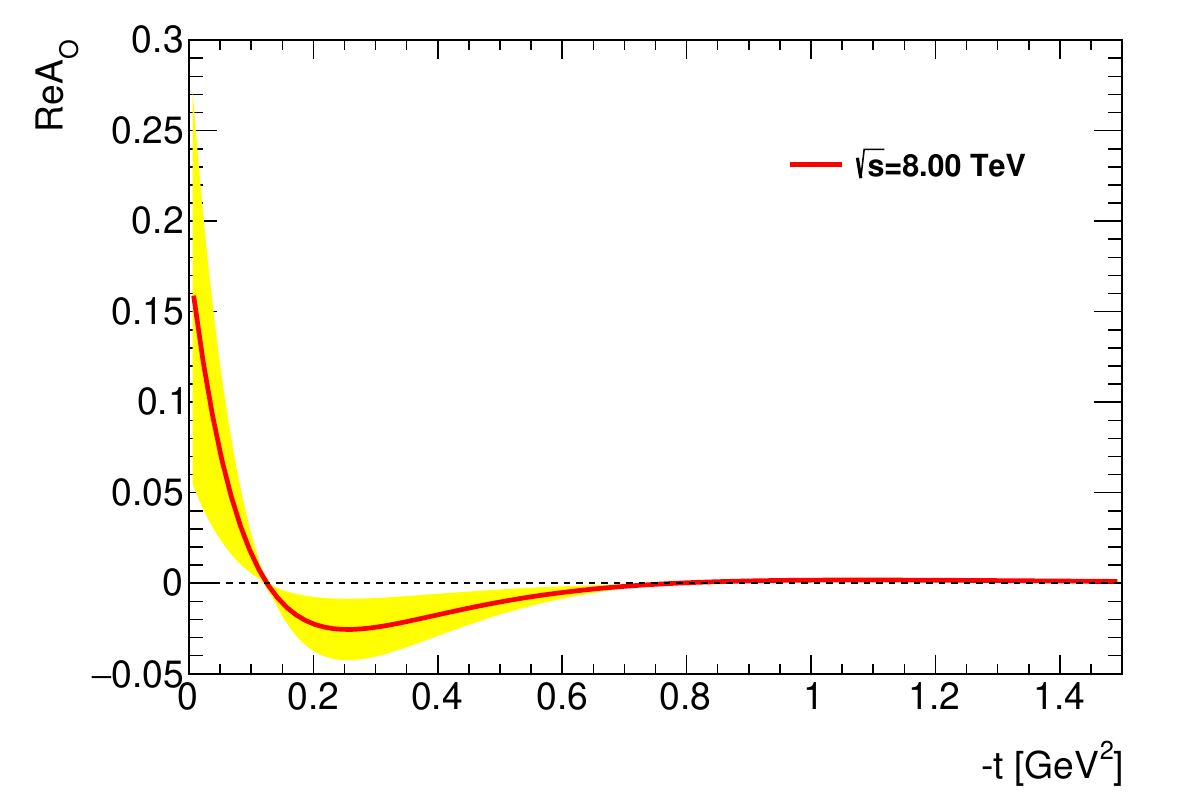}%
	}
\caption{Real and imaginary parts of the pomeron (crossing-even) and odderon (crossing-odd) contributions to the scattering amplitude at $\sqrt{s}=8$ TeV as calculated from the ReBB model using the energy calibration of the fit parameters shown in \cref{fig:reBB_model_log_lin_Rq}, \cref{fig:reBB_model_log_lin_Rd}, \cref{fig:reBB_model_log_lin_Rqd},  and \cref{fig:reBB_model_log_lin_alpha}. The yellow band indicates the estimated systematic errors of the calculation. 
}
\label{fig:ImRe_even_odd}
\end{figure}

The results for the real and imaginary parts of the ReBB model pomeron and odderon amplitudes at $\sqrt{s}=8$ TeV in the impact parameter representation are displayed in \cref{fig:ImRe_even_odd}. These results are based on the energy calibration of the fit parameters shown in \cref{fig:reBB_model_log_lin_Rq}, \cref{fig:reBB_model_log_lin_Rd}, \cref{fig:reBB_model_log_lin_Rqd},  and \cref{fig:reBB_model_log_lin_alpha}. At small values of $|t|$, as one naturally expects, the most dominant contribution comes from the imaginary part of the pomeron amplitude (see \cref{sec:eik}): the real part of the pomeron amplitude is about an order of magnitude smaller than its imaginary part, and both the real and imaginary parts of the odderon amplitude are smaller than the real part of the pomeron amplitude. At low-$|t|$ values, the imaginary parts of the odderon amplitude are about an order of magnitude smaller than its real part. Thus, at low-$|t|$, the odderon amplitude is small and dominantly real. 

Note that the $s$-dependence of the ReBB model pomeron and odderon amplitudes happens through its energy-dependent parameters. Based on \cref{eq:pomeron-amplitude}, \cref{eq:odderon-amplitude}, \cref{eq:even-amplitude} and \cref{eq:odd-amplitude}, we can write:
\begin{equation}\label{eq:pomeron-amplitude-in-terms-of-G}
    T_{\rm el}^{\mathbb P}(s,t)   \equiv T_{\rm el}^{+}(s,t) \, =
          G(R_q^{pp}(s), R_d^{pp}(s), R_{qd}^{pp}(s),\alpha^{pp}_R(s), \alpha^{p\bar p}_R(s);t), 
\end{equation}
\begin{equation}\label{e:odderon-amplitude-in-terms-of-H}
         T_{\rm el}^{\mathbb O}(s,t)   \equiv      T_{\rm el}^{-}(s,t) \, =
          H(R_q^{pp}(s), R_d^{pp}(s), R_{qd}^{pp}(s),\alpha^{pp}_R(s), \alpha^{p\bar p}_R(s);t), 
\end{equation}
where $G$ and $H$ are symbolic short-hand notations showing how the ReBB model pomeron and odderon amplitudes depend on $s$ through the $s$-dependent parameters of the model.

As discussed in \cref{sec:oddintro}, if the elastic differential cross section of $pp$ scattering differs from that of  $p\bar p$ scattering at the same kinematic ($s$, $t$) domain, the odderon contribution to the scattering amplitude is nonzero. It is clear from the above results, and the relation between $\rho_0$ and $\alpha_R$ as discussed in \cref{sec:rebb_endep_TeV}, that, within the ReBB model, the elastic differential cross section of $pp$ scattering differs from that of $p\bar p$ scattering at the same kinematic ($s$, $t$) domain, $i.e.$, the odderon contribution is nonzero in that domain if \mbox{$\alpha^{pp}_R \neq \alpha^{p\bar p}_R$} or equivalently if  $\rho_0^{pp} \neq \rho_0^{p\bar p}  $. For $\sqrt{s}\geq1$ TeV, we can write this in formulas as
     \begin{eqnarray}
        \frac{d\sigma^{pp}}{dt}   \neq   \frac{d\sigma^{p\bar p}}{dt} 
        &\iff& 
                T_{\rm el}^O(s,t) \neq 0  \, \nonumber \\  &\iff&  \rho_0^{pp}(s) \neq \rho_0^{p\bar p}(s) \nonumber \\
                & \iff &  \alpha^{pp}(s) \neq \alpha^{p\bar p}(s) \nonumber.
    \end{eqnarray}

Since $\alpha^{pp}_R(s) \neq \alpha^{p\bar p}_R(s)$  (see \cref{sec:rebb_endep_TeV}), there is a non-vanishing odderon contribution as shown in \cref{fig:ImRe_even_odd}. In this chapter, I calculate the significance of this odderon contribution by comparing the $pp$ and $p\bar p$ differential cross sections at the same kinematic ($s$, $t$) domain. I compare ReBB model extrapolations to experimental data.   While in \cref{chap:oddTD0}, I presented preliminary results on the ReBB model odderon analysis, in the present Chapter, I discuss the final results of the odderon observation within the ReBB model.

\section{Test of the ReBB model description}\label{sec:rebb_test}

Before comparing ReBB model curves to measured datasets to conclude about the signal of the $t$-channel odderon exchange contribution, I test in this section the validity of the ReBB model description obtained in \cref{chap:rebbdesc}. 

In \cref{chap:rebbdesc}, I determined the energy dependencies of the ReBB model parameters, $R_q(s)$, $R_d(s)$, $R_{qd}(s)$, $\alpha^{pp}_R(s)$, and $\alpha^{p\bar p}_R(s)$ as shown in \cref{fig:reBB_model_log_lin_Rq}, \cref{fig:reBB_model_log_lin_Rd}, \cref{fig:reBB_model_log_lin_Rqd},  and \cref{fig:reBB_model_log_lin_alpha}. These energy dependencies are based on altogether 16 parameter values obtained from fitting the available $p\bar p$ differential cross section data at $\sqrt{s}= 546$ GeV and 1.96 TeV, and the $pp$ differential cross section data at $\sqrt{s}= 2.76$ TeV and 7 TeV in the squared four-momentum transfer range of 0.38~GeV$^2$~$\lesssim -t\lesssim1.2$~GeV$^2$.  In this section, I show that all available $pp$ and $p\bar p$ differential cross section data are described by the ReBB model in a statistically acceptable manner, $i.e.$, with a $CL>$ 0.1\% in the kinematic range of 0.38~GeV$^2$~$\leq -t\leq1.2$~GeV$^2$ and 0.546~TeV~$\leq\sqrt{s}\leq 8$~TeV. Note that the $t$ range is the same as the $t$ range of the analyzed existing data in \cref{chap:rebbdesc} while the $\sqrt{s}$ range extends up to 8 TeV.

Fixing the ReBB model parameters, $R_q$, $R_d$, $R_{qd}$, $\alpha^{pp}_R$, and $\alpha^{p\bar p}_R$, at the values given by the energy dependence trends shown in \cref{fig:reBB_model_log_lin_Rq}, \cref{fig:reBB_model_log_lin_Rd}, \cref{fig:reBB_model_log_lin_Rqd},  and \cref{fig:reBB_model_log_lin_alpha} I compare the ReBB model curves to the differential cross section data by using the $\chi^2$ function of \cref{eq:chi2_refind}. $\lambda=1/2$ and $A_{qq}=1$ are kept fixed, too, as always. This means that all the physical model parameters are fixed. The only free parameters are the $\epsilon$ parameters of the type $b$ and type $c$ errors, $\epsilon_b$ and $\epsilon_c$, present in the $\chi^2$ function of \cref{eq:chi2_refind}.

\begin{figure}[hbt!]
	\centering
	\includegraphics[width=0.8\linewidth]{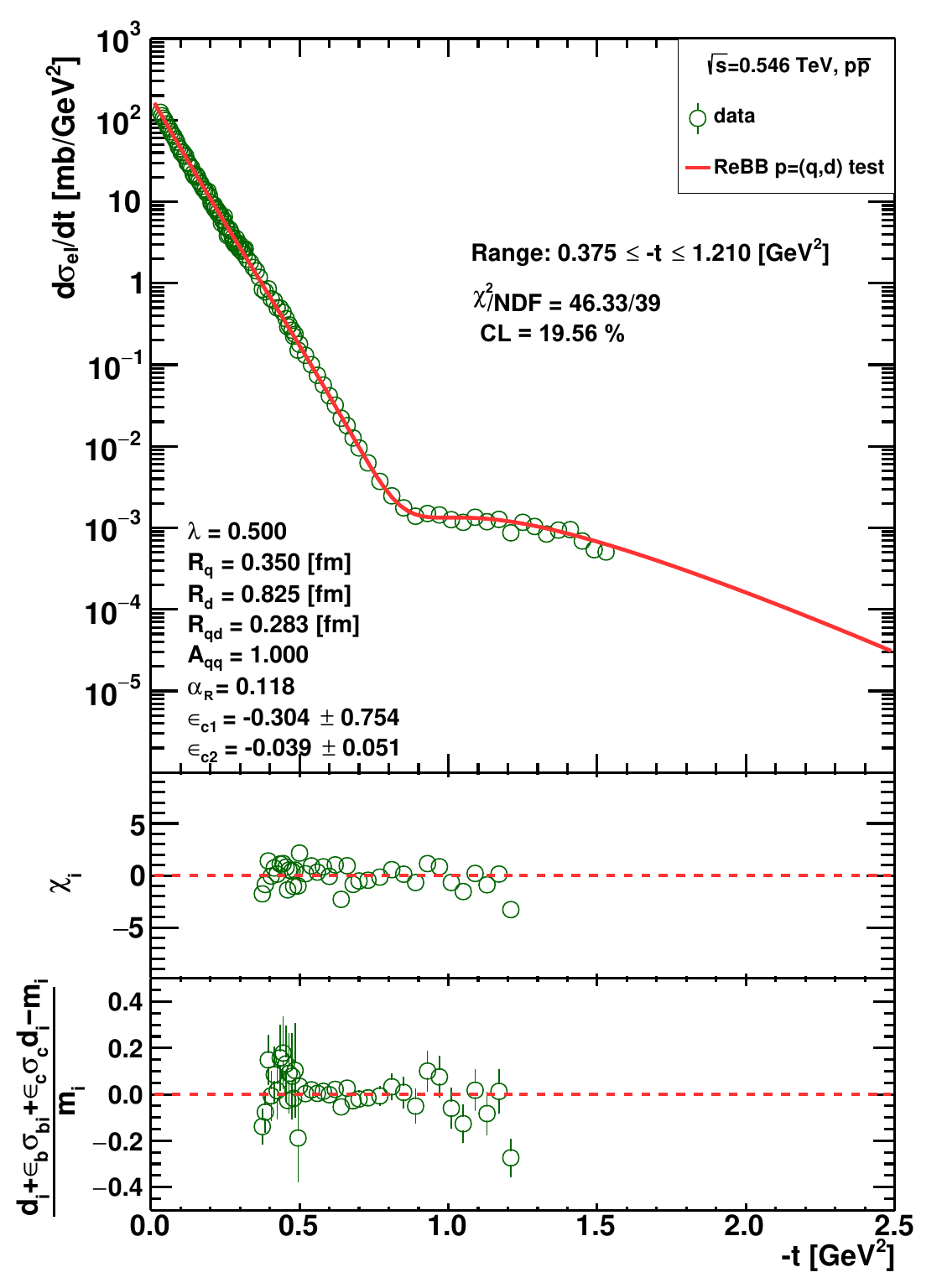}
	\caption{Test of the ReBB model description using the UA4 SPS $\sqrt{s}=$ 0.546 TeV $p\bar p$ elastic differential cross section data \cite{UA4:1983mlb,UA4:1984uui,UA4:1985oqn} in the squared four-momentum transfer range of 0.375 ${\rm GeV^2}\leq-t\leq1.210$ GeV$^2$. Only type $a$ vertical errors of the data points are shown. This sanity test was performed as a fit during which the model parameters $R_q$, $R_d$, $R_{qd}$ and $\alpha_R$ were fixed at their values based on the energy dependence trends as given in \cref{fig:reBB_model_log_lin_Rq}, \cref{fig:reBB_model_log_lin_Rd}, \cref{fig:reBB_model_log_lin_Rqd},  and \cref{fig:reBB_model_log_lin_alpha}, while the $\epsilon$ parameters in the $\chi^2$ definition of \cref{eq:chi2_refind} were free parameters.}
	\label{fig:reBB_model_test_0_546_GeV}
 \vspace{-5mm}
\end{figure}

\begin{figure}[hbt!]
	\centering
	\includegraphics[width=0.8\linewidth]{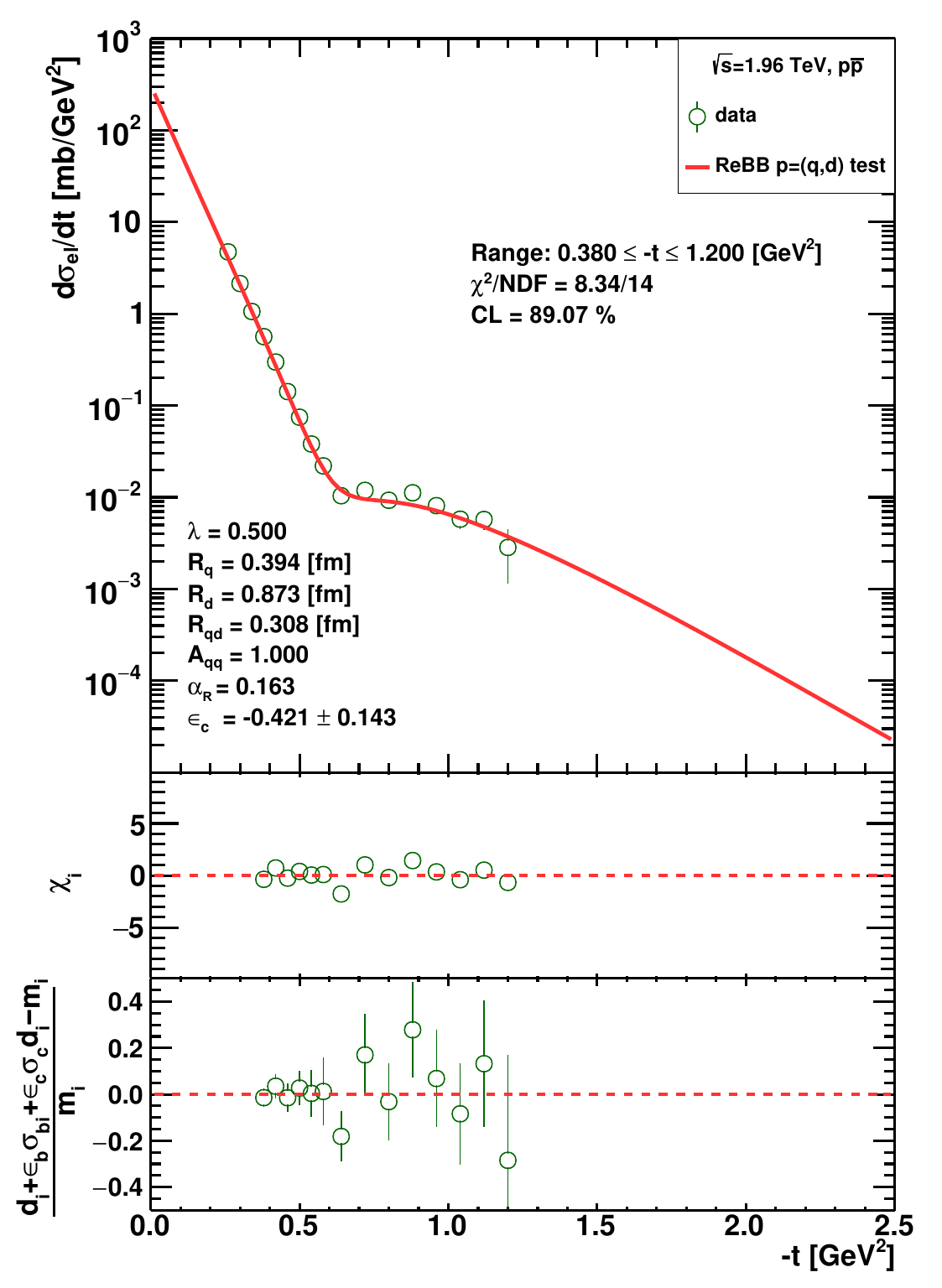}
	\caption{Same as \cref{fig:reBB_model_test_0_546_GeV} but using the Tevatron D0 $\sqrt{s}=$ 1.96 TeV $p\bar p$ elastic differential cross section data \cite{D0:2012erd}.}
	\label{fig:reBB_model_test_1_96_TeV}
 \vspace{-5mm}
\end{figure}

\begin{figure}[hbt!]
	\centering
	\includegraphics[width=0.8\linewidth]{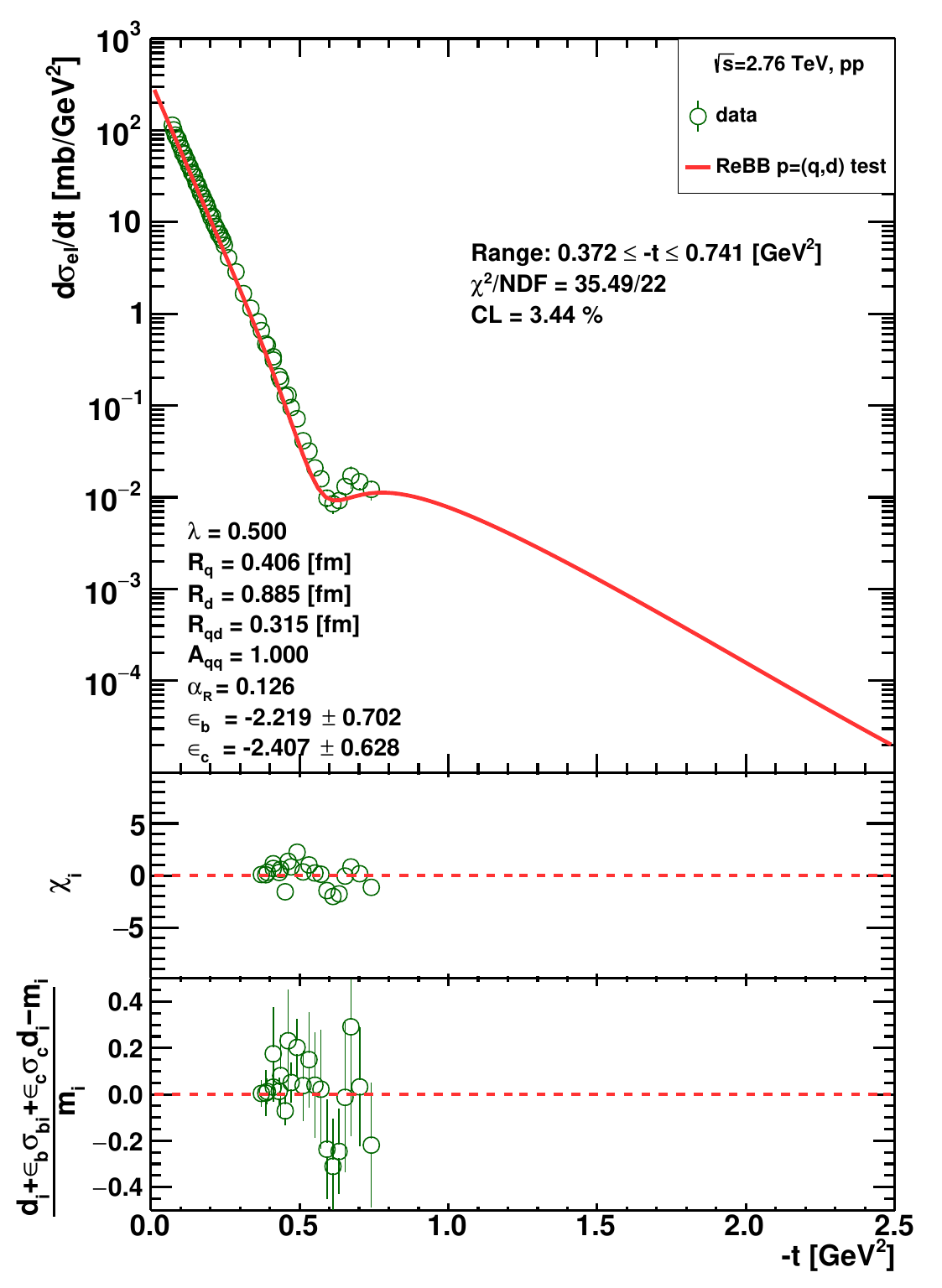}
	\caption{Same as \cref{fig:reBB_model_test_0_546_GeV} but using the LHC TOTEM $\sqrt{s}=$ 2.76 TeV $pp$ elastic differential cross section data \cite{TOTEM:2018psk}. 
	}
	\label{fig:reBB_model_test_2_76_TeV}
 \vspace{-5mm}
\end{figure}

\begin{figure}[hbt!]
	\centering
	\includegraphics[width=0.8\linewidth]{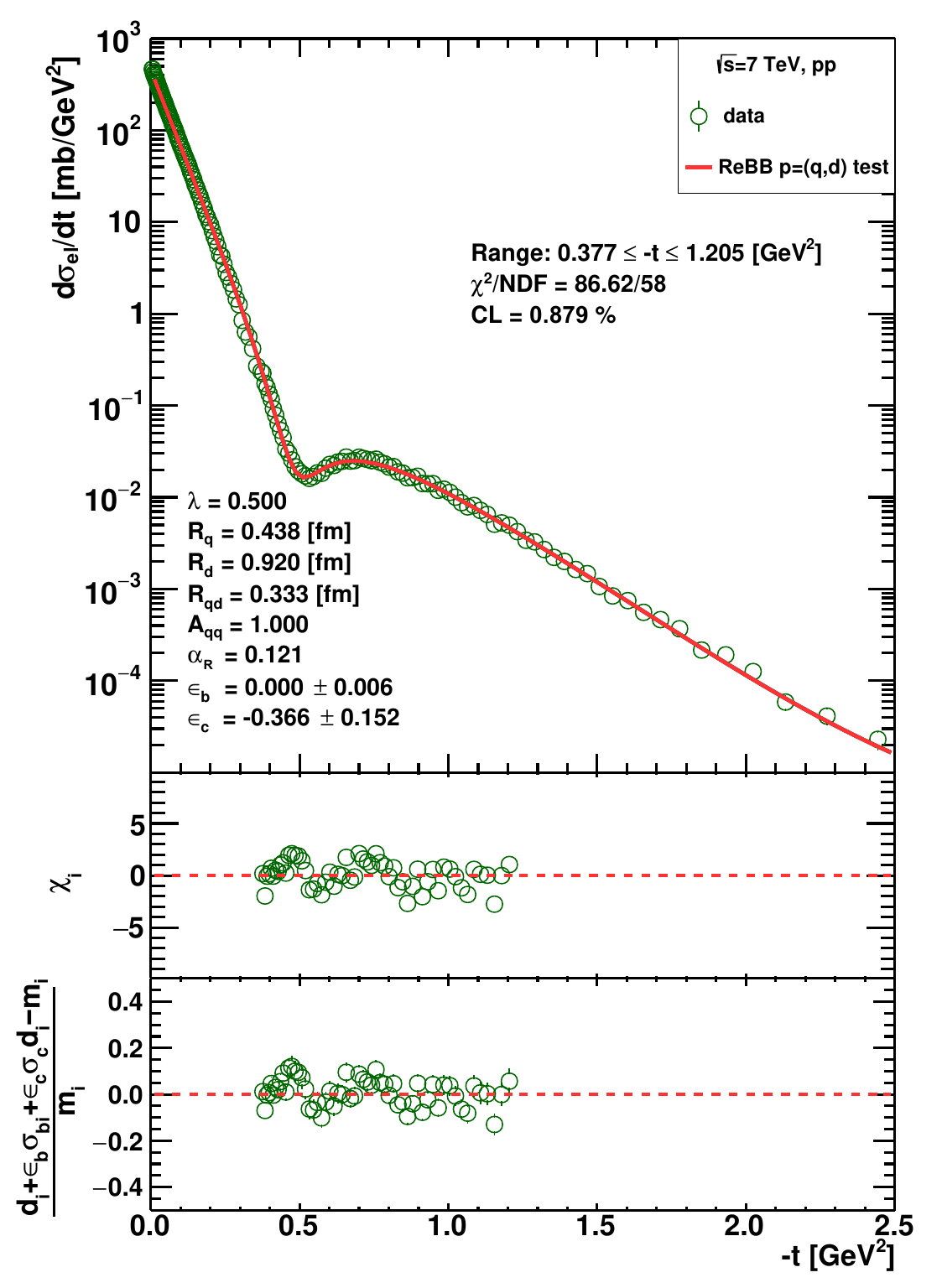}
	\caption{Same as \cref{fig:reBB_model_test_0_546_GeV} but using the LHC TOTEM $\sqrt{s}=$ 7 TeV $pp$ elastic differential cross section data \cite{TOTEM:2013lle}. 
	}
	\label{fig:reBB_model_test_7_TeV}
 \vspace{-5mm}
\end{figure}

To test the validity of the ReBB model description, I used data that were and were not utilized in determining the energy dependence trends of the ReBB model parameters. \cref{fig:reBB_model_test_0_546_GeV}, \cref{fig:reBB_model_test_1_96_TeV}, \cref{fig:reBB_model_test_2_76_TeV}, and \cref{fig:reBB_model_test_7_TeV} show the test on the $pp$ and $p\bar p$ elastic differential cross section data at $\sqrt{s}= 546$ GeV, 1.96 TeV, 2.76 TeV, and 7 TeV in the squared four-momentum transfer range of 0.38~GeV$^2$~$\leq -t\leq1.2$~GeV$^2$. These are the data that were considered during the determinations of the energy dependence trends of the ReBB model parameters. In all of the cases, we find a description with a $CL$ value higher than 0.1\%. The other details of these fits are the same as in \cref{sec:rebb_desc_refind_TeV}.

\begin{figure}[hbt!]
	\centering
	\includegraphics[width=0.8\linewidth]{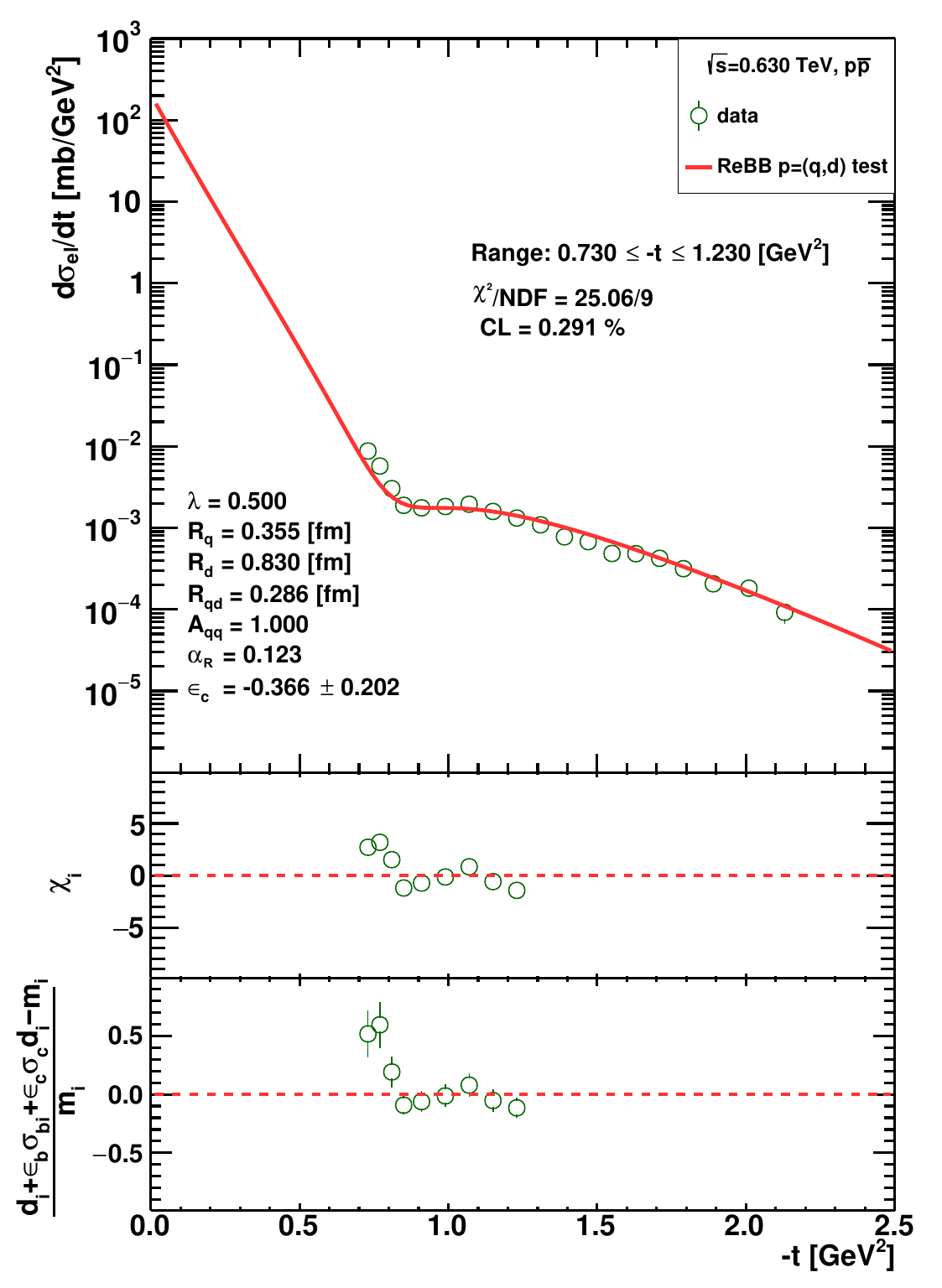}
	\caption{Same as \cref{fig:reBB_model_test_0_546_GeV} but using the SPS UA4 $\sqrt{s}=$ 0.63 TeV $p\bar p$ elastic differential cross section data \cite{UA4:1986cgb}. }
 \vspace{-0.5 cm}
	\label{fig:reBB_model_test_0_630_GeV}
\end{figure}

\begin{figure}[hbt!]
	\centering
	\includegraphics[width=0.8\linewidth]{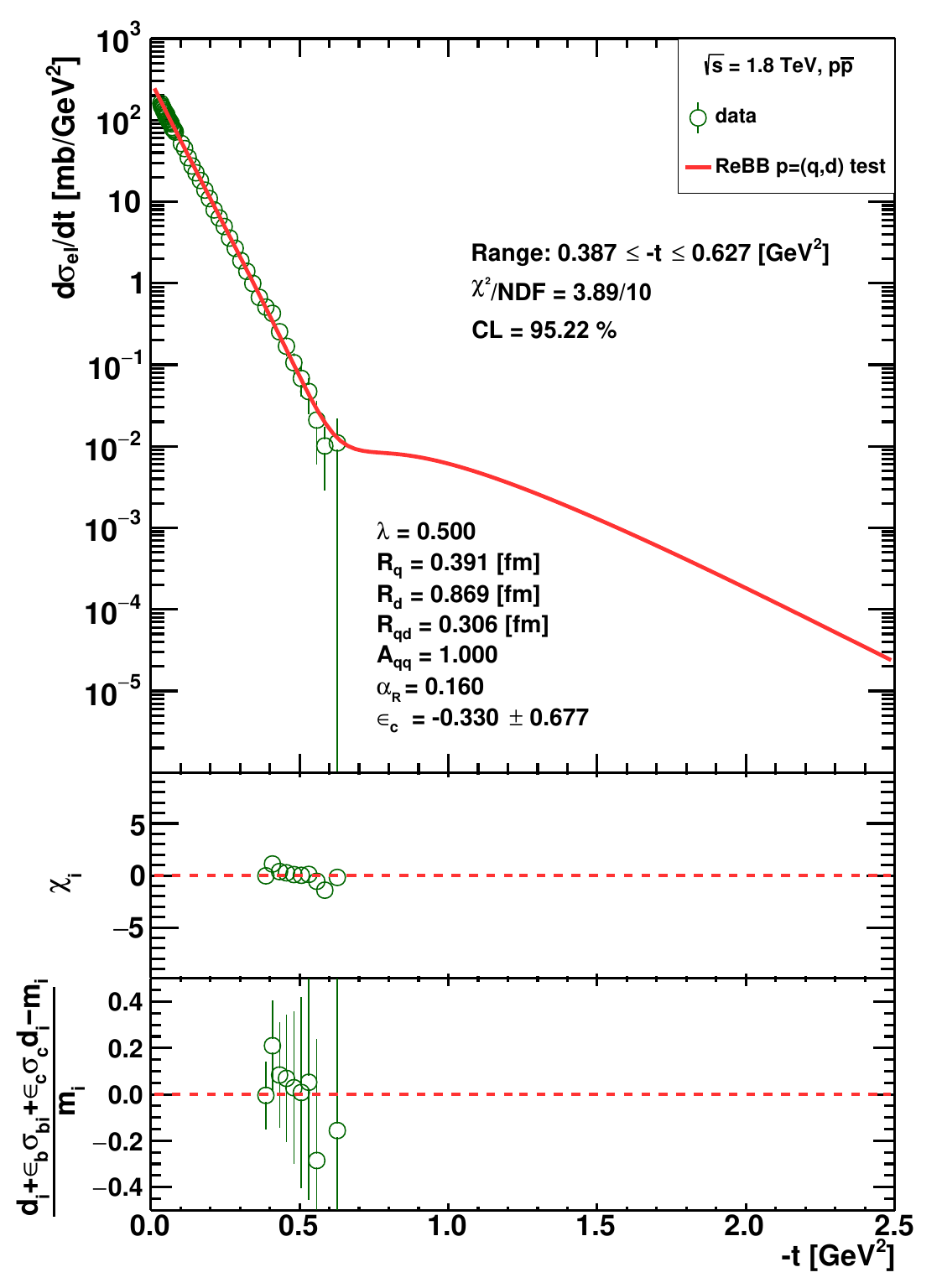}
	\caption{Same as \cref{fig:reBB_model_test_0_546_GeV} but with the Tevatron E-710 $\sqrt{s}=$ 1.8 TeV $p\bar p$ elastic differential cross section data \cite{E-710:1990vqb}. 
	}
	\label{fig:reBB_model_test_1_8_TeV}
\end{figure}

\begin{figure}[hbt!]
	\centering
	\includegraphics[width=0.8\linewidth]{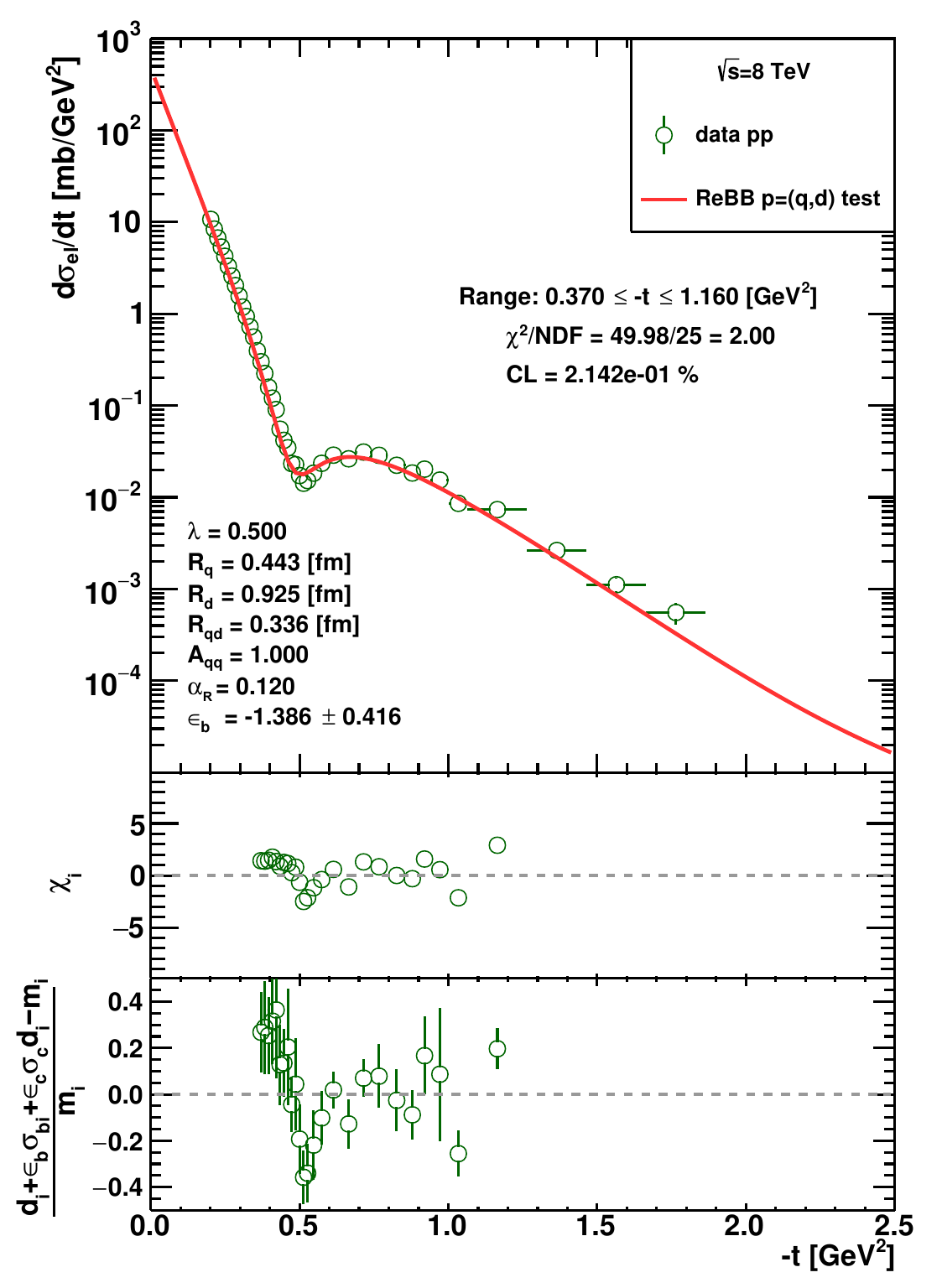}
	\caption{Same as \cref{fig:reBB_model_test_0_546_GeV} but with the LHC TOTEM $\sqrt{s}=$ 8 TeV $pp$ elastic differential cross section data \cite{TOTEM:2021imi}. 
}
	\label{fig:reBB_model_test_8_TeV}
\end{figure}

\cref{fig:reBB_model_test_0_630_GeV}, \cref{fig:reBB_model_test_1_8_TeV}, and \cref{fig:reBB_model_test_8_TeV} show the test of the ReBB model description on the $pp$ and $p\bar p$ elastic differential cross section data at $\sqrt{s}= 630$ GeV, 1.8 TeV, and 8 TeV. These are the data that were not considered during the determination of the energy dependence trends of the ReBB model parameters. In the case of the $\sqrt{s}= 630$ GeV and 1.8 TeV data, the limited acceptance did not allow to reliably determine the values of the ReBB model fit parameters, while the $\sqrt{s}= 8$ TeV data was not published in the time of determining the energy dependence trends of the ReBB model parameters. Despite the fact that these datasets were not used before, the ReBB model describes them with a $CL$ value higher than 0.1\%.

Since the ReBB model fits to $pp$ and $p\bar p$ elastic differential cross section data at \mbox{$\sqrt{s}=630$ GeV}, 1.8 TeV, and 8 TeV with the $\chi^2$ definition of \cref{eq:chi2_refind} were not performed and detailed in \cref{sec:rebb_desc_refind_TeV}, in the following part I describe the details of \mbox{these fits.}

The result of the ReBB model test using the SPS 630 GeV elastic $p\bar p$ differential cross section data \cite{UA4:1986cgb} is shown in \cref{fig:reBB_model_test_0_630_GeV}. This dataset has a type $c$ error of \mbox{$\sigma_{c}$~=~0.15.}  In addition to the type $c$ error, only type $a$ vertical errors ($\sigma_{ai}$) are published and consequently used in the fit. The $CL$ of the fit is 0.29\%.

The result of the ReBB model test using the Tevatron 1.8 TeV elastic $p\bar p$ differential cross section data \cite{E-710:1990vqb} is shown in \cref{fig:reBB_model_test_1_8_TeV}. This dataset has a type $c$ error of \mbox{$\sigma_{c}$ = 0.09.}  In addition to the type $c$ error, only type $a$ vertical errors ($\sigma_{ai}$) are published and consequently used in the fit. The $CL$ of the fit is 95.22\%.

Finally, the result of the ReBB model test using the TOTEM 8 TeV elastic $pp$ differential cross section data \cite{TOTEM:2021imi} is shown in \cref{fig:reBB_model_test_8_TeV}. No type $c$ error is given in Ref.~\cite{TOTEM:2021imi} separately. This type $c$ error is included in the systematic errors which I consider as type $b$ errors. In addition to the type $b$ error, type $a$ vertical errors ($\sigma_{ai}$) are published and consequently used in the fit. There is an additional subtlety. No representative $t$ values are given in Ref.~\cite{TOTEM:2021imi}, only $t$ bin sizes. I consider the bin center as a representative $t$ value and the half-bin-size as its error (horizontal type $a$, $\delta_a t_{i}$). In Ref.~\cite{TOTEM:2020zzr}, the same procedure was applied, and it was noted that the half-bin-size in $|t|$ is comparable to the $|t|$-resolution. The $CL$ of the resulting fit is 0.21\%.

\begin{figure}[hbt!]
	\centering
	\includegraphics[width=0.8\linewidth]{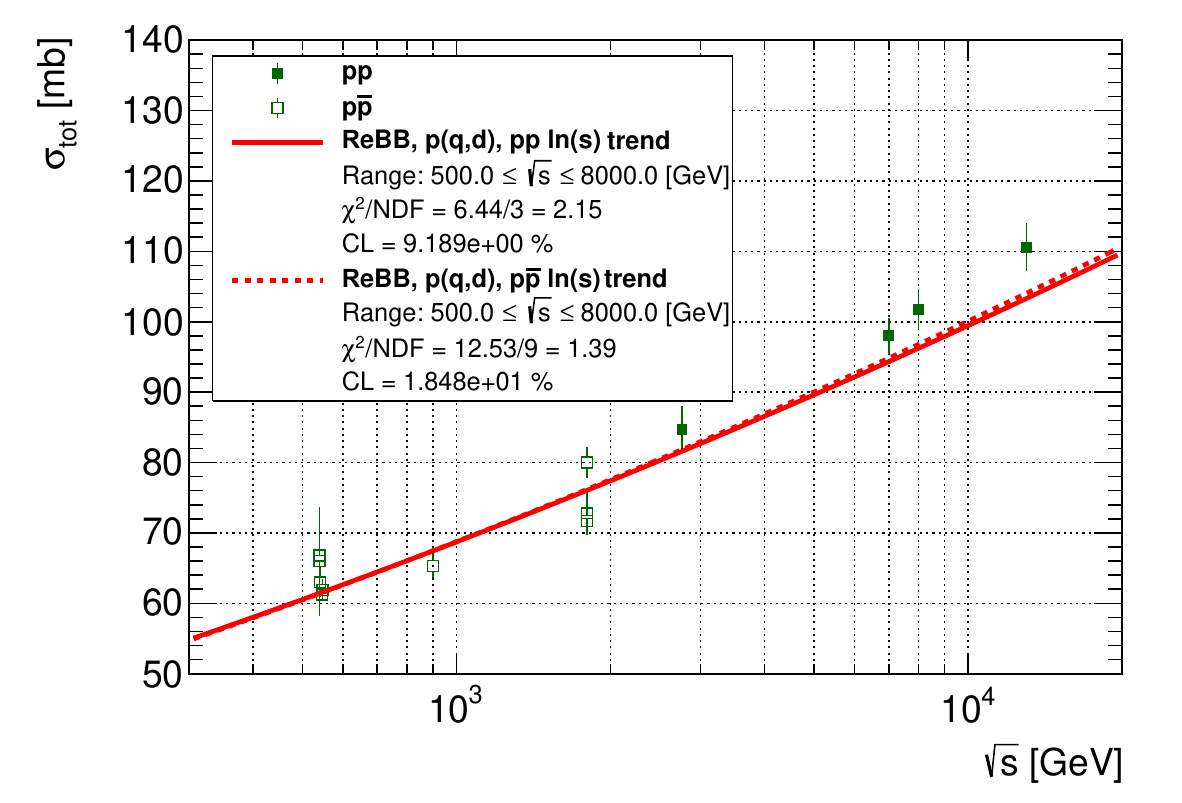}
	\caption{Test of the ReBB model description with the $pp$ \cite{TOTEM:2013vij,TOTEM:2012oyl,TOTEM:2017asr} and $p\bar p$ \cite{ParticleDataGroup:2018ovx} total cross section data. This test was performed as a fit during which the model parameters $R_q$, $R_d$, $R_{qd}$ and $\alpha_R$ were fixed at their values based on the energy dependence trends as given in \cref{fig:reBB_model_log_lin_Rq}, \cref{fig:reBB_model_log_lin_Rd}, \cref{fig:reBB_model_log_lin_Rqd},  and \cref{fig:reBB_model_log_lin_alpha}. }
	\label{fig:reBB_model_test_sig_tot}
\end{figure}

\begin{figure}[hbt!]
	\centering
	\includegraphics[width=0.8\linewidth]{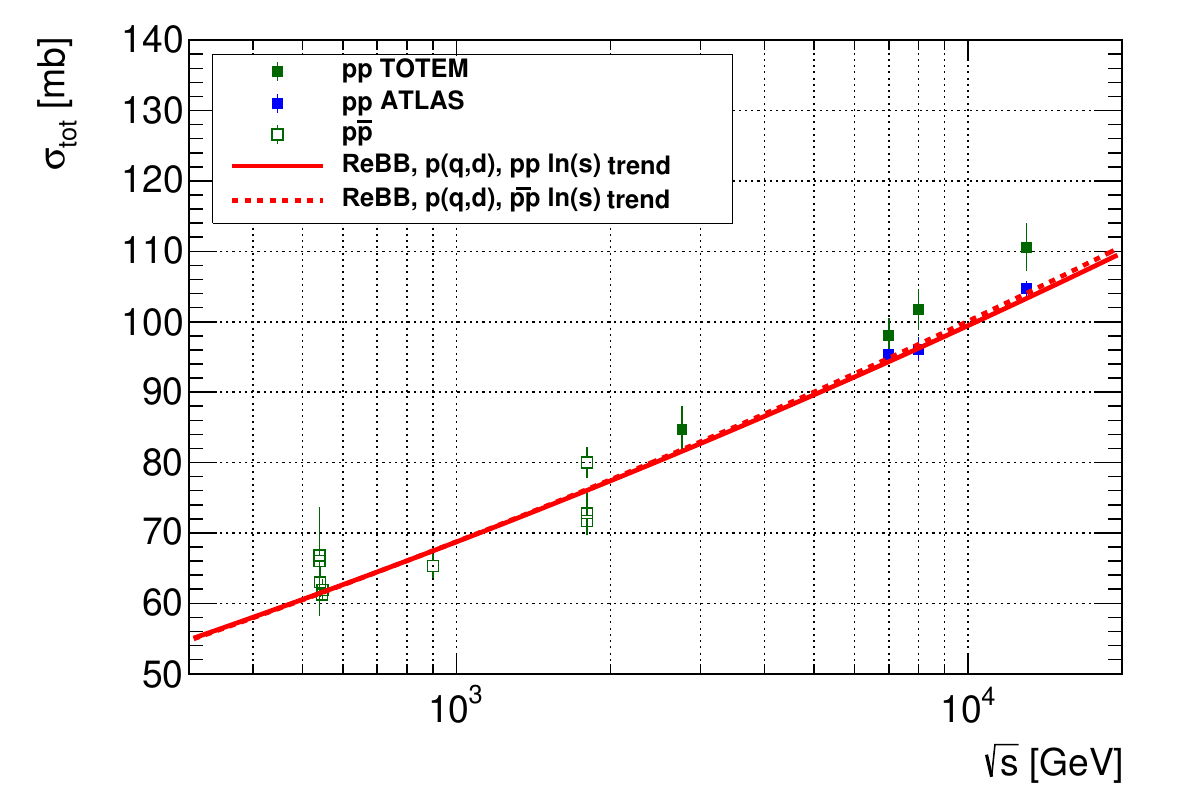}
	\caption{Same as \cref{fig:reBB_model_test_sig_tot} but with the ATLAS $pp$ data \cite{ATLAS:2014vxr,ATLAS:2016ikn, ATLAS:2022mgx} included.
}
	\label{fig:reBB_model_test_tot_A}
\end{figure}

\begin{figure}[hbt!]
	\centering
	\includegraphics[width=0.8\linewidth]{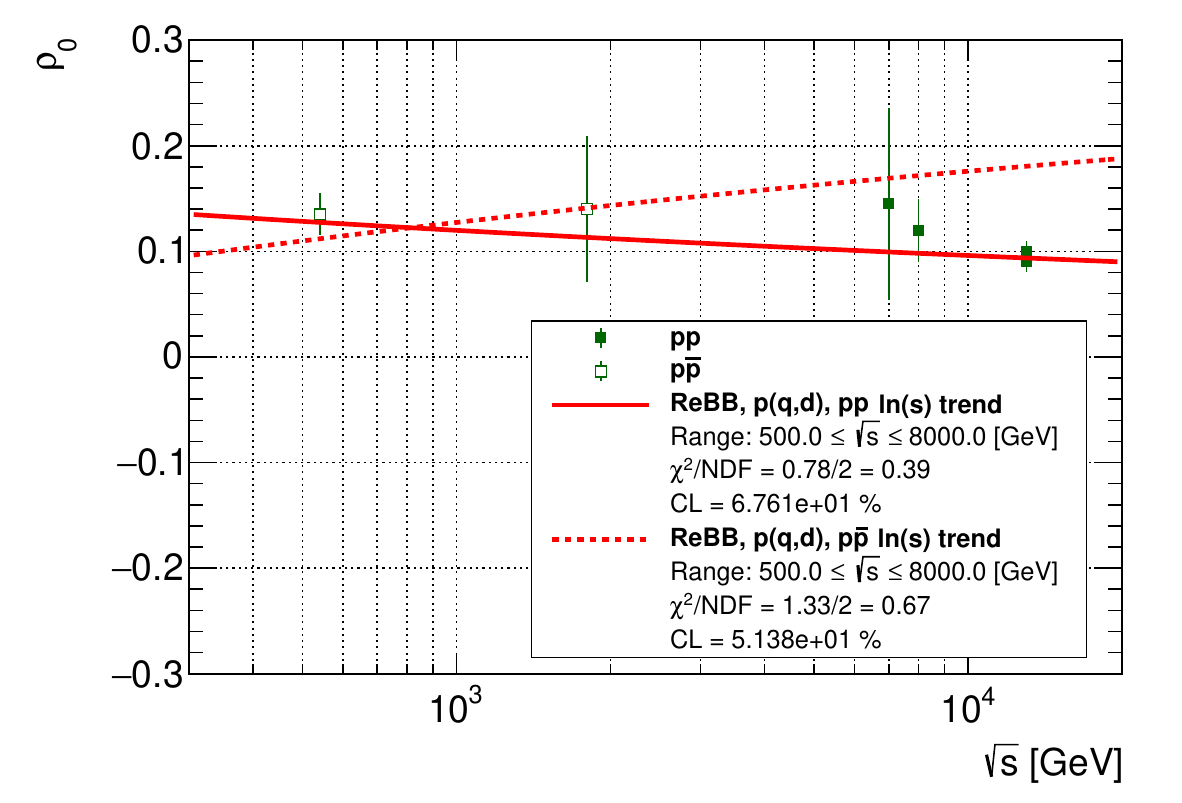}
	\caption{Same as \cref{fig:reBB_model_test_sig_tot} but with the $pp$ \cite{TOTEM:2013vij,TOTEM:2016lxj,TOTEM:2017sdy} and $p\bar p$ \cite{ParticleDataGroup:2018ovx}  $\rho_0$ data.
}
	\label{fig:reBB_model_test_rho}
\end{figure}

As an additional validity test, I cross-checked if the ReBB model with the energy dependence trends as given in \cref{fig:reBB_model_log_lin_Rq}, \cref{fig:reBB_model_log_lin_Rd}, \cref{fig:reBB_model_log_lin_Rqd},  and \cref{fig:reBB_model_log_lin_alpha}, describes the $pp$ and $p\bar p$ total cross section and $\rho_0$ data in the c.m. energy range of 0.546~TeV~$\leq\sqrt{s}\leq 8$~TeV, or not. The answer is yes, it does describe these data in the mentioned c.m. energy range. 

~

\cref{fig:reBB_model_test_sig_tot} shows the results for the total cross section data: the $CL$ of the $pp$ data description is 9.18\%, while the $CL$ of the $p\bar p$ data description is 18.48\%. \cref{fig:reBB_model_test_tot_A} is the same as \cref{fig:reBB_model_test_sig_tot} but with the ATLAS $pp$ data \cite{ATLAS:2014vxr,ATLAS:2016ikn, ATLAS:2022mgx} included. It was noticed only after the publication of the results of the ReBB model analysis that the ReBB model calibrated to the SPS UA4 $p\bar p$, Tevatron D0 $p\bar p$, and LHC TOTEM $pp$ \mbox{higher-$|t|$} data predicts a $pp$ $\sigma_{\rm tot}$ energy evolution that very nicely describes the LHC ATLAS measurements that are systematically below the LHC TOTEM measurements\footnote{Studies of the ATLAS-TOTEM discrepancy that also include the analysis of the data in the low-$|t|$ domain are in progress but these studies are outside the scope of this dissertation.}. 
\cref{fig:reBB_model_test_rho} shows the results for the $\rho_0$ data: the $CL$ of the $pp$ data description is 67.61\%, while the $CL$ of the $p\bar p$ data description is 51.38\%.

The final conclusion of this section is that the ReBB model with the energy dependence as determined in \cref{sec:rebb_endep_TeV} describes all the experimentally measured differential cross section datasets in the kinematic range of 0.546~TeV~$\leq\sqrt{s}\leq 8$~TeV and \mbox{0.38~GeV$^2$~$\leq -t\leq1.2$~GeV$^2$} in a statistically acceptable manner, $i.e.$, with $CL$ $>$ 0.1\%. In addition, this calibrated ReBB model also describes $pp$ and $p\bar p$ total cross section and $\rho_0$ data in the c.m. energy range of 0.546~TeV~$\leq\sqrt{s}\leq 8$~TeV in a statistically acceptable manner.

\clearpage

\section{Observation of the $t$-channel odderon}\label{sec:rebbextraps}

The validity of the ReBB model in the kinematic range of 0.546~TeV~$\leq\sqrt{s}\leq 8$~TeV and 0.38~GeV$^2$~$\leq -t\leq1.2$~GeV$^2$ even at those datasets not included in the calibration of the model indicates the reliability of extrapolations and interpolations made by using this model. In this section, in the $\sqrt{s}>1$ TeV kinematic domain, I calculate the elastic $pp$ differential cross section at those energies where the elastic $p\bar p$ differential cross section is measured and vice versa, I calculate the elastic $p\bar p$ differential cross section at those energies where the elastic $pp$ differential cross section is measured. After this, in the validated kinematic range of the ReBB model, I compare the model curves with the measured datasets using the $\chi^2$ definition of \cref{eq:chi2_refind} to calculate the statistical significance of the difference between the $pp$ and $p \bar p$ elastic differential cross sections, $i.e.$, the statistical significance of the $t$-channel odderon exchange signal.

I calculate the $pp$ differential cross section at $\sqrt s$ = 1.96 TeV and compare it to the Tevatron $p\bar p$ differential cross section dataset measured at $\sqrt s$ = 1.96 TeV. Then I compute the $p\bar p$ differential cross sections at $\sqrt s$ = 2.76 TeV, 7 TeV, and 8~TeV and compare them to the TOTEM $pp$ differential cross section measurements at \mbox{$\sqrt s$ = 2.76 TeV}, 7 TeV, and 8~TeV. 
I emphasize that I compare the $pp$ and $p\bar p$ differential cross sections at exactly the same energies in the $\sqrt{s}>1$ TeV energy domain and in the squared four-momentum transfer domain 0.38~GeV$^2$~$\leq -t\leq1.2$~GeV$^2$ tested in \cref{sec:rebb_test} based on the available experimental data.  
This way, any statistically significant difference between them can be attributed to the effect of the odderon exchange. If the differential cross sections are not compared at exactly the same energies, an additional difference could arise from ignoring the energy evolution. 

According to the results of \cref{sec:rebb_endep_TeV}, the values of the ReBB model scale parameters, $R_q$, $R_d$, and $R_{qd}$ are compatible with the same energy dependence in $pp$ and $p\bar p$ scattering, only the energy dependence of the opacity parameter, $\alpha_R$, differs in the two processes. Thus, when computing the $pp$ differential cross section at $\sqrt s$ = 1.96 TeV and the $p\bar p$ differential cross sections at \mbox{$\sqrt s$ = 2.76 TeV} and 7 TeV, I use the same values of the scale parameters as obtained from the direct fits to the data presented in \cref{fig:reBB_model_fit_1_96_TeV}, \cref{fig:reBB_model_fit_2_76_TeV}, and \cref{fig:reBB_model_fit_7_TeV}. I take only the values of the $\alpha_R^{pp}$ and $\alpha_R^{p\bar p}$ parameters from extrapolations based on their energy dependence trends presented in \cref{fig:reBB_model_log_lin_alpha}. At \mbox{$\sqrt s$ = 8 TeV},  I take also the values of the scale parameters from the energy dependence trends presented in \cref{fig:reBB_model_log_lin_Rq}, \cref{fig:reBB_model_log_lin_Rd}, and \cref{fig:reBB_model_log_lin_Rqd}. As I detail below, it turns out that this procedure gives a more conservative estimate for the signal of the odderon exchange at \mbox{$\sqrt s$ = 8 TeV} than the case when I use the same values of the scale parameters as obtained from the direct fit to the \mbox{$\sqrt s$ = 8 TeV} data. 

\begin{figure}[hbt!]
	\centering
	\includegraphics[width=0.8\linewidth]{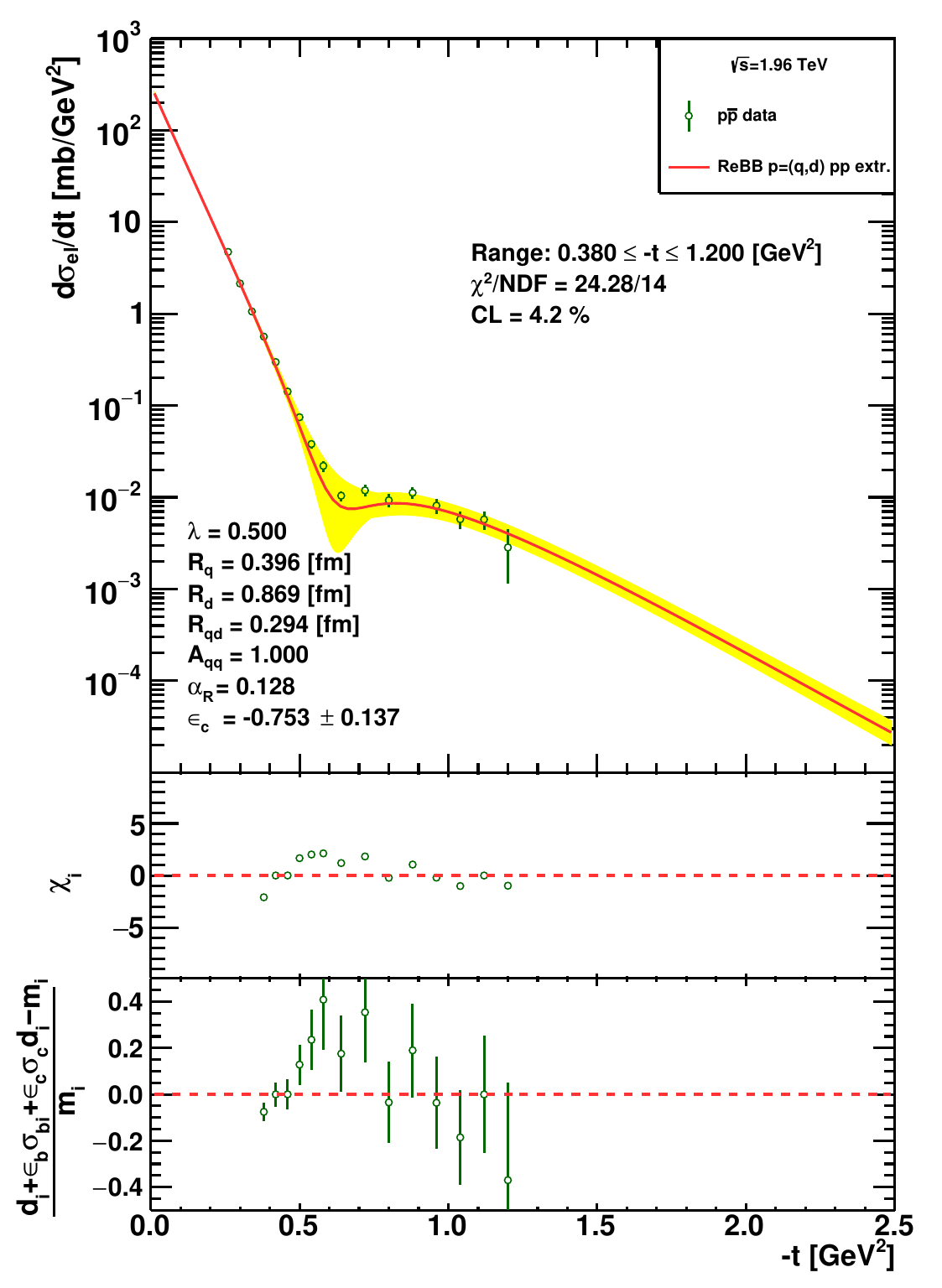}
	\caption{Comparison of the ReBB model $pp$ differential cross section curve at \mbox{$\sqrt{s}=1.96$~TeV} to the Tevatron D0 $p\bar p$
 differential cross section data \cite{D0:2012erd} measured at the same energy. The yellow band is the estimated uncertainty of the theoretical calculation.}
	\label{fig:reBB_model_extr_1_96_TeV}
\end{figure}

\begin{figure}[hbt!]
	\centering
	\includegraphics[width=0.8\linewidth]{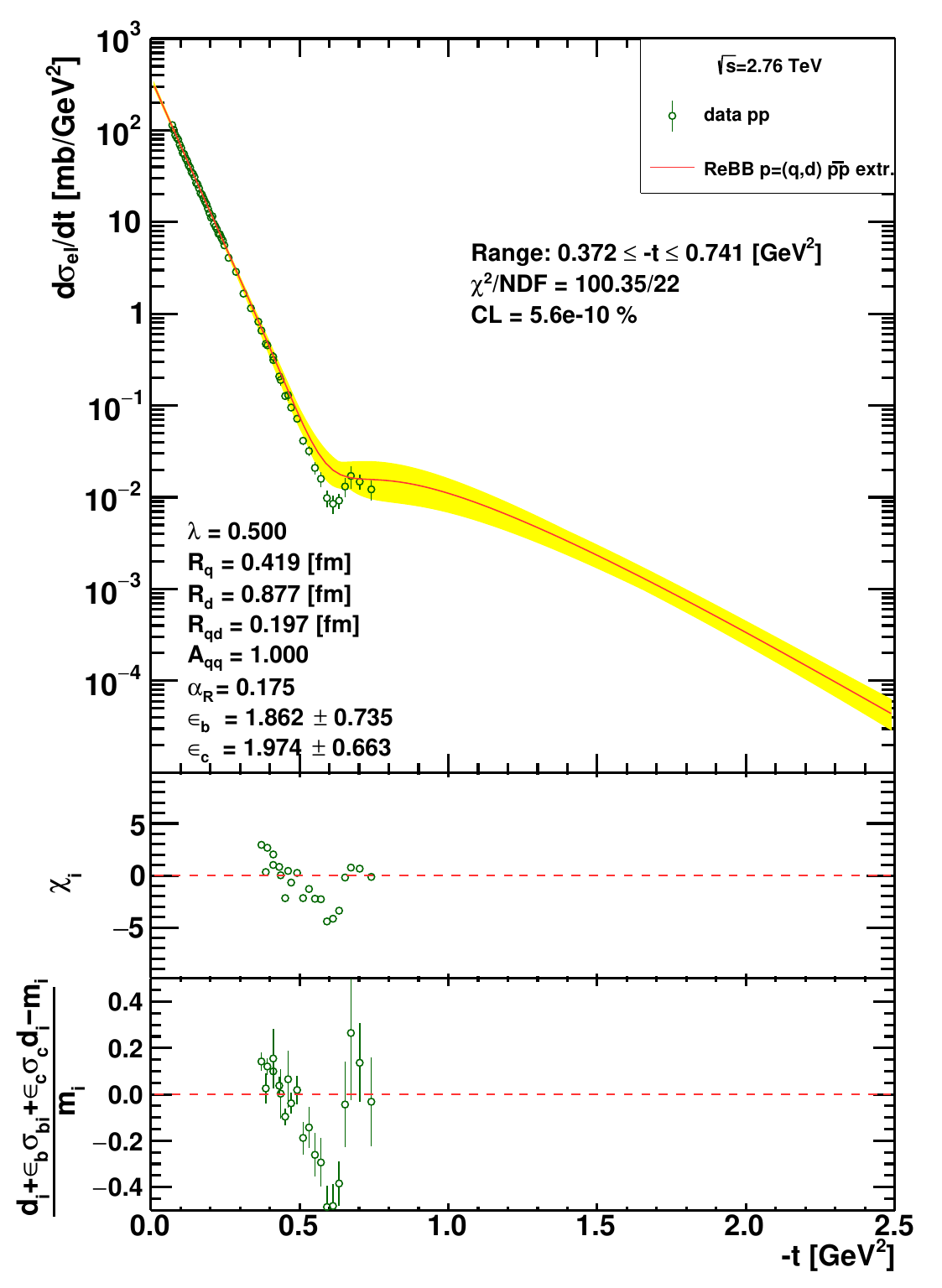}
	\caption{Comparison of the ReBB model $p\bar p$ differential cross section curve at \mbox{$\sqrt{s}=2.76$~TeV} to the LHC TOTEM $pp$
 differential cross section data \cite{TOTEM:2018psk} measured at the same energy. Only type $a$ vertical errors of the data points are shown. The yellow band is the estimated uncertainty of the theoretical calculation.}
	\label{fig:reBB_model_extr_2_76_TeV}
\end{figure}

\begin{figure}[hbt!]
	\centering
    \includegraphics[width=0.8\linewidth]{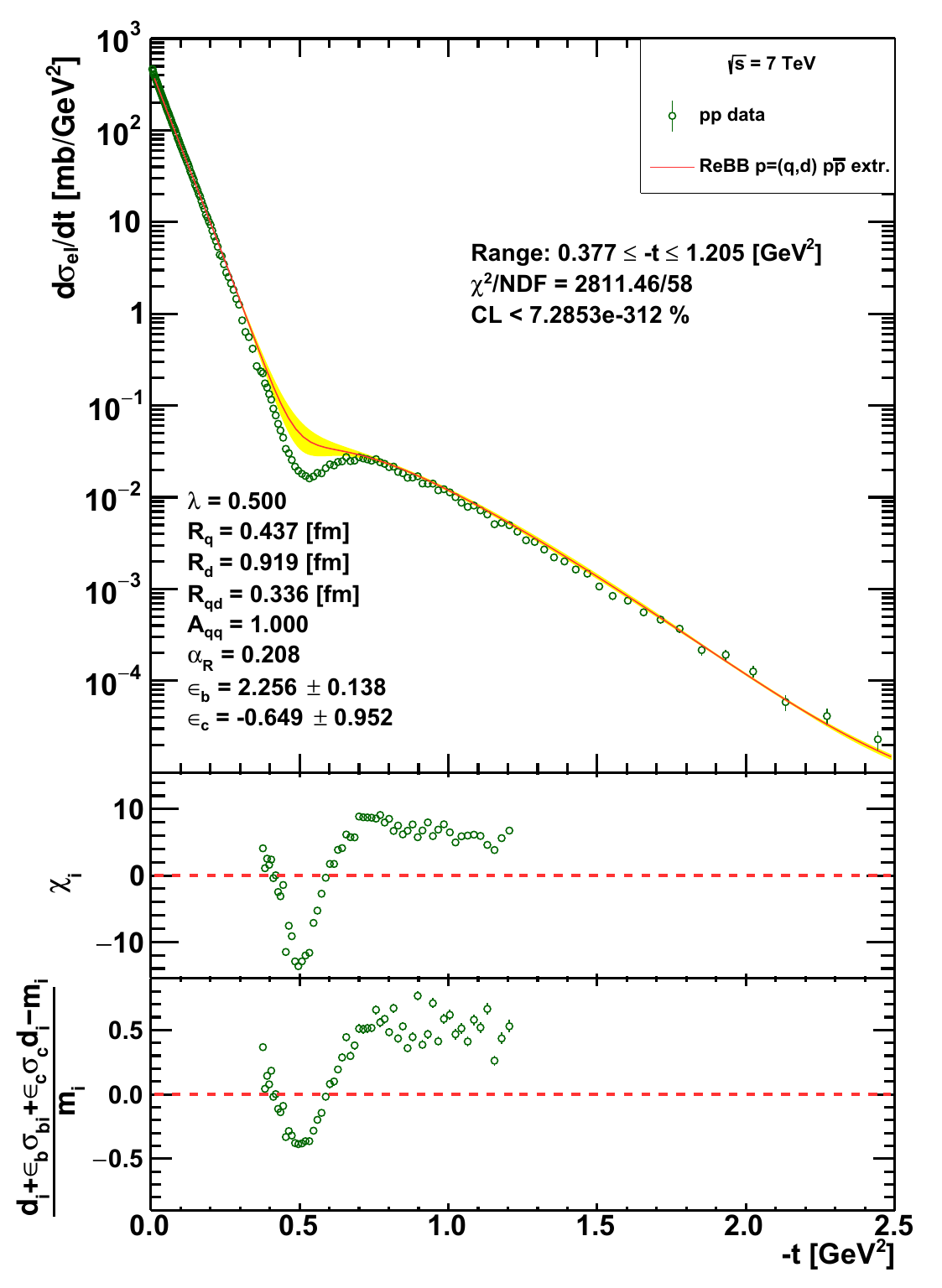}
	\caption{Comparison of the ReBB model $p\bar p$ differential cross section curve at \mbox{$\sqrt{s}=7$~TeV} to the LHC TOTEM $pp$
 differential cross section data \cite{TOTEM:2013lle} measured at the same energy. The yellow band is the estimated uncertainty of the theoretical calculation.}
	\label{fig:reBB_model_extr_7_TeV}
\end{figure}

\begin{figure}[hbt!]
	\centering
    \includegraphics[width=0.8\linewidth]{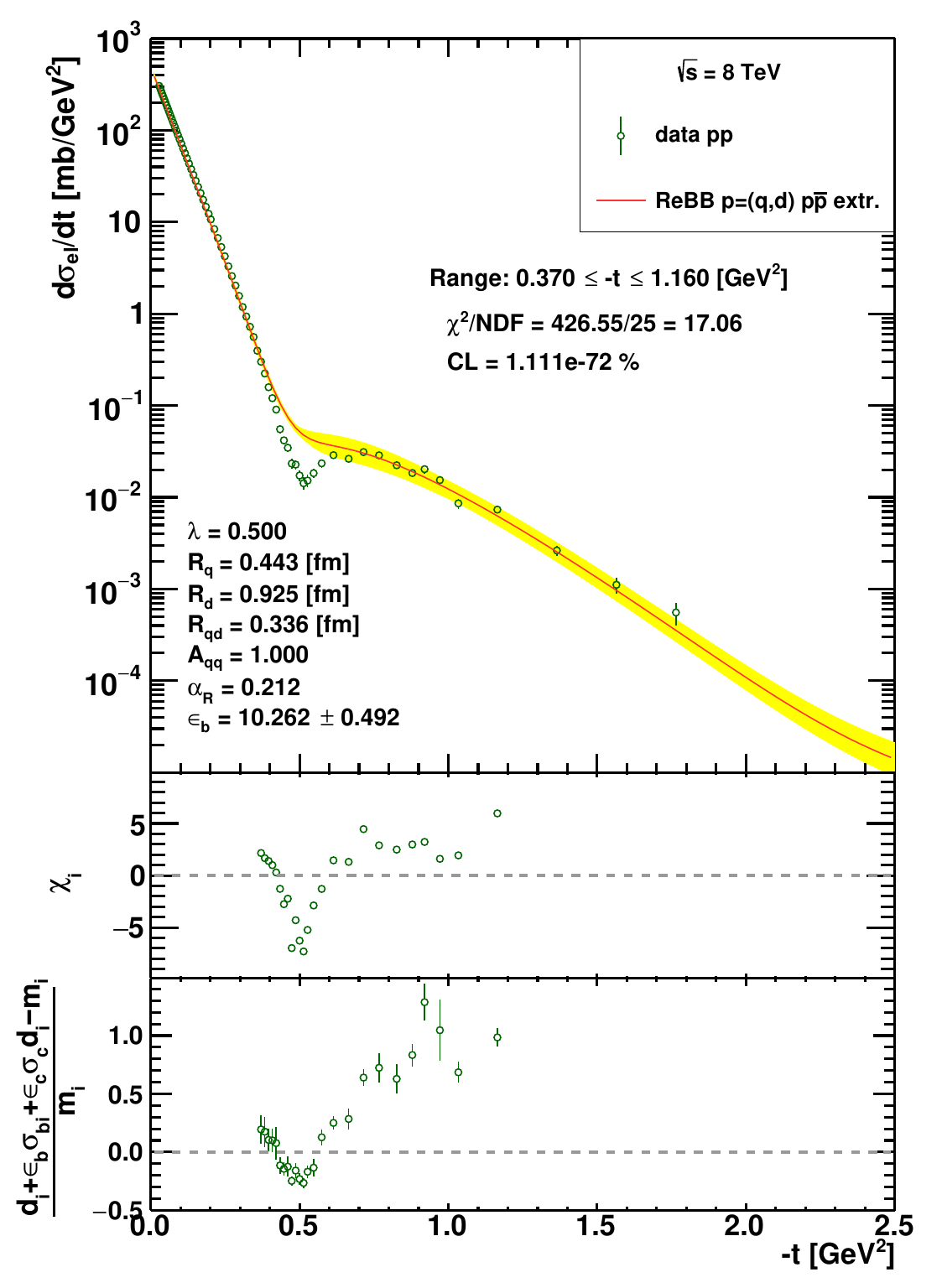}
	\caption{Comparison of the ReBB model $p\bar p$ differential cross section curve at \mbox{$\sqrt{s}=8$~TeV} to the LHC TOTEM $pp$
 differential cross section data \cite{TOTEM:2021imi} measured at the same energy. The yellow band is the estimated uncertainty of the theoretical calculation. 
}
	\label{fig:reBB_model_extr_8_TeV}
\end{figure}

The result of the comparison of the $pp$ and $p\bar p$ differential cross sections at the c.m. energy of \mbox{1.96 TeV} is shown in \cref{fig:reBB_model_extr_1_96_TeV}. Only the value of the $\chi^2$ parameter, $\epsilon_c$ is optimized. Using the $\chi^2$ definition of \cref{eq:chi2_refind}, the comparison results in a $CL$ value of 4.2\%. One can conclude that no statistically significant odderon effect is observed when comparing the calculated $pp$ differential cross section with the measured $p\bar p$ differential cross section at $\sqrt{s}=1.96$ TeV. $CL$ = 4.2\% corresponds to an odderon signal with a statistical significance of only 2$\sigma$. This significance is two times smaller than that obtained in \cref{chap:oddTD0} by comparing the ReBB $pp$ extrapolation to the $p\bar p$ measurement at $\sqrt{s}=1.96$ TeV utilizing the $\chi^2$ definition of \cref{eq:chi_Cj++}. We can conclude that the analysis with the $\chi^2$ definition of \cref{eq:chi2_refind} presented in the present Chapter gives a more conservative result for the odderon significance.

The estimated theoretical systematic error bands are also shown around the calculated curves at $\sqrt{s}=1.96$ TeV and higher energies. These error bands are based on the envelope of the curves obtained by shifting up and down the ReBB model parameter values, or the parameter values that determine the energy dependencies of the ReBB model parameters, by their uncertainties. For example, the error band around the $\sqrt{s}=1.96$ TeV curve is the envelope of 10 curves: 6 curves are obtained from shifting up and down the values of the ReBB model scale parameters at  $\sqrt{s}=1.96$ TeV with their uncertainties while additional 4 curves are obtained by shifting by their uncertainties the values of the $p_0$ and $p_1$ parameters that determine the energy dependence of the opacity parameter $\alpha_R^{pp}$.

The result of the comparison of the $pp$ and $p\bar p$ differential cross sections at \mbox{$\sqrt{s}=2.76$ TeV} is shown in \cref{fig:reBB_model_extr_2_76_TeV}. Only the values of the $\chi^2$ parameters, $\epsilon_b$ and $\epsilon_c$, are optimized. At \mbox{$\sqrt{s}=2.76$ TeV}, the hypothesis that the $p\bar p$ curve describes the $pp$ data has a $CL$ value of 5.6$\times$10$^{-10}$\% as determined by using the $\chi^2$ definition of \cref{eq:chi2_refind}. This $CL$ value corresponds to an odderon signal with a statistical significance of 6.8$\sigma$. Given that in high-energy particle physics, the discovery threshold corresponds to a 5$\sigma$ effect, this result already means a model-dependent $t$-channel odderon exchange observation. Looking at \cref{fig:reBB_model_extr_2_76_TeV} we can see that the odderon signal is coming from the diffractive minimum region: while there is a prominent dip in the elastic differential cross section of $pp$ scattering, this dip fills in in the elastic differential cross section of $p\bar p$ scattering. My result qualitatively confirms the prediction of Ref.~\cite{Donnachie:1983hf} that the dip fills in in high-energy elastic $p\bar p$ collisions because of the three-gluon exchange mechanism associated with the $t$-channel odderon exchange in QCD (see \cref{sec:amplitude} and \cref{sec:oddintro}).

The result of the comparison of the $pp$ and $p\bar p$ differential cross sections at \mbox{$\sqrt{s}=7$ TeV} is shown in \cref{fig:reBB_model_extr_7_TeV}. As in the previous cases, only the values of the $\chi^2$ parameters, $\epsilon_b$ and $\epsilon_c$ are optimized. The resulting $CL$ value is smaller than 7.3$\times$10$^{-310}$ \%, corresponding to an odderon signal with a statistical significance greater than 37.7$\sigma$. The odderon signal with the highest statistical significance is obtained at $\sqrt s = 7$ TeV, where the data has a relatively large acceptance in $t$. By looking at the middle and bottom panels of \cref{fig:reBB_model_extr_7_TeV}, we see that not only the dip fills in in $p\bar p $ scattering but also in the bump region the $p\bar p $ $t$ distribution goes below the $pp$ $t$-distribution in a similar manner as observed in the model-independent analysis of Ref.~\cite{Csorgo:2019ewn}.

Finally, the result of the comparison of the $pp$ and $p\bar p$ differential cross sections at \mbox{$\sqrt{s}=8$ TeV} is shown in \cref{fig:reBB_model_extr_8_TeV}.
Only the value of the $\chi^2$ parameter, $\epsilon_b$, is optimized. The resulting $CL$ value is 1.1$\times$10$^{-72}$ \%, corresponding to an odderon signal with a statistical significance greater than or equal to 18.2$\sigma$. 

I checked also the case when the values of the ReBB model parameters, $R_q$, $R_d$, $R_{qd}$, and $\alpha_R$, are optimized to the \mbox{$\sqrt{s}=8$ TeV} $pp$ elastic differential cross section data. When the ReBB model parameters are optimized to the \mbox{$\sqrt{s}=8$ TeV} $pp$ elastic differential cross section data, the $CL$ of the description by the $pp$ ReBB model curve to the $pp$ data at $\sqrt{s}=8$ TeV improves to 14.95\%. The extrapolated and the fitted values for $R_q$ and $R_{qd}$ are the same within triple errors but for $R_d$ within errors. The odderon significance at \mbox{$\sqrt{s}=8$ TeV} grows by calculating the $p\bar p$ ReBB model differential cross section curve using the scale parameter values optimized to the \mbox{$\sqrt{s}=8$ TeV} $pp$ elastic differential cross section data.  One would expect, in the latter case, the odderon signal to decrease, but it is not so because the ReBB model scale parameter values change in a way making the difference between $pp$ and $p\bar p$ scattering bigger. 
Thus, the final result for the odderon significance at \mbox{$\sqrt{s}=8$ TeV} is the more conservative one, $i.e.$, lower value of 18.2$\sigma$.
 
The obtained confidence levels and the corresponding odderon signal significances at different energies are summarised in \cref{tab:odderon_signif}.

\begin{table}[!hbt]
    \centering
    \begin{tabular}{cccccc}
   \hline\hline\noalign{\smallskip}
        $\sqrt{s}$ (TeV) & $\chi^2$ & $NDF$ & $CL$ & significance ($\sigma$)   \\ \noalign{\smallskip}\hline \hline\noalign{\smallskip}
        1.96 & 24.283 & 14 & 0.0423 & 2.0  \\ 
        2.76 & 100.347 & 22 & 5.6093 $\times10^{-12}$ & 6.8  \\ 
        7 & 2811.46 & 58 & $\textless$ 7.2853$\times 10^{-312} $ & $\textgreater$37.7  \\ 
        8 & 426.553 & 25 & 1.1111$\times 10^{-74}$ & $\geq$18.2   \\ \hline\hline
    \end{tabular}
    \caption{Odderon signal significances in the ReBB model analysis at different energies. Significances higher than 8$\sigma$ were calculated using an analytical approximation as detailed in the text.}\label{tab:odderon_signif}
    \vspace{-5mm}
\end{table}

Significances higher than 8$\sigma$ as obtained at $\sqrt s = $ 7 TeV and 8 TeV were estimated using an analytical approximation which allows one to calculate a lower limit for the significance.  In the following I present the details of this approximation.

Defining the Gaussian probability density function
with mean $x_0$ and variance $\sigma^2$ as
\begin{equation}
    G(x) = \frac{1}{\sqrt{2\pi\sigma^2}}{\rm e}^{-\frac{(x-x_0)^2}{2\sigma^2}},\,\,\,\int_{-\infty}^{\infty}dx\,G(x)=1,
\end{equation}
the $CL$ corresponding to $n\sigma$ significance is calculated as \cite{ParticleDataGroup:2022pth}

\begin{equation}
    {CL} = 2\int_{x_0+n\sigma}^{\infty}G(x)dx,
\end{equation}
Applying a variable change, $$x\rightarrow x' = x-x_0-n\sigma,$$ we have that
$$
    {CL} = \frac{2}{\sqrt{2\pi\sigma^2}}\int_{0}^{\infty}{\rm e}^{-\frac{(x'+n\sigma)^2}{2\sigma^2}}dx' \leq \frac{2}{\sqrt{2\pi\sigma^2}}{\rm e}^{-\frac{n^2}{2}}\int_{0}^{\infty}{\rm e}^{-\frac{x'n}{\sigma}}dx' = \sqrt{\frac{2}{\pi}}\frac{1}{n}{\rm e}^{-\frac{n^2}{2}}.
$$
Thus, one obtains an upper limit for the $CL$ that corresponds to a significance $n$ in $\sigma$-s:
\begin{equation}\label{eq:CLlim}
{CL} \leq \sqrt{\frac{2}{\pi n^2}}{\rm e}^{-\frac{n^2}{2}} .
\end{equation}
\cref{eq:CLlim} is then used to associate to a given $CL$ value a lower limit of the significance $n$ in $\sigma$-s.
\cref{eq:CLlim} is useful when the significance is too large (larger than $8\sigma$) to be quantified with the standard numerical packages of CERN Root, Wolfram Mathematica, or Microsoft Excel.

As a next step, I calculated the combined odderon significances by combining the significances obtained at different energies and summarized in \cref{tab:odderon_signif}. I proceeded step by step: first, I combined the result at the two lowest c.m. energies, at $\sqrt s = 1.96$ TeV and \mbox{2.76 TeV,} then I included the result at $\sqrt s = 7$ TeV or 8 TeV; finally, I included the result at all the analyzed energies. To combine the significances, I applied two different methods, as detailed below. 


One of the methods consists in adding the individual $\chi^2$ and $NDF$ values to calculate the combined $CL$ and significance values. The results are summarized in \cref{tab:odsum-combo_a}. The second method I applied is Stouffer's method utilized in Ref.~\cite{TOTEM:2020zzr} by the TOTEM and D0 Collaborations for combining odderon significances. Using Stouffer's method, the combined significances are obtained by summing the significances and dividing this sum by the square root of the number of summed significances. My results obtained by using Stouffer's method are summarized in \cref{tab:odsum-combo_b}.

\begin{table}[!hbt]
     \begin{subtable}[h]{0.99\textwidth}
        \centering
    \begin{tabular}{ccccc}
    \hline\hline\noalign{\smallskip}
        $\sqrt{s}$  of combined data (TeV) & $\chi^2$ & $NDF$ & $CL$ & \Centerstack{ combined\\significance ($\sigma$)\\ by $\chi^2$ \& $NDF$ sum \\ method}\\ \noalign{\smallskip}\hline
        1.96 \& 2.76 & 124.63 & 36 & 1.07$\times$ $10^{-11}$   & 6.7\\ 
        1.96 \& 2.76 \& 7 & 2936.09 & 94 & $\textless$ 9.13$\times$ $10^{-312}$    & $\textgreater$37.7\\ 
        1.96 \& 2.76 \& 8 & 551.183 & 61 & 4.63$\times$ $10^{-80}$  & $\textgreater$18.9  \\ 
        1.96 \& 2.76 \& 7 \& 8 & 3362.64 & 119 & $\textless$8.07$\times$ $10^{-312}$ & $\textgreater$37.7 \\ \hline\hline
    \end{tabular}
           \caption{\label{tab:odsum-combo_a}}
       \label{tab:sub1}
    \end{subtable}
    \vfill
    \vspace{0.3cm}
    \begin{subtable}[h]{0.99\textwidth}
        \centering
    \begin{tabular}{cc}
    \hline\hline\noalign{\smallskip}
         $\sqrt{s}$  of combined data (TeV) & \Centerstack{ combined\\significance ($\sigma$)\\ by Stouffer's method}  \\ \noalign{\smallskip}\hline
        1.96 \& 2.76  & 6.3  \\ 
        1.96 \& 2.76 \& 7  & $\textgreater$26.9   \\ 
        1.96 \& 2.76 \& 8  & $\textgreater$15.7   \\ 
        1.96 \& 2.76 \& 7 \& 8  & $\textgreater$32.4   \\ \hline\hline
    \end{tabular}
           \caption{\label{tab:odsum-combo_b}}
       \label{tab:sub2}
    \end{subtable}
    \vspace{-1mm}
    \caption{Combined odderon signal significances in the ReBB model analysis (a) by adding the individual $\chi^2$ and the individual $NDF$ values and (b) by Stouffer's method. }\label{tab:odsum-combo}
    \vspace{-5mm}
\end{table}

Looking at the values in \cref{tab:odsum-combo}, we can conclude that Stouffer's method gives a bit lower, more conservative estimate for the combined significances. Nevertheless, we see that combining the results for the odderon significances obtained at the two lowest energies, $\sqrt s = 1.96$ TeV and 2.76 TeV, independently from the method applied, the resulting combined significance is greater than $6\sigma$ exceeding the discovery limit of $5\sigma$. Including the results obtained at $\sqrt s = 7$ TeV or 8 TeV, the combined odderon significance exceeds $15\sigma$. Finally, when I include the results at all the four analyzed energies,  \mbox{$\sqrt s = 1.96$ TeV}, \mbox{2.76 TeV,} 7 TeV and 8 TeV,  the resulting combined odderon signal exceeds even a $32\sigma$ statistical significance. Thus, in practical terms, within the framework of the ReBB model analysis, the existence of the $t$-channel odderon exchange \mbox{is not a probability but a certainty.}

Based on the $H(x)$ scaling property of the $pp$ elastic differential cross section, a statistically significant, at least $6.26\sigma$ model-independent odderon observation was reported beforehand in Ref.~\cite{Csorgo:2019ewn} and in an anonymously refereed/peer-reviewed ISMD 2019 conference proceedings, Ref.~\cite{Csorgo:2020msw}. This ReBB model odderon analysis aimed to calculate the statistical significance of the $t$-channel odderon exchange with a different, model-dependent method. 
Afterwards, an at least $5.2\sigma$ odderon effect was reported in the joint paper of the CERN LHC TOTEM and FNAL Tevatron D0 Collaborations, Ref.~\cite{TOTEM:2020zzr}. This result of Ref.~\cite{TOTEM:2020zzr} is based on the comparison of D0 measured $\rm{p\bar p}$ ${\rm d}\sigma_{\rm el}/{\rm d}t$ with D0-TOTEM extrapolated pp $d\sigma_{\rm el}/dt$ at $\sqrt{s} = 1.96$ TeV (utilizing also new, beforehand unpublished TOTEM data) and the comparison of TOTEM measured $\rho_0$ parameter and total cross section at \mbox{$\sqrt{s} = 13$ TeV} with a set of models not containing odderon contribution.

The authors of Refs.~\cite{Donnachie:2019ciz,Donnachie:2022aiq} presented a Regge-motivated model that explains the low-$|t|$ data measured by TOTEM at $\sqrt{s} = 13$ TeV without an Odderon contribution; however, the Odderon contribution at large-|t|, in the diffractive
minimum-maximum/shoulder region is not questioned, and hence the criticism does not affect the results obtained using the $H(x)$ scaling property of elastic $pp$ scattering \cite{Csorgo:2019ewn,Csorgo:2020msw} or using the ReBB \mbox{model \cite{Csorgo:2020wmw,Szanyi:2022ezh}.}  Studies based on the $H(x)$ scaling and the ReBB model report discovery level
odderon signal observation by re-analyzing only already published large-$|t|$ datasets.

\vspace{0.5cm}
\textbf{Summary}
\vspace{0.2cm}

In this Chapter, I demonstrated that the ReBB model of elastic $pp$ and $p\bar p$ scattering is valid in the kinematic range of \mbox{0.546~TeV~$\leq\sqrt{s}\leq 8$~TeV} and \mbox{0.38~GeV$^2$~$\leq -t\leq1.2$~GeV$^2$} by showing that the ReBB model with the energy dependence as determined in \cref{sec:rebb_endep_TeV} describes all the measured elastic $pp$ and $p\bar p$ differential cross section datasets in the mentioned kinematic domain in a statistically acceptable manner, $i.e.$, with $CL$ $>$ 0.1\%. Using this validated ReBB model description, I compared the $pp$ and $p\bar p$ elastic differential cross sections in common squared four-momentum transfer ranges at exactly the same energies. I found that the $pp$ and $p\bar p$ elastic differential cross sections differ with a statistical significance of at least 6.3$\sigma$ when significances obtained at the two lowest energies -- at \mbox{$\sqrt s$ = 1.96 TeV} and 2.76 TeV -- are combined. Then I found by combining the significances obtained at all the four analyzed energies, $\sqrt s$ = 1.96~TeV, 2.76 TeV, 7 TeV and \mbox{8 TeV,} that the statistical significance of the $t$-channel odderon exchange signal within the ReBB model analysis is higher than $30\sigma$.

\clearpage
\chapter{Interpretation and test of $H(x)$  scaling at higher values of $|t|$}\label{chap:Hxvalidity}

The $H(x)$ scaling property of elastic $pp$ scattering allows us to scale out the energy dependence of the elastic differential cross section data in a certain $(s,t)$ range as discussed in \cref{sec:Hx}. This $H(x)$ scaling property of $pp$ elastic scattering was used to compare the TOTEM measured $pp$ and the D0 measured $p\bar p$ differential cross sections at the TeV $\sqrt s$ domain resulting in a model-independent, data-driven odderon signal with a statistical significance of at least 6.26$\sigma$ \cite{Csorgo:2019ewn,Csorgo:2023rzm,Csorgo:2020rlb,Csorgo:2020msw}. My model-dependent results, presented in this Chapter, indicate that this comparison is made in the domain of validity of the $H(x)$ scaling.

In the present Chapter, in \cref{sec:Hxmodels}, I show that models of elastic scattering with an impact parameter distribution depending on $b$ only via a dimensionless scaling variable, \mbox{$\tilde\xi=b/R(s)$,} where $R(s)$ is an internal scale and can be given in terms of the nuclear slope parameter, $R(s)\equiv\sqrt{B_0(s)}$, manifest scaling behavior at all values of $t$. By this, I provide an interpretation for the observed $H(x)$ scaling of the data at higher $|t|$ values, including the region of the diffractive minimum-maximum structure. In \cref{sec:HxReBB}, I identify the $H(x)$ scaling limit of the ReBB model. In \cref{sec:scfun},
utilizing the $H(x)$ scaling limit of the ReBB model, I test model-dependently the $H(x)$  scaling of $pp$ scattering against the $pp$ and $p\bar p$ differential cross section data in the TeV energy domain. 


This chapter is based on Refs.~\cite{Csorgo:2019ewn,Szanyi:2022ezh,Csorgo:2023rzm,Csorgo:2020rlb,Csorgo:2020msw}.

\vfill

\section{Models with scaling behavior}\label{sec:Hxmodels}

\vspace{-0.2cm}

The $H(x)$ scaling behavior was found to be present not only in the exponentially decreasing low-$|t|$ domain but also in the dip-bump region and at high $|t|$ values (see \cref{sec:Hx}). In this section, I introduce a special class of models containing an $s$-dependent internal scale, $R(s)$, and show that this class of models manifests scaling behavior not only at low~$|t|$ values but at all values of $|t|$. By identifying $R(s)$ with $\sqrt{B_0(s)}$,  I demonstrate that (i)~this scaling is the $H(x)$ scaling and (ii)~the $H(x)$ scaling function can have more complex shapes, other than $e^{-x}$.

Let us assume that the impact parameter representation of the scattering amplitude can be written in a factorized form, \begin{equation}\label{eq:amplstr}
    \widetilde T_{\rm el} (s,b) =  C(s)\widetilde E(\tilde \xi(b,s)),
\end{equation}
where $C(s)$ is some, in general,  $s$-dependent and complex normalization factor, and  $\widetilde E(\tilde \xi)$ is, in general, a complex function of any shape that describes the $b$-dependence of the scattering amplitude through the dimensionless scaling variable,
\begin{eqnarray}
       \tilde \xi \equiv \tilde \xi(s,b) =  \frac{b}{R(s)},
\end{eqnarray}
where $R(s)$ is some real, $s$-dependent internal scale parameter. 

\cref{eq:amplstr} via \cref{eq:relPWtoeik_2} yields
\begin{equation}\label{eq:tdepamp}
         T_{el}(s, t) =  2\pi R^2(s)C(s)E(\xi(t,s)),
\end{equation}
where $E(\xi)$ describes the $t$-dependence via the dimensionless variable
\begin{eqnarray}
    \xi \equiv \xi (t,s)=q(t)R(s),
\end{eqnarray}
with $q(t)=\sqrt{-t}$.
The function $E(\xi)$ arises by Fourier transforming $\widetilde E(\tilde \xi)$:
\begin{equation}\label{eq:Efunc}
    E\left(q(t)R(s)\right)=\frac{1}{2\pi R^2(s)}\int {\rm d}^2b e^{i\vec q\cdot\vec b}\widetilde E\left(b/R(s)\right).
\end{equation}

\cref{eq:tdepamp} gives the following result for the differential cross section:
\begin{equation}\label{eq:dsigdtscaling}
    \frac{d\sigma_{\rm el}}{dt}(s,t) = \pi R^4(s) |C(s)|^2 | E(\xi(t,s))|^2 \,.
\end{equation}
The optical point reads
\begin{equation}\label{eq:normscala}
    \frac{d\sigma_{\rm el}}{d t}\bigg|_{t=0}  = \pi R^4(s)|C(s)|^2 | E(0)|^2 \,.
\end{equation}
Then, the scaling relation is obtained as 
\begin{equation} 
    \frac{d\sigma_{\rm el}}{d t}\bigg/\frac{d\sigma_{\rm el}}{d t}\bigg|_{t=0} = \frac{| E(q(t)R(s))|^2}{| E(0)|^2}
\end{equation}
based on which I define the scaling function 
\begin{equation}\label{eq:Hx-general}
     \mathcal{H}(x)=\frac{| E(\sqrt{x})|^2}{| E(0)|^2},
\end{equation}
where
\begin{equation}
    x=q^2(t)R^2(s)=-tR^2(s).
\end{equation}
Then
\begin{equation} \label{eq:dsigma-Hx_general}
    \frac{d\sigma_{\rm el}}{d t}\bigg/\frac{d\sigma_{\rm el}}{d t}\bigg|_{t=0} = \mathcal{H}(x)\bigg|_{x=-tR^2(s)}.
\end{equation}
It is clear that that $\mathcal{H}(0)=1$, 
while 
the integral of $\mathcal{H}(x)$, $\int_{0}^{\infty} dx\mathcal{H}(x)$ is not necessarily unity.

Let us apply the transformation on the differential cross section used to obtain the $H(x)$ scaling function in \cref{sec:Hx}. For this, the first step is to calculate the elastic cross section and the slope parameter. 

The elastic cross section is calculated using \cref{eq:elastic_cross_section} with introducing the variable
\begin{equation}
    \vec{q}\,' = \vec{q} R(s).
\end{equation}
Thus, the elastic cross section reads
\begin{equation}\label{eq:elxsecscaling}
\sigma_{\rm el} = 2\pi R^2(s)|C(s)|^2\int_{0}^{\infty}{\rm d}q'\,q'|E(q')|^2.
\end{equation}

The slope parameter is calculated based on \cref{eq:Bst0}. Taking into account \cref{eq:dsigma-Hx_general}, we have 
\begin{equation}\label{eq:slopescaling}
    B_0(s)=\lim_{t\to 0}\frac{\rm d}{{\rm d}t}\ln \mathcal{H}(-tR^2(s))=-R^2(s)\mathcal{H}'(0)
\end{equation}
where 
\begin{equation}
\mathcal{H}'(0) \equiv \frac{\rm d}{{\rm d}x}\mathcal{H}(x)\bigg|_{x\to0}.
\end{equation}
The identification, $R^2(s)\equiv B_0(s)$, fixes  $\mathcal{H}'(0)=-1$. Now using \cref{eq:dsigdtscaling}, \cref{eq:Hx-general}, \cref{eq:elxsecscaling}, \cref{eq:slopescaling}, and 
\begin{equation}
    q'=\sqrt{x},
\end{equation}
we have
\begin{equation}\label{eq:Hxscaleddxsec}
    H(x)=\frac{1}{B_0(s)\,\sigma_{\rm el}(s)} \frac{d\sigma_{\rm el}}{d t}  \bigg|_{t=-x/B_0(s)}  \equiv \frac{\mathcal{H}(x)}{\int_{0}^\infty {\rm d}x'\mathcal{H}(x')}
\end{equation}
with $H(0)=-H'(0)$ and  $\int_{0}^\infty {\rm d}xH(x) = 1$.

\cref{eq:Hxscaleddxsec} has the $H(x)$ scaling property of the differential cross section: the differential cross section manifests $H(x)$ scaling by applying the same transformations used to obtain the scaling function in \cref{sec:Hx} in the case of a differential cross section with a purely exponential shape. The $H(x)$ scaling function given in  \cref{eq:Hxscaleddxsec} is not necessarily an exponential
scaling function, it can have a more general shape as determined via \cref{eq:Hx-general} by the function $E(\sqrt{x})$ given in \cref{eq:Efunc}. Furthermore, it is clear that if $\int_{0}^\infty {\rm d}x'\mathcal{H}(x')=1$,
\begin{equation}\label{eq:HH}
     H(x) \equiv \mathcal{H}(x).
\end{equation}

Actually, $H(x)$ is very close to $\mathcal{H}(x')$: numerical calculations using the $H(x)$ scaling limit of the ReBB model detailed in \cref{sec:HxReBB} of this dissertation gives $$H(0) = -H'(0)= 0.998$$ and  $$\int_{0}^\infty {\rm d}x'\mathcal{H}(x') = 1.002.$$

The shapes of the $pp$ and $p\bar p$ differential cross sections are nearly exponentially falling at low-$|t|$. The integral $\int_{0}^\infty {\rm d}xH(x)$ gets the leading contribution from the nearly exponentially falling part. For example, the numerical calculations using  the $H(x)$ scaling limit of the ReBB model, give: $\int_{0}^{5} {\rm d}xH(x)=0.994$, $\int_{0}^{10} {\rm d}xH(x)=0.999$, and $\int_{0}^{50}{\rm d}xH(x)=1.000$. 



Every model of elastic scattering with an impact parameter representation amplitude of the form given by \cref{eq:amplstr} manifests scaling behavior, and the dependence of the scaling variable $x$ is given in terms of the modulus squared of the Fourier-transform of the function $\tilde E(\tilde{\xi})$ that determines the $b$-dependence of the elastic scattering amplitude only through the dimensionless combination $\tilde\xi = \frac{b}{R(s)}$. In such a scenario, the scaling that transforms out the $s$-dependence is directly connected to the impact parameter dependence of the elastic amplitude, and scaling violations may happen in an $\sqrt{s}$ domain where the \mbox{$s$-dependence} of the function $\tilde E$ can not be described in terms of a single energy-dependent scale parameter, $R(s)$, via the dimensionless variable $\tilde\xi$. 

It was shown in Ref.~\cite{Csorgo:2019egs} that $\tilde\sigma_{\rm in}(s,b)$, which is related to the scattering amplitude in the impact parameter representation (see \cref{eq:unitb_2a}), has a different $b$-dependent shape at $\sqrt s$ = 13 TeV at low $b$ values as compared to the results obtained at lower energies. This is the so-called hollowness effect. The statistical significance of this effect at 13 TeV is higher than $5\sigma$ \cite{Csorgo:2019egs}. This indicates that, at 13 TeV, there is a noticeable change in the $b$-dependence of the elastic scattering amplitude, leading to scaling violations.

$H(x)$ scaling shows some similarities with the so-called geometrical \mbox{scaling \cite{Jenkovszky:1987yd,Matthiae:1994uw,DiasdeDeus:1977af}} introduced in the 1970s. Geometrical scaling relates the differential cross section to the imaginary and real parts of the amplitude in a specific way. This relation takes the form of a non-linear first-order differential equation. Moreover, geometrical scaling requires the condition $\Omega(s,b) = \Omega\left(b/R(s)\right)$ which after the identification, $R^2(s)\equiv B_0(s)$, is a sufficient condition for $H(x)$ scaling (see \cref{eq:amplstr} and \cref{eq:impact_ampl_eik_sol}). 
Scaling properties of recent LHC TOTEM data on $pp$ elastic scattering were analyzed also by the authors of Ref.~\cite{Baldenegro:2022xrj}.

The main conclusion of this section is that the physical reason for the $H(x)$ scaling is that the impact parameter dependence of the scattering amplitude can be described by a function of a dimensionless variable written in terms of a single $s$-dependent internal scale parameter that can be identified with $\sqrt{B_0(s)}$.

\section{H(x) scaling limit of the ReBB model}\label{sec:HxReBB}


The full ReBB model, as demonstrated in \cref{chap:odderon}, describes the available elastic $pp$ and $p\bar p$ differential cross section, total cross section and $\rho_0$ datasets in the kinematic range of 0.546~TeV~$\leq\sqrt{s}\leq 8$~TeV and 0.38~GeV$^2$~$\leq -t\leq1.2$~GeV$^2$ in a statistically acceptable manner with $CL$ $\ge $ 0.1\%.  In this section, I identify the $H(x)$ scaling limit of this ReBB model.  The $H(x)$ scaling limit of the ReBB model is a particular example of the general class of scaling models introduced in the previous section.

In the ReBB model, the $b$-dependence of the scattering amplitude is described by \cref{eq:ReBB_b_ampl} via $\tilde{\sigma}_{\rm in}(s,b)$. $\tilde{\sigma}_{\rm in}(s,b)$ has the form (see \cref{eq:tilde_sigma_inel}, \cref{eq:sig_tilde_b} of \cref{chap:levy}, and \mbox{\hyperref[sec:app_BBcalc]{Appendix B}):}
\begin{equation}
    \tilde{\sigma}_{\rm in}(s,b) = \sum_i c_i(s) \tilde f_i\left(\frac{b}{R_i(s)}\right),  
\end{equation}
where $c_i(s)$ are dimensionless functions of the energy-dependent ReBB model scale parameters $R_q(s)$, $R_d(s)$, and $R_{qd}(s)$; 
\begin{eqnarray}
    \tilde f_i\left(\frac{b}{R_i(s)}\right) = \exp{-\left(\frac{b}{R_i(s)}\right)^2};
\end{eqnarray}
and $R_i(s)$ are  length dimension functions of $R_q(s)$, $R_d(s)$, and $R_{qd}(s)$.

The structure of the ReBB\, model allows for\, a\, scaling behavior\, given that\, the \mbox{$s$-dependencies} of the ReBB model scale parameters is determined by the same, factorizable function denoted as $r(s)$, $i.e.$, 
   \begin{equation}
R_q(s)  =  r(s) R_q(s_0) \,, \label{eq:Hx-Rq}
   \end{equation}
      \begin{equation}
R_d(s)  =  r(s) R_d(s_0) \,, \label{eq:Hx-Rd}
   \end{equation}
      \begin{equation}
R_{qd}(s)  =  r(s) R_{qd}(s_0) \,,\label{eq:Hx-Rqd}
   \end{equation}
where $R_q(s_0)$, $R_d(s_0)$, and $R_{qd}(s_0)$ are the values of the ReBB model scale parameters at some reference energy, $s_0$. 

We saw in \cref{chap:rebbdesc} that the values of the parameters $\lambda$ and $A_{qq}$ are energy independent, $i.e.$,
\begin{equation}
  \lambda(s)  =  \lambda(s_0),\,\,\,  A_{qq}(s)  =  A_{qq}(s_0) \,. \label{eq:Hx-lamA}
\end{equation}
Let us assume also that the opacity parameter $\alpha_R$ is energy independent, $i.e.$,
   \begin{equation}
             \alpha_R(s)  =  \alpha_R(s_0) \,. \label{eq:Hx-alpha}
     \end{equation}
As we saw in \cref{sec:rebb_endep_TeV}, the opacity parameter, $\alpha_R$, for elastic $pp$ scatting practically has an energy independent value in the c.m. energy range of 0.546~TeV~$\leq\sqrt{s}\leq 8$~TeV. This means that the requirement of \cref{eq:Hx-alpha} is satisfied for $pp$ scattering. This also explains why the $H(x)$ scaling is not present in elastic $p\bar p$ scattering: as we saw in \cref{sec:rebb_endep_TeV}, the opacity parameter, $\alpha_R$ for elastic $p\bar p$ scattering has an energy evolution in the c.m. energy range of 0.546~TeV~$\leq\sqrt{s}\leq 8$~TeV. It is true, however, that a scaling behavior is still possible if $\alpha_R$ depends on $s$ but only via the variable $b/R(s)$.

Sticking to the requirements as given by \cref{eq:Hx-Rq}, \cref{eq:Hx-Rd}, \cref{eq:Hx-Rqd}, \cref{eq:Hx-lamA}, and \cref{eq:Hx-alpha}, \cref{eq:ReBB_b_ampl} takes the form 
\begin{equation}\label{eq:scalingb}
    \widetilde T_{\rm el} (s,b) =i\left(1-e^{i\, \alpha_R\, \tilde\sigma_{\rm in}(b/R(s))}\sqrt{1-\tilde\sigma_{\rm in}(b/R(s))}\right) \equiv \widetilde E(b/R(s)),
\end{equation}
with
\begin{equation}\label{eq:Rsr0}
R(s) = r(s)r_0,
\end{equation}
where $r_0$ is some length dimension and $s$-independent scale parameter.
\cref{eq:scalingb} clearly has a scaling structure as defined by \cref{eq:amplstr}. \cref{eq:scalingb} via \cref{eq:relPWtoeik_2} and \cref{eq:dsigma-Hx_general} yields
\begin{equation}\label{eq:rebbhx_1}
    \mathcal{H}(x') = \frac{| T_{\rm el}(s,t)|^2}{| T_{\rm el}(s,t=0)|^2}\bigg|_{t=-x'/ R^2(s)} = \frac{1}{\frac{d\sigma_{\rm el}}{d t}\big|_{t=0}} \frac{d\sigma_{\rm el}}{d t}  \bigg|_{t=-x'/ R^2(s)},
\end{equation}
where $T_{\rm el}(s,t)$ is the Fourier transformed of \cref{eq:scalingb}. 
Now based on \cref{eq:rebbhx_1} and \cref{eq:Hxscaleddxsec}, we can construct the $H(x)$ scaling function from the scaling ReBB model amplitude given by \cref{eq:scalingb}. The resulting $H(x)$ scaling function is not a simple exponential: $H(x)\neq e^{-x}$. 

Because of the structure of the ReBB model, $B_0(s)$ must be given in terms of a length squared dimension function of $R_q(s)$, $R_d(s)$, and $R_{qd}(s)$. Since $R_q(s)$, $R_d(s)$, and $R_{qd}(s)$ evolve with the same $s$-dependent factor, $r(s)$, in the scaling limit, the slope parameter must evolve in $s$ with $r^2(s)$, $i.e.$,
\begin{equation}\label{eq:scaleslope}
    B_0(s) =  r^2(s) B_0(s_0).
\end{equation}
Again, a dimensional analysis gives that, in the $H(x)$ scaling limit, the elastic differential cross section, the elastic and the total cross section must evolve in $s$ with $r^4(s)$ or  $r^2(s)$, $i.e.$,
\begin{equation}
\frac{d\sigma_{\rm el}}{dt}(s,t) = r^4(s) \frac{d\sigma_{\rm el}}{dt}\left(s_0, r^2(s)t\right),
\end{equation}
\begin{equation}\label{eq:scaleel}
\sigma_{\rm el}(s) =  r^2(s) \sigma_{\rm el}(s_0),
\end{equation}
\begin{equation}\label{eq:scaletot}
\sigma_{\rm tot}(s) =  r^2(s) \sigma_{\rm tot}(s_0).
\end{equation}
With the identification, $R^2(s)\equiv B_0(s)$, and based on \cref{eq:Rsr0} and \cref{eq:scaleslope}, we can write
\begin{equation}
    r_0=\sqrt{B_0(s_0)}.
\end{equation}


\section{H(x) scaling ReBB model and the data}\label{sec:scfun}

In this section, I test the $H(x)$ scaling of $pp$ scattering model-dependently by comparing the $H(x)$ scaling version of the ReBB model to the experimental data using the refind $\chi^2$ definition of \cref{eq:chi2_refind}. In the $H(x)$  scaling limit of the ReBB model, the scale parameters evolve in energy by a factorizable function $r(s)$ as given by \cref{eq:Hx-Rq}, \cref{eq:Hx-Rd}, and \cref{eq:Hx-Rqd}, while all the other parameters are energy independent.  

Considering the ReBB model $H(x)$ scaling condition of \cref{eq:Hx-alpha}, the $H(x)$ scaling could be present in elastic $pp$ scattering in the $\sqrt{s}$ domain of \mbox{0.546~TeV~$\leq\sqrt{s}\leq 8$~TeV} where $\alpha_{R}^{pp}$ is compatible with a constant value. However, in $p\bar p$ elastic scattering the $H(x)$ scaling is clearly violated in the $\sqrt{s}$ domain of \mbox{0.546~TeV~$\leq\sqrt{s}\leq 8$~TeV} because of the energy evolution of the $\alpha_{R}^{p\bar p}$ parameter. Surprisingly, the test of the H(x) scaling of $pp$ scattering against the $p\bar p$ scattering data is possible since the values of the ReBB model scale parameters $R_q$, $R_d$, and $R_{qd}$ are the same in $pp$ and $p\bar p$ scattering within errors. Though $pp$ and $p\bar p$ elastic differential cross section data are not measured at the same energies, the energy evolution of each of the parameters $R_q$, $R_d$, and $R_{qd}$ is consistent within errors with the same linear-logarithmic rise in $pp$ and $p\bar p$ scattering (see \cref{sec:rebb_endep_TeV}). The $H(x)$ scaling condition for the opacity parameter as given in \cref{eq:Hx-alpha} is violated in $p\bar p$ scattering, however, by letting the $\alpha_R$ parameter to be free but constraining the values of the scale parameters with the scaling condition during the ReBB model fits to the data, the $H(x)$ scaling conditions given in \cref{eq:Hx-Rq}, \cref{eq:Hx-Rd}, and  \cref{eq:Hx-Rqd} can be tested against the $p\bar p$ elastic differential cross section data. Thus, in the model-data comparisons below, when I use the $p\bar p$ data, I require only the fulfillment of the $H(x)$ scaling conditions for the scale parameters as given in \cref{eq:Hx-Rq}, \cref{eq:Hx-Rd}, and  \cref{eq:Hx-Rqd}. However, when I use $pp$ data, I require also the fulfillment of the $H(x)$ scaling condition for the opacity parameter as given in \cref{eq:Hx-alpha}. In both cases, the $H(x)$ scaling conditions for $\lambda$ and $A_{qq}$ as given in \cref{eq:Hx-lamA} is trivially fulfilled since these parameters are always kept on the same fixed values at all energies (see \cref{chap:rebbdesc}). All model-data comparison below is done in the $-t$ range of 0.38~GeV$^2$~$\lesssim -t\lesssim1.2$~GeV$^2$, where the ReBB model description of the data is validated (see \cref{chap:odderon}).

 
The values of ReBB model parameters are most precisely determined at \mbox{$\sqrt{s}=7$ TeV} (see \cref{sec:rebb_desc_refind_TeV}). For this reason, I chose this energy as the reference, $i.e.$, \mbox{$\sqrt{s_0}=7$ TeV.} The values of the ReBB model scale and opacity parameters at the reference energy of $\sqrt{s_0}=7$ TeV are: $R_q(s_0) = 0.438$ fm, $R_d(s_0) = 0.920$ fm, $R_{qd}(s_0) = 0.333$ fm, and \mbox{$\alpha_{R}^{pp}(s_0)$ = 0.125.} The uncertainties of these parameters are presented in \cref{fig:reBB_model_fit_7_TeV} and in \cref{tab:rebb_fit_parameters}.

\begin{figure}[htb!]
	\centering 
    \includegraphics[width=0.8\linewidth]{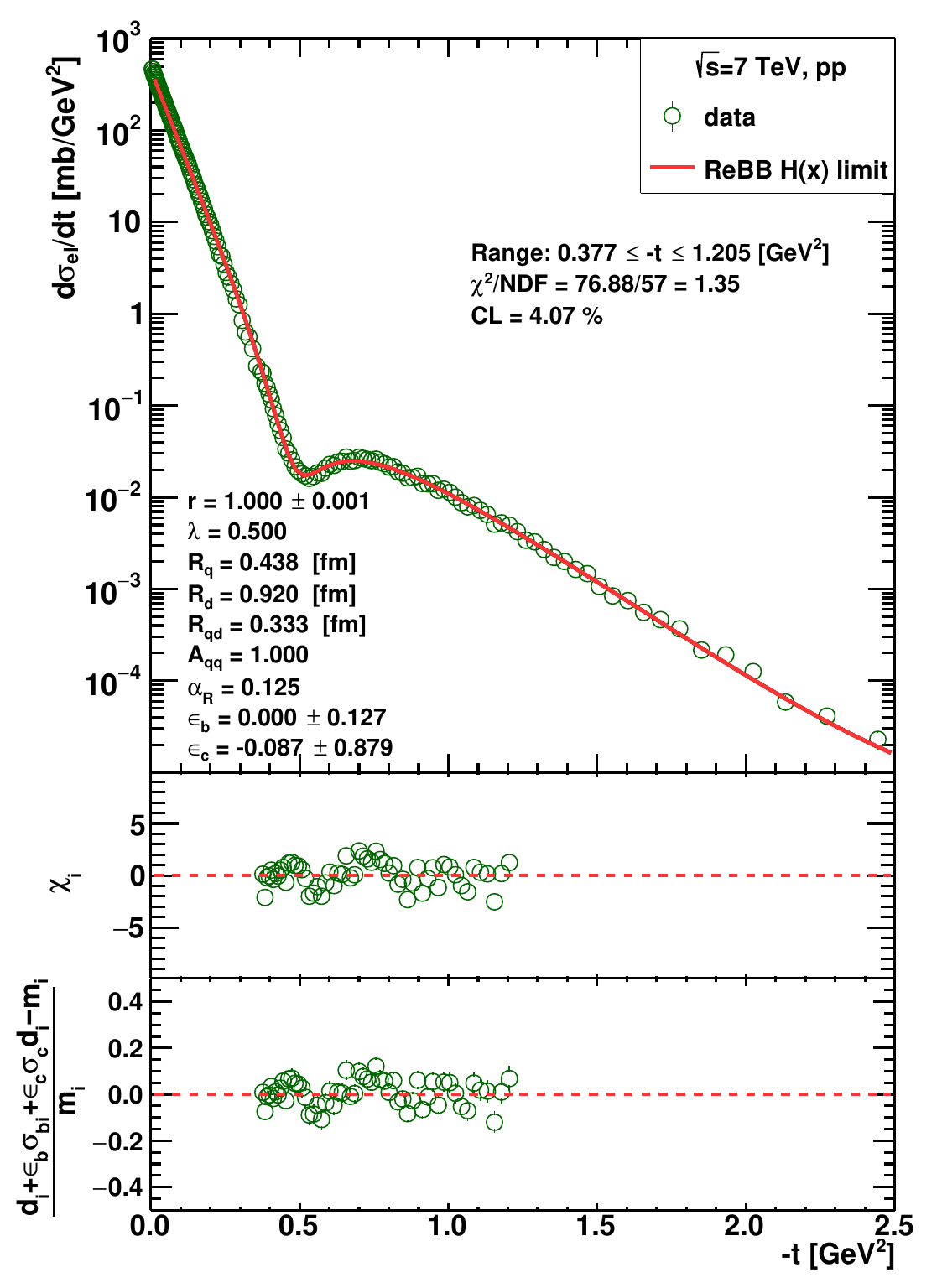}
	\caption{H(x) scaling ReBB model fit to the LHC TOTEM $\sqrt{s}=7$ TeV $pp$ differential cross section data \cite{TOTEM:2013lle} in the $-t$ range of 0.377~GeV$^2$~$\leq -t\leq1.205$~GeV$^2$.}
	\label{fig:reBB_model_hx_s-cross-check-at-7-TeV}
\end{figure}

The fit to the $\sqrt{s}=7$ TeV $pp$ differential cross section data \cite{TOTEM:2013lle} with the $H(x)$  scaling ReBB model in the $-t$ domain of \mbox{0.377~GeV$^2$~$\leq -t\leq1.205$~GeV$^2$} is presented in \cref{fig:reBB_model_hx_s-cross-check-at-7-TeV}. Only the value of $r$, $\epsilon_b$, and $\epsilon_c$ are optimized. The $CL$ of the fit is 4.07\%. At \mbox{$\sqrt{s}=7$ TeV,} the value of $r$ that determines the common energy dependence of the ReBB model scale parameters according to \cref{eq:Hx-Rq}, \cref{eq:Hx-Rd}, and \cref{eq:Hx-Rqd} is 1.0 $\pm $ 0.001. Since the reference energy is $\sqrt{s_0}=7$ TeV, it is expected based on \cref{eq:Hx-Rq}, \cref{eq:Hx-Rd}, and  \cref{eq:Hx-Rqd} that $r$ at $\sqrt{s}=7$ TeV is unity within errors.

\begin{figure}[htb!]
	\centering 
    \includegraphics[width=0.8\linewidth]{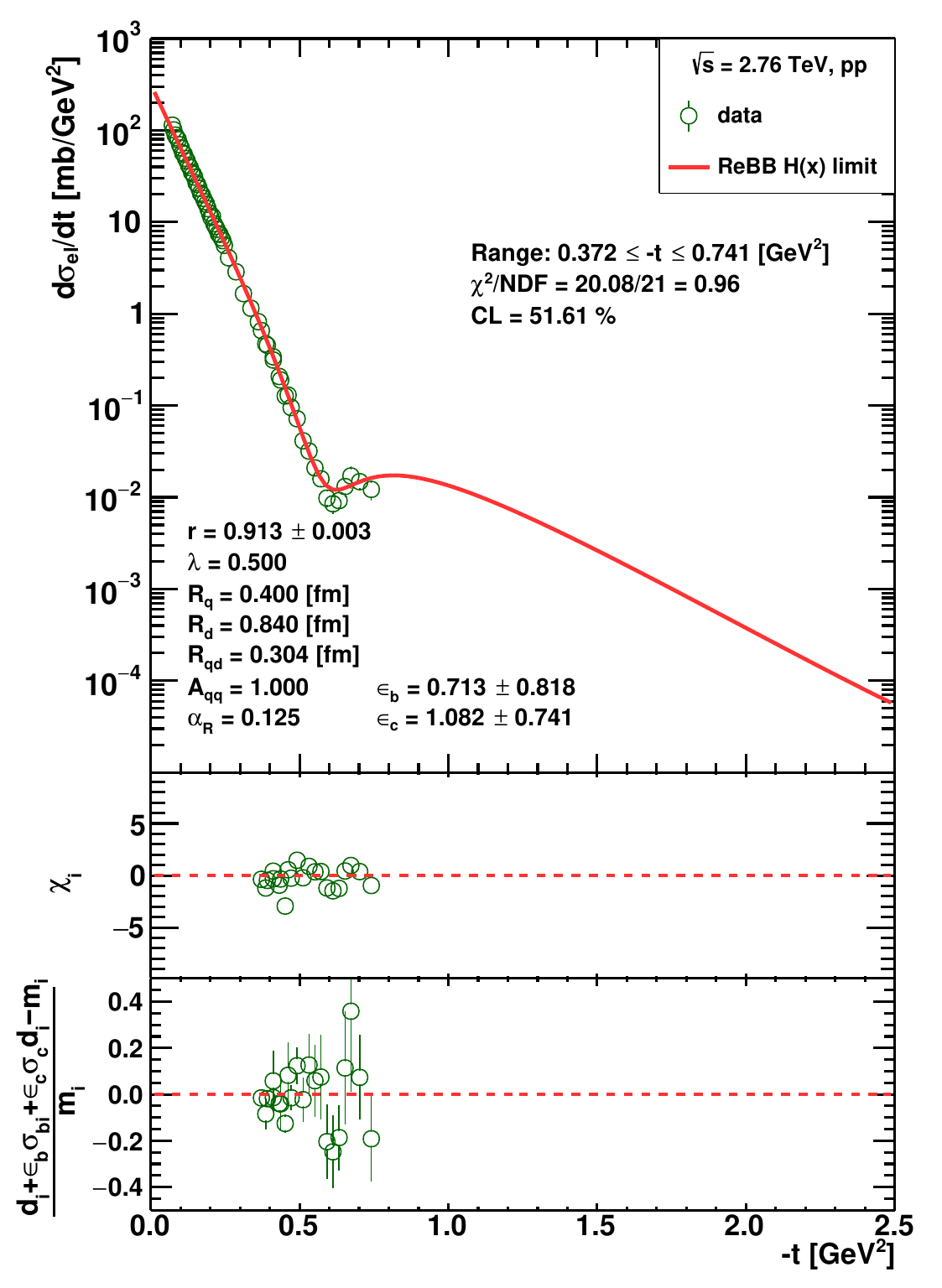}
	\caption{H(x) scaling ReBB model fit to the LHC TOTEM $\sqrt{s}=2.76$ TeV $pp$ differential cross section data \cite{TOTEM:2018psk} in the $-t$ range of 0.372~GeV$^2$~$\leq -t\leq0.741$~GeV$^2$.}
	\label{fig:reBB_model_hx_s-cross-check-at-2.76-TeV}
\end{figure}

The fit to the $\sqrt{s}=2.76$ TeV $pp$ differential cross section data \cite{TOTEM:2018psk} with the $H(x)$  scaling ReBB model in the $-t$ range of \mbox{0.372~GeV$^2$~$\leq -t\leq0.741$~GeV$^2$} is presented in \cref{fig:reBB_model_hx_s-cross-check-at-2.76-TeV}. Only the value of $r$, $\epsilon_b$, and $\epsilon_c$ are optimized. The $CL$ of the fit is 51.61\%. At $\sqrt{s}=2.76$ TeV, the value of $r$ is 0.913 $\pm $ 0.003. 

\begin{figure}[htb!]
	\centering 
    \includegraphics[width=0.8\linewidth]{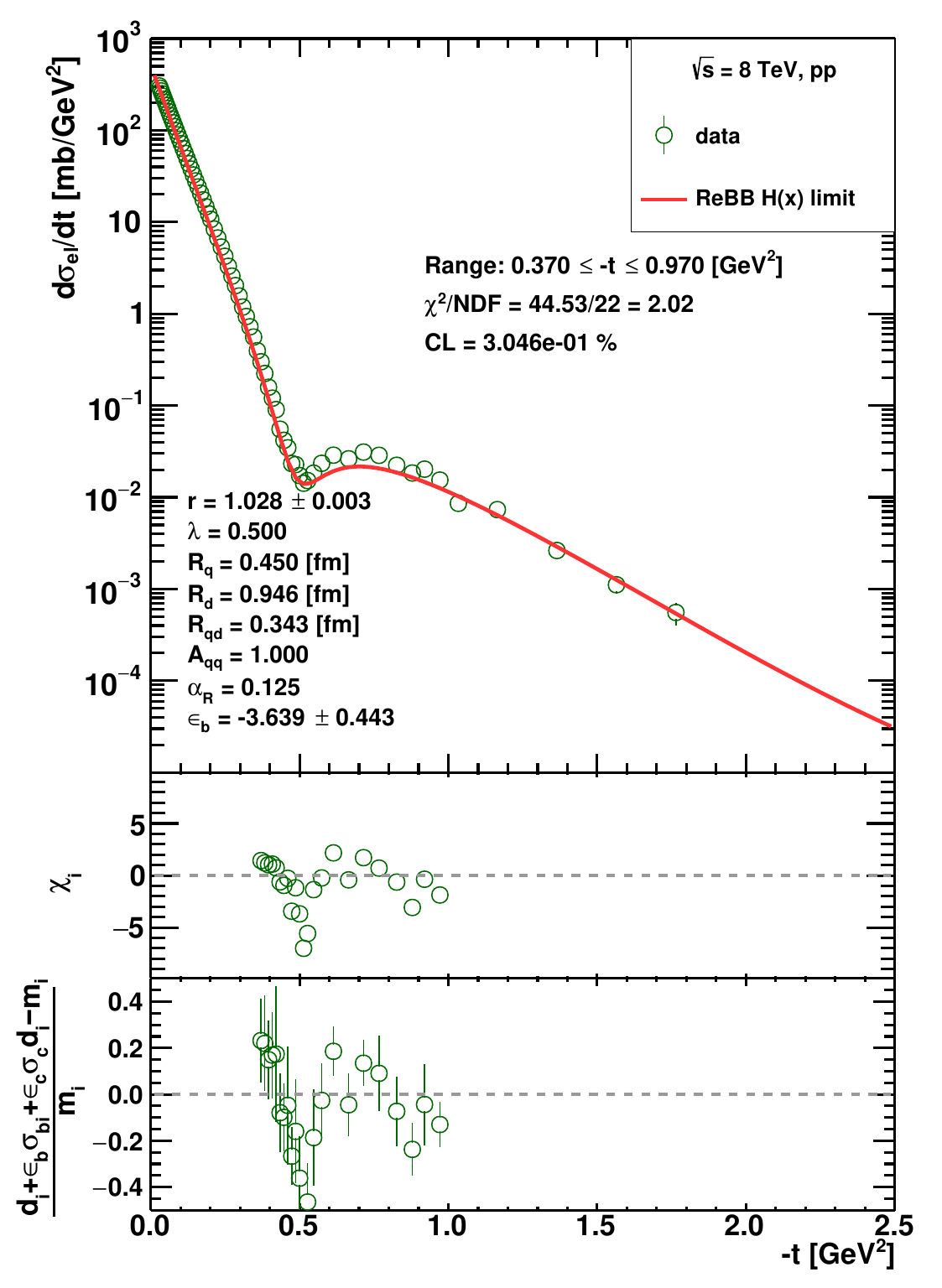}
	\caption{H(x) scaling ReBB model fit to the LHC TOTEM $\sqrt{s}=8$ TeV $pp$ differential cross section data \cite{TOTEM:2021imi} in the $-t$ range of 0.370~GeV$^2$~$\leq -t\leq0.970$~GeV$^2$.}
	\label{fig:reBB_model_hx_s-cross-check-at-8-TeV}
\end{figure}

The fit to the $\sqrt{s}=8$ TeV $pp$ differential cross section data \cite{TOTEM:2021imi} with the $H(x)$  scaling ReBB model in the $-t$ range of \mbox{0.37~GeV$^2$~$\leq -t\leq0.97$~GeV$^2$} is presented in \cref{fig:reBB_model_hx_s-cross-check-at-8-TeV}. Only the value of $r$, $\epsilon_b$ are optimized. The $CL$ of the fit is 0.3\%. At $\sqrt{s}=8$ TeV, the value of $r$ is 1.028 $\pm $ 0.003. Extending the fit range up to \mbox{$-t=1.2$ GeV$^2$,} the value of the $CL$ is less than 0.1\% unless an additional at least $\sigma_c=0.045$ overall normalization error\footnote{This particular value of $\sigma_c$ is obtained by gradually increasing the overall normalization error until the $CL$ reaches 0.1\%.} is considered and $\epsilon_c$ is optimized. However, no overall normalization error is separately given in Ref.~\cite{TOTEM:2021imi}; it is included in the point-to-point varying systematic errors of the data.

\begin{figure*}[htb!]
	\centering 
    \includegraphics[width=0.8\linewidth]{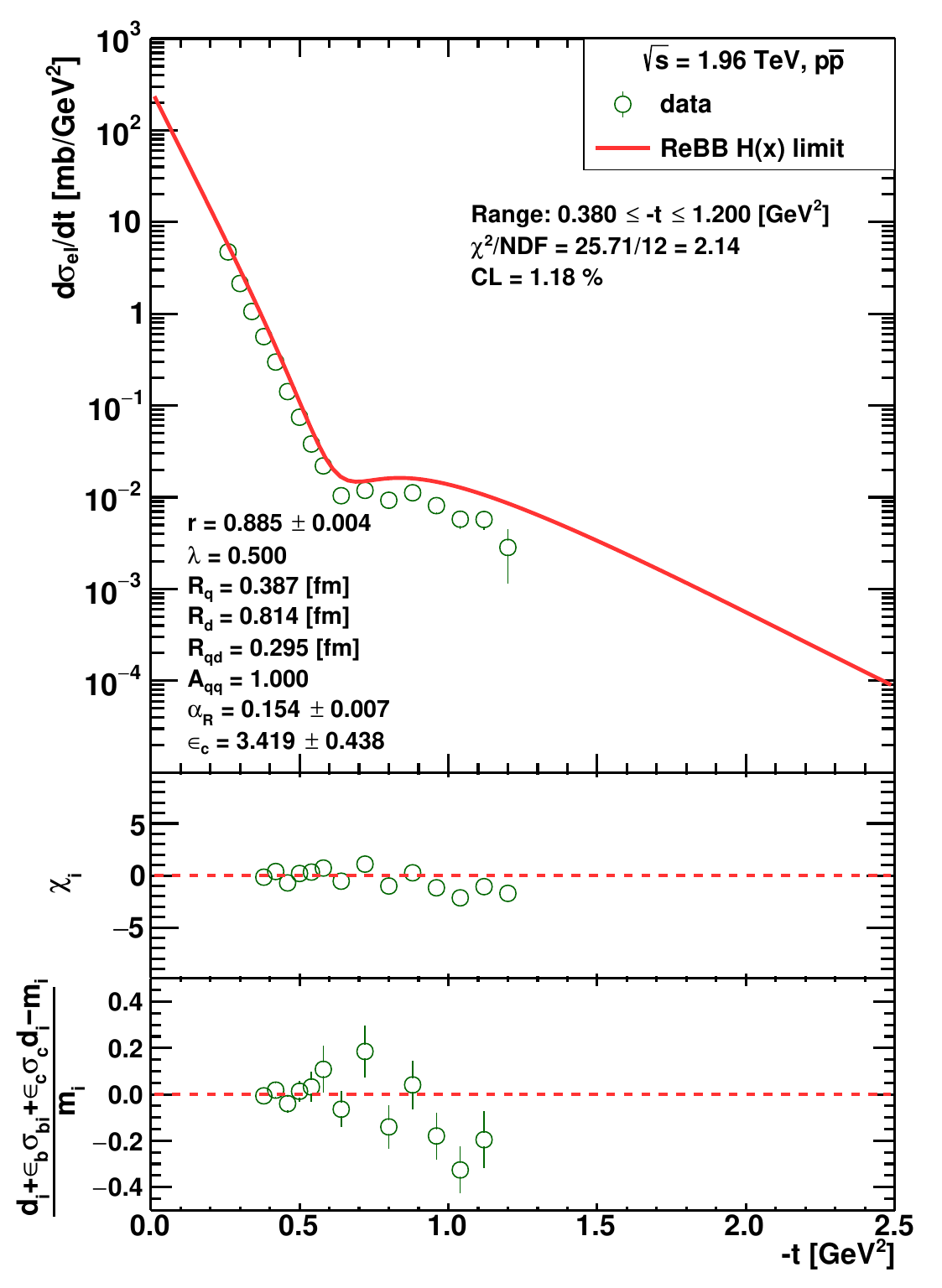}
	\caption{H(x) scaling ReBB model fit to the Tevatron D0 $\sqrt{s}=1.96$ TeV $p\bar p$ differential cross section data \cite{D0:2012erd} in the $-t$ range of 0.38~GeV$^2$~$\leq -t\leq1.2$~GeV$^2$.
	Note that the value of $\epsilon_c$ is rather large, outside its usual range of $(-1,1)$. Since the data points are not rescaled by $\epsilon_c$, in the upper panel of the plot, the data points are visibly below the best-fit curve.
	The good quality agreement between the rescaled data and the $H(x)$  limit ReBB model curve is visible in the middle and the bottom panels of the plot.}
	\label{fig:App-D-20.2-reBB_model_hx_s-cross-check-at-1.96-TeV}
\end{figure*}

\begin{figure}[htb!]
	\centering 
    \includegraphics[width=0.8\linewidth]{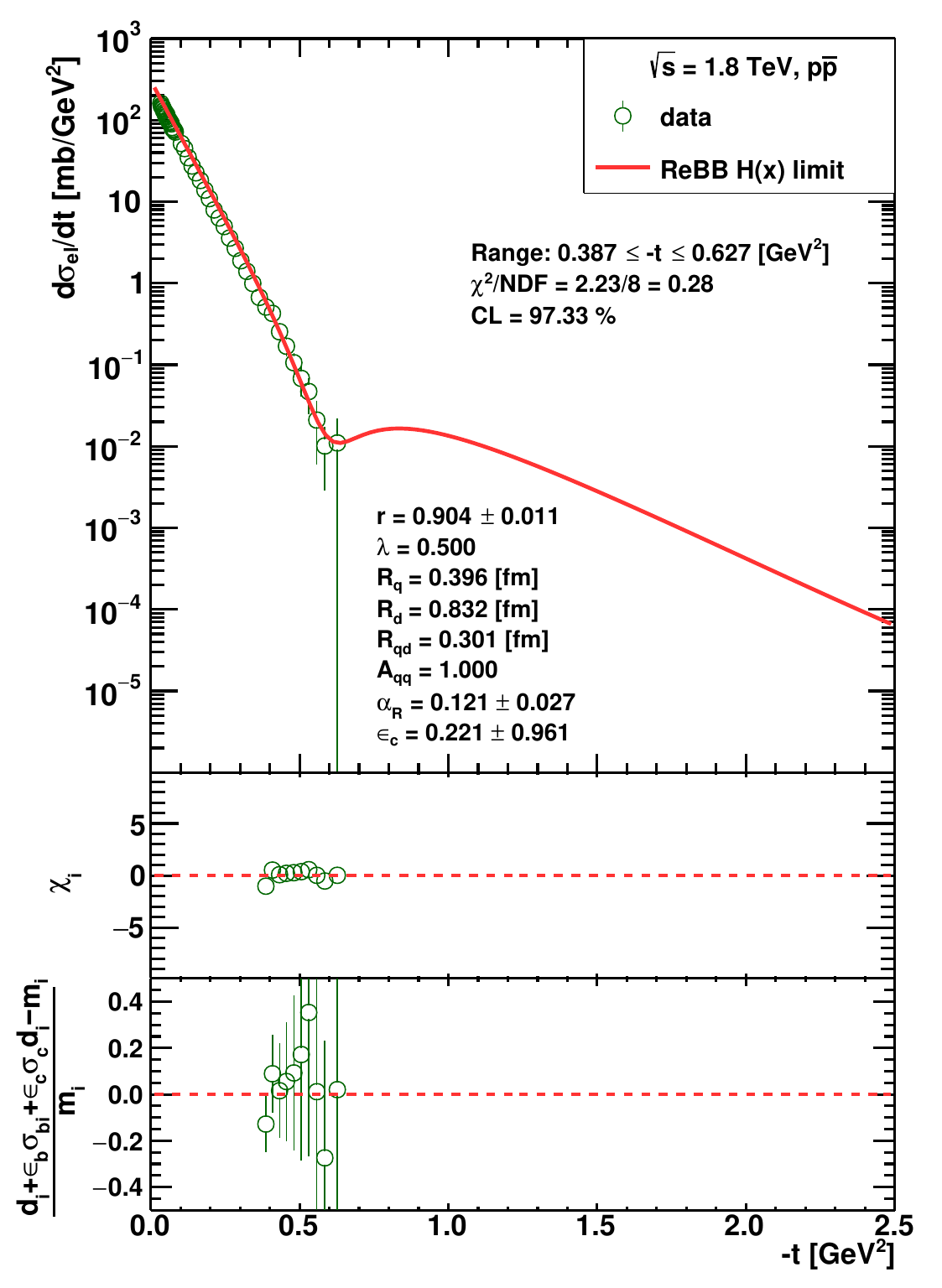}
	\caption{H(x) scaling ReBB model fit to the Tevatron E-710 $\sqrt{s}=1.8$ TeV $p\bar p$ differential cross section data \cite{E-710:1990vqb} in the $-t$ range of 0.387~GeV$^2$~$\leq -t\leq0.627$~GeV$^2$.}
  \vspace{-0.5cm}
	\label{fig:reBB_model_hx_s-cross-check-at-1.8-TeV}
\end{figure}

The fit of the $\sqrt{s}=1.96$ TeV $p\bar p$ differential cross section data \cite{D0:2012erd} with the $H(x)$  scaling ReBB model in the squared four-momentum transfer range of 0.38~GeV$^2$~$\leq -t\leq1.2$~GeV$^2$ is presented in \cref{fig:App-D-20.2-reBB_model_hx_s-cross-check-at-1.96-TeV}. Only the value of $r$, $\alpha_R$, and $\epsilon_c$ are optimized. The $CL$ of the fit is 1.18\%. At $\sqrt{s}=1.96$ TeV, the value of $r$ is 0.885 $\pm $ 0.004. According to this result, at $\sqrt{s} = 1.96$ TeV, 
the domain of validity of the $H(x)$ scaling in $pp$ scattering can not be excluded up to $-t \leq 1.2$ GeV$^2$, which corresponds to $x=-tB_0 = 20.2 $. 

One may notice that the value of $\epsilon_c$ at $\sqrt{s}=1.96$ TeV is rather large, outside its usual range of $(-1,1)$. Because of the rather large $\epsilon_c$ value, the data points are visibly below the best-fitted curve, as in the upper panel of \cref{fig:App-D-20.2-reBB_model_hx_s-cross-check-at-1.96-TeV} the data points were not rescaled by $\epsilon_c$. However, the good
quality agreement between the rescaled data and the solid red line is indicated on the two lower panels of \cref{fig:App-D-20.2-reBB_model_hx_s-cross-check-at-1.96-TeV} and by the value of the confidence level ($CL=1.18$\%).

Finally, the fit of the $\sqrt{s}=1.8$ TeV $p\bar p$ differential cross section data \cite{E-710:1990vqb} with the $H(x)$  scaling ReBB model in the $-t$ range of 0.387~GeV$^2$~$\leq -t\leq0.627$~GeV$^2$ is presented in \cref{fig:reBB_model_hx_s-cross-check-at-1.8-TeV}. Only the value of $r$, $\alpha_R$, and $\epsilon_c$ are optimized. The $CL$ of the fit is 97.33\%. At $\sqrt{s}=1.8$ TeV, the value of $r$ is 0.904 $\pm $ 0.011. 

The above results show that the ReBB model $H(x)$ scaling conditions \cref{eq:Hx-Rq}, \cref{eq:Hx-Rd}, and \cref{eq:Hx-Rqd}, are fulfilled on the available $pp$ and $p\bar p$ differential cross section data in the c.m. energy range of \mbox{2.76 TeV $\leq \sqrt{s} \leq$ 7 TeV} and in the $-t$ range of \mbox{0.38 GeV$^2$ $ \leq-t\leq1.2$ GeV$^2$} (this squared four-momentum transfer range corresponds to the tested validity range of the ReBB model; see \cref{chap:odderon}). We also saw that the ReBB model $H(x)$ scaling condition \cref{eq:Hx-alpha} is fulfilled on the available $pp$ differential cross section data in the c.m. energy range of \mbox{1.8 TeV $\leq \sqrt{s} \leq$ 7 TeV} and in the $-t$ range of \mbox{0.38 GeV$^2$ $ \leq-t\leq1.2$ GeV$^2$}. At $\sqrt{s} = 8$ TeV, the $H(x)$  scaling ReBB model describes the $pp$ differential cross section data in a bit more limited squared four-momentum transfer range of \mbox{0.37 GeV$^2$ $ \leq-t\leq0.97$ GeV$^2$} with \mbox{$CL>0.1\%$.}

In the following, I proceed with determining the energy dependence of the general scale parameter $r$. 

From the above fits, the value of $r$ is known at $\sqrt{s}=1.8$ TeV, 1.96 TeV, 2.76 TeV, \mbox{7 TeV,} and 8 TeV, since, from \cref{eq:Hx-Rq}, \cref{eq:Hx-Rd}, and \cref{eq:Hx-Rqd} one has

\begin{equation}\label{eq:rsfromReBBscales}
r(s) = \frac{R_q(s)}{R_q(s_0)}=\frac{R_d(s)}{R_d(s_0)}=\frac{R_{qd}(s)}{R_{qd}(s_0)}.
\end{equation}

From \cref{eq:scaleslope}, \cref{eq:scaleel}, and \cref{eq:scaletot} one has

\begin{equation}\label{eq:sloper}
r(s) = \sqrt \frac{B_0(s)}{B_0(s_0)} =\sqrt \frac{\sigma_{\rm el}(s)}{\sigma_{\rm el}(s_0)}= \sqrt \frac{\sigma_{\rm tot}(s)}{\sigma_{\rm tot}(s_0)}.
\end{equation}
Thus, one can calculate the function $r(s)$ directly from the experimental data on $B_0$, $\sigma_{\rm tot}$ and $\sigma_{\rm el}$, too.  

\begin{figure}[htb!]
	\centering
    \includegraphics[width=0.85\linewidth]{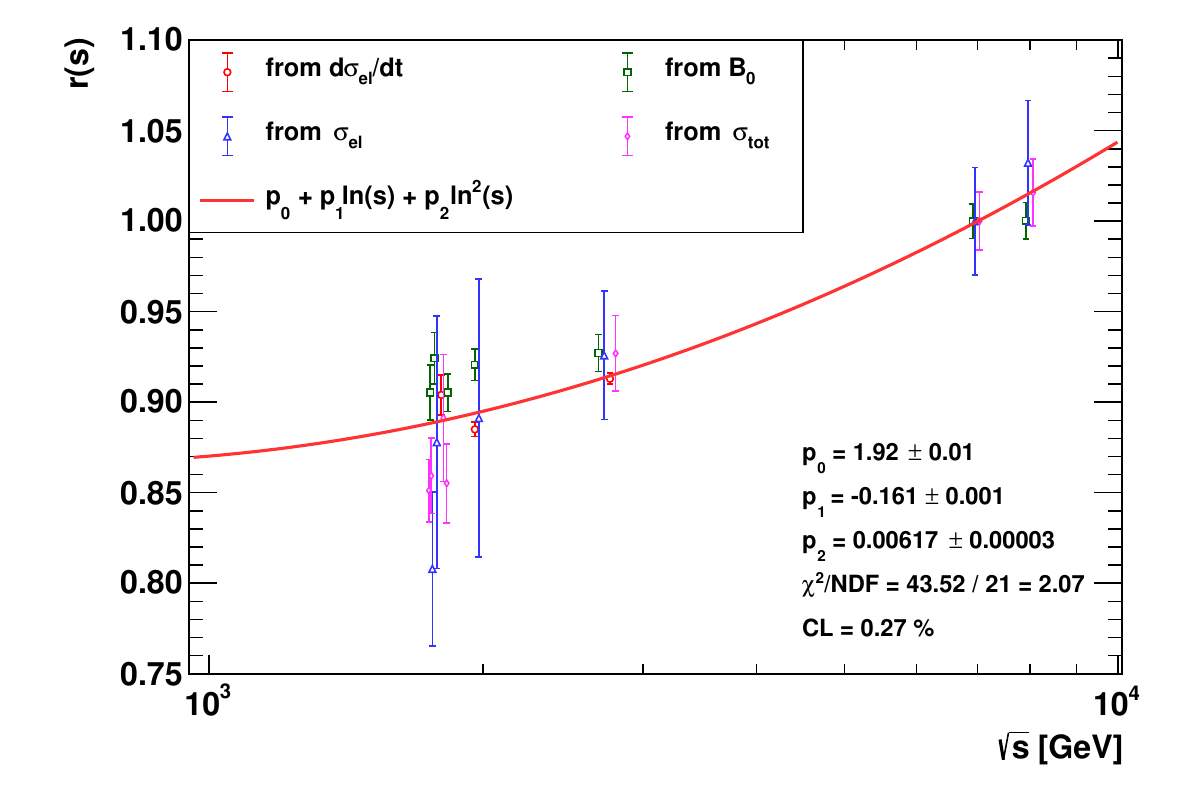}
    \vspace{-0.4cm}
	\caption{The function $r(s)$ as determined from $pp$ and $p\bar p$ $B_0$, $\sigma_{\rm el}$, and $\sigma_{\rm tot}$ data with the help of \cref{eq:sloper}, and from the $pp$ and $p\bar p$ $d\sigma_{\rm el}/dt$ data with the help of \cref{eq:rsfromReBBscales}. }
	\label{fig:rs}
\end{figure}

Keeping the reference energy of $\sqrt{s_0}=$ 7 TeV, \cref{fig:rs} shows the $r$ values as obtained from the experimental $B_0$, $\sigma_{\rm el}$, and $\sigma_{\rm tot}$ values \cite{E710:1989vnd,E-710:1990xdw,E-710:1990vqb,E710:1991bcl,E-811:2002iqx,D0:2012erd,Csorgo:2019ewn,TOTEM:2018psk,Nemes:2017gut,TOTEM:2017asr,TOTEM:2013lle,TOTEM:2012oyl} based on \cref{eq:sloper} as well as from the above fits to the differential cross section \mbox{data \cite{E-710:1990vqb,D0:2012erd,TOTEM:2018psk,TOTEM:2013lle,TOTEM:2021imi}} at different energies based on \cref{eq:rsfromReBBscales}. As presented in \cref{fig:rs}, unless the value of $r$ at $\sqrt{s}=8$ TeV as obtained from the $d\sigma_{\rm el}/dt$ data fit is included, the function $r(s)$ is compatible with a squared-logarithmic shape\footnote{The  squared-logarithmic energy evolution of $r$ violates the Froissart bound on the asymptotic growth of the total cross section (see \cref{sec:Regge}), however, this is not a problem since the ReBB model itself is valid only in a limited energy range certainly below 13 TeV.},
\begin{equation}\label{eq:squaredlog}
P(s)=p_0+p_1\ln{(s/s_0)}+p_2\ln^2{(s/s_0)},\,\,\,s_0=1~{\rm GeV}^2,
\end{equation}
with $p_0=1.92 \pm 0.01$, $p_1=-0.161 \pm 0.001$ and $p_2=0.00617 \pm 0.00003$. The $CL$ of the description is 0.27\%. If the value of $r$ at $\sqrt{s}=8$ TeV as obtained from the $d\sigma_{\rm el}/dt$ data fit is included, the $CL$ of a fit by \cref{eq:squaredlog} goes below the 0.1\% threshold value. There are two possible explanations for this phenomenon. Since we know that the $H(x)$ scaling is violated at $\sqrt{s}=13$ TeV \cite{Csorgo:2019ewn}, it is possible that some scaling violations start to arise already at $\sqrt{s}=8$ TeV. The other explanation is that the uncertainties of the differential cross section data at $\sqrt{s}=8$ TeV are published in an irregular way: no representative $|t|$ values are given, only $t$ bin sizes are given, and the overall normalization error is contained in the point-to-point varying systematic errors. A unified treatment of the uncertainties may improve the compatibility of the $r$ value at 8 TeV with the $r$ values at lower energies within a squared-logarithmic energy evolution trend.

I tested that all the available $pp$ and $p\bar p$ 
differential cross section data in the c.m. energy range of 1.8 TeV $\leq \sqrt{s} \leq$ 7 TeV and in the squared four-momentum transfer range of 0.38 GeV$^2$ $ \leq-t\leq1.2$ GeV$^2$ can be described with \mbox{$CL>0.1\%$}, when the $r$ values are taken from the determined squared-logarithmic shape as shown in \cref{fig:rs}. The same test fails at $\sqrt{s}=8$ TeV. The possible reasons are the same as detailed above.

\vspace{0.5cm}
\textbf{Summary}
\vspace{0.2cm}


In this Chapter, I showed that the $H(x)$ scaling of the data in the non-exponential higher-$|t|$ region can be interpreted with the fact that the $b$-dependence of the elastic scattering amplitude is determined by a function of a dimensionless variable, \mbox{$\tilde\xi=b/R(s)$,} where $R(s)$ is an internal scale and it can be identified with $\sqrt{B_0(s)}$.

I identified the $H(x)$ scaling limit of the ReBB model and demonstrated that this model manifests $H(x)$ scaling if its scale parameters evolve in energy with the same factorizable function $r(s)$ and all the other fit parameters are energy-independent. Based on experimental and phenomenological inputs, I concluded that the $r(s)$ function is compatible with a squared-logarithmic rise in $s$ in the c.m. energy domain of \mbox{1.8 TeV $\leq \sqrt{s} \leq$ 7 TeV.} 

Utilizing the $H(x)$ scaling limit of the ReBB model, I tested the $H(x)$ scaling conditions against the available $pp$ and $p\bar p$ elastic differential cross section data in the kinematic range from 8 TeV down to 1.8 TeV in the validated squared four-momentum transfer range of the original ReBB model. My results show that the $H(x)$ scaling in $pp$ scattering is valid in the squared four-momentum transfer range 0.38 GeV$^2$ $ \leq-t\leq1.2$ GeV$^2$ from \mbox{$\sqrt{s}=$ 7 TeV} down to 1.96 TeV or even down to 1.8 TeV (the acceptance of the measurement at $\sqrt{s}=$ 1.8 TeV in $-t$ terminates at $-t=0.627$ GeV$^2$).

The $H(x)$ scaling property of $pp$ elastic scattering was used to compare the TOTEM measured $\sqrt{s}=7$ TeV $pp$ and the D0 measured $\sqrt{s}=1.96$ TeV $p\bar p$ elastic differential cross section data 
resulting in a model-independent, data-driven odderon signal with a statistical significance of at least 6.26$\sigma$ \cite{Csorgo:2019ewn,Csorgo:2023rzm,Csorgo:2020rlb,Csorgo:2020msw}. 
My model-dependent results indicate that this comparison is made in the domain of validity of the $H(x)$ scaling in $pp$ scattering\footnote{Later on, in Ref.~\cite{Csorgo:2023rzm}, it was shown that, outside the region of the odderon signal, the experimental $pp$ $H(x)$ scaling functions at $\sqrt s =$ 7 TeV and 8 TeV agree with the experimental $p\bar p$ $H(x)$ scaling function at $\sqrt s =$ 1.96 TeV. This is a model-independent indication of the validity of the $H(x)$ scaling of elastic $pp$ scattering in the c.m. energy region of \mbox{1.96 TeV $\leq \sqrt{s} \leq$ 8 TeV.}}.

\newpage
\thispagestyle{empty}
\chapter{Lévy $\alpha$-stable model for elastic scattering} \label{chap:levy}

Precise measurements at the LHC show that the low-$|t|$ $pp$ differential cross section has a strong non-exponential behavior. I show in this chapter that this non-exponential behavior can be easily handled by generalizing the Gaussian model of elastic scattering to a Lévy $\alpha$-stable model. 
 
In \cref{sec:levy_simp_gen}, I generalize the Gaussian impact parameter amplitude to a Lévy \mbox{$\alpha$-stable} impact parameter amplitude to obtain a simple Lévy \mbox{$\alpha$-stable} model for the low-$|t|$ elastic $pp$ ($p\bar p$) differential cross section. Then, in \cref{sec:simp_levy_desc}, I use this model to describe experimental data on elastic $pp$ and $p\bar p$ scattering, and I determine the energy dependencies of the model's parameters. Motivated by the nice descriptions obtained by the simple Lévy $\alpha$-stable model in the non-exponential low-$|t|$ domain, where the ReBB model does not give a quantitative description, in \cref{sec:rebb_levy_gen}, I work out the real extended Lévy $\alpha$-stable generalized Bialas--Bzdak (LBB) model. I do this by generalizing the Gaussian distributions in the ReBB model to Lévy $\alpha$-stable distributions. By approximations that are valid in the low-$|t|$ domain, I deduce from the LBB model the simple Lévy $\alpha$-stable model and relate the parameters of this simple model to the parameters of the LBB model.


This chapter is based on Refs.~\cite{Csorgo:2023pdn,Csorgo:2023rbs}.

\newpage

\section{A simple Lévy $\alpha$-stable model for low-$|t|$ $d\sigma_{\rm el}/dt$}\label{sec:levy_simp_gen}

The exponential differential cross section as given by \cref{eq:dsdt-exp} with \cref{eq:norma0} results from a scattering amplitude of the form 
\begin{equation}\label{eq:ampl_Gauss}
     T_{\rm el}(s,t) = \frac{i+\rho_0(s)}{2} \sigma_{\rm tot}(s)\,{\rm e}^{\frac{tB_0(s)}{2}}.
\end{equation}
Utilizing \cref{eq:invFourier}, the impact parameter amplitude that corresponds to \cref{eq:ampl_Gauss} is
\begin{equation}\label{eq:ampl_b_Gauss}
    \widetilde T_{\rm el}(s,b) = \frac{i+\rho_0(s)}{4\pi}\frac{\sigma_{\rm tot}(s)}{B_0(s)}\,{\rm e}^{-\frac{b^2}{2 B_0(s)}}.
\end{equation}
\cref{eq:ampl_b_Gauss} is a Gaussian model of elastic hadron-hadron scattering \cite{Block:2006hy} that via \cref{eq:relPWtoeik_2} results the exponential differential cross section of \cref{eq:dsdt-exp}.

However, it was experimentally observed that the low-$|t|$ differential cross section is non-exponential. It was first observed at ISR \cite{Barbiellini:1972ua} and later confirmed at LHC by the TOTEM Collaboration \cite{TOTEM:2015oop,TOTEM:2017sdy,TOTEM:2018hki} that the $pp$ elastic differential cross section at low values of squared four-momentum transfer does not have an exponential, $Ae^{-B_0|t|}$ structure: the slope of the differential cross section, $B_0$, changes at around $|t|=0.1$ GeV$^2$. At $\sqrt{s} = 8$ TeV, the TOTEM Collaboration excluded an exponential $pp$ elastic differential cross section in the range of $0.027~{\rm GeV}^2\lesssim|t|\lesssim0.2~{\rm GeV}^2$ with a statistical significance greater than $7\sigma$~\cite{TOTEM:2015oop}. In $p\bar p$ scattering, a change in the slope was observed by the UA4 Collaboration at SPS at $\sqrt{s} = 540$ GeV and 546 GeV around \mbox{$|t|=0.15$ GeV$^2$} \cite{UA4:1983mlb,UA4:1984skz}.

\cref{eq:ampl_b_Gauss} can be easily and rather naturally generalized to a Levy $\alpha$-stable model, where the Gaussian model is recovered in the $\alpha_L=2$ special case. In the generalized, L\'evy $\alpha$-stable model of low-$|t|$ elastic $pp$ ($p\bar p$) scattering the amplitude in the impact parameter representation reads: 
\begin{equation}\label{eq:gray_Levy}
     \widetilde T_{\rm el} (s,b)  =  \frac{i+\rho_0(s)}{8\pi^2}\sigma_{\rm tot}(s) \int d^2\vec q  {\rm e}^{-i\vec q\cdot\vec b}
                         {\rm e}^{-\frac{1}{2}\big|q^2B_L(s)\big|^{\alpha_L(s)/2}}.
\end{equation}
See \hyperref[sec:app_biGaussLevy]{Appendix C} for details on bivariate symmetric Lévy-$\alpha$ stable distributions. \cref{eq:gray_Levy} via \cref{eq:relPWtoeik_2} yields the amplitude in the momentum representation,
\begin{equation}\label{eq:Levy_t_ampl}
T_{\rm el}(s,t)   =  \frac{i+\rho_0(s)}{2} \sigma_{\rm tot}(s) 
                {\rm e}^{-\frac{1}{2}\big|tB_L(s)\big|^{\alpha_L(s)/2}},    
\end{equation}
and the corresponding differential cross section is calculated by \cref{eq:Tdsigma} yielding
\begin{equation}\label{eq:SL0app}
\frac{{\rm d}\sigma_{\rm el}}{{\rm d} t}(s,t) = 
\frac{1+\rho^2_0(s)}{16 \pi} \sigma^2_{\rm tot}(s) 
                {\rm e}^{-\big|tB_L(s)\big|^{\alpha_L(s)/2}}.
\end{equation}
Clearly, for $\alpha_L=2$, $B_L(s)\equiv B_0(s)$ and \cref{eq:gray_Levy} gives \cref{eq:ampl_b_Gauss}, \cref{eq:Levy_t_ampl} gives \cref{eq:ampl_Gauss}, and \cref{eq:SL0app} gives \cref{eq:dsdt-exp}.

Using \cref{eq:norma0}, \cref{eq:SL0app} can be rewritten in the form: 
\begin{equation}\label{eq:SLmodel}
\frac{{\rm d}\sigma_{\rm el}}{{\rm d} t}(s,t) = 
a(s) {\rm e}^{-\big|tB_L(s)\big|^{\alpha_L(s)/2}}.
\end{equation}
I define the simple L\'evy $\alpha$-stable model by \cref{eq:SLmodel}, where I consider the optical point parameter $a(s)$, the L\'evy slope parameter $B_L$ and the L\'evy index of stability $\alpha_L(s)$ as free parameters to be fitted to the $pp$ and $p\bar p$ differential cross section data at a given energy.

Note that an $H_L(x_L)=e^{-x_L}$ scaling function can be constructed for the SL model differential cross section, \cref{eq:SLmodel}, by introducing the variable $x_L=\left(B_L(s)|t|\right)^{\alpha_L(s)/2}$: 
\begin{equation}\label{eq:SLmodelHxL}
H_L(x_L)=\frac{1}{a(s)}\frac{{\rm d}\sigma_{\rm el}}{{\rm d} t}\bigg|_{|t|=x_L^{2/\alpha_L(s)}/B_L(s)} \equiv e^{-x_L}.
\end{equation}
\cref{eq:SLmodelHxL} is the generalization of the $H(x)$ scaling function of the exponential differential cross section (see \cref{sec:Hx}) for a non-exponential $t$ distribution of the shape given \mbox{by \cref{eq:SLmodel}.}

The essential difference between Gaussian and  L\'evy $\alpha$-stable distributions is that the latter have tails behaving asymptotically as a power law. L\'evy $\alpha$-stable distributions are heavy-tailed distributions \cite{nolan2020univariate} (see \hyperref[sec:app_biGaussLevy]{Appendix C} and particularly \cref{fig:Levy-lengthdep}). This causes the variance of these distributions to be infinite for $\alpha_L<2.$  Stable distributions do not have simple analytical forms, but they have simple characteristic functions \mbox{(see \hyperref[sec:app_biGaussLevy]{Appendix C}).} 

\begin{figure}[hbt!]
	\centering
 \vspace{-0.2cm}
\includegraphics[width=0.8\linewidth]{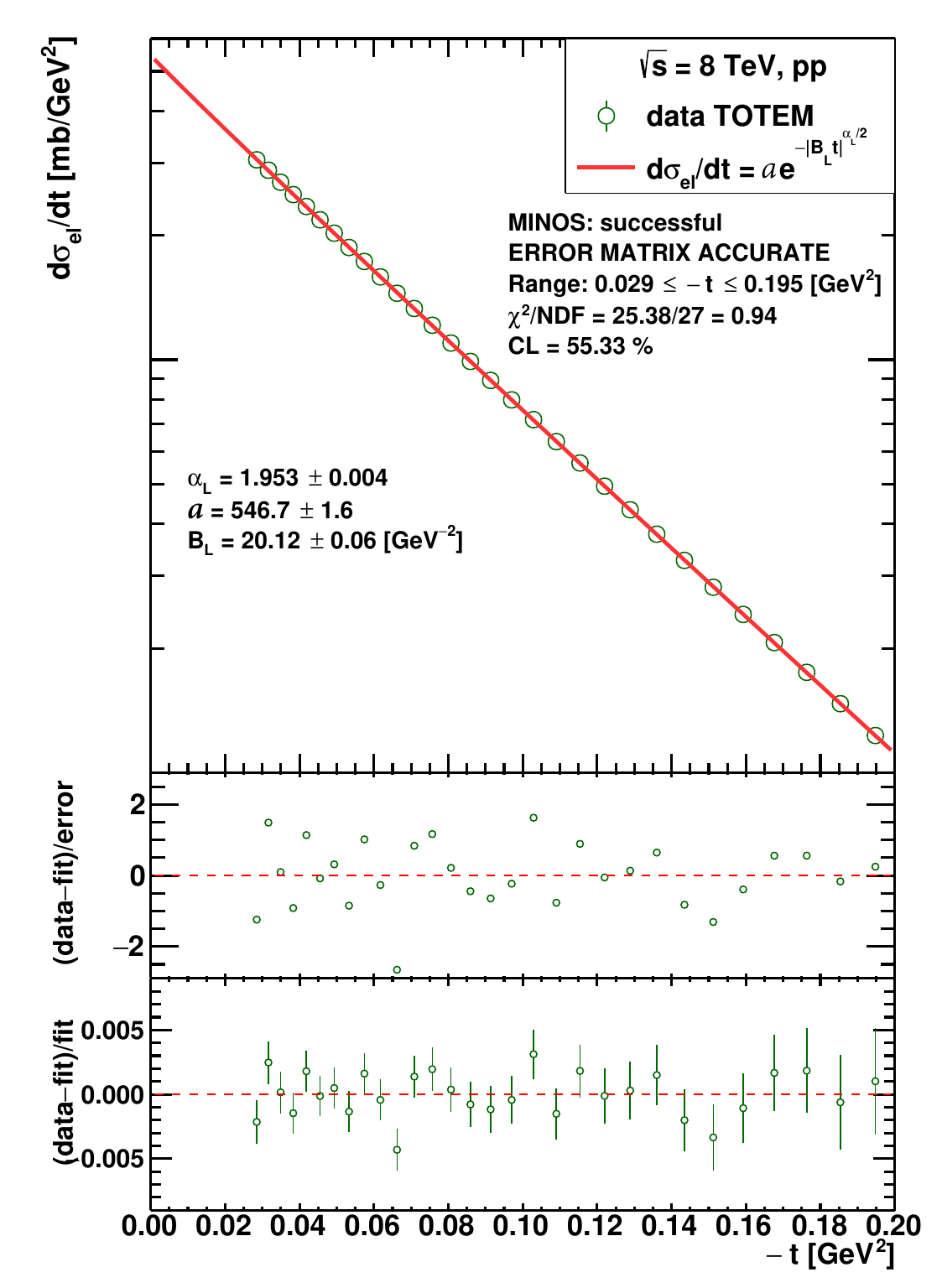}
\vspace{-0.5cm}
	\caption{Simple L\'evy $\alpha$-stable model fit to the 8 TeV low-$|t|$ $pp$ elastic differential cross section data \cite{TOTEM:2015oop}.}
 \vspace{-0.4cm}
	\label{fig:SLfit8TeV}
\end{figure}

I demonstrate the power of the simple L\'evy $\alpha$-stable model of \cref{eq:SLmodel} by fitting its free parameters to the 8 TeV low-$|t|$ $pp$ elastic differential cross section data \cite{TOTEM:2015oop} where the TOTEM Collaboration observed a non-exponential behavior with a statistical significance greater than $7\sigma$. The result of the fit is shown in \cref{fig:SLfit8TeV}. The $CL$ of the fit is 55.33\% and the value of $\alpha_L$ is 1.953$\pm$0.004. This value is very close to 2 but its uncertainty is very small. Thus, we can say that $\alpha_L$ is significantly less than 2, indicating the necessity of a non-exponential model for elastic $pp$ scattering. When I fix $\alpha_L=2$, the $CL$ of the resulting fit is $1.06\times10^{-24}$\%. This result clearly excludes the low-$|t|$ Gaussian model of elastic $pp$ scattering.

\begin{figure}[hbt!]
	\centering
 \vspace{-0.2cm}
\includegraphics[width=0.8\linewidth]{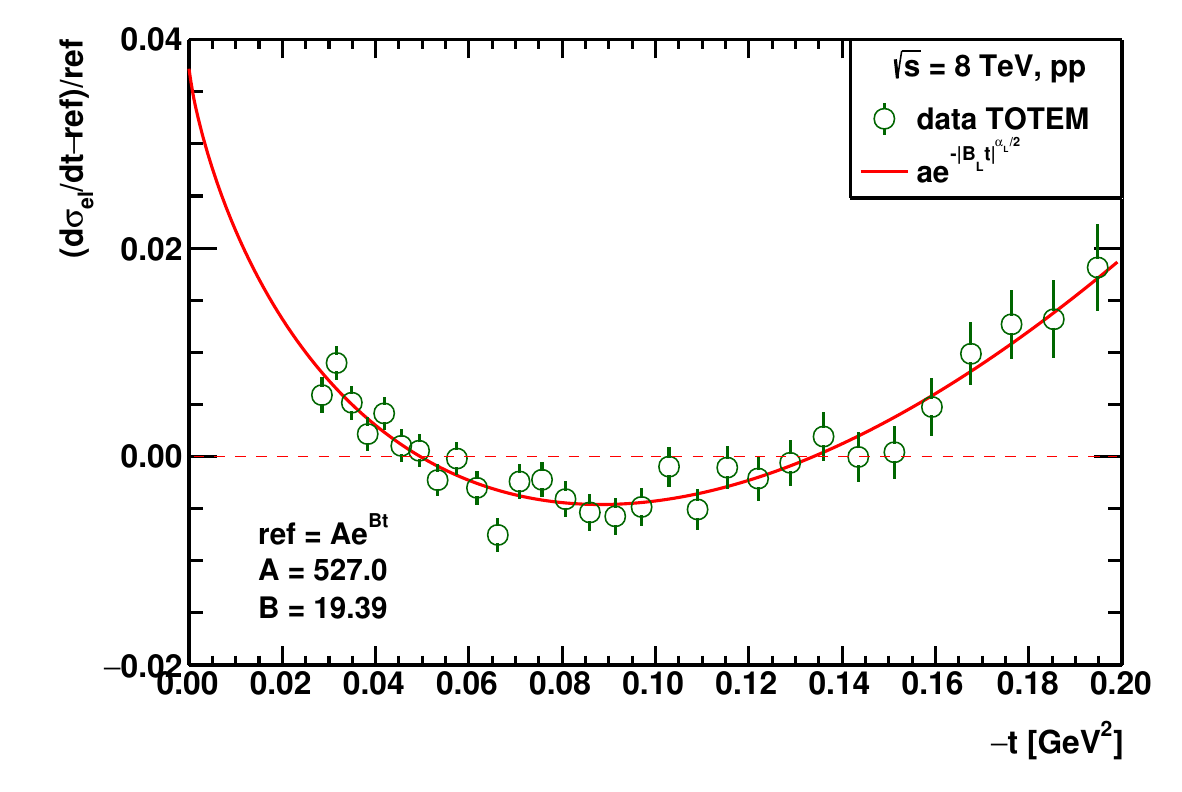}
\vspace{-0.4cm}
	\caption{Simple L\'evy $\alpha$-stable model fit to the 8 TeV low-$|t|$ $pp$ elastic differential cross section data \cite{TOTEM:2015oop} compared to a reference exponential shape, ${\rm ref} = A{\rm e}^{B t}$.}
 \vspace{-0.4cm}
	\label{fig:SLfit8TeVratio}
\end{figure}

To make visible the non-exponential behavior of the data, I calculated the ratio
\begin{equation}
    R = \frac{{\rm d}\sigma_{\rm el}/{\rm d} t-{\rm ref}}{\rm ref},
\end{equation}
where the reference function has an exponential shape,
\begin{eqnarray}
{\rm ref} = A{\rm e}^{B t}.
\end{eqnarray}
The values of $A$ and $B$ are obtained from a fit to the data. This $R$ ratio is presented in \cref{fig:SLfit8TeVratio}. The ratio $R$ crosses zero two times but is nonzero otherwise. This indicates that the differential cross section deviates from an exponential shape.

Authors of  Refs.~\cite{Kohara:2019qoq, Kohara:2018wng} discuss that the interplay between the Coulomb and nuclear amplitudes may effect the low-$|t|$ behavior of the elastic $pp$ differential cross section up to $-t\simeq$ 0.05 GeV$^2$ but the differential cross section must be measured with a precision
better than 0.1\% to draw more concrete conclusions. Also, TOTEM  concludes in Ref.~\cite{TOTEM:2017sdy} that Coulomb effects modify the nuclear cross section by less than 1\% for $|t|\gtrsim0.007$ GeV$^2$ and since the experimental errors are around 5\%, in the range $|t|\gtrsim0.007$ GeV$^2$, the differential cross section can be analyzed independently from the modeling of the Coulomb effects.

In the framework of the Regge approach, the non-exponential behavior of the elastic differential cross section was related to the $4m_\pi^2$ branch point of $t$-channel scattering amplitude and, hence, is explained as the manifestation of $t$-channel \mbox{unitarity  \cite{Jenkovszky:2017efs,Szanyi:2019kkn,Cohen-Tannoudji:1972gqd,Anselm:1972ir, Tan:1974gd,Khoze:2000wk,Jenkovszky:2014yea,Fagundes:2015vva, Jenkovszky:2017dox,Jenkovszky:2017pqs}.} Refs.~\cite{Jenkovszky:2017efs,Szanyi:2019kkn, Jenkovszky:2017dox,Jenkovszky:2017pqs, } contain my contribution to these results. 

The emergence of Lévy $\alpha$-stable distributions can be expected based on the generalizations of the Central Limit Theorem (CLT). According to the classical CLT, the normalized sum of independent and identically distributed random variables with finite variance converges to a Gaussian distribution. If the finite variance assumption is dropped, we get a generalized CLT: the normalized sum of independent and identically distributed random variables with infinite variance converges to a stable distribution (see more precise mathematical definitions in the book of J. P. Nolan \cite{nolan2020univariate}). The first complete proof of such a generalized CLT was given by Paul Lévy in 1937 \cite{levy1937theorie}, and stable distributions were named after him. There are even such CLT versions in which, if some other conditions are satisfied, the random variables need not necessarily be identically distributed (see more detail in Ref.~\cite{chung2001course}).

The application of Lévy $\alpha$-stable distributions is not new in the field of high-energy physics. Let me mention some examples. The $\alpha_L=1$ case gives the Cauchy-Lorentz or Breit-Wigner distribution used to model unstable particles \cite{ParticleDataGroup:2022pth}. In Ref.~\cite{Wilk:1999dr}, the power-law tails in the transverse momentum spectra of particles were explained by stable distributions. Stable distributions were utilized to describe an intermittent behavior in the quark-gluon plasma-hadron gas phase transition \cite{Brax:1990jv} and to describe anisotropic dynamical fluctuations in multiparticle production dynamics \cite{Zhang:1995uf}. The application of stable distributions is widespread in high energy heavy ion physics \cite{Csorgo:2003uv,Csanad:2024hva}. Several experiments in high-energy particle and nuclear physics analyze their data utilizing stable distributions, \textit{e.g.}, ATLAS~\cite{Schegelsky:2018tit} and CMS~\cite{CMS:2023xyd} experiments at LHC, PHENIX~\cite{Lokos:2018dqq} and STAR~\cite{Kincses:2024sin} experiments at RHIC,  and the NA61/SHINE experiment at \mbox{CERN SPS~\cite{Porfy:2019scf,Porfy:2023yii}.}

A model-independent L\'evy expansion technique was introduced in Ref.~\cite{Novak:2016cyc} to test possible deviations of distributions from a Lévy $\alpha$-stable shape. This technique was successfully applied to describe elastic $pp$ and $p\bar p$ $t$-distributions in wide ranges of $t$ where significant deviations were found \cite{Csorgo:2019egs, Csorgo:2018uyp}. The leading order of this expansion formula of Ref.~\cite{Csorgo:2018uyp} coincides with the SL model of \cref{eq:SLmodel}.

In\, my\, work,\, I\, applied\, the\, SL\, model\, of\, \cref{eq:SLmodel}\, to\, describe\, elastic\, $pp$\, and\, $p\bar p$ \mbox{$t$-distributions}\, in\, a\, relatively\, narrow\, range,\, specifically\, in\, the\, low-$|t|$\, domain of \mbox{0.02 GeV$^2$ $\leq -t\leq0.15$ GeV$^2$.} As it turns out in the following section, this SL model perfectly describes the data in the specified $t$-domain.

\section{SL model analysis of the low-$|t|$ data} \label{sec:simp_levy_desc}

To describe the low-$|t|$ $pp$ differential cross section data, the TOTEM Collaboration utilized the parametrization 
 \cite{TOTEM:2015oop}:

\begin{equation}\label{eq:expnonlin}
   \frac{d\sigma_{\rm el}}{dt}(s,t)=a(s)e^{b_1(s)t+b_2(s)t^2},
\end{equation}
where $a$, $b_1$, and $b_2$ are free parameters to be determined at a given c.m. energy. In \cref{eq:expnonlin}, a non-exponential behavior is generated by the non-vanishing quadratic term in the exponent, while in the SL model, a non-exponential behavior is generated by an $\alpha_L$ parameter value which is less than 2. Since the SL model is the generalization of the $\alpha_L=2$ Gaussian model to the $\alpha_L\leq2$ Lévy $\alpha$-stable model, it may be more natural to use \cref{eq:SLmodel} instead of \cref{eq:expnonlin} to describe the experimental data.

In this section, I analyze the low-$|t|$ $pp$ and $p\bar p$ elastic differential cross section data in the c.m. energy range of $546~{\rm GeV}<\sqrt s<13~{\rm TeV}$ and in the squared four-momentum transfer range of \mbox{0.02 GeV$^2$ $\leq -t\leq0.15$ GeV$^2$} 
using the SL model of \cref{eq:SLmodel} and the $\chi^2$ definition of \cref{eq:chi2_refind} without the terms for $\sigma_{\rm tot}$ and $\rho_0$. I analyzed eleven $pp$ and $p\bar p$ differential cross section datasets. The data sources, the values of the fitted SL model parameters at different energies, and the confidence levels of the description are summarized in \cref{tab:anLevy}. The $CL$ values range from  8.8\% to 96\%, implying that the SL model describes the data in a statistically acceptable manner.
 
\begin{table}[!hbt]
    \centering
    \begin{tabular}{cccccc}
   \hline\hline\noalign{\smallskip}
        $\sqrt{s}$ [GeV] &data from& $\alpha_L$ & $a$ [mb/GeV$^2$] & $B_L$ [GeV$^{-2}$] & $CL$ (\%)   \\ \noalign{\smallskip}\hline\noalign{\smallskip}
546& UA4 \cite{UA4:1984skz}&1.93 $\pm$ 0.09&209 $\pm$ 15&15.8 $\pm$ 0.9&18.1 \\
1800& E-710 \cite{E-710:1990vqb}&2.0 $\pm$ 1.5&270 $\pm$ 24&16.2 $\pm$ 0.2&77.1\\
2760&TOTEM \cite{TOTEM:2018psk}&1.6 $\pm$ 0.3&637 $\pm$ 252&28 $\pm$ 11&20.5\\
7000 &TOTEM \cite{TOTEM:2013lle}&1.95 $\pm$ 0.01&535 $\pm$ 30&20.5 $\pm$ 0.2&8.8\\
7000 &ATLAS \cite{ATLAS:2014vxr} &1.97 $\pm$ 0.01&463 $\pm$ 13&19.8 $\pm$ 0.2&96.0\\
8000 &TOTEM \cite{TOTEM:2015oop}&1.955 $\pm$ 0.005&566 $\pm$ 31&20.09 $\pm$ 0.08&43.9\\
8000 &TOTEM \cite{TOTEM:2016lxj}&1.90 $\pm$ 0.03&582 $\pm$ 33&20.9 $\pm$ 0.4&19.6\\
8000 &ATLAS \cite{ATLAS:2016ikn}&1.97 $\pm$ 0.01&480 $\pm$ 11&19.9 $\pm$ 0.1&55.8\\
13000 
 &TOTEM \cite{TOTEM:2017sdy}&1.959 $\pm$ 0.006&677 $\pm$ 36&20.99 $\pm$ 0.08&76.5\\
13000 &TOTEM \cite{TOTEM:2018hki}&1.958 $\pm$ 0.003&648 $\pm$ 10&21.06 $\pm$ 0.05&89.1\\
13000 &ATLAS \cite{ATLAS:2022mgx}&1.968 $\pm$ 0.006&569 $\pm$ 17&20.84 $\pm$ 0.07&29.7\\ \hline\hline
    \end{tabular}
    \vspace{-1mm}
    \caption{
Values of the SL model parameters at different energies from half TeV up to 13~TeV as obtained from fits to the data in the four-momentum transfer range of \mbox{0.02 GeV$^2$ $\leq -t\leq0.15$ GeV$^2$} using the $\chi^2$ definition of \cref{eq:chi2_refind} without the terms for $\sigma_{\rm tot}$ and $\rho_0$. The last column shows the confidence level of the description. 
    }\label{tab:anLevy}
    \vspace{-6mm}
\end{table}

\begin{figure}[hbt!]
	\centering
 \vspace{-0.2cm}
\includegraphics[width=0.8\linewidth]
{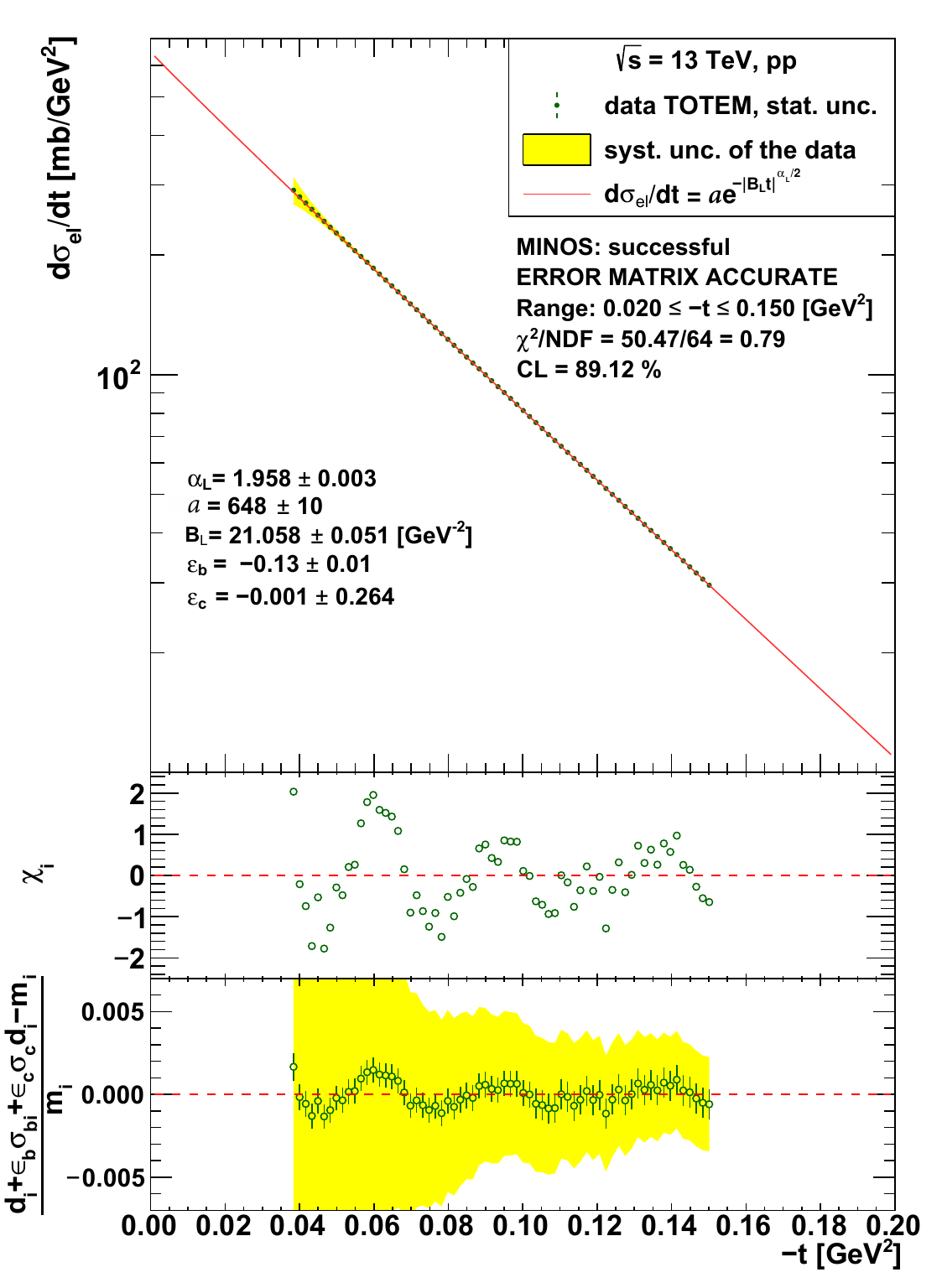}
\vspace{-0.4cm}
	\caption{SL model fit to the 13 TeV low-$|t|$ $pp$ elastic differential cross section data \cite{TOTEM:2018hki} using the $\chi^2$ definition of \cref{eq:chi2_refind} without the terms for $\sigma_{\rm tot}$ and $\rho_0$.}
 \vspace{-0.4cm}
	\label{fig:SLfit13TeV}
\end{figure}

\begin{figure}[hbt!]
	\centering
 \vspace{-0.2cm}
\includegraphics[width=0.8\linewidth]{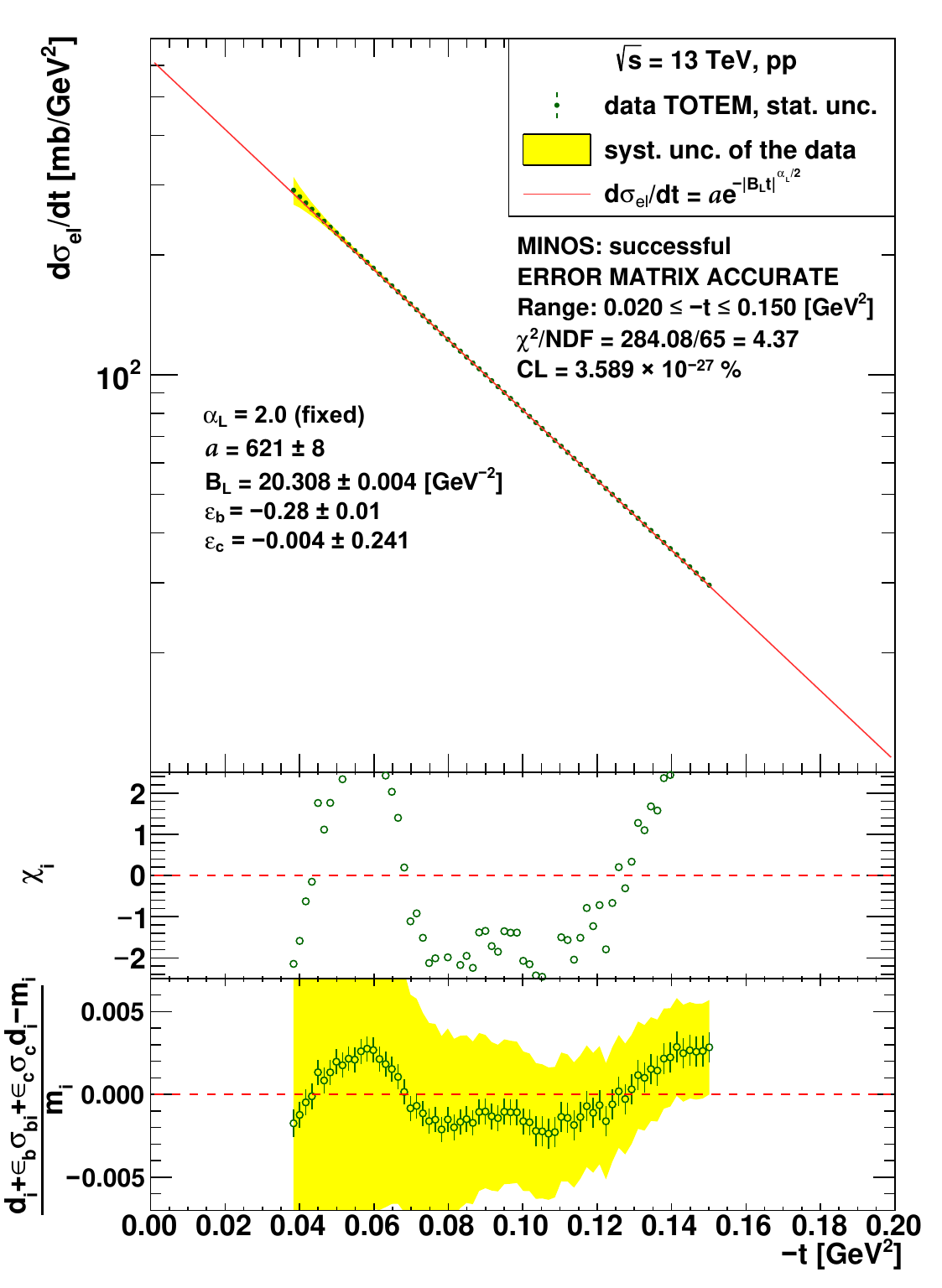}
\vspace{-0.4cm}
	\caption{SL model fit to the 13 TeV low-$|t|$ $pp$ elastic differential cross section data \cite{TOTEM:2018hki} with $\alpha_L=2$ fixed using the $\chi^2$ definition of \cref{eq:chi2_refind} without the terms for $\sigma_{\rm tot}$ and $\rho_0$.}
 \vspace{-0.4cm}
	\label{fig:SLfit13TeVfix}
\end{figure}

\begin{figure}[hbt!]
	\centering
 \vspace{-0.2cm}
\includegraphics[width=0.8\linewidth]{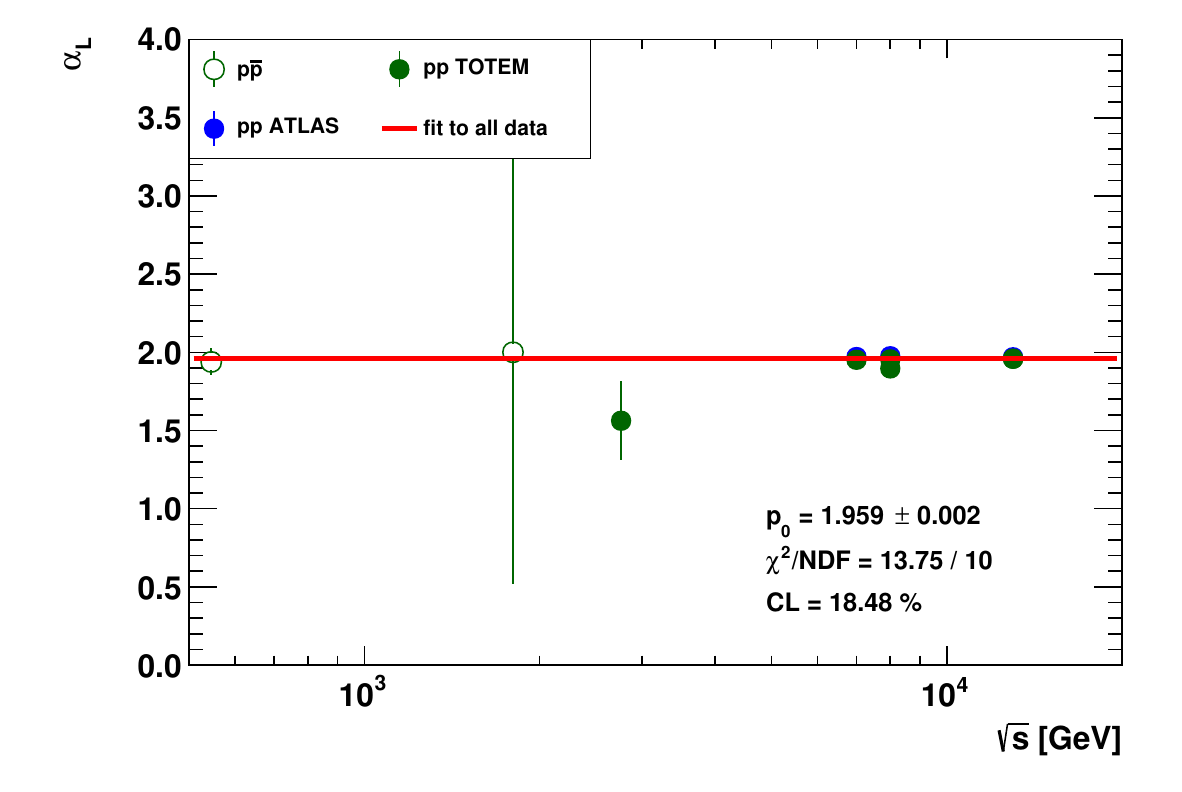}
\vspace{-0.4cm}
	\caption{The values of the $\alpha_L$ parameter of the SL model as determined from fits to $pp$ and $p\bar p$ elastic differential cross section data in the c.m. energy range of \mbox{$546~{\rm GeV}<\sqrt s<13~{\rm TeV}$} and in the squared four-momentum transfer range of \mbox{0.02 GeV$^2$ $\leq -t\leq0.15$ GeV$^2$} using the $\chi^2$ definition of \cref{eq:chi2_refind} without the terms for $\sigma_{\rm tot}$ and $\rho_0$.}
 \vspace{-0.4cm}
	\label{fig:SL_alpha_endep}
\end{figure}

\begin{figure}[hbt!]
	\centering
\includegraphics[width=0.8\linewidth]{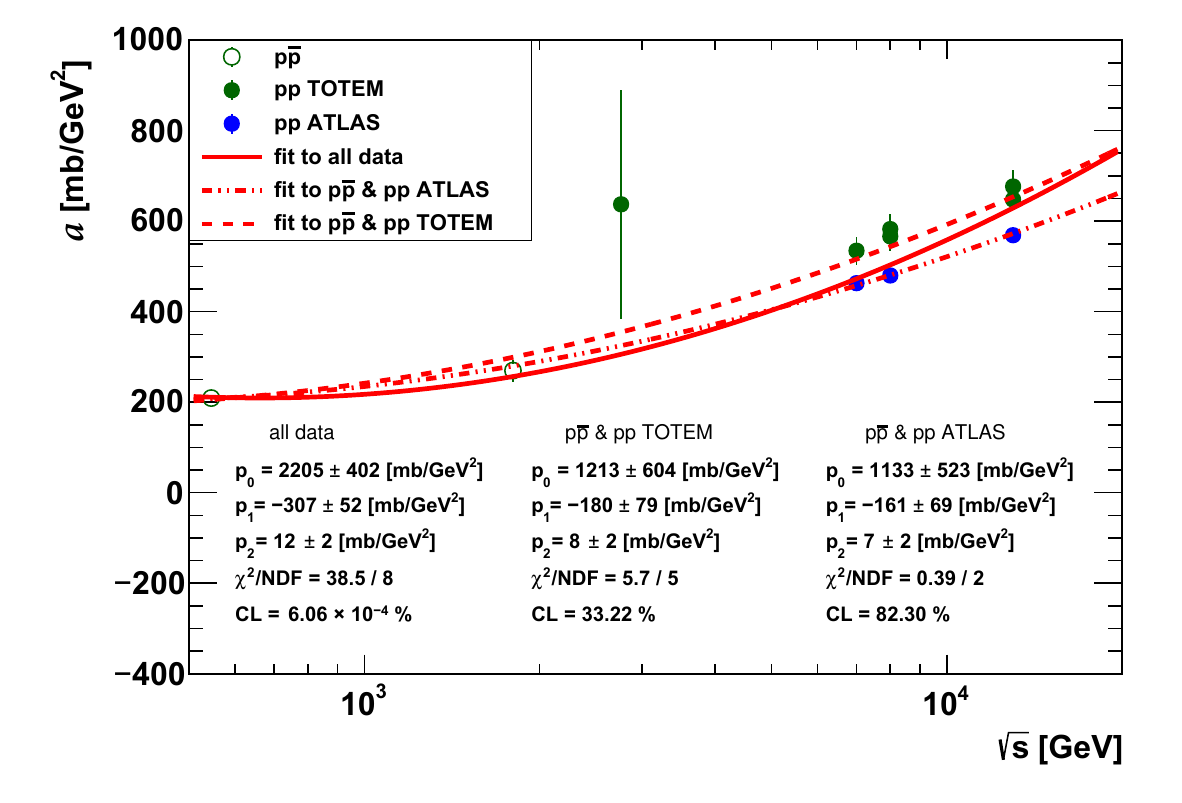}
 \vspace{-0.4cm}
	\caption{Same as \cref{fig:SL_alpha_endep} but for the $a$ parameter of the SL model.}
   \vspace{-0.4cm}
	\label{fig:SL_a_endep}
\end{figure}

\begin{figure}[hbt!]
	\centering
 \vspace{-0.2cm}
\includegraphics[width=0.8\linewidth]{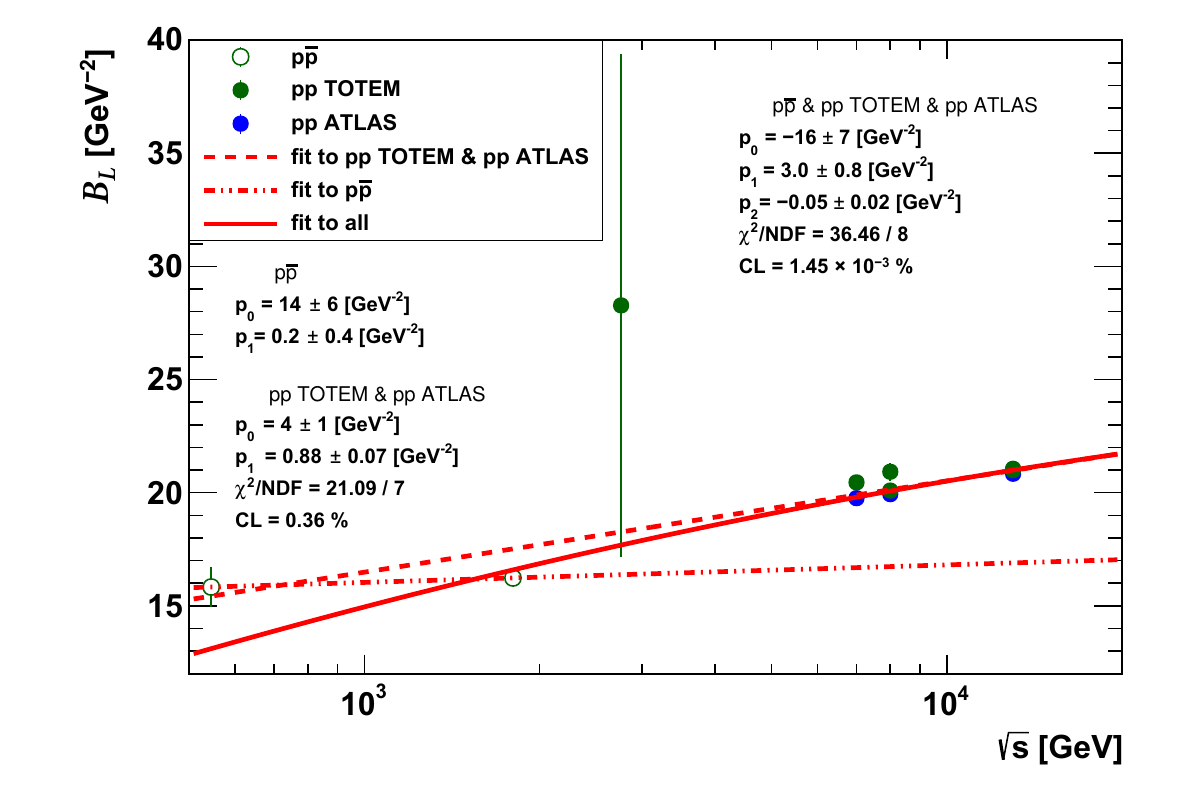}
 \vspace{-0.4cm}
	\caption{Same as \cref{fig:SL_alpha_endep} but for the $B_L$ parameter of the SL model.}
  \vspace{-0.4cm}
	\label{fig:SL_BL_endep}
\end{figure}

\cref{fig:SLfit13TeV} shows the result of the SL model fit to the most precise TOTEM data measured at $\sqrt s =$ 13 TeV \cite{TOTEM:2018hki}. The $CL$ of the description is 89.12\% with $\alpha_L=1.958\pm0.003$.  \cref{fig:SLfit13TeVfix} shows the result of the SL model fit to the same data with $\alpha_L=2$ fixed. In this case, the $CL$ of the description is 3.6 $\times~10^{-27}$\%. The need for a model that reproduces a proper non-exponential behavior is undoubted.

Now I proceed with determining the energy dependencies of the SL model parameters using the values of the parameters at different energies as given in \cref{tab:anLevy}.

The values of the $\alpha_L$ parameter at different energies are shown in \cref{fig:SL_alpha_endep}. These values can be fitted with a constant  1.959 $\pm$ 0.002, resulting in a $CL$ value of 18.48\%. This $\alpha_L = 1.959 \pm 0.002$ constant value is 
close to the Gaussian $\alpha_L = 2$ case, but its error is small, and thus, the value of $\alpha_L$ is significantly less than 2 in a statistical sense. 

The values of the optical point parameter $a$ in the SL model at different energies are shown in \cref{fig:SL_a_endep}. I found that the energy dependence of this parameter is compatible with a squared-logarithmic shape as given by \cref{eq:squaredlog}. The subtlety is that the $a$ parameter values obtained using TOTEM data and obtained using ATLAS data are not compatible. For $p\bar p$ and ATLAS data, the values of the parameters in \cref{eq:squaredlog} are \mbox{$p_0=1213\pm604$~mb/GeV$^2$,} $p_1=-180\pm79$ mb/GeV$^2$, and $p_2=8\pm2$ mb/GeV$^2$, resulting in a confidence level of $33.22\%$. For $p\bar p$ and TOTEM data, the parameter values are $p_0=1133\pm523$ mb/GeV$^2$, $p_1=-161\pm69$ mb/GeV$^2$, and $p_2=7\pm2$ mb/GeV$^2$, giving a confidence level of $82.30~\%$. A simultaneous fit to all $a$ parameter values obtained from $p\bar p$, ATLAS, and TOTEM data gives a $CL$ value of 6.06$~\times~10 ^{-4}\%$ corresponding to a statistically unacceptable description.  The observed discrepancy might be related to the fact that the TOTEM and ATLAS experiments use different methods to measure luminosity, $i.e.$, to obtain the absolute normalization for the differential cross section measurements \cite{ATLAS:2022mgx}.

The values of the Lévy slope parameter $B_L$ in the SL model at different energies are shown in \cref{fig:SL_BL_endep}. I found that for  ATLAS and TOTEM $pp$ data, the energy dependence of $B_L$ is compatible with a linearly logarithmic shape as given by \cref{eq:linlog}. For ATLAS and TOTEM data, the values of the parameters in \cref{eq:linlog} are $p_0=4\pm1$ GeV$^{-2}$ and $p_1=0.88\pm0.07$ GeV$^{-2}$ giving a $CL$ value of $0.36\%$. The $B_L$ values for $p\bar p$ data lie on a different linear curve with $p_0=14\pm6$ GeV$^{-2}$ and $p_1=0.2\pm0.4$ GeV$^{-2}$. The fit of all $B_L$ values ($p\bar p$, ATLAS, and TOTEM), even with the quadratic parametrization of \cref{eq:squaredlog}, is statistically not acceptable since the $CL$ value of the description is $1.45\times 10^{-3}\%$. There are two alternative interpretations for the incompatibility of the  $pp$ and $p\bar p$ $B_L$ parameter trends. As discussed by the TOTEM Collaboration \cite{TOTEM:2017asr}, there is a jump in the energy dependence of the $B_0$ slope parameter somewhere in the energy domain of \mbox{3~TeV $\lesssim \sqrt s \lesssim$ 4 TeV.} One of the possibilities is that the same jump is seen in my analysis, preventing the lower energy $p\bar p$ data from lying on the same curve with the higher energy LHC ATLAS and TOTEM data. The second possible interpretation is that the difference between the $pp$ and $p\bar p$ slope is an odderon effect (see \cref{sec:oddintro}). New $pp$ low-$|t|$ measurements at LHC at $\sqrt s =$ 1.8 TeV and/or 1.96 TeV may bring a deeper understanding.

In Ref.~\cite{Petrov:2023lho}, the authors analyzed ATLAS and TOTEM data together and obtained similar results: TOTEM and ATLAS differential cross sections differ only in their normalization, but their shapes (slopes) are consistent.

\vspace{-2mm}
\section{Lévy $\alpha$-stable generalization of the ReBB model}\label{sec:rebb_levy_gen}
\vspace{-2mm}

Motivated by the success of the Lévy $\alpha$-stable model of low-$|t|$ elastic $pp$ and $p\bar p$ scattering, in this section, I generalize the Gaussian shapes in the ReBB model to Lévy $\alpha$-stable shapes resulting the real extended Lévy $\alpha$-stable generalized Bialas--Bzdak model or LBB model for short. 

In the original BB model and in its extended version, in the ReBB model, the transverse density of the proton or, in other words, the quark-diquark distribution in the proton and the inelastic differential cross sections for the collision of two constituents have Gaussian forms as given by \cref{eq:quark_diquark_distribution} and \cref{eq:inelastic_cross_sections}, respectively.

\cref{eq:quark_diquark_distribution} assumes a Gaussian distribution for the square root of the sum of the constituent's squared transverse position vectors. Now, I assume Gaussian distribution for the relative position of the constituents in a single proton. Thus, introducing the normalized bivariate Gaussian distribution (see \hyperref[sec:app_biGaussLevy]{Appendix C}),

\begin{equation}
    G(\vec x\, | R_G)= \frac{1}{(2\pi)^2}\int d^2 q e^{i{\vec q}\cdot \vec x}e^{-\frac{1}{2}q^2R_G^2}=\frac{1}{2\pi R_G^2}e^{-\frac{x^2}{2 R_G^2}},
\end{equation}
we have
\begin{equation}\label{eq:quark-diquark_dist}
    D(\vec s_q, \vec s_d) = (1+\lambda)^2 G\left(\vec s_{q} - \vec s_d | R_{qd}/\sqrt{2}\right)\delta^2(\vec s_d+\lambda \vec s_q).
\end{equation}
\cref{eq:quark-diquark_dist} satisfies the normalisation condition of \cref{eq:densnorm}:
\begin{eqnarray} 
 \int d^2\vec s_d D(\vec s_{q},\vec s_{d}) & =& G\left(\vec s_{q} |R_{qd*}/\sqrt{2}\right), \\
 \int d^2\vec s_q d^2\vec s_d D(\vec s_{q},\vec s_{d}) & =& 1,
\end{eqnarray}
where
\begin{equation} \label{eq:Rqdresc}
R_{qd*}=\frac{R_{qd}}{1+\lambda}.
\end{equation}
\cref{eq:quark-diquark_dist} allows interpreting the $R_{qd}$ parameter of the model as the relative distance between the constituents in a single proton: $R_{qd}\equiv d_{qd}$ (see \cref{sec:obb}). Moreover, the form of \cref{eq:quark-diquark_dist} 
prepares the ground for the Lévy $\alpha$-stable generalization. \cref{eq:quark_diquark_distribution} is given in terms of a product of two Gaussian shapes, however, the product of two Lévy $\alpha$-stable shapes with $\alpha_L<2$ is not easy to deal with when calculating integrals. The form of \cref{eq:quark-diquark_dist} avoids dealing with products of Lévy $\alpha$-stable shapes in $D(\vec s_{q},\vec s_{d})$.


The inelastic differential
cross sections of the binary constituent-constituent collisions, \cref{eq:inelastic_cross_sections}, in terms of a normalized bivariate Gaussian distribution can be rewritten as
\begin{equation}\label{eq:inelastic_cross_sections_NG}
     \sigma^{ab}_{\rm in}(\vec s\,) =  A_{ab} \pi S_{ab}^{2} G\left(\vec s\,|S_{ab}/\sqrt{2}\right),\,\,\,S^2_{ab}=R_a^2+R_b^2,\,\,\,a,b\in\{q,d\}\,.
 \end{equation}
\cref{eq:inelastic_cross_sections_NG} can be obtained in terms of a convolution of the constituent's parton distributions. The convolution represents an overlap between the constituent's parton distributions. Assuming that the constituent quark and the constituent diquark have Gaussian parton distributions, $G(\vec s_q|R_q/\sqrt{2})$ and $G(\vec s_d |R_d/\sqrt{2})$, one recovers \cref{eq:inelastic_cross_sections_NG} as follows:
\begin{align}\label{eq:dsig_inab}
         \sigma^{ab}_{\rm in}(\vec s) &= A_{ab} \pi S_{ab}^{2}  \int d^2  s_a G\left(\vec s_a |R_a/\sqrt{2}\right) G\left(\vec s - \vec s_a | R_b/\sqrt{2} \right) \\\nonumber
         &\equiv  A_{ab} \pi S^{2}_{ab} G\left(\vec s\,| S_{ab}/\sqrt{2}\right).
\end{align}
\cref{eq:dsig_inab} is calculated by using the convolution theorem
(see \cref{eq:CT0} and \cref{eq:CT} below). The scale parameters of the convoluted normalized Gaussian distributions add up quadratically.

Requiring the condition given in \cref{eq:ratiosforsigma}, the inelastic constituent-constituent differential cross sections read: 
\begin{equation}\label{eq:sigin_qq_G}
   \sigma_{\rm in}^{qq}(\vec s_{q},\vec s_{q}^{\,\prime}%
;\vec b)= 2\pi A_{qq} R_{q}^2 G(\vec b+\vec s_q^{~\prime}-\vec s_q|  R_{q}),
\end{equation}
\begin{equation}\label{eq:sigin_qd_G}
 \sigma_{\rm in}^{qd}(\vec s_{q},\vec s_{d}^{\,\prime}%
;\vec b)= 4\pi A_{qq}R_{q}^2 G\Bigg(\vec b+\vec s_d^{~\prime}-\vec s_q\Bigg|\sqrt{\frac{R_{q}^2+R_{d}^2}{2}}\Bigg),
\end{equation}
\begin{equation}\label{eq:sigin_dq_G}
\sigma_{\rm in}^{dq}(\vec s_{q}^{\,\prime},\vec s_{d}%
;\vec b\,) = 4\pi A_{qq}R_{q}^2 G\Bigg(\vec b+\vec s_q^{~\prime}-\vec s_d\Bigg|\sqrt{\frac{R_{q}^2+R_{d}^2}{2}}\Bigg),
\end{equation}
and
\begin{equation}\label{eq:sigin_dd_G}
\sigma_{\rm in}^{dd}(\vec s_{d},\vec s_{d}^{\,\prime}%
;\vec b\,) = 8\pi A_{qq}R_{q}^2 G(\vec b+\vec s_d^{~\prime}-\vec s_d|R_{d}).
\end{equation}
Substituting \cref{eq:sigin_qq_G}, \cref{eq:sigin_qd_G}, \cref{eq:sigin_dq_G} and \cref{eq:sigin_dd_G} into \cref{eq:tilde_sigma_inel} we obtain $\tilde\sigma_{\rm in}(\,\vec b\,)$ in terms of eleven different integral expressions,
\begin{align}\label{eq:sig_tilde_b}
\tilde\sigma_{\rm in}(\,\vec b\,)&=\tilde\sigma^{qq}_{\rm in}(\,\vec b\,)+2\tilde\sigma^{qd}_{\rm in}(\,\vec b\,)+\tilde\sigma^{dd}_{\rm in}(\,\vec b\,)-\left[2\tilde\sigma^{qq,qd}_{\rm in}(\,\vec b\,) +\tilde\sigma^{qd,dq}_{\rm in}(\,\vec b\,)+\tilde\sigma^{qq,dd}_{\rm in}(\,\vec b\,) + \right.\\ \nonumber
&\left. + 2\tilde\sigma^{qd,dd}_{\rm in}(\,\vec b\,)\right]
+\left[\tilde\sigma^{qq,qd,dq}_{\rm in}(\,\vec b\,)+2\tilde\sigma^{qq,qd,dd}_{\rm in}(\,\vec b\,)+\tilde\sigma^{dd,qd,dq}_{\rm in}(\,\vec b\,)\right]-\tilde\sigma^{qq,qd,dq,dd}_{\rm in}(\,\vec b\,),
\end{align}
where, after making use of the presence of the Dirac $\delta$ function in \cref{eq:quark-diquark_dist}, the last term reads:
\begin{align}\label{eq:tildesigma_general}
\tilde\sigma^{qq,qd,dq,dd}_{\rm in}(\,\vec b\,) &= \int d^2s_qd^2s'_qG(\vec s_{q} |R_{qd*}/\sqrt{2})G(\vec s_{q}^{~\prime} |R_{qd*}/\sqrt{2}) \times \\\nonumber &\times \sigma_{\rm in}^{qq}(\vec s_{q},\vec s_{q}^{~\prime}%
;\vec b\,)
\sigma_{\rm in}^{qd}(\vec s_{q}, -\lambda \vec s_{q}^{~\prime}%
;\vec b\,) \sigma_{\rm in}^{dq}(\vec s_{q}^{~\prime},-\lambda\vec s_{q}%
;\vec b\,)\sigma_{\rm in}^{dd}(-\lambda \vec s_{q},-\lambda\vec s_{q}^{~\prime}%
;\vec b\,).
\end{align}
Each term in the integrand of \cref{eq:tildesigma_general} gives a Gaussian-shaped factor. The integral \cref{eq:tildesigma_general} can be obtained explicitly as a function of the
parameters of the various Gaussians. The other terms in \cref{eq:sig_tilde_b} can be obtained by
setting these parameters to suitable values (see \hyperref[sec:app_BBcalc]{Appendix B}). After computing all the integrals, we obtain $\tilde\sigma_{\rm in}(\,\vec b\,)$ in terms of eleven different Gaussian-shaped expressions given in \hyperref[sec:app_BBcalc]{Appendix B}.


The L\'evy $\alpha$-stable generalization of the BB model lies in generalizing both (i) the quark-diquark distribution in the proton and (ii) the inelastic differential cross sections for the collision of two constituents from the $\alpha_L=2$ Gaussian shapes to $\alpha_L\leq2$  L\'evy $\alpha$-stable shapes. 


I use the normalized bivariate symmetric L\'evy $\alpha$-stable distribution in its common form (see \hyperref[sec:app_biGaussLevy]{Appendix C}), 
\begin{equation}
L(\vec x\,|\alpha_L, R_L) = \frac{1}{(2\pi)^2}\int d^2 \vec q e^{i{\vec q}\cdot \vec x}e^{-\left|q^2R_L^2\right|^{\alpha_L/2}},
\end{equation}
which, for $\alpha_L=2$, after the substitution $R_L=R_G/\sqrt{2}$, gives exactly the bivariate Gaussian distribution: 
\begin{equation}
    L(\vec x\,|\alpha_L=2, R_L=R_G/\sqrt{2})\equiv G(\vec x | R_G).
\end{equation}
I work here with symmetric L\'evy $\alpha$-stable distribution. Hence, the skewness and shift parameters of the distribution, $\beta$ and $\delta$, respectively, are implicit and have vanishing values. 

In the L\'evy $\alpha$-stable generalized BB model, the quark-diquark distribution in a single proton reads:
\begin{equation}\label{eq:quark-diquark_levy}
    D(\vec s_q, \vec s_d) = (1+\lambda)^2 L\left(\vec s_{q} - \vec s_d | \alpha_L, R_{qd}/2\right) \delta^2(\vec s_d+\lambda \vec s_q)
\end{equation}
and satisfies the normalisation condition of \cref{eq:densnorm}:
\begin{eqnarray} 
 \int d^2\vec s_d D(\vec s_{q},\vec s_{d}) & =& L\left(\vec s_{q} |\alpha_L,R_{qd*}/2\right), \\
 \int d^2\vec s_q d^2\vec s_d D(\vec s_{q},\vec s_{d}) & =& 1.
\end{eqnarray}
Assuming that the constituent quark and the constituent diquark have L\'evy $\alpha$-stable parton distributions, $L(\vec s_q|R_q/2)$ and $L(\vec s_d |R_d/2)$, the inelastic differential cross sections for the collision of two constituents read:
\begin{align}
         \sigma^{ab}_{\rm in}(\vec s) &= A_{ab} \pi S_{ab}^{2}   \int d^2  s_a L(\vec s_a |\alpha_L,R_a/2) L(\vec s - \vec s_a | \alpha_L, R_b/2 ) \\\nonumber
         &=  A_{ab} \pi S_{ab}^{2} L\left(\vec s\,|\alpha_L, S_{ab}/2\right),
\end{align}
where now
\begin{equation}
    S_{ab}= \left(R_a^{\alpha_L}+R_b^{\alpha_L}\right)^{1/\alpha_L},
\end{equation}
since, after utilizing the convolution theorem, in this case, the radii add up not quadratically but at the power of $\alpha_L$ (see \cref{eq:CT0}, \cref{eq:CT}, and \cref{eq:levyadd} below).

Again, requiring the condition given in \cref{eq:ratiosforsigma}, the inelastic constituent-constituent differential cross sections now read: 
\begin{equation}\label{eq:qqlevy}
\sigma_{\rm in}^{qq}(\vec s_{q},\vec s_{q}^{\,\prime}%
;\vec b\,)= \pi A_{qq} \left(2R_{q}^{\alpha_L}\right)^{2/\alpha_L} L\left(\vec b+\vec s_q^{~\prime}-\vec s_q\,|\,\alpha_L,  \left(2 R_{q}^{\alpha_L}\right)^{1/\alpha_L}/2\right),
\end{equation}
\begin{equation}\label{eq:qdlevy}
\sigma_{\rm in}^{qd}(\vec s_{q},\vec s_{d}^{\,\prime}%
;\vec b\,)= 2\pi A_{qq}\left(2R_{q}^{\alpha_L}\right)^{2/\alpha_L} L\Bigg(\vec b+\vec s_d^{~\prime}-\vec s_q\,\Bigg|\,\alpha_L,\left(R_q^{\alpha_L}+R_d^{\alpha_L}\right)^{1/\alpha_L}/2\Bigg),
\end{equation}
\begin{equation}\label{eq:dqlevy}
\sigma_{\rm in}^{dq}(\vec s_{q}^{\,\prime},\vec s_{d}%
;\vec b\,) = 2\pi A_{qq}\left(2R_{q}^{\alpha_L}\right)^{2/\alpha_L} L\Bigg(\vec b+\vec s_q^{~\prime}-\vec s_d\,\Bigg|\,\alpha_L,\left(R_q^{\alpha_L}+R_d^{\alpha_L}\right)^{1/\alpha_L}/2\Bigg),
\end{equation}
and
\begin{equation}\label{eq:ddlevy}
\sigma_{\rm in}^{dd}(\vec s_{d},\vec s_{d}^{\,\prime}%
;\vec b\,) = 4\pi A_{qq}\left(2R_{q}^{\alpha_L}\right)^{2/\alpha_L} L\left(\vec b+\vec s_d^{~\prime}-\vec s_d\,|\,\alpha_L,\left(2 R_{d}^{\alpha_L}\right)^{1/\alpha_L}/2\right).
\end{equation}

\cref{eq:tilde_sigma_inel} with \cref{eq:elprob}, \cref{eq:quark-diquark_levy}, \cref{eq:qqlevy}--\cref{eq:ddlevy} define the L\'evy \mbox{$\alpha$-stable} generalized Bialas--Bzdak (LBB) model for $\tilde\sigma_{\rm in}(\,\vec b\,)$. Now, \cref{eq:sig_tilde_b} is a sum of integrals of products of normalized L\'evy $\alpha$-stable distributions. Integrals of products of \mbox{L\'evy $\alpha$-stable} distributions can be easily calculated if the integral can be written in a convolution form as in the case of the first three terms in \cref{eq:sig_tilde_b}. Then, utilizing the convolution theorem, the calculation is straightforward.

According to the convolution theorem, the Fourier transform of a convolution of two functions is the product of their Fourier transforms,
\begin{equation}\label{eq:CT0}
    \mathcal{F}\left[\int d\tau g(\tau)f(x-\tau) \right]= \mathcal{F}\left[f(x)\right]\mathcal{F}\left[g(x)\right],
\end{equation}
implying that
\begin{equation}\label{eq:CT}
    \int d\tau g(\tau)f(x-\tau) = \mathcal{F}^{-1}\left[\mathcal{F}\left[f(x)\right]\mathcal{F}\left[g(x)\right]\right],
\end{equation}
where $\mathcal{F}$ denotes Fourier transformation and $\mathcal{F}^{-1}$ denotes the inverse Fourier transformation.

The Fourier transformed of a L\'evy $\alpha$-stable distribution is its characteristic function. The product of two L\'evy $\alpha$-stable characteristic functions with scale parameters $R_1$ and $R_2$ is another L\'evy $\alpha$-stable characteristic function with scale parameter 
\begin{equation}\label{eq:levyadd}
|R|=\left(|R_1|^{\alpha_L}+|R_2|^{\alpha_L}\right)^{\frac{1}{\alpha_L}}.
\end{equation}
The inverse Fourier transformation of a L\'evy $\alpha$-stable characteristic function gives L\'evy $\alpha$-stable distribution. 
Thus, utilizing \cref{eq:CT}, the first three terms in \cref{eq:sig_tilde_b} in the LBB model are:
\begin{align}\label{eq:LBBsqqb}
&\tilde\sigma_{\rm in}^{qq}(\,\vec b\,)=  \pi A_{qq} \left(2R_{q}^{\alpha_L}\right)^{2/\alpha_L}\times\\ \nonumber &\times\int d^2s_qd^2s'_qL(\vec s_{q} |\,\alpha_L, R_{qd*}/2)L(\vec s_q^{~\prime} |\,\alpha_L,R_{qd*}/2)L\left(\vec b+\vec s_q^{~\prime}-\vec s_q|\left(2 R_{q}^{\alpha_L}\right)^{1/\alpha_L}/2\right) \\ \nonumber 
&=\pi A_{qq} \left(2R_{q}^{\alpha_L}\right)^{2/\alpha_L} L\left(\vec b\,\Big|\,\alpha_L,\left(2R_{qd*}^{\alpha_L}+2R_{q}^{\alpha_L}\right)^{1/\alpha_L}/2\right),
\end{align}
\begin{align}\label{eq:LBBsqdb}
&\tilde\sigma_{\rm in}^{qd}(\,\vec b\,)=2\pi A_{qq} \left(2R_{q}^{\alpha_L}\right)^{2/\alpha_L}\times\\ \nonumber &\times\int d^2s_qd^2s'_qL(\vec s_{q} |\,\alpha_L,R_{qd*}/2)L(\vec s_{q}^{~\prime} |\,\alpha_L,R_{qd*}/2)L\left(\vec b-\lambda\vec s_q^{~\prime}-\vec s_q\Bigg|\,\alpha_L,\left(R_{q}^{\alpha_L}+R_{d}^{\alpha_L}\right)^{1/\alpha_L}/2\right)\\ \nonumber 
&= 
2\pi A_{qq} \left(2R_{q}^{\alpha_L}\right)^{2/\alpha_L} L\left(\vec b\,\Big|\,\alpha_L,\left((1+\lambda^{\alpha_L})R_{qd*}^{\alpha_L}+ R_{q}^{\alpha_L}+R_{d}^{\alpha_L}\right)^{1/\alpha_L}/2\right),
\end{align}
\begin{align}\label{eq:LBBsddb}
&\tilde\sigma_{\rm in}^{dd}(\,\vec b\,)
=4\pi A_{qq} \left(2R_{q}^{\alpha_L}\right)^{2/\alpha_L}\times\\ \nonumber &\times\int d^2s_qd^2s'_qL(\vec s_{q} |\,\alpha_L,R_{qd*}/2)L(\vec s_q^{~\prime} |\,\alpha_L,R_{qd*}/2)L\left(\vec b+\lambda(\vec s_q-\vec s_q^{~\prime})|\,\alpha_L,\left(2 R_{d}^{\alpha_L}\right)^{1/\alpha_L}/2\right) \\ \nonumber
&=4\pi A_{qq} \left(2R_{q}^{\alpha_L}\right)^{2/\alpha_L} L\left(\vec b\,\Big|\,\alpha_L,\left(2\lambda^{\alpha_L} R_{qd*}^{\alpha_L}+2R_{d}^{\alpha_L}\right)^{1/\alpha_L}/2\right).
\end{align}
We see that the results of these three integrals, constituting the leading order terms in \cref{eq:sig_tilde_b}, are in terms of L\'evy $\alpha$-stable distributions. The remaining eight integrals, however, can not be written in convolution forms. The results of these remaining integrals are yet to be computed in terms of analytic formulas.

It is not easy to do numerical calculations either. Gaussian distributions have closed forms in terms of elementary functions, however, L\'evy $\alpha$-stable distributions do not. L\'evy $\alpha$-stable\, distributions\, can\, be\, expressed\, in\, terms\, of\, special\, functions,\, namely \mbox{Fox H-functions} (see \hyperref[sec:app_biGaussLevy]{Appendix C}). Since products of Fox H-functions can be rewritten as a single Fox H-function \cite{doi:10.1137/0133036}, every integral term in \cref{eq:sig_tilde_b} will be an integral of a single \mbox{Fox H-function} over all possible $\vec s_q$ and $\vec s_q^{\,\prime}$. But even the values of a Fox H-function is calculated by numerical integration (see \hyperref[sec:app_biGaussLevy]{Appendix C}). Thus, to get the scattering amplitude in momentum representation, one has to calculate numerically an integral (see \cref{eq:relPWtoeik_2}) with an integrand (see \cref{eq:ReBB_b_ampl}) given in terms of a sum of numerically calculated four-dimensional integrals (see \cref{eq:sig_tilde_b}) with integrands being again numerically calculated integrals (computation of the value of the Fox H-function). This means a sequence of three integral calculations where the integrand of the second integral contains the result of the first integral, and the result of the second integral is contained by the integrand of the third integral. Thus, a relatively high computing capacity or improved analytic insight will be needed in the future in other to fit the LBB model parameters to the experimental data.

Interestingly, the SL model as introduced in \cref{sec:levy_simp_gen} can be obtained from the LBB model by (i) approximating $\tilde\sigma_{\rm in}(s,b)$ by a single Lévy $\alpha$-stable shape in $b$ and (ii) sticking only to the applicability in the low-$|t|$ domain. Scattering of protons characterized by small $|t|$ values corresponds to scattering resulting from collisions with high values of $b$. At high $b$ values $\tilde\sigma_{\rm in}(s,b)$ is small. The leading order expression in the Taylor expansion of \cref{eq:ReBB_b_ampl} is 
\begin{equation}\label{eq:full_amplitude_expanded}
    \widetilde T_{\rm el}(s, b) = \left(\alpha_R+\frac{i}{2}\right)\tilde\sigma_{\rm in}(s, b).
\end{equation}
\cref{eq:full_amplitude_expanded} is considered as a low-$|t|$ approximation of the scattering amplitude of \cref{eq:ReBB_b_ampl}. In this approximation: 
\begin{equation}\label{eq:rho0_ReBB}
    \rho_{0}(s)=\frac{{\rm Re}T_{\rm el}(s,t=0)}{{\rm Im}T_{\rm el}(s,t=0)}=2\alpha_R(s).
\end{equation}
Then \cref{eq:full_amplitude_expanded} can be rewritten as
\begin{equation}\label{eq:full_amplitude_expanded_rewrite}
    \widetilde T_{\rm el}(s, b) = \frac{1}{2}\left(i+\rho_0(s)\right)\tilde\sigma_{\rm in}(s, b).
\end{equation}
Comparing \cref{eq:full_amplitude_expanded_rewrite} to \cref{eq:gray_Levy}, it is clear that choosing a Lévy $\alpha$-stable form for $\tilde\sigma_{\rm in}(s, b)$, as
\begin{equation}\label{eq:sigin_Levy}
     \tilde\sigma_{\rm in}(s, b)  =  \frac{\sigma_{\rm tot}(s)}{4\pi^2} \int d^2\vec q  {\rm e}^{i\,\vec q\cdot\vec b}
                         {\rm e}^{-\frac{1}{2}\big|q^2B_L(s)\big|^{\alpha_L(s)/2}},
\end{equation}
we recover the Lévy $\alpha$-stable model of elastic scattering as given by \cref{eq:gray_Levy}. Alternatively, by choosing a  Gaussian shape for $\tilde\sigma_{\rm in}(s, b)$, as
\begin{equation}\label{eq:sigin_Levy}
     \tilde\sigma_{\rm in}(s, b)  =  \frac{1}{2\pi}\frac{\sigma_{\rm tot}(s)}{B_{0}(s)} \,{\rm e}^{-\frac{b^2}{2 B_0(s)}},
\end{equation}
we recover the Gaussian model of elastic scattering as given by \cref{eq:ampl_b_Gauss}.

Now, I relate first the parameters of the Gaussian model of low-$|t|$ elastic scattering to the ReBB model parameters. Then, I relate the parameters of the Lévy $\alpha$-stable model of  low-$|t|$ elastic scattering to the LBB model parameters.

I utilize the approximation as given by \cref{eq:full_amplitude_expanded} and I consider only the leading order terms in the ReBB model $\tilde\sigma_{\rm in}(s, b)$, i.e., 
\begin{equation}\label{eq:sigbqq}
\tilde\sigma^{qq}_{\rm in}(s,b) = \frac{A_{qq}R_q^2(s)}{R_{qd*}^2(s)+R_q^2(s)}\,{\rm e}^{-\frac{1}{2}\frac{b^2}{R_{qd*}^2(s)+R_q^2(s)}}, 
\end{equation}
\begin{equation}\label{eq:sigbqd}
\tilde\sigma^{qd}_{\rm in}(s,b)=\frac{4A_{qq}R_q^2(s)}{(1+\lambda^2)R_{qd*}^2(s)+R_q^2(s)+R_d^2(s)}\,{\rm e}^{-\frac{b^2}{(1+\lambda^2)R_{qd*}^2(s)+R_q^2(s)+R_d^2(s)}},
\end{equation}
and
\begin{equation}\label{eq:sigbdd}
\tilde\sigma^{dd}_{\rm in}(s,b)= \frac{4 A_{qq}R_q^2(s)}{\lambda^2R_{qd*}^2(s)+R_d^2(s)}\,{\rm e}^{-\frac{1}{2}\frac{b^2}{\lambda^2R_{qd*}^2(s)+R_d^2(s)}}. 
\end{equation}
These terms give the dominant contribution when $t\to0$. Substituting \cref{eq:sigbqq}, \cref{eq:sigbqd} and \cref{eq:sigbdd} into \cref{eq:full_amplitude_expanded_rewrite} and utilizing \cref{eq:relPWtoeik_2} we get the amplitude in momentum space. Then, the differential cross section is calculated via \cref{eq:Tdsigma}. The optical point parameter is calculated via \cref{eq:optpoint} resulting
\begin{equation}\label{eq:op_ReBB}
    a(s)=\frac{81}{4} \pi R_q^4(s) \left(1 + 4 \alpha^2_R(s)\right).
\end{equation}
The slope parameter is calculated via \cref{eq:Bst0} yielding
\begin{equation}\label{eq:B_ReBB}
   B_0(s)=\frac{2}{9}R_{qd}^2(s)+\frac{2}{3}R_{d}^2(s)+\frac{1}{3}R_{q}^2(s).
\end{equation}


Now I consider the leading order terms in the LBB model $\tilde\sigma_{\rm in}(s, b)$, as given by \cref{eq:LBBsqqb}, \cref{eq:LBBsqdb}, and \cref{eq:LBBsddb}. 
Then, the result for the optical point is 
\begin{equation} \label{eq:op_LBB}
   a(s)=\frac{81}{16} \pi \left(2 R_q^{\alpha_L}(s)\right)^{4/\alpha_L} \left(1 + 4 \alpha^2_R(s)\right).
\end{equation}

The expression for $B_L$ is easily obtained by regarding the differential cross section as a function of $\tilde t = -|t|^{\alpha_L/2}$. The result is:
\begin{align}\label{eq:B_LBB}
   B_L(s)=\frac{1}{36}\left(\frac{4}{3}\right)^{2/\alpha_L}\left(\left(2+2^{\alpha_L}\right)R_{qd}^{\alpha_L}(s)+3^{\alpha_L}\left(2R_d^{\alpha_L}(s)+R_q^{\alpha_L}(s)\right)\right)^{2/\alpha_L}.
\end{align}

Thus, the relation of the parameters of the Gaussian model of low-$|t|$ elastic scattering to the ReBB model parameters are given in \cref{eq:op_ReBB} and \cref{eq:B_ReBB} while the relation of the parameters of the Lévy $\alpha$-stable model of  low-$|t|$ elastic scattering to the LBB model parameters are given in \cref{eq:op_LBB} and \cref{eq:B_LBB}. Clearly, for $\alpha_L=2$, \cref{eq:op_LBB} reduces to \cref{eq:op_ReBB} and \cref{eq:B_LBB} reduces to \cref{eq:B_ReBB}.

Using the low-$|t|$ approximation of the scattering amplitude, \cref{eq:full_amplitude_expanded_rewrite}, and only the first three, leading order terms in $\tilde\sigma_{\rm in}(s, b)$ as above, the ReBB model total cross section, calculated by \cref{eq:Ttot}, is
\begin{equation}\label{eq:sigtot_ReBB}
   \sigma_{tot}(s)=18\pi R_q^2(s),
\end{equation}
while the LBB model total cross section is
\begin{equation}\label{eq:sigtot_LBB}
   \sigma_{tot}(s)=9\pi \left(2R_q^{\alpha_L}(s)\right)^{2/\alpha_L}.
\end{equation}
Since $R_q$ has the same value in $pp$ and $p\bar p$ scattering at a given c.m. energy, the approximate results \cref{eq:sigtot_ReBB} and \cref{eq:sigtot_LBB} support the strong version of the Pomeranchuk theorem stating that the total cross section for particle-particle and particle-antiparticle scattering become equal at asymptotically high energies (see \cref{sec:oddintro}).

\newpage

\textbf{Summary}
\vspace{0.2cm}


In this Chapter, I generalized the Gaussian, $\alpha_L=2$ special case model of low-$|t|$ elastic $pp$ and $p\bar p$ scattering to a Lévy \mbox{$\alpha$-stable,} $0<\alpha_L\leq2$ model. I demonstrated that the simple Lévy \mbox{$\alpha$-stable} model describes the 8 TeV low-$|t|$ non-exponential $pp$ differential cross section data with $CL$=55.33\% (the Gaussian model describes the same data with $CL$=1.06$\times10^{-24}$\%). Then, utilizing the simple Lévy $\alpha$-stable model, I analyzed the low-$|t|$ $pp$ and $p\bar p$ elastic differential cross section data in the c.m. energy range of $546~{\rm GeV}<\sqrt s<13~{\rm TeV}$ and in the squared four-momentum transfer range of \mbox{0.02 GeV$^2$ $\leq -t\leq0.15$ GeV$^2$}. The analysis revealed that the value of $\alpha_L$ is $1.958\pm0.003$, $i.e.$, significantly less than two. This result excludes the Gaussian model of low-$|t|$ elastic scattering.

I found that the energy dependence of the optical point parameter $a$ in the SL model is compatible with a squared-logarithmic shape. I observed the known incompatibility between TOTEM and ATLAS data regarding their overall normalization \cite{ATLAS:2022mgx}.

I found that the energy dependence of the Lévy slope parameter, $B_L$, for  ATLAS and TOTEM $pp$ data and separately for $p\bar p$ data, is compatible with linear-logarithmic shapes; however, the $pp$ and $p\bar p$ $B_L$ parameter trends are incompatible. This can be interpreted with the jump in the energy dependence of the $B_0$ slope parameter somewhere in the energy domain of 3~TeV $\lesssim \sqrt s \lesssim$ 4 TeV \cite{TOTEM:2017asr} or as an odderon effect. 

Finally, motivated by the success of the simple Lévy $\alpha$-stable model of low-$|t|$ elastic $pp$ and $p\bar p$ scattering, I generalized the Gaussian shapes in the ReBB model to Lévy $\alpha$-stable shapes resulting in the LBB model. I related the parameters of the simple Lévy $\alpha$-stable model to the parameters of the LBB model. However, a relatively high computing capacity or improved analytic insight will be needed to apply the full LBB model to describe experimental data. 



\newpage


\chapter*{\vspace{-1.1cm}Summary}\label{chap:summary}
\addcontentsline{toc}{chapter}{\protect\numberline{}Summary}
\markboth{Summary}{Summary}

In this dissertation, aiming to conclude about possible odderon effects, I analyzed pp and $\rm{p\bar p}$ elastic scattering data measured in collider experiments in the c.m. energy range from half TeV up to 13 TeV in the ``soft'' domain, \textit{i.e.}, in the squared four-momentum transfer region $|t|\lesssim$ 3 GeV$^2$. I investigated also the $H(x)$ scaling of elastic pp scattering and the low-$|t|$ non-exponential behavior of the elastic pp differential cross section.

In \cref{chap:ReBBpbarp}, I generalized the ReBB model of elastic pp scattering to elastic $\rm{p\bar p}$ scattering. It was a crucial step towards the ReBB model odderon analysis. By fitting the ReBB model to the $p\bar p$ elastic differential cross section data, I found that the energy dependencies of the geometrical parameters of the model, $R_q$, $R_d$, and $R_{qd}$, for $\rm{p\bar p}$ scattering, are compatible with linear-logarithmic curves form ISR energies up to LHC energies.

In \cref{chap:oddTD0}, I studied the possibilities of
closing the energy gap between $\rm pp$ and $\rm p\bar p$ elastic differential cross section measurements by presenting my results obtained during my participation in the joint project of the CERN LHC TOTEM and FNAL Tevatron D0 experimental collaborations on the odderon search. Closing the energy gap between $\rm pp$ and $\rm p\bar p$ elastic scattering is essential for studying odderon effects in these processes. Utilizing the ReBB model and a Regge model with dipole pomeron and odderon exchanges, I extrapolated the TOTEM measurements on $\rm pp$ ${\rm d}\sigma_{\rm el}/{\rm d}t$ to $\sqrt{s}=1.96$ TeV and determined that the effect of the odderon exchange manifests mainly in generating a prominent diffractive minimum in pp  ${\rm d}\sigma_{\rm el}/{\rm d}t$ and filling in this minimum in the $\rm{p\bar p}$ ${\rm d}\sigma_{\rm el}/{\rm d}t$. I gave twelve internal presentations within ten months on my progress in this work \cite{talk1,talk2,talk3,talk4,talk5,talk6,talk7,talk8,talk9,talk10,talk11,talk12}. My results served as a guide during the joint D0-TOTEM analysis that finally led to the observation of the signal of the odderon exchange with a statistical significance of at \mbox{least 5.2$\sigma$ \cite{TOTEM:2020zzr}.}  


In \cref{chap:rebbdesc}, I completed a refined ReBB model analysis of elastic pp and $\rm p\bar p$ scattering data. I found that each of the geometrical parameters of the model, $R_q$, $R_d$, and $R_{qd}$, has an energy dependence compatible with a single, monotonically increasing linear-logarithmic curve in pp and $\rm{p\bar p}$ collisions implying that the values of these parameters within errors are the same in pp and $\rm{p\bar p}$ reactions. I found, however, that the energy dependence of the opacity parameter of the model, $\alpha_R$, is incompatible with a single curve in pp and $\rm{p\bar p}$ reactions, implying that the value of this parameter is different in pp and $\rm{p\bar p}$ scattering. In this refined analysis, the kinematic range of the quantitative ReBB model description, \textit{i.e.}, the description with $CL$~$\geq$~0.1\%, is: 0.546~TeV~$\leq\sqrt{s}\leq 7$~TeV and 0.38~GeV$^2$~$\leq -t\leq1.2$~GeV$^2$ \cite{Csorgo:2020wmw}. 

In \cref{chap:odderon}, I validated the ReBB model analysis of elastic pp and $\rm{p\bar p}$ scattering in the kinematic domain \mbox{0.546~TeV~$\leq\sqrt{s}\leq 8$~TeV} and 0.38~GeV$^2$~$\leq -t\leq1.2$~GeV$^2$. Based on this validated analysis, I closed the energy gap between elastic pp and $\rm{p\bar p}$ scattering and compared pp and $\rm{p\bar p}$ differential cross sections at exactly the same energies in the TeV c.m. energy range. I found that the pp and $\rm{p\bar p}$ differential cross sections differ with a statistical significance of at least 6.3$\sigma$ when significances obtained at 1.96 TeV and 2.76 TeV are combined. By combining the significances obtained at all the four analyzed energies, 1.96~TeV, 2.76 TeV, 7 TeV and 8 TeV, the statistical significance of the difference between pp and $\rm{p\bar p}$ differential cross sections, \textit{i.e.}, the statistical significance of the $t$-channel odderon exchange signal is higher than $30\sigma$ \cite{Szanyi:2022ezh,Szanyi:2022qgx}.  

In \cref{chap:Hxvalidity}, I interpreted the $H(x)$ scaling of elastic pp scattering:
I showed that models of elastic scattering with an impact parameter distribution depending on $b$ only via a dimensionless variable, \mbox{$\tilde\xi=b/R(s)$,} where $R(s)$ is an internal scale and can be identified with $\sqrt{B_0(s)}$, $i.e.$ the square root of the slope parameter of ${\rm d}\sigma_{\rm el}/{\rm d}t$, manifest $H(x)$ scaling at all values of $t$. 
I identified the $H(x)$ scaling limit of the ReBB model and compared it to the pp and $\rm p\bar p$  differential cross section data to test the domain of validity of the $H(x)$ scaling law of elastic scattering in $s$ and $t$. Based on this comparison, I concluded that the $H(x)$ scaling in elastic pp scattering is valid from \mbox{$\sqrt s =$ 7 TeV} down to \mbox{1.96 TeV} in the $t$ range where the original ReBB model is validated. 
The $H(x)$ scaling property was used to compare the TOTEM measured pp $\sqrt{s}=$ 7 TeV and the D0 measured $\rm p\bar p$ $\sqrt{s}=$ 1.96 TeV elastic differential cross section data 
resulting in a model-independent, data-driven odderon signal with a statistical significance of at least 6.26$\sigma$  \cite{Csorgo:2019ewn,Csorgo:2020rlb,Csorgo:2023rzm,Csorgo:2020msw}.  My results indicate that this comparison is made in the domain of validity of the $H(x)$ scaling of elastic pp scattering \cite{Csorgo:2019ewn}.

In \cref{chap:levy}, I generalized the Gaussian, $\alpha_L=2$ special case model of low-$|t|$ elastic pp and $\mathbf{\rm p\bar p}$ scattering to a Lévy $\alpha$-stable, $0<\alpha_L\leq2$ model and demonstrated that the generalized model describes the strongly non-exponential behavior of the precise LHC low-$|t|$ data with $\alpha_L=1.959\pm0.002$ and $CL\geq8.8$ \% in the kinematic range of \mbox{0.02 GeV$^2$ $\leq -t\leq0.15$ GeV$^2$} \cite{Csorgo:2023rbs}. Motivated by the success of the simple Lévy $\alpha$-stable model and aiming to achieve a statistically acceptable description in a wider kinematic range in an improved BB model framework, I formulated the real extended Lévy $\alpha$-stable generalized Bialas--Bzdak (LBB) model by generalizing the Gaussian distributions in the ReBB model to Lévy $\alpha$-stable distributions and consequently introducing a new free parameter, $\alpha_L$, the Lévy index of stability \cite{Csorgo:2023pdn}. The studies presented in \cref{chap:levy} and published in Refs.~\cite{Csorgo:2023rbs,Csorgo:2023pdn} constitute the first applications of Lévy $\alpha$-stable distributions with $0 < \alpha_L \leq 2$ in physical models of elastic hadronic scattering.

In conclusion, the signal of the odderon exchange with a statistical significance higher than 5$\sigma$ is observed using three different methods and the results are published in four refereed papers \cite{Csorgo:2019ewn,Csorgo:2020wmw,Szanyi:2022ezh,TOTEM:2020zzr} to which I contributed significantly. The discovery of the odderon exchange deepens our understanding of strong interactions by providing evidence for a leading color-neutral $t$-channel exchange with negative spatial and charge parity which gives an observable contribution in the c.m. energy domain dominated by the pomeron exchange. The odderon discovery was announced to the public, among others, by MATE \cite{mate}, ELTE \cite{elte}, Lund University \cite{lund}, \mbox{CERN \cite{cerncourier, cernnews}} and Fermilab \cite{fermilabnews} and was mentioned as a highlighted result at CERN in the first place in 2021 \cite{cernhighlights}. The science portal of the European Commission, CORDIS, considered the odderon discovery 
as a new milestone in particle physics \cite{cordismilestone}. 

\enlargethispage{1\baselineskip} 
Some properties of the odderon and its role in the physics of elastic hadron-hadron scattering can be studied with higher precision by developing theoretical models that describe all the available experimental data on elastic $\rm pp$ and $\rm p\bar p$ scattering in a statistically acceptable way, $i.e.$, with $CL$ $\geq0.1\%$. I took some of the first steps in this new upcoming direction by determining certain properties of the odderon exchange in the framework of the ReBB model that gives a statistically acceptable description to the data in a limited kinematic range \cite{Csorgo:2020wmw} and by formulating the real extended Lévy $\alpha$-stable generalized Bialas--Bzdak (LBB) model~\cite{Csorgo:2023rbs,Csorgo:2023pdn}.

\clearpage
\thispagestyle{empty}
\chapter*{Appendices}
\addcontentsline{toc}{chapter}{\protect\numberline{}Appendices}
\markboth{Appendices}{Appendices}


\setcounter{equation}{0}\renewcommand\theequation{A.\arabic{equation}}

\setcounter{figure}{0}\renewcommand\thefigure{A.\arabic{figure}}

\renewcommand\theHequation{A.\theequation}

\section*{Appendix A: Multiple scattering approach}\label{sec:app_multscatt}
\addcontentsline{toc}{section}{\protect\numberline{}Appendix A: Multiple scattering approach}
\markboth{Appendix A: Multiple scattering approach}{Appendix A: Multiple scattering approach}

The\, diffractive\, multiple-scattering\, theory\, was\, developed\, by\, Roy\, J.\, Glauber\, in\, the \mbox{1950s \cite{glauber1959lectures}} and used later on to describe high-energy hadron-nucleus \mbox{scattering \cite{Glauber:1970jm}.}  The approach was extended to high-energy nucleus-nucleus scattering \cite{Czyż1971} and later to high-energy hadron-hadron scattering \cite{Glauber:1984su,Glauber:186995,pike2002scattering}. In this Appendix, first, I discuss the high-energy eikonal approximation of the non-relativistic potential scattering problem. Then I discuss the theory of multiple collisions or
diffractive multiple-scattering theory which is the generalization of the eikonal approximation of high energy non-relativistic potential scattering to the collision of a composite few-body system of nucleons with a single-body hadron \cite{glauber1959lectures,pike2002scattering}. Finally, I discuss a probabilistic approach to multiple collisions and the problem of the collision of two composite few-body systems of nucleons.



\subsection*{Potential scattering and eikonal approximation}

The basic theory of potential scattering describes the simplest type of quantum scattering problem, the elastic interaction of two spinless particles \cite{Barone:2002cv, Collins:1977jy, glauber1959lectures}. Potential scattering is a two-body problem; however, 
we can describe the interaction of the two particles with masses $m_1$, $m_2$, and position vectors $r_1$, $r_2$, as the scattering of a single particle in a static force field represented by a short-range potential. This potential, $V(\vec{r})$, depends only on the relative positions of the particles, 
\begin{equation}
    \vec r = \vec r_1 - \vec r_2.
\end{equation}
The mass of the single particle emerging in the treatment of the problem is the reduced mass of the two-particle system,
\begin{equation}
    \mu=\frac{m_1 m_2}{m_1+m_2}.
\end{equation}
Then the energy of the incident particle is 
\begin{equation}
    E= \frac{\hbar^2 k^2}{2\mu},
\end{equation}
where $\hbar$ is the reduced Planck constant, $k^2=|\vec k|^2$,  and $\vec k$ is the propagation vector of the incident particle wave. A particle in quantum mechanics is represented by a wavepacket $\Psi(x,t)$. When the wavepacket has well-defined energy and momentum, it may be approximated by a plane wave: $\Psi(x,t)= \psi(\vec{r})e^{-iE t/\hbar}$. In the standard theory of potential scattering, the incoming plane wave $\psi(\vec{r})$  continuously scatters on the potential and a part of it propagates further in the form of a spherical wave originating from the scattering center. The problem is thus to solve the time-independent Schrödinger equation, 
\begin{equation}\label{eq:Sch}
    \left(\nabla^2 + k^2\right) \psi(\vec{r})=\frac{2\mu}{\hbar^2}V(\vec{r})\psi(\vec{r}),
\end{equation}
with the boundary condition
\begin{equation}\label{eq:bc}
    \psi(\vec{r})\underset{r \to \infty}{\sim} e^{i\vec k \cdot \vec r} + f(\vec k, \vec k^{\,\prime}) \frac{e^{i\vec k^{\,\prime}\cdot \vec r}}{r},
\end{equation}
where $r=\left|\vec r\,\right|$ furthermore $\vec r$ and $\vec k^{\,\prime}$ are parallel. \cref{eq:bc} represents the result of a three-dimensional scattering process of a plane wave against a fixed scatterer: asymptotically far from the short-range potential, the scattered wave function is the sum of the incoming plane wave with wave vector $\vec k$ and an outgoing spherical wave originating from the scattering center and characterized by the wave vector $\vec k^{\,\prime}$. Energy conservation requires that $|\vec k|=|\vec k^{\,\prime}|\equiv k$. The quantum mechanical non-relativistic scattering amplitude is denoted as $f(\vec k, \vec k')$, and its absolute value squared determines the observable differential cross section,
\begin{equation}\label{eq:QMdiffxsec}
    \frac{d\sigma}{d\Omega} = |f(\vec k, \vec k^{\,\prime})|^2,
\end{equation}
interpreted as the measure of probability that the incoming particle scatters in the infinitesimal solid angle range
$d\Omega$ with the outgoing wave vector $\vec k^{\,\prime}$. 


The soloution of \cref{eq:Sch} with the boundary condition \cref{eq:bc} is given by the Lippman-Schwinger integral equation:
\begin{equation}\label{eq:LSsol}
   \psi(\vec{r})=e^{i\vec k \cdot \vec r}-\frac{2\mu}{4\pi\hbar^2}\int \frac{e^{ik | \vec r-\vec r^{\,\prime}|}}{| \vec r-\vec r^{\,\prime}|}V(\vec{r}^{\,\prime})\psi(\vec{r}^{\,\prime})d\vec{r}^{\,\prime}.
\end{equation}
The $\vec{r}^{\,\prime}$ variable runs over the domain where $V(\vec{r}^{\,\prime})\neq 0$ and the integration domain in \cref{eq:LSsol}  includes the region where the potential is non-vanishing. At asymptotic distances, $i.e$, when $|\vec r|\rightarrow \infty$ (or equivalently when  $|\vec r|\gg a$, where $a$ is the width of the potential outside which the potential vanishes),  \cref{eq:LSsol} takes the form \cite{glauber1959lectures,davydov1976quantum}:
\begin{equation}\label{eq:LSsolasymp}
   \psi(\vec{r})=e^{i\vec k \cdot \vec r}-\frac{2\mu}{4\pi\hbar^2}\frac{e^{ikr}}{r}\int e^{-i\vec k^{\,\prime} \cdot \vec{r}^{\,\prime}}V(\vec{r}\,')\psi(\vec{r}\,')d\vec{r}\,'.
\end{equation}
Comparing \cref{eq:bc} to \cref{eq:LSsolasymp} one finds that the scattering amplitude is

\begin{equation} \label{eq:eiksolampl}
    f(\vec k, \vec k')=-\frac{2\mu}{4\pi\hbar^2}\int e^{-i\vec k^{\,\prime} \cdot \vec r^{\,\prime}}V(\vec{r}^{\,\prime})\psi(\vec{r}^{\,\prime})d\vec{r}^{\,\prime}.
\end{equation}
\cref{eq:eiksolampl} implies that we only need to know the wave function inside the potential to calculate the measurable cross section.

We are interested in the high energy limit, \textit{i.e.}, when the energy of the incident particle greatly exceeds the absolute magnitude of the potential, $V/E\ll 1$, and this energy is large enough that the wavelength of the particle is much smaller than the width of the potential (interaction range) $a$, $2\pi a/\lambda=ka\gg 1$ \cite{Barone:2002cv,glauber1959lectures,glauber1970theory}. The latter condition means that the distance over which the potential varies appreciably is large compared to the wavelength of the incident particle. Under such conditions back-scattering is negligible and the solution of \cref{eq:Sch}, to a very good accuracy, can be given by the incident plane wave itself but modulated by an unknown function:
\begin{equation}
    \label{eq:eiksol}
    \psi(\vec{r})=e^{i\vec k \cdot \vec r}\varphi(\vec{r}),
\end{equation}
where $\varphi(\vec{r})$ is a function that varies slowly over a distance corresponding to the wavelength of the particle, $1/k$. This is the eikonal approximation for the potential scattering problem. The direction of propagation of the incident plane wave is chosen to be the direction of the $z$ axis. Since \cref{eq:eiksol} must reduce to the incident plane wave at $z=-\infty$, the boundary condition 
\begin{equation}
    \label{eq:bc2}
    \varphi(x,y,-\infty)=1
\end{equation}
holds. Inserting now \cref{eq:eiksol} into \cref{eq:Sch} and neglecting the term $\nabla^2\varphi$ since $\varphi$ is slowly varying, the solution of the resulting equation
\begin{equation}
    2i\left(\vec k \cdot \vec \nabla \right)\varphi (\vec{r})= \frac{2\mu}{\hbar^2}V(\vec{r})\varphi(\vec{r})
\end{equation}
is
\begin{equation}
    \label{eq:soleik}
    \varphi(x,y,z)=e^{-\frac{i\mu}{\hbar^2k}\int_{-\infty}^zV(x,y,z')dz'},
\end{equation}
and the full wavefunction is 
\begin{equation}
    \label{eq:soleikfull}
    \psi(x,y,z)=e^{ikz-\frac{i\mu}{\hbar^2k}\int_{-\infty}^zV(x,y,z')dz'}.
\end{equation}
 \cref{eq:soleikfull} does not describe a spherical outgoing wave but holds within the volume occupied by the potential.


When we evaluate the scattering amplitude it is useful to resolve the vector $\vec r$ into two components:
\begin{equation}\label{eq:vecomp}
    \vec r = \vec b + \vec \kappa z,
\end{equation}
where the vector $\vec b$ is lying in a plane perpendicular to  $\vec \kappa = \vec k / k$, the unit vector in the direction of propagation of the incoming wave. Furthermore, we consider the case that the potential is centered at the origin and the distance $b=|\vec b|$ then can be interpreted as the impact parameter. The impact parameter is the perpendicular distance between the path of the incoming particle and the center of the potential field.

Then \cref{eq:soleikfull} can be rewritten as
\begin{equation}
    \label{eq:soleikfull_2}
    \psi(\vec r)=e^{i\vec k \cdot \vec r-\frac{i\mu}{\hbar^2k}\int_{-\infty}^zV(\vec b,z')dz'}.
\end{equation}
Inserting \cref{eq:soleikfull_2} into \cref{eq:eiksolampl} and using that $ d^3\vec r =d^2\vec b d z$ one finds 
\begin{equation} \label{eq:eiksolampl_3}
    f(\vec k, \vec k')=-\frac{2\mu}{4\pi\hbar^2}\int_{-\infty}^{+\infty}dz\int d^2\vec b e^{-i(\vec k'-\vec k) \cdot (\vec b-\vec \kappa z)}V(\vec{b},z)e^{-\frac{i\mu}{\hbar^2k}\int_{-\infty}^{z}V(\vec{b},z')dz'}.
\end{equation}

Since energy conservation requires that $|\vec k|=|\vec k'|$, for small scattering angles, the  momentum transfer vector $\vec q=\vec k '-\vec k$ is nearly orthogonal to $\vec k$: $\vec q \underset{\theta\to 0}{\simeq} \vec k'_\bot$, where 
\begin{equation}
    \vec k'_\bot=\vec k'\sin\theta
\end{equation}
is the component of the outgoing three-momentum transverse to the $z$-direction. Thus $(\vec k-\vec k')\cdot \vec \kappa \simeq 0$ and the $z$ integration can be performed in \cref{eq:eiksolampl_3}, yielding
\begin{equation} \label{eq:eiksolampl_4}
    f(\vec q)=\frac{i k}{2\pi}\int d^2\vec b e^{i\vec q \cdot \vec b}\left(1-e^{i\chi(\vec b)}\right),
\end{equation}
where
\begin{equation}
    \chi(\vec b)=-\frac{\mu}{\hbar^2k}\int_{-\infty}^{+\infty}V(\vec{b},z')dz'
\end{equation}
is called the \textit{eikonal function} \cite{Barone:2002cv,Collins:1977jy}. In case of a potential with azimuthal symmetry the angular integration can be performed and  \cref{eq:eiksolampl_4} can be rewritten as \cite{Collins:1977jy}
\begin{equation} \label{eq:eiksolampl_5}
    f(\vec q)=i k\int_0^\infty db bJ_0(qb)\left(1-e^{i\chi(b)}\right),
\end{equation}
where $q=|\vec q|=2k\sin{\left(\theta/2\right)}$ and $J_0$ is the zeroth order Bessel function of the first kind. \cref{eq:eiksolampl_4} and also \cref{eq:eiksolampl_5} are referred to as the eikonal representation of the scattering amplitude.

We can understand the meaning of the eikonal function by taking the $z\rightarrow\infty$ limit of \cref{eq:soleikfull_2}, 
\begin{equation}
    \psi(\vec r) \underset{z\rightarrow\infty}{\sim} e^{i\chi(\vec b)}e^{i\vec k \cdot \vec r}.
\end{equation}
One can see that $\chi(\vec b)$ represents the phase shift of the incident wave due to its passage through the scattering potential.

The quantum mechanical results for high energy elastic scattering are analogous to Kirchhoff's results for Fraunhofer diffraction by an absorbing opaque obstacle obtained by applying Maxwell's equations \cite{Barone:2002cv}. 
The quantity 
\begin{equation}
    \Gamma(\vec b) = 1-e^{i\chi(\vec b)}
\end{equation}
is called as the profile function in analogy to optics \cite{Barone:2002cv}. 
The scattering amplitude, as given by \cref{eq:eiksolampl_4}, is a two-dimensional Fourier transform of the profile function:
\begin{equation} \label{eq:eiksolampl_6}
    f(\vec q)=\frac{i k}{2\pi}\int d^2\vec b e^{i\vec q \cdot \vec b}\Gamma(\vec b).
\end{equation}
By inverting the Fourier transform in \cref{eq:eiksolampl_6} we get the profile function in terms of the scattering amplitude:
\begin{equation} \label{eq:eiksolampl_7} 
    \Gamma(\vec b)=\frac{1}{2\pi ik}\int d^2\vec b e^{-i\vec q \cdot \vec b}f(\vec q\,).
\end{equation}

These formulas are derived here for non-relativistic scattering; however, they can be derived independently from the Schrödinger equation directly from the partial-wave expansion for relativistic scattering \cite{Barone:2002cv}  (see \cref{sec:eik} of this dissertation).

\subsection*{The theory of multiple collisions}

The diffractive multiple-scattering theory was introduced as a generalization of the eikonal approximation of high energy non-relativistic potential scattering to the collision of a composite few-body system of nucleons with a single-body hadron \cite{glauber1959lectures,pike2002scattering}. 

In hadron-nucleus collisions, the target, \textit{i.e.}, the nucleus is a composite $N$-body system of nucleons and the projectile is a single-body hadron. The problem is to solve the Schrödinger equation of an $N+1$ body system. When a high-energy particle passes through a nucleus, it is not very much deflected, and there is not enough time for any nuclear rearrangement to take place. The projectile is much more energetic than any nucleon in the nucleus; thus, essentially, when it collides with individual nucleons, the others can be considered to be stationary. This simplifies the problem, allowing the use of the adiabatic approximation \cite{pike2002scattering}. Let $\vec R$ be the projectile position and let $\{\vec r_j\}$ be the positions of the $N$ target constituents. Let us assume also that the projectile interacts with the constituents of the target through central two-body potentials, $V_j(\vec R - \vec r_j)$, and the total interaction between the target and the projectile is given by the sum
 \begin{equation}\label{eq:multipot}
     U(\{\vec R - \vec r_j\}) = \sum_{j=1}^NV_j(\vec R - \vec r_j).
 \end{equation}
In the scattering process the target nucleus goes from an initial state $\ket{i}$ to a final state $\ket{f}$ and the projectile that had initial wave vector $\vec k$ will have a final wave vector $\vec k '$. The scattering amplitude that describes this scattering process in the high-energy eikonal approximation can be written as \cite{glauber1959lectures}
\begin{align}\label{eq:nbampl_2}
    F_{fi}(\vec q)   =\frac{ik}{2\pi}\int e^{i\vec q \cdot \vec b}\bra{f} \Gamma(\vec b, \vec s_1, ..., \vec s_N) \ket{i} d^2\vec b,
\end{align}
where $\vec q = \vec k-\vec k'$ is the momentum transfer, $\vec s_j = \vec r_j - \vec \kappa(\vec \kappa\cdot \vec r_j)$ is the component of $\vec r_j$ perpendicular to $\vec k$, and
\begin{equation}\label{eq:Gamma}
    \Gamma(\vec b, \vec s_1, ..., \vec s_N) = 1 - e^{i\chi(\vec b, \vec s_1, ..., \vec s_N)} 
\end{equation}
is the profile function with phase shift \cite{Czyż1971}
\begin{equation}\label{eq:psmany}
    \chi(\vec b, \vec s_1, ..., \vec s_N) = \sum_{j=1}^N \chi_j(\vec b-\vec s_j) \sim-\sum_{j=1}^N\int_{-\infty}^{+\infty}V_j(\vec b  - \vec s_j + \vec \kappa z)dz.
\end{equation}
In \cref{eq:nbampl_2}, $\bra{f} \Gamma(\vec b, \vec s_1, ..., \vec s_N) \ket{i}$ means an averaging over the constituent positions inside the nucleus. \cref{eq:psmany} says that when the incident particle interacts with the constituents of the target by means of two-body forces, the phase shift \cref{eq:psmany} is just the sum of the phase shifts resulting from the contribution of the individual constituents. The vector $\vec b-\vec s_j$ is the relative transverse position of the projectile particle and the $j$th constituent of the nucleus. 


Using \cref{eq:psmany} we can rewrite \cref{eq:Gamma} as
\begin{equation}\label{eq:Gamma_2}
    \Gamma(\vec b, \vec s_1, ..., \vec s_N) = 1 - \prod_{j=1}^N e^{i\chi_j(\vec b-\vec s_j)} = 1- \prod_{j=1}^N \left(1-\Gamma_j(\vec b-\vec s_j)\right),
\end{equation}
where 
\begin{equation}
    \Gamma_j(\vec b-\vec s_j) = 1- e^{i\chi_j(\vec b-\vec s_j)}
\end{equation}
are the individual constituent profile functions.

\cref{eq:Gamma_2} can be expanded as
\begin{align}\label{eq:Gamma_3}
    \Gamma(\vec b, \vec s_1, ..., \vec s_N) &= \sum_j \Gamma_j (\vec b-\vec s_j) - \sum_{j< l} \Gamma_j (\vec b-\vec s_j)\Gamma_l (\vec b-\vec s_l) + ... \nonumber \\&+(-1)^N\prod_{j=1}^N\Gamma_j (\vec b-\vec s_j),
\end{align}
and one can use the relation given by \cref{eq:eiksolampl_7}
to express the individual constituent profile functions in terms of the individual scattering amplitudes $f_j(\vec q)$ describing the collision of the projectile with the $j$-th constituent of the nucleus. Then, using \cref{eq:Gamma_3}, we can rewrite the nuclear scattering amplitude \cref{eq:nbampl_2} in ascending powers of amplitudes $f_j(\vec q)$. This way, we obtain the so-called multiple scattering expansion developed by the Nobel laureate Roy J. Glauber \cite{glauber1959lectures}. The terms containing a single function $f_j(\vec q)$ describe the single scattering contributions due to the interaction of the projectile with each of the constituents. The further terms provide multiple (double, triple, etc.) scattering corrections. The maximal order of multiple scattering is $N$ since the use of the high-energy (eikonal) approximation restricts the scattering angle to have small values prohibiting backscattering. The incident particle that once interacted with a particular constituent of the target and scattered in the forward direction will not meet again with this particular constituent.  Actually, the first approximation, \textit{i.e.}, the sum of the individual amplitudes includes too much. Multiple scattering corrections with alternating signs (see \cref{eq:Gamma_3}) account for ``shadowing'' effects (when the constituents are in the ``shadow'' cast by one another) or for not counting multiple scattering processes multiple times.  

\subsection*{Probabilistic approach to multiple collisions}

\cref{eq:Gamma_2} says that the profile function for hadron-nucleus scattering can be written in terms of individual constituent profiles due to the additivity of the phase shifts resulting from the individual collisions of the projectile with the target constituents. The same profile function can be obtained by computing the probabilities of all possible multiple collisions (with backscattering prohibited) by considering the individual profiles as probability densities related to the scattering probability for collision in some impact parameter range \cite{Czyz:1969jg,Czyż1971}. The probability density interpretation of the profile is reasonable since, for high-energy small-angle scattering, the amplitude is dominantly imaginary, and thus, the profile is dominantly real (see \cref{sec:eik} of this dissertation).

The idea of the probabilistic approach is as follows. Let us first consider the case when a projectile is a single unit, and the target consists of two subunits with profiles $\Gamma_1(\vec b-\vec s_1)$ and $\Gamma_2(\vec b-\vec s_2)$. The argument of $\Gamma_j$ is the relative transverse position of the projectile and the $j$-th constituent of the target. Then, from probability calculus, the total profile is:
\begin{align}\label{eq:probprof}
    \Gamma(\vec b,\vec s_1,\vec s_2) &= \Gamma_1(\vec b-\vec s_1)\left[1-\Gamma_2(\vec b-\vec s_2)\right] +\Gamma_2(\vec b-\vec s_2)\left[1-\Gamma_1(\vec b-\vec s_1)\right] \nonumber\\&+ \Gamma_1(\vec b-\vec s_1)\Gamma_2(\vec b-\vec s_2).
\end{align}
\cref{eq:probprof} thus gives the sum of the probabilities of all possible collisions in the following interpretation: the projectile collides with constituent 1 and does not collide with constituent 2 or the projectile collides with constituent 2 and does not collide with constituent 1 or the projectile collides with both constituents.

\begin{figure}[h!]
	\centering
\includegraphics[width=0.75\linewidth]{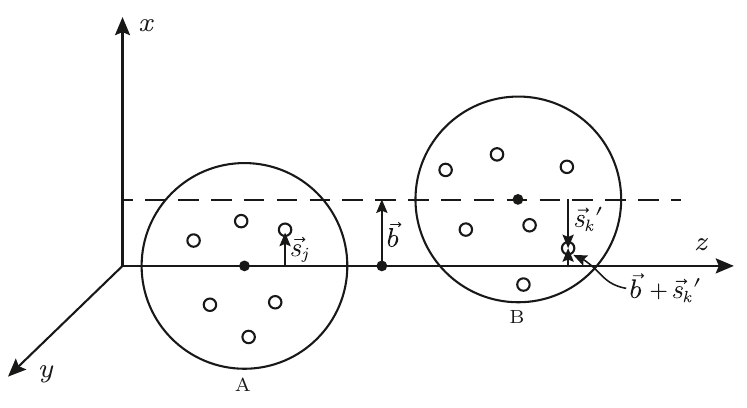}
	\caption{Geometry of the collision of two composite objects and definitions of the coordinates used in the cross section formulae.}
	\label{fig:scattgeom}
\end{figure}

\cref{eq:probprof} can be rewritten as 
\begin{equation}\label{eq:probprof_b}
    \Gamma(\vec b,\vec s_1,\vec s_2) = \Gamma_1(\vec b-\vec s_1) +\Gamma_2(\vec b-\vec s_2) - \Gamma_1(\vec b-\vec s_1)\Gamma_2(\vec b-\vec s_2).
\end{equation}
The generalization of \cref{eq:probprof} to $N$ constituents in the target gives exactly \cref{eq:Gamma_3}, the overall profile function. We can further generalize this result to the case of two composite objects, $A$ and $B$, with $N$ and $N'$ constituents, respectively. Let us denote the profile function of the collision of two constituents $j$ and $k$ as $\Gamma_{jk}(\vec b-\vec s_j+\vec s_k')$, where $\vec s_j$ is the transverse position of the $j$-th constituent in object $A$ and  $\vec s_k'$ is the transverse position of the $k$-th constituent in object $B$. Definitions of the coordinates are shown in \cref{fig:scattgeom}. The argument of $\Gamma_{jk}$ is the relative transverse position of the $j$-th constituent of the object $A$ and the $k$-th constituent of the object $B$. The resulting overall profile function for the collision of two composite objects is:
\begin{equation}\label{eq:Gamma_comp}
    \Gamma(\vec b, \vec s_1, ..., \vec s_N,\vec s_1^{\,\prime}, ..., \vec s_{N'}^{\,\prime}) = 1- \prod_{j=1}^N\prod_{k=1}^{N'} \left[1-\Gamma_{jk}(\vec b-\vec s_j+\vec s\,'_k)\right].
\end{equation}
The scattering amplitude is calculated from \cref{eq:Gamma_comp} using the formula
\begin{align}\label{eq:nbampl_comp}
    F_{fi}(\vec q) =\frac{ik}{2\pi}\int e^{i\vec q \cdot \vec b}\bra{f} \Gamma(\vec b, \vec s_1, ..., \vec s_N,\vec s_1^{\,\prime}, ..., \vec s_{N'}^{\,\prime}) \ket{i} d^2\vec b,
\end{align}
where $\ket{i}$ and $\ket{f}$ are the initial and final states of the system of the two composite particles. The $\vec q$ momentum transfer vector is defined as $\vec q=\vec k - \vec k'$ where $\vec k$ is the relative momentum of $A$ and $B$ before the collision and $\vec k'$ is the relative momentum of $A$ and $B$ after the collision.

\clearpage

\section*{Appendix B: Calculation of $\tilde\sigma_{\rm in }$ in the ReBB model}\label{sec:app_BBcalc}
\addcontentsline{toc}{section}{\protect\numberline{}Appendix B: Calculation of $\tilde\sigma_{\rm in }$ in the ReBB model}
\markboth{Appendix B: Calculation of $\tilde\sigma_{\rm in }$ in the ReBB model}{Appendix B: Calculation of $\tilde\sigma_{\rm in }$ in the ReBB model}
\setcounter{equation}{0}\renewcommand\theequation{B.\arabic{equation}}

In this appendix, I detail the computation of the ReBB model $\tilde\sigma_{\rm in }$ defined by \cref{eq:tilde_sigma_inel} with \cref{eq:elprob}, \cref{eq:quark_diquark_distribution}, \cref{eq:inelastic_cross_sections}, and \cref{eq:Apars}. I discuss also the case when instead of \cref{eq:quark_diquark_distribution}, \cref{eq:quark-diquark_dist} is used.

Substituting \cref{eq:elprob} into \cref{eq:tilde_sigma_inel} we obtain $\tilde\sigma_{\rm in}(\, \vec b\,)$ in terms of eleven different integral expressions as given in \cref{eq:sig_tilde_b}. After making use of the presence of the \mbox{Dirac $\delta$ function} in \cref{eq:quark_diquark_distribution}, the last term in \cref{eq:sig_tilde_b} is given by \cref{eq:tildesigma_general}. \cref{eq:tildesigma_general} can be rewritten to make all the Gaussian integrals explicit \cite{Bialas:2006qf,Nemes:2015iia} 
    \begin{align}
	\tilde\sigma^{qq,qd,dq,dd}_{\rm in}(\,\vec b\,)=-\frac{4v^{2}A}{\pi^2}\int\limits^{+\infty}_{-\infty}\int\limits^{+\infty}_{-\infty}{{\rm d}^2\vec s_q {\rm d}^2\vec s_q^{\,\prime}\,e^{-2v\left(s_q^2+s_q'^2\right)}
            \prod_{k}\prod_{l}e^{-c_{kl}\left(\vec{b}-\vec{s}_{k}+\vec{s}_{l}^{\,\prime}\right)^2}}\,,
            \label{master_formula_2}
    \end{align}
	where $A=A_{qq}A_{qd}A_{dq}A_{dd}$, $v=(1+\lambda^{2})/(2 R_{qd}^{2})$, $c_{kl}=S_{kl}^{-2}$, $S_{kl}^2=R_k^2+R_l^2$, $k,l\in\{q,d\}$. 
After evaluating the Gaussian integrals in \cref{master_formula_2}, we have
    \begin{align}
		\tilde\sigma^{qq,qd,dq,dd}_{\rm in}(\,\vec b\,)=-\frac{4v^{2}A}{B}e^{-{\vec b\,}^2\frac{\Gamma}{B}}\,,
            \label{master_formula}
    \end{align}
	where
    \begin{align}
        B&=C_{qd,dq}\left(v+c_{qq} + \lambda^2 c_{dd}\right)+\left(1-\lambda\right)^2 D_{qd,dq}\,,\notag\\
        \Gamma&=C_{qd,dq}D_{qq,dd} + C_{qq,dd}D_{qd,dq}\,,
            \label{master_formula1}
    \end{align}
	and
    \begin{align}
	C_{kl,mn}&=4v + \left(1+\lambda\right)^2\left(c_{kl}+c_{mn}\right)\,,\notag\\
	D_{kl,mn}&=v \left(c_{kl}+c_{mn}\right)+\left(1+\lambda\right)^2c_{kl}c_{mn}\,.
	            \label{master_formula2}
    \end{align} 
	Each term in \cref{eq:sig_tilde_b} can be obtained from \cref{master_formula}, by setting one or more coefficients to
	zero,~$c_{kl}=0$, and the corresponding amplitude to one,~$A_{kl}=1$.

To write out the result of each integral, I define a rescaled $R_{qd}$ parameter, $R_{qd*}$. If $D(\vec s_{q},\vec s_{d})$ is defined by \cref{eq:quark_diquark_distribution}
\begin{equation}
    R_{qd*}=R_{qd}\frac{1}{\sqrt{1+\lambda^2}}.
\end{equation}
If $D(\vec s_{q},\vec s_{d})$ is defined by \cref{eq:quark-diquark_dist}, the rescaled parameter, $ R_{qd*}$ is given by \cref{eq:Rqdresc}.

Defining the quantities 
\begin{equation}
    \sigma^{qq}_0=2\pi A_{qq} R_{q}^2,
\end{equation}
\begin{equation}
\sigma^{
qd}_0=\sigma^{dq}_0=4\pi A_{qq}R_{q}^2,
\end{equation}
and
\begin{equation}
    \sigma^{dd}_0 =8\pi A_{qq}R_{q}^2,
\end{equation}
the first three integrals in \cref{eq:sig_tilde_b} given in terms of normalized bivariate Gaussian distributions (see \hyperref[sec:app_biGaussLevy]{Appendix C} for the notation) are
\begin{align}
&\tilde\sigma_{\rm in}^{qq}(\,\vec b\,)= \int d^2s_qd^2s'_qG\left(\vec s_{q} |R_{qd*}/\sqrt{2}\right)G\left(\vec s_q^{~\prime} |R_{qd*}/\sqrt{2}\right)\sigma^{qq}_{\rm in}\left(\vec b,\vec s_q,\vec s_q^{~\prime}\right) \\ \nonumber 
&= \sigma^{qq}_0\int d^2s_qd^2s'_qG\left(\vec s_{q} |R_{qd*}/\sqrt{2}\right)G\left(\vec s_q^{~\prime} |R_{qd*}/\sqrt{2}\right)G\left(\vec b+\vec s_q^{~\prime}-\vec s_q| R_{q}\right) \\ \nonumber 
&= \sigma^{qq}_0\int d^2s_qG\left(\vec s_{q} |R_{qd*}/\sqrt{2}\right)G\left(\vec b-\vec s_q| \sqrt{R_{qd*}^2/2+R_{q}^2}\right) \\ \nonumber
&=\sigma^{qq}_0G\Big(\vec b\,\Big|\sqrt{R_{qd*}^2+ R_{q}^2}\Big),
\end{align}
\begin{align}
&\tilde\sigma_{\rm in}^{qd}(\,\vec b\,)=\int d^2s_qd^2s'_qG\left(\vec s_{q} |R_{qd*}/\sqrt{2}\right)G\left(\vec s_q^{~\prime} |R_{qd*}/\sqrt{2}\right)\sigma^{qd}_{\rm in}\left(\vec b,\vec s_q,\vec s_q^{~\prime}\right)\\ \nonumber 
&=\sigma^{qd}_0\int d^2s_qd^2s'_qG\left(\vec s_{q} |R_{qd*}/\sqrt{2}\right)G\left(\vec s_q^{~\prime} |R_{qd*}/\sqrt{2}\right)G\Bigg(\vec b-\lambda\vec s_q^{~\prime}-\vec s_q\Bigg|\sqrt{\frac{R_{q}^2+R_{d}^2}{2}}\Bigg)\\ \nonumber 
&=\sigma^{qd}_0\int d^2s_qd^2s''_qG\left(\vec s_{q} |R_{qd*}/\sqrt{2}\right)G\left(\vec s_{q}^{~\prime\prime} |\lambda R_{qd*}/\sqrt2\right)G\left(\vec b-\vec s_q^{~\prime\prime}-\vec s_q\Bigg|\sqrt{\frac{R_{q}^2+R_{d}^2}{2}}\right)\\ \nonumber 
&= 
\sigma^{qd}_0G\left(\vec b\,\Bigg|\sqrt{\frac{(1+\lambda^2)R_{qd*}^2+ R_{q}^2+R_{d}^2}{2}}\right),
\end{align}
and
\begin{align}
&\tilde\sigma_{\rm in}^{dd}(\,\vec b\,)=\int d^2s_qd^2s'_qG\left(\vec s_{q} \,\Big|R_{qd*}/\sqrt{2}\right)G\left(\vec s_q^{~\prime} \,\Big|R_{qd*}/\sqrt{2}\right)\sigma^{dd}_{\rm in}\left(\vec b,\vec s_q,\vec s_q^{~\prime}\right) \\ \nonumber
&=\sigma^{dd}_0\int d^2s_qd^2s'_qG\left(\vec s_{q} \,\Big|R_{qd*}/\sqrt{2}\right)G\left(\vec s_q^{~\prime} \,\Big|R_{qd*}/\sqrt{2}\right)G\left(\vec b+\lambda(\vec s_q-\vec s_q^{~\prime})\,\Big|R_{d}\right) \\ \nonumber
&=\sigma^{dd}_0\int d^2s''_qd^2s'''_qG\left(\vec s_{q}^{~\prime\prime} \,\Big|\lambda R_{qd*}/\sqrt{2}\right)G\left(\vec s_q^{~\prime\prime\prime} \,\Big|\lambda R_{qd*}/\sqrt{2}\right)G\left(\vec b+\vec s_q^{~\prime\prime}-\vec s_q^{~\prime\prime\prime}\,\Big|R_{d}\right) \\ \nonumber
&=\sigma^{dd}_0G\Big(\vec b\,\Big|\sqrt{\lambda^2R_{qd*}^2+R_{d}^2}\Big),
\end{align}
where the calculations are done by using the convolution theorem, \textit{i.e.}, \cref{eq:CT}. The same results can be obtained by using \cref{master_formula_2}.  The remaining eight integrals in \cref{eq:sig_tilde_b} are calculated utilizing \cref{master_formula_2} and fixing the appropriate amplitudes $A_{kl}$ at unity and the appropriate coefficients $c_{kl}$ at zero. The results are
\enlargethispage{3\baselineskip}
\begin{align}
&\tilde\sigma_{\rm in}^{qq,qd}(\,\vec b\,)= \int d^2s_qd^2s'_qG\left(\vec s_{q} |R_{qd*}/\sqrt{2}\right)G\left(\vec s_q^{~\prime} |R_{qd*}/\sqrt{2}\right)\sigma^{qq}_{\rm in}\left(\vec b,\vec s_q,\vec s_q^{~\prime}\right)\sigma^{qd}_{\rm in}\left(\vec b,\vec s_q,\vec s_q^{~\prime}\right) \nonumber\\ 
&=\sigma^{qq}_0\sigma^{qd}_0\int d^2s_qd^2s'_qG\left(\vec s_{q} |R_{qd*}/\sqrt{2}\right)G\left(\vec s_q^{~\prime} |R_{qd*}/\sqrt{2}\right)G\left(\vec b+\vec s_q^{~\prime}-\vec s_q| R_{q}\right)\times   
\end{align}
\begin{align}
\nonumber&\times G\left(\vec b+\vec s_q^{~\prime}+\lambda\vec s_q\Bigg|\sqrt{\frac{R_{q}^2+R_{d}^2}{2}}\right) 
=\frac{\sigma^{qq}_0\sigma^{qd}_0}{\pi\Big(3R_{q}^2+R_{d}^2+(1+\lambda)^2R_{qd*}^2\Big)}\times \\
\nonumber&\times G\left(\vec b\,\Bigg|\sqrt{\frac{R_{d}^2 \left(R_{qd*}^2+R_{q}^2\right)+(\lambda +1)^2 R_{qd*}^4/2+ \left(\lambda ^2+2\right) R_{qd*}^2 R_{q}^2+R_{q}^4}{R_{d}^2+3 R_{q}^2+(\lambda +1)^2 R_{qd*}^2}}\right),
\end{align}
\begin{align}
&\tilde\sigma_{\rm in}^{qd,dq}(\,\vec b\,)=\int d^2s_qd^2s'_qG\left(\vec s_{q} |R_{qd*}/\sqrt{2}\right)G\left(\vec s_q^{~\prime} |R_{qd*}/\sqrt{2}\right)\sigma^{qd}_{\rm in}\left(\vec b,\vec s_q,\vec s_q^{~\prime}\right)\sigma^{dq}_{\rm in}\left(\vec b,\vec s_q,\vec s_q^{~\prime}\right)\nonumber\\ 
&=\sigma^{qd}_0\sigma^{qd}_0\int d^2s_qd^2s'_qG\left(\vec s_{q} |R_{qd*}/\sqrt{2}\right)G\left(\vec s_q^{~\prime} |R_{qd*}/\sqrt{2}\right)\times \\ \nonumber &\times G\left(\vec b-\lambda\vec s_q^{~\prime}-\vec s_q\Bigg|\sqrt{\frac{R_{q}^2+R_{d}^2}{2}}\right) G\left(\vec b+\vec s_q^{~\prime}+\lambda\vec s_q\Bigg|\sqrt{\frac{R_{q}^2+R_{d}^2}{2}}\right)\\ \nonumber
&=\frac{\sigma^{qd}_0\sigma^{qd}_0}{2\pi\Big(R_{q}^2+R_{d}^2+(1+\lambda)^2R_{qd*}^2\Big)}G\left(\vec b\,\Bigg|\sqrt{\frac{R_{d}^2+R_{q}^2+(\lambda-1)^2R_{qd*}^2}{4}}\right),
\end{align}
\begin{align}
&\tilde\sigma_{\rm in}^{qq,dd}(\,\vec b\,)=\int d^2s_qd^2s'_q\,G\left(\vec s_{q} |R_{qd*}/\sqrt{2}\right)G\left(\vec s_q^{~\prime} |R_{qd*}/\sqrt{2}\right)\sigma^{qq}_{\rm in}\left(\vec b,\vec s_q,\vec s_q^{~\prime}\right)\sigma^{dd}_{\rm in}\left(\vec b,\vec s_q,\vec s_q^{~\prime}\right)\nonumber\\ 
&=\sigma^{qq}_0\sigma^{dd}_0\int d^2s_qd^2s'_qG\left(\vec s_{q} |R_{qd*}/\sqrt{2}\right)G\left(\vec s_q^{~\prime} |R_{qd*}/\sqrt{2}\right)\times \\ \nonumber &\times G\left(\vec b+\vec s_q^{~\prime}-\vec s_q| R_{q}\right)G\left(\vec b+\lambda(\vec s_q-\vec s_q^{~\prime})|R_{d}\right)\\ \nonumber
&=\frac{\sigma^{qq}_0\sigma^{dd}_0}{2\pi\Big(R_{q}^2+R_{d}^2+(1+\lambda)^2R_{qd*}^2\Big)}G\left(\vec b\,\Bigg|\sqrt{\frac{R_{d}^2\left(R_q^2+R_{qd*}^2\right)+R_q^2R_{qd*}^2\lambda^2}{R_d^2+R_q^2+(1+\lambda)^2R_{qd*}^2}}\right),
\end{align}
\enlargethispage{3\baselineskip} 
\begin{align}
&\tilde\sigma_{\rm in}^{dd,dq}(\,\vec b\,)=\int d^2s_qd^2s'_q\,G\left(\vec s_{q} |R_{qd*}/\sqrt{2}\right)G\left(\vec s_q^{~\prime} |R_{qd*}/\sqrt{2}\right)\sigma^{dd}_{\rm in}\left(\vec b,\vec s_q,\vec s_q^{~\prime}\right)\sigma^{dq}_{\rm in}\left(\vec b,\vec s_q,\vec s_q^{~\prime}\right)\nonumber\\ 
&=\sigma^{dd}_0\sigma^{qd}_0\int d^2s_qd^2s'_qG\left(\vec s_{q} |R_{qd*}/\sqrt{2}\right)G\left(\vec s_q^{~\prime} |R_{qd*}/\sqrt{2}\right)\times \\ \nonumber
&\times G\left(\vec b+\lambda(\vec s_q-\vec s_q^{~\prime})| R_{d}\right)G\left(\vec b+\vec s_q^{~\prime}+\lambda\vec s_q\Bigg|\sqrt{\frac{R_{q}^2+R_{d}^2}{2}}\right)\\
\nonumber
&=\frac{\sigma^{dd}_0\sigma^{qd}_0}{\pi\Big(3R_{d}^2+R_{q}^2+(1+\lambda)^2R_{qd*}^2\Big)}\times \\ \nonumber
&\times G\left(\vec b\,\Bigg|\sqrt{\frac{ R_{d}^2 \left(R_{qd*}^2(1+2\lambda^2)+R_{q}^2\right)+R_{qd*}^2\lambda^2\left(2R_q^2+R_{qd*}^2(1+\lambda)^2\right)/2+ R_{d}^4}{3R_{d}^2+ R_{q}^2+(\lambda +1)^2 R_{qd*}^2}}\right),
\end{align}
\begin{align}
&\tilde\sigma^{qq,qd,dq}_{\rm in}(\,\vec b\,)=\int d^2s_qd^2s'_qG\left(\vec s_{q} |R_{qd*}/\sqrt{2}\right)G\left(\vec s_q^{~\prime} |R_{qd*}/\sqrt{2}\right)\sigma^{qq}_{\rm in}\left(\vec b,\vec s_q,\vec s_q^{~\prime}\right)\times 
\end{align}
\begin{align}
 \nonumber
&\times \sigma^{qd}_{\rm in}\left(\vec b,\vec s_q,\vec s_q^{~\prime}\right)\sigma^{dq}_{\rm in}\left(\vec b,\vec s_q,\vec s_q^{~\prime}\right)\\\nonumber &=\sigma^{qq}_0\sigma^{qd}_0\sigma^{dq}_0\int d^2s_qd^2s'_qG\left(\vec s_{q} |R_{qd*}/\sqrt{2}\right)G\left(\vec s_q^{~\prime} |R_{qd*}/\sqrt{2}\right)G\left(\vec b+\vec s_q^{~\prime}-\vec s_q| R_{q}\right)\times \\ \nonumber
&\times G\left(\vec b-\lambda\vec s_q^{~\prime}-\vec s_q\Bigg|\sqrt{\frac{R_{q}^2+R_{d}^2}{2}}\right)G\left(\vec b+\vec s_q^{~\prime}+\lambda\vec s_q\Bigg|\sqrt{\frac{R_{q}^2+R_{d}^2}{2}}\right) \\\nonumber 
&=\frac{\sigma^{qq}_0\sigma^{qd}_0\sigma^{dq}_0}{\pi^2 (R_d^2 + R_q^2 + R_{qd*}^2 (1 + \lambda)^2) (R_d^2 + 
   5 R_q^2 + R_{qd*}^2 (1 + \lambda)^2)}\times\\
\nonumber 
&\times G\left(\vec b\, \Bigg | \sqrt{\frac{R_q^4 + R_d^2 (R_q^2 + R_{qd*}^2) + 
 R_q^2 R_{qd*}^2 (2 - 2 \lambda + \lambda^2)}{R_d^2 + 5 R_q^2 + R_{qd*}^2 (1 + \lambda)^2}}\right),
\end{align}
\begin{align}
&\tilde\sigma^{qq,qd,dd}_{\rm in}(\,\vec b\,)=\int d^2s_qd^2s'_q\,G\left(\vec s_{q} |R_{qd*}/\sqrt{2}\right)G\left(\vec s_q^{~\prime} |R_{qd*}/\sqrt{2}\right)\sigma^{qq}_{\rm in}\left(\vec b,\vec s_q,\vec s_q^{~\prime}\right)\times \\ \nonumber
&\times \sigma^{qd}_{\rm in}\left(\vec b,\vec s_q,\vec s_q^{~\prime}\right)\sigma^{dd}_{\rm in}\left(\vec b,\vec s_q,\vec s_q^{~\prime}\right)\\ \nonumber
&=\sigma^{qq}_0\sigma^{qd}_0\sigma^{dd}_0\int d^2s_qd^2s'_qG\left(\vec s_{q} |R_{qd*}/\sqrt{2}\right)G\left(\vec s_q^{~\prime} |R_{qd*}/\sqrt{2}\right)G\left(\vec b+\vec s_q^{~\prime}-\vec s_q| R_{q}\right) \times \\ \nonumber &\times G\left(\vec b-\lambda\vec s_q^{~\prime}-\vec s_q\Bigg|\sqrt{\frac{R_{q}^2+R_{d}^2}{2}}\right)G\left(\vec b+\lambda(\vec s_q-\vec s_q^{~\prime})|R_{d}\right)
\\\nonumber &=\frac{\sigma^{qq}_0\sigma^{qd}_0\sigma^{dd}_0}{\pi^2 R_2^4}
    G\left(\vec b\, \Bigg | \sqrt{\frac{R_1^6}{R_2^4}}\right),
\end{align}
\enlargethispage{1\baselineskip} 
\begin{align}
&\tilde\sigma^{dd,qd,dq}_{\rm in}(\,\vec b\,)=\int d^2s_qd^2s'_qG\left(\vec s_{q} |R_{qd*}/\sqrt{2}\right)G\left(\vec s_q^{~\prime} |R_{qd*}/\sqrt{2}\right)\sigma^{dd}_{\rm in}\left(\vec b,\vec s_q,\vec s_q^{~\prime}\right)\times \\ \nonumber
&\times \sigma^{qd}_{\rm in}\left(\vec b,\vec s_q,\vec s_q^{~\prime}\right)\sigma^{dq}_{\rm in}\left(\vec b,\vec s_q,\vec s_q^{~\prime}\right)\\ \nonumber
&=\sigma^{dd}_0\sigma^{qd}_0\sigma^{dq}_0\int d^2s_qd^2s'_qG\left(\vec s_{q} |R_{qd*}/\sqrt{2}\right)G\left(\vec s_q^{~\prime} |R_{qd*}/\sqrt{2}\right)G\left(\vec b+\lambda(\vec s_q-\vec s_q^{~\prime})|R_{d}\right) \times \\
\nonumber & \times  G\left(\vec b-\lambda\vec s_q^{~\prime}-\vec s_q\Bigg|\sqrt{\frac{R_{q}^2+R_{d}^2}{2}}\right)G\left(\vec b+\vec s_q^{~\prime}+\lambda\vec s_q\Bigg|\sqrt{\frac{R_{q}^2+R_{d}^2}{2}}\right)
\\
\nonumber &=\frac{\sigma^{dd}_0\sigma^{qd}_0\sigma^{dq}_0}{\pi^2 (R_d^2 + R_q^2 + R_{qd*}^2 (1 + \lambda)^2) (5 R_d^2 +
    R_q^2 + R_{qd*}^2 (1 + \lambda)^2)} \times \\\nonumber & \times G\left(\vec b\, \Bigg |\sqrt{\frac{R_d^4 + R_q^2 R_{qd*}^2 \lambda^2 + 
  R_d^2 (R_q^2 + R_{qd*}^2 (1 - 2 \lambda + 2 \lambda^2))}{5 R_d^2 + R_q^2 + R_{qd*}^2 (1 + \lambda)^2}}\right),
\end{align}
\begin{align}
&\tilde\sigma^{qq,qd,dq,dd}_{\rm in}(\,\vec b\,)=\int d^2s_qd^2s'_qG\left(\vec s_{q} |R_{qd*}/\sqrt{2}\right)G\left(\vec s_q^{~\prime} |R_{qd*}/\sqrt{2}\right)\sigma^{qq}_{\rm in}\left(\vec b,\vec s_q,\vec s_q^{~\prime}\right)\times \\ \nonumber
&\times \sigma^{qd}_{\rm in}\left(\vec b,\vec s_q,\vec s_q^{~\prime}\right)\sigma^{dq}_{\rm in}\left(\vec b,\vec s_q,\vec s_q^{~\prime}\right)\sigma^{dd}_{\rm in}\left(\vec b,\vec s_q,\vec s_q^{~\prime}\right)
\end{align}
\begin{align}
\nonumber
&=\sigma^{qq}_0\sigma^{qd}_0\sigma^{dq}_0\sigma^{dd}_0\int d^2s_qd^2s'_qG\left(\vec s_{q} |R_{qd*}/\sqrt{2}\right)G\left(\vec s_q^{~\prime} |R_{qd*}/\sqrt{2}\right)G\left(\vec b+\vec s_q^{~\prime}-\vec s_q| R_{q}\right) \times \\\nonumber &\times G\left(\vec b-\lambda\vec s_q^{~\prime}-\vec s_q\Bigg|\sqrt{\frac{R_{q}^2+R_{d}^2}{2}}\right) G\left(\vec b+\vec s_q^{~\prime}+\lambda\vec s_q\Bigg|\sqrt{\frac{R_{q}^2+R_{d}^2}{2}}\right)G\left(\vec b+\lambda(\vec s_q-\vec s_q^{~\prime})|R_{d}\right)
\\
\nonumber &=\frac{\sigma^{qq}_0\sigma^{qd}_0\sigma^{dq}_0\sigma^{dd}_0}{ 2\pi^3 (R_d^2 + 
   R_q^2 + (1 + \lambda)^2 R_{qd*}^2) R_3^4} \times
   \\\nonumber 
   &\times G\left(\vec b\, \Bigg |\sqrt{ \frac{R_q^4 \lambda^2 R_{qd*}^2 + 
 R_d^4 (R_q^2 + R_{qd*}^2) + 
 R_d^2 (R_q^4 + 2 R_q^2 (1 - \lambda + \lambda^2) R_{qd*}^2)}{R_3^4}}\right),
\end{align}
where
\begin{align}
    R_1^6 &= 2 R_d^4 (R_q^2 + R_{qd*}^2) + 
 R_q^2 R_{qd*}^2 \lambda^2 (2 R_q^2 + 
    R_{qd*}^2 (1 + \lambda)^2) + \\ \nonumber &+
 R_d^2 (2 R_q^4 + R_{qd*}^4 (1 + \lambda)^2 + 
    4 R_q^2 R_{qd*}^2 (1 + \lambda^2)),
\end{align}
\begin{equation}
R_2^4 =  2 R_d^4 + 2 R_q^4 + 4 R_q^2 R_{qd*}^2 (1 + \lambda)^2 + 
 R_{qd*}^4 (1 + \lambda)^4 + 
 4 R_d^2 (2 R_q^2 + R_{qd*}^2 (1 + \lambda)^2),
\end{equation}
and
\begin{equation}
  R_3^4 =   R_d^4 + 2 R_d^2 (3 R_q^2 + (1 + \lambda)^2 R_{qd*}^2) + 
 R_q^2 (R_q^2 + 2 (1 + \lambda)^2 R_{qd*}^2).
\end{equation}

\clearpage

\setcounter{equation}{0}\renewcommand\theequation{C.\arabic{equation}}

\setcounter{figure}{0}\renewcommand\thefigure{C.\arabic{figure}}

\section*{Appendix C: Bivariate Gaussian and stable distributions}\label{sec:app_biGaussLevy}
\addcontentsline{toc}{section}{\protect\numberline{}Appendix C: Bivariate Gaussian and stable distributions}
\markboth{Appendix C: Bivariate Gaussian and stable distributions}{Appendix C: Bivariate Gaussian and stable distributions}

Let $\vec{X}$ be a $2$-dimensional random vector. The characteristic function (CF) of $\vec{X}$, denoted as $\tilde D(\vec{q}\,)$, is the expectation value of $e^{i\vec q\cdot \vec X}$ where $\vec q\in\mathbb{R}^2$. Let  $D(\vec{x})$ be the probability density function (PDF) of $\vec{X}$, then the corresponding CF is given by the Fourier transform of the PDF:
\begin{equation}
    \tilde D(\vec{q}\,) = \int_{-\infty}^{+\infty}d^2\vec xe^{i\vec{q}\cdot\vec{x}}D(\vec{x}).
\end{equation}
The inverse Fourier transform gives the PDF if the CF is known:
\begin{equation}
     D(\vec{x}) = \frac{1}{(2\pi)^2}\int_{-\infty}^{+\infty}d^2\vec q\,e^{-i\vec{q}\cdot\vec{x}}\tilde D(\vec{q}\,).
\end{equation}
While the PDF is real-valued, the CF is a complex-valued function. 

The bivariate Gaussian PDF,  $G(\vec{x}|\vec{\mu},\mathbf{\Sigma})$, is defined as 
\begin{equation}
    G(\vec{x}|\vec{\mu},\mathbf{\Sigma})= \frac{1}{2\pi\sqrt{{\rm det}\mathbf{\Sigma}}}e^{-\frac{1}{2}\left(\vec{x}-\vec\mu\right)^{\rm T}\mathbf{\Sigma}^{-1}\left(\vec{x}-\vec\mu\right)}.
\end{equation}
The corresponding Gaussian CF is 
\begin{equation}
    \widetilde G(\vec{q}\,|\vec{\mu},\mathbf{\Sigma})= e^{i\,\vec{q}\cdot\vec{\mu}-\frac{1}{2}\vec{q}^T\mathbf{\Sigma}\vec{q}}.
\end{equation}

Regarding this work, the interesting case is when $\vec\mu =\vec0$ and  
\begin{equation}
\mathbf{\Sigma} \equiv \mathbf{\Sigma}(\sigma) =   \left( \begin{matrix}
        \sigma^2 & 0 \\
        0 & \sigma^2 
    \end{matrix}\right).
\end{equation}
I define the notations:
\begin{equation}
    G(\vec{x}\,|\sigma)\equiv G\left(\vec{x}\,|\vec{\mu}=\vec0,\mathbf{\Sigma}(\sigma)\right)= \frac{1}{2\pi\sigma^2}e^{-\frac{1}{2}\frac{\vec x^2}{\sigma^2}}
\end{equation}
and
\begin{equation}
    \widetilde G(\vec{q}\,|\sigma)\equiv \widetilde G(\vec{q}\,|\vec{\mu}=\vec0,\mathbf{\Sigma}(\sigma))= e^{-\frac{1}{2}\vec{q}^2\sigma^2}.
\end{equation}

The general form of a bivariate Lévy-stable CF is complicated. We are interested in a bivariate Lévy $\alpha$-stable CF, which gives the Gaussian CF if $\alpha_L=2$. This is the CF of the so-called sub-Gaussian symmetric $\alpha$-stable distribution and is defined as: 
\begin{equation}
\widetilde L(\vec{q}\,|\alpha_L,\beta=0,\vec{\mu},\mathbf{\Gamma})=  e^{i \vec{q}\cdot\vec{\mu}-\left(\vec{q}^{\rm T}\mathbf{\Gamma}\vec{q}\right)^{\alpha_L/2}}
\end{equation}
with
\begin{equation}
\mathbf{\Gamma} \equiv \mathbf{\Gamma}(\gamma) =   \left( \begin{matrix}
        \gamma^2 & 0 \\
        0 & \gamma^2 
    \end{matrix}\right).
\end{equation}
It is obvious that 
\begin{equation}
    \widetilde G(\vec{q}\,|\vec{\mu},\mathbf{\Sigma}(\sigma))\equiv\tilde L\left(\vec{q}\,|\alpha_L=2,\beta=0,\vec{\mu},\mathbf{\Gamma}(\gamma=\sigma/\sqrt{2})\right).
\end{equation}
$\alpha_L$ is called the index of stability. $\beta$ is called the skewness parameter, and if $\beta\neq 0$ the distribution is not symmetric. The $\alpha$-stable parameters can take values in the following domains: $0<\alpha_L\leq 2$, $-1 \leq \beta \leq 1$, $\gamma>0$, and $\vec{\mu}\in\mathbb{R}^2$. We will be interested in the case when $\vec\mu=\vec0$ and $\beta=0$. I define the notations:
\begin{equation}
    \widetilde L(\vec{q}\,|\alpha_L,\gamma)\equiv \widetilde L(\vec{q}\,|\alpha_L,\beta=0,\vec{\mu}=\vec0,\mathbf{\Gamma}(\gamma))=  e^{-\big|\vec{q}^{2}\gamma^2\big|^{\alpha_L/2}},
\end{equation}
and
\begin{equation}
    L(\vec{x}\,|\alpha_L,\gamma)\equiv L\left(\vec{x}\,|\alpha_L,\beta=0,\vec{\mu}=\vec0,\mathbf{\Gamma}(\gamma)\right)= \frac{1}{(2\pi)^2}\int_{-\infty}^{+\infty}d^2\vec q\,e^{-i\vec{q}\cdot\vec{x}}\,e^{-\big|\vec{q}^{2}\gamma^2\big|^{\alpha_L/2}}.
\end{equation}
Note that 
\begin{equation}
    L(\vec{x}\,|\alpha_L=2,\gamma=\sigma/\sqrt 2)\equiv G(\vec{x}\,|\sigma).
\end{equation}

For our purposes, it is enough to know the PDF-s only as a function of the length of $\vec x$, \textit{i.e.}, as a function of $x=|\vec{x}|$. The bivariate $\vec \mu = 0$ sub-Gaussian symmetric Lévy $\alpha$-stable PDF as a function of $x$ can be expressed as \cite{uchaikin2011chance}: 
\begin{equation}\label{eq:Levy-lengthdep}
L(x|\alpha_L,\gamma) = \frac{1}{2\pi \gamma^2}\int_{0}^{\infty}dte^{-t^{\alpha_L}}tJ_0\left(\gamma^{-1}xt\right),
\end{equation}
where $J_0$ is the zeroth order Bessel functions of the first kind.

\begin{figure}[hbt!]
	\centering
\includegraphics[width=0.8\linewidth]{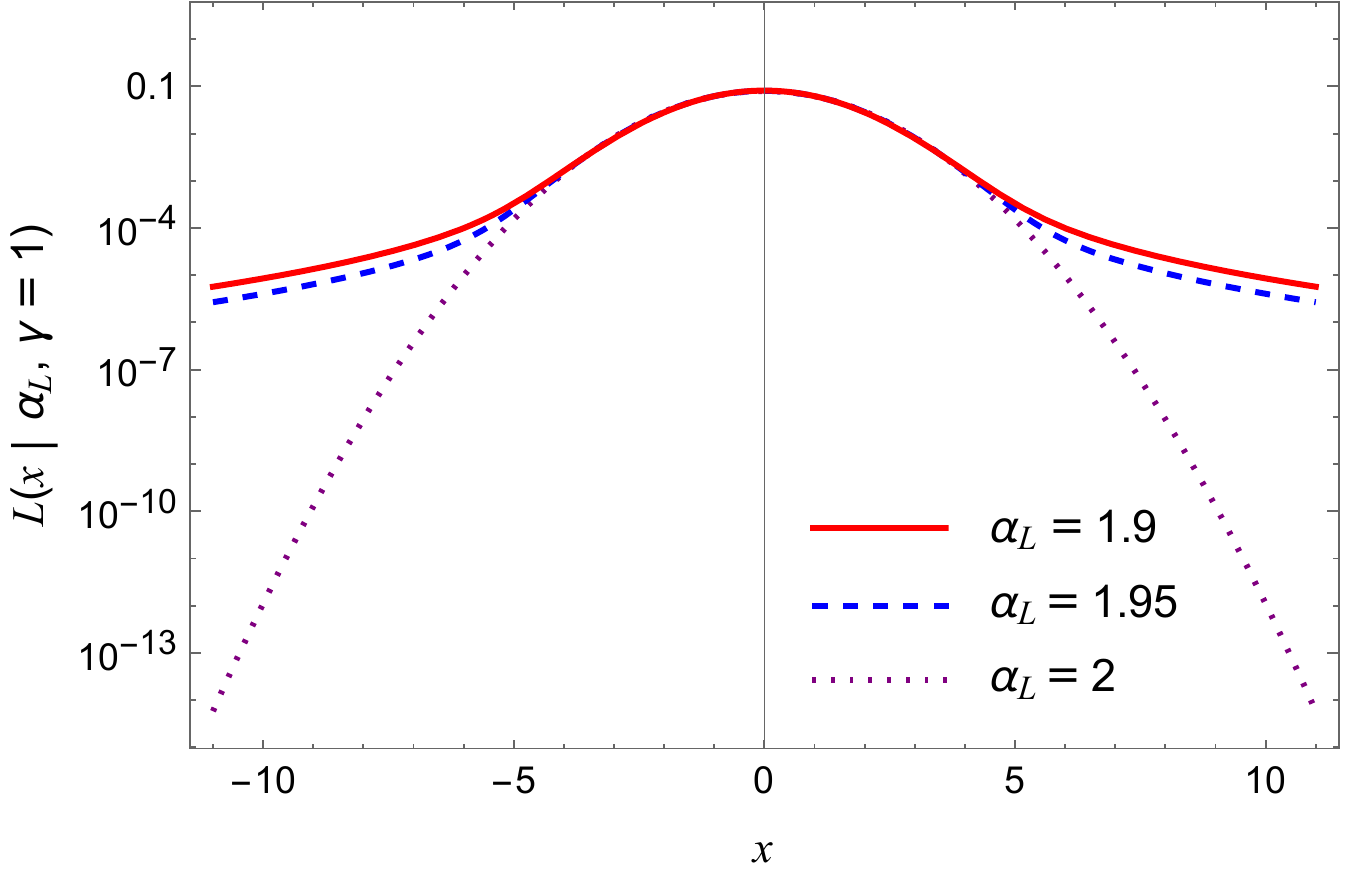}
	\caption{Sub-Gaussian symmetric, bivariate Lévy $\alpha$-stable PDF as a function of $x=|\vec x|$ as given by \cref{eq:Levy-lengthdep} for $\gamma=1$ and for different $\alpha_L$ values.}
	\label{fig:Levy-lengthdep}
\end{figure}

\cref{eq:Levy-lengthdep} is plotted in \cref{fig:Levy-lengthdep} for $\gamma=1$ and for different $\alpha_L$ values. The $\alpha_L=2$ case gives the Gaussian distribution for $\sigma=\sqrt 2$ while for $\alpha_L<2$, heavy-tailed distributions show up. 

$L(x\,|\alpha_L,\gamma)$ can be expressed also in terms of Fox H-functions \cite{uchaikin2011chance}. Based on \mbox{Ref.~\cite{uchaikin2011chance},} where the Mellin transform of $$x\times L\left(x\big|\alpha_L,\gamma=1\right)$$ is calculated for $\vec{x}$ being an $N$-dimensional vector, one can conclude that the $N$-variate sub-Gaussian symmetric Lévy $\alpha$-stable PDF as a function of $x$ can be expressed in terms of a univariate Fox-H function as: 
\begin{align}\label{eq:LsymN}
&L\left(x\big|\alpha_L,\beta=0,\vec{\mu}=\vec{0},\mathbf{\Gamma}(\gamma)\right) = \frac{1}{(4\pi)^{N/2}}\frac{1}{\sqrt{\pi}} \frac{1}{\alpha_L\gamma^{N-1}} \frac{1}{x} \times\\\nonumber &\times H^{1,3}_{3,3}\Bigg[\frac{x}{\gamma}\Bigg|\begin{matrix} \left(1, \frac{1}{2}\right),\left(\frac{1}{2},\frac{1}{2}\right),\left(1-\frac{N-1}{\alpha_L},\frac{1}{\alpha_L}\right) \\ \left(\frac{1}{2},\frac{1}{2}\right),\left(1,1\right),\left(1-\frac{N-1}{2},\frac{1}{2}\right) \end{matrix}\Bigg].
\end{align}
For the interesting $N=2$ case, \cref{eq:LsymN} simplifies to 
\begin{equation}
L\left(x\big|\alpha_L,\beta=0,\vec{\mu}=0,\mathbf{\Gamma}(\gamma)\right) = \frac{1}{4\pi^{3/2}}\frac{1}{\alpha_L \gamma} \frac{1}{x} H^{1,2}_{2,2}\Bigg[\frac{x}{\gamma}\Bigg|\begin{matrix} \left(1, \frac{1}{2}\right),\left(1-\frac{1}{\alpha_L},\frac{1}{\alpha_L}\right) \\ \left(\frac{1}{2},\frac{1}{2}\right),\left(1,1\right) \end{matrix}\Bigg].
\end{equation}

The general Fox H-function is defined as \cite{uchaikin2011chance}: 
\begin{align}
H^{m,n}_{p,q}\Bigg[z\Bigg|\begin{matrix} (a_1,A_1),...,(a_p,A_p) \\ (b_1,B_1),...,(b_q,B_q) \end{matrix} \Bigg] =\frac{1}{2\pi i}\int_{L}ds\theta(s)z^{s},
\end{align}
where 
$$\theta(s)=\frac{\prod_{j=1}^{m}\Gamma(b_j-B_j s)\prod_{j=1}^{n}\Gamma(1-a_j+A_j s)}{\prod_{j=m+1}^{q}\Gamma(1-b_j+B_j s)\prod_{j=n+1}^{p}\Gamma(a_j-A_j s)},$$
furthermore $m$, $n$, $p$, and $q$ are non-negative integers; $0\leq n\leq p$; $1\leq m\leq q$; $a_i,b_j\in\mathbb{R}$ or $\mathbb{C}$; $A_i$ and $B_j$ are positive non-zero reals; $i=1,...,p$; $j=1,...,q$; $z^{-s}=e^{s(\ln |z|+i\arg z)}$, $z\neq0$ and $\arg z$ is not necessarily the principal value; empty products of gamma functions are taken to be 1; the integration path $L$ is an infinite contour which separates poles of the Gamma functions. This $L$ integration contour exists if: $A_l(b_j+\nu)\neq B_j(a_l-\mu-l)$; $j=1,...,m$; $l=1,...,n$; $\nu=0,1,2,...$; $\mu=0,1,2,...$. See more details in Refs.~\cite{mathai2009h,uchaikin2011chance}. 

The Fox H-Function is implemented in Wolfram Mathematica \cite{Mathematica} and available for symbolic and numerical manipulations.

\clearpage
\chapter*{\vspace{-0.3cm}Thesis points}\label{chap:TP}
\addcontentsline{toc}{chapter}{\protect\numberline{}Thesis points}
\markboth{Thesis points}{Thesis points}


\textbf{\circled{1} 
I interpreted the $\mathbf{H(x)}$ scaling of elastic $\mathbf{pp}$ scattering, identified the $\mathbf{H(x)}$ scaling limit of the ReBB model and 
tested the domain of validity of the $\mathbf{H(x)}$ scaling law.
}\\[2pt] 
I showed that models of elastic scattering with an impact parameter distribution depending on $b$ only via a dimensionless variable, \mbox{$\tilde\xi=b/R(s)$,} where $R(s)$ is an internal scale and can be identified with $\sqrt{B_0(s)}$, $i.e.$ the square root of the slope parameter of ${\rm d}\sigma_{\rm el}/{\rm d}t$, manifest $H(x)$ scaling at all values of $t$. 
By comparing the $H(x)$ scaling limit of the ReBB model to the pp and $\rm p\bar p$  ${\rm d}\sigma_{\rm el}/{\rm dt}$ data, I showed, model-dependently, that the scaling is valid in pp scattering in the kinematics range \mbox{1.96~TeV~$\leq\sqrt{s}\leq 7$~TeV} and 0.38~GeV$^2$~$\leq -t\leq1.2$~GeV$^2$. The $H(x)$ scaling property was used to compare the TOTEM measured pp, and the D0 measured ${\rm p\bar p}$ ${\rm d}\sigma_{\rm el}/{\rm d}t$-s resulting in an at least 6.26$\sigma$ model-independent odderon signal \cite{Csorgo:2019ewn,Csorgo:2020rlb,Csorgo:2020msw,Csorgo:2023rzm}.  My results indicate that this comparison is done in the domain of validity of the $H(x)$ scaling \cite{Csorgo:2019ewn}.

\noindent\textbf{\circled{2} I generalized the\,ReBB\,model\,of\,elastic\,$\mathbf{pp}$\,scattering\,to\,elastic\,$\mathbf{p\bar p}$\,scattering.} \\ [2pt]
The ReBB model was proposed for describing elastic pp ${\rm d}\sigma_{\rm el}/{\rm d}t$ data. I~used the ReBB model to describe not only elastic pp but also elastic $\rm{p\bar p}$ ${\rm d}\sigma_{\rm el}/{\rm d}t$ data.~I~found that the geometrical parameters of the model, $R_q$, $R_d$, and $R_{qd}$, for pp and $\rm{p\bar p}$ collisions are the same within errors; however, its opacity parameter, $\alpha_R$, differs in pp and $\rm{p\bar p}$ reactions at the same energy. The kinematic range of the quantitative description, \textit{i.e.}, the description with confidence~level~(CL)~$\geq$~0.1\%, is: 0.546~TeV~$\leq\sqrt{s}\leq 8$~TeV and \mbox{0.38~GeV$^2$~$\leq -t\leq1.2$~GeV$^2$ \cite{Csorgo:2020wmw}.} 

\noindent\textbf{\circled{3} I found discovery level odderon exchange signals by comparing the ReBB model extrapolations to experimental $\mathbf{pp}$ and $\mathbf{p\bar p}$ elastic scattering data.} \\ [2pt]
Using the results of the ReBB model analysis of elastic pp and $\rm{p\bar p}$ scattering data, I compared pp and $\rm{p\bar p}$ ${\rm d}\sigma_{\rm el}/{\rm d}t$-s at exactly the same energies in the TeV center-of-mass energy range. I found that the pp and $\rm{p\bar p}$ \mbox{${\rm d}\sigma_{\rm el}/{\rm d}t$-s} differ with a statistical significance of at least 6.3$\sigma$ when significances obtained at 1.96 TeV and 2.76 TeV are combined. By combining the significances obtained at all the four analyzed energies, 1.96~TeV, \mbox{2.76 TeV,} \mbox{7 TeV,} and 8 TeV, the statistical significance of the difference between pp and $\rm{p\bar p}$ \mbox{${\rm d}\sigma_{\rm el}/{\rm d}t$-s,} \textit{i.e.}, the statistical significance of the $t$-channel odderon exchange signal is higher than $30\sigma$ \cite{Csorgo:2020wmw,Szanyi:2022ezh,Szanyi:2022qgx}.

\newpage

\noindent\textbf{\circled{4} I showed that the TOTEM measured $\mathbf{pp}$ $\mathbf{{\rm d}\sigma_{\rm el}/{\rm dt}}$ can be extrapolated to \mbox{$\sqrt{s}=1.96{\rm~TeV}$}  and I  predicted the effects of the odderon exchange at this c.m. energy contributing to the joint D0-TOTEM odderon observation.
} \\ [2pt]
I participated in the joint project of the CERN LHC TOTEM and FNAL Tevatron D0 experimental collaborations on the odderon search. Utilizing the ReBB model and a phenomenological model based on Regge theory, I extrapolated the TOTEM measured pp ${\rm d}\sigma_{\rm el}/{\rm d}t$-s to \mbox{$\sqrt{s}=1.96$ TeV} and determined that the effect of the odderon exchange manifests mainly in generating a prominent diffractive minimum in pp  ${\rm d}\sigma_{\rm el}/{\rm d}t$ and filling in this minimum in the $\rm{p\bar p}$ ${\rm d}\sigma_{\rm el}/{\rm d}t$. I gave twelve internal presentations within ten months on my progress in this work \cite{talk1,talk2,talk3,talk4,talk5,talk6,talk7,talk8,talk9,talk10,talk11,talk12}. My results served as a guide during the joint D0-TOTEM analysis that, utilizing also new data, finally led to an odderon observation with a statistical significance of at \mbox{least 5.2$\sigma$ \cite{TOTEM:2020zzr}.}   

\noindent\textbf{\circled{5} I generalized the Gaussian model of low-$|t|$ elastic hadronic scattering to a Lévy $\alpha$-stable model and I formulated the real extended Lévy $\alpha$-stable generalized Bialas--Bzdak (LBB) model.}\\ [2pt]
I generalized the Gaussian, $\alpha_L=2$ special case model of elastic pp and $\mathbf{\rm p\bar p}$ scattering to a Lévy $\alpha$-stable, $0<\alpha_L\leq2$ model and demonstrated that the generalized model describes the strongly non-exponential behavior of the precise LHC data with \mbox{$\alpha_L=1.959\pm0.002$} and $CL\geq8.8$ \% in the kinematic range of \mbox{0.02 GeV$^2$ $\leq -t\leq0.15$ GeV$^2$} \cite{Csorgo:2023rbs}. Motivated by the success of the simple Lévy $\alpha$-stable model, I worked out the LBB model by generalizing the Gaussian distributions in the ReBB model to Lévy \mbox{$\alpha$-stable} distributions and consequently introducing a new free parameter, $\alpha_L$, the Lévy index of stability \cite{Csorgo:2023pdn}.

\clearpage
\chapter*{\vspace{-1.1cm}Publications}\label{chap:publications}
\addcontentsline{toc}{chapter}{\protect\numberline{}Publications}
\markboth{Publications}{Publications}


	\section*{I. My publications in refereed journals related to my thesis points}
	\begin{enumerate}
		\setstretch{1.0}
		\item \fullcite{Csorgo:2023rbs} {\color{gray}}
		\item \fullcite{Csorgo:2023pdn} {\color{gray}} 
		\item \fullcite{Szanyi:2022ezh} {\color{gray}}
		\item \fullcite{Csorgo:2020wmw} {\color{gray}}
		\item \fullcite{TOTEM:2020zzr} {\color{gray}}    
		\item \fullcite{Csorgo:2019ewn} {\color{gray}}  
	\end{enumerate}

 \section*{II. My publications in refereed journals not related to my thesis points}
 
	\begin{enumerate}
		\setstretch{1.0}
		\item \fullcite{Jenkovszky:2024} {\color{gray}}
		\item \fullcite{Szanyi:2023ano} {\color{gray}}     
		\item \fullcite{Jenkovszky:2022bkw} {\color{gray}}   
		\item \fullcite{Szanyi:2019kkn} {\color{gray}}          
		\item \fullcite{Jenkovszky:2018itd} {\color{gray}}
		\item \fullcite{Szanyi:2018pew} {\color{gray}}
		\item \fullcite{Broniowski:2018xbg} {\color{gray}}    
		\item \fullcite{Jenkovszky:2017efs} {\color{gray}} 
		\item \fullcite{Jenkovszky:2017pqs} {\color{gray}} 
		\item \fullcite{Jenkovszky:2017dox} {\color{gray}}
	\end{enumerate}

	\section*{III. My book chapter and my book}
	\begin{enumerate}
		\setstretch{1.0}
		\item \fullcite{Csorgo:2020rlb}
		\item Jenkovszky László, Spenik Sándor, Szanyi István, Turóci-Sütő Jolán. Rugalmas és diffraktív szórás az LHC korában: a pomeron, az odderon és gluonlabdák (in Hungarian). Uzhhorod, ``AUTDOR-SHARK", 2021, 152 pages.
	\end{enumerate}
	
	\section*{IV. My conference papers related to my thesis points}
	\begin{enumerate}
		\setstretch{1.0}
		\item \fullcite{Csorgo:2023rzm}
		\item \fullcite{Szanyi:2022qgx}
		\item \fullcite{Csorgo:2020msw}
	\end{enumerate}
	
	\section*{V. My conference papers not related to my thesis points}

 \begin{enumerate}
		\setstretch{1.0}
        \item \fullcite{Lengyel:2023jst}
		\item \fullcite{Szanyi:2019mtf}
		\item \fullcite{Szanyi:2019pye}
		\item \fullcite{Bence:2018cby}
		\item \fullcite{Jenkovszky:2018pcm}
		\item \fullcite{Bence:2017rcp}
		\item \fullcite{Szanyi:2017uoi}
		\item \fullcite{Jenkovszky:2017vnh}
	\end{enumerate}

\newpage
 \thispagestyle{empty}
\chapter*{\vspace{-1.1cm}Talks}\label{chap:talks}
\addcontentsline{toc}{chapter}{\protect\numberline{}Talks}
\markboth{Talks}{Talks}

	\section*{I. My talks at international scientific events}
	\begin{enumerate}
		\setstretch{1.0}
		\item ``High-energy, low-|t| elastic proton-proton scattering'' at the 16th Zimányi Winter School on Heavy Ion Physics, Budapest, Hungary, 5-9 December 2016
		\item ``Structures in the diffraction cone: the "break" and "dip" in high-energy proton-proton scattering'' at the 17th conference on Elastic and Diffractive Scattering (EDS Blois 2017), Prague, Czech Republic, 26-30 June  2017
		\item ``Proton-proton scattering at the LHC'' at the 10th Bolyai-Gauss-Lobachevsky Conference~ on~ Non-Euclidean geometry and its Application, Gyöngyös, Hungary, \mbox{21-25 August 2017} (awarded ``Best student talk of Universe prize'' given by the MDPI Universe journal) 
		\item ``The Coulomb-nuclear interference region and forward physics of proton-proton scattering at high energies'' at the EMMI workshop “Challenges in Photon Induced Interactions”, Cracow, Poland, 5-8 September 2017
		\item ``Elastic pp scattering at the LHC: the non-exponential low-|t| diffraction cone and the energy dependence of the slope'' at the Wilhelm and Else Heraeus Physics School "QCD - Old Challenges and New Opportunities", Bad Honnef, Germany, \mbox{24-30 September 2017} (awarded best student talk prize)
		\item ``Forward proton-proton scattering at high energies'' at the 17th Zimányi-COST Winter School on Heavy Ion Physics, Budapest, Hungary, 4-8 December 2017
		\item ``New physics from TOTEM's recent measurements of elastic and total cross sections'' at the Diffraction and low-x 2018  workshop, Reggio Calabria, Italy, 26 August -- 1 September 2018
		\item ``Comparing the ReBB model predictions with recent TOTEM data at 2.76 and \mbox{13 TeV}'' at the Day of Femtoscopy 2018, Gyöngyös, Hungary, 30 October 2018
		\item ``Study of pp and $\rm{p\bar p}$ scattering at LHC energies using an extended Bialas--Bzdak model'' at the 18th Zimányi Winter School on Heavy Ion Physics, Budapest, \mbox{Hungary}, 3-7 December 2018
		\item ``The shape of the interaction region of colliding protons in a Regge model'' at the New Trends in High-Energy Physics international conference, Odessa, Ukraine, \mbox{12-18 May 2019} (awarded best student talk prize)
		\item ``Status of the Odderon search using the ReBB model'' at the Day of Femtoscopy 2019, Gyöngyös, Hungary, 31 October 2019
		\item ``Results of the Odderon search with the Real Extended Bialas--Bzdak model'' at the 19th\, Zimányi\, Winter\, School\, on\, Heavy\, Ion\, Physics,\, Budapest,\, Hungary,\, \mbox{2-6 December 2019}
		\item ``Observation of Odderon Effects at LHC energies - A Real-Extended Bialas--Bzdak
Model Study'' at the 17th International Scientific Days, Femtoscopy Session -- Online Conference, Gyöngyös, Hungary, 5 June 2020
		\item ``New results on Odderon search using the Bialas--Bzdak model'' at the Day of Femtoscopy 2020, Gyöngyös, Hungary, 29 October 2020 
		\item ``New results on Odderon search using the Bialas--Bzdak model'' at the 20th Zimányi Winter School on Heavy Ion Physics, Budapest, Hungary, 7-11 December 2020 (online contribution)
		\item ``Observation of Odderon Effects at LHC energies -- A Real Extended Bialas Bzdak Model Study'' at the ELFT Winter School "Physics beyond the Standard Model: Modern Approaches", Budapest, Hungary, 1-5 February 2021 (online contribution)
		\item ``Observation of Odderon Effects at LHC energies -- A Real Extended \mbox{Bialas--Bzdak} Model Study'' at~ the~ Low-x~ 2021~ workshop,~ Island Elba,~ Italy,~ 26 September \mbox{-- 1 October 2021}
		\item ``Model-dependent results on the energy dependence of the optical point on elastic pp scattering'' at the Day of Femtoscopy 2021, Gyöngyös, Hungary, 28 October 2021 
		\item ``H(x) scaling and the pp and ${\rm p \bar p}$ slope B(s,t)'' at the 21th Zimányi Winter School on Heavy Ion Physics, Budapest, Hungary, 6-10 December 2021
		\item ``Real Extended Bialas--Bzdak Model at 8 TeV'' at the Seminarium HEP Białasówka, Kraków, Poland, 1 April 2022
		\item ``Real Extended Bialas--Bzdak Model and the Odderon at 8 TeV'' at the 18th International Scientific Days, Gyöngyös, Hungary, 5 May 2022 
		\item ``The ReBB model at 8 TeV: Odderon exchange is a certainty'' at the Diffraction and low-x 2022 workshop, Corigliano Calabro, Italy, 24-30 September 2022
		\item ``Pomeron and Odderon from the ReBB model'' at the Day of Femtoscopy 2022, Gyöngyös, Hungary, 15 November 2022 
		\item ``The ReBB model at 8 TeV: Odderon exchange is a certainty'' at the 22th Zimányi Winter School on Heavy Ion Physics, Budapest, Hungary, 5-9 December 2022
		\item ``Lévy alpha-stable model for the non-exponential low-|t| proton-proton differential cross section'' at the 52nd International Symposium on Multiparticle Dynamics, Gyöngyös, Hungary, 21-26 August 2023
		\item ``Lévy alpha-stable model for the non-exponential low-|t| proton-proton differential cross section'' at the Low-x 2023 workshop, Leros Island, Greece, \mbox{4-8 September 2023}
		\item ``Dip-bump structures in pp elastic scattering and single diffractive dissociation'' at the EMMI workshop on Forward Physics in ALICE 3, Heidelberg, Germany, \mbox{18-20 October 2023}
		\item ``Dip-bump structures in elastic pp scattering and single diffractive dissociation'' and ``Levy alpha-stable model for the non-exponential low $-t$ region of elastic pp scattering'' at the Day of Femtoscopy 2023, Gyöngyös, Hungary, 30-31 October 2023 
		\item ``Dip-bump structure in pp single diffraction'' at the 23rd Zimányi School, Budapest, Hungary, 4-8 December 2023
  \item ``Lévy $\alpha$-stable generalization of the ReBB model'' at the Diffraction and gluon saturation at the LHC and the EIC ECT* workshop, Trento, Italy, 10-14 June 2024
  \item ``Lévy $\alpha$-stable generalization of the ReBB model'' at the 19th International Scientific Days, Gyöngyös, Hungary, 26 June 2024
  \item ``Simple Levy-alpha stable model analysis of the low-|t| elastic pp and ${\rm p \bar p}$ data'' at the LHC Forward Physics workshop, 15-16 July 2024 (online contribution)
  
	\end{enumerate}

	\section*{II. My  presentations in the TOTEM experiment}
	\begin{enumerate}
		\setstretch{1.0}
		\item \fullcite{talk1}
		\item \fullcite{talk2}
		\item \fullcite{talk3}
		\item \fullcite{talk4}
		\item \fullcite{talk5}
		\item \fullcite{talk6}
		\item \fullcite{talk7}
		\item \fullcite{talk8}
		\item \fullcite{talk9}
		\item \fullcite{talk10}
		\item \fullcite{talk11}
		\item \fullcite{talk12}
	\end{enumerate}

\newpage
\thispagestyle{empty}
\chapter*{\vspace{-0.3cm}List of acronyms}\label{chap:LA}
\addcontentsline{toc}{chapter}{\protect\numberline{}List of acronyms}
\markboth{List of acronyms}{List of acronyms}

\begin{itemize}
    \item \textbf{ATLAS} A Toroidal LHC Apparatus
    \item \textbf{CERN} European Organization for Nuclear Research (in French: Conseil Européen pour la Recherche Nucléaire)
    \item \textbf{CMS} Compact Muon Solenoid
    \item \textbf{FNAL} Fermi National Accelerator Laboratory
    \item \textbf{ISR} Intersecting Storage Rings
    \item \textbf{LHC} Large Hadron Collider
    \item \textbf{NA61/SHINE} North Area 61 / SPS Heavy Ion and Neutrino Experiment
    \item \textbf{PHENIX} Pioneering High Energy Nuclear Interaction eXperiment
    \item \textbf{RHIC} Relativistic Heavy Ion Collider
    \item \textbf{SPS} Super Proton Synchrotron
    \item \textbf{STAR} Solenoidal Tracker at RHIC
    \item \textbf{TOTEM} TOTal Elastic and diffractive cross section Measurement
    \item \textbf{UA4} Underground Area 4

    
\end{itemize}

\clearpage
\thispagestyle{empty}
\addcontentsline{toc}{chapter}{\protect\numberline{}Bibliography}
\printbibliography[notkeyword=nocite]
\label{chap:bib}
\clearpage
{\protect\numberline{}}
\markboth{}{}
\thispagestyle{empty}
\clearpage
{\protect\numberline{}}
\markboth{}{}
\thispagestyle{empty}
\vspace{-1.8cm}
\begin{center}
Short summary of my PhD Dissertation

{\bf Odderon Exchange in Elastic Proton-Proton and\\ Proton-Antiproton Scattering at TeV Energies}

\vspace{0.2cm}
{István Szanyi}
\vspace{0.2cm}

{ Supervisor: Tamás Csörgő, D.Sc.,\\ Member of Academia Europaea}

{ Co-supervisor: Máté Csanád, D.Sc.}
\end{center}

\enlargethispage{1\baselineskip} 


\noindent Elastic\, proton-proton\, ($pp$)\,\, scattering\,\, measurements\,\, at\, CERN's\, LHC\, and\, elastic \mbox{proton-antiproton ($p\bar{p}$)}\, scattering\, measurements\, at\, FNAL's\, Tevatron,\, performed at \mbox{TeV-scale} center-of-mass energies ($\sqrt{s}$) and over wide ranges of squared four-momentum transfer~($t$), gave new opportunities to study the physics of elastic hadronic processes. In the framework of T. Regge's theory of complex angular momenta, the scattering amplitude of hadronic process is specified by $t$-channel exchanges of Regge trajectories.  
The pomeron trajectory exchange, characterized by the quantum numbers of vacuum, was invented in the 1960s to account for hadronic cross sections that appeared at first to be constant and were later found to be increasing with $\sqrt{s}$. 
In 1973, the negative spatial and charge parity counterpart of the pomeron, the so-called odderon was proposed by L.~Lukaszuk and B.~Nicolescu. Any statistically significant difference between the measurable quantities of pp and proton-antiproton ($\rm{p\bar p}$)~elastic scattering in the TeV energy region is an indication for the existence of odderon exchange. For 48 years, there was no definitive answer to the question about the existence of the odderon exchange. The joint analysis of the LHC pp and the SPS and Tevatron $\rm{p\bar p}$ elastic scattering data allowed us to do comparative studies of elastic pp and $\rm{p\bar p}$ scattering in the same kinematic \mbox{range ($s$, $t$)} and reveal the existence of the $t$-channel odderon exchange.  My dissertation details my contribution to the odderon discovery and related studies. In my dissertation, I analyzed the pp and $\rm{p\bar p}$ elastic scattering data in the $\sqrt{s}$ range from half TeV up to \mbox{13 TeV} in the domain of ``soft'' interactions (\mbox{$|t|\lesssim$ 3 GeV$^2$}). I generalized the ReBB model of elastic ${\rm pp}$ scattering to elastic $\rm p\bar p$ scattering. I showed that the TOTEM measured pp ${\rm d}\sigma_{\rm el}/{\rm dt}$ can be extrapolated to $\sqrt{s}=1.96{\rm~TeV}$  and I predicted the effects of the odderon exchange at $\sqrt{s}=1.96{\rm~TeV}$. Based on a refined ReBB model analysis of experimental data, I found discovery level odderon exchange signals by comparing the ReBB model extrapolations to experimental pp and ${\rm p\bar p}$ elastic scattering data. I investigated the so-called H(x) scaling of elastic $\rm pp$ scattering: I interpreted the H(x) scaling, identified the H(x) scaling limit of the ReBB model, and tested the H(x) scaling conditions against the data. \mbox{I investigated} the \mbox{low-$|t|$} non-exponential behavior of the pp elastic differential cross section and generalized the Gaussian model of \mbox{low-$|t|$} elastic hadronic scattering to a Lévy $\alpha$-stable model. \mbox{I formulated} the real extended Lévy $\alpha$-stable generalized Bialas--Bzdak (LBB) model, which may be used in the future for a more detailed analysis of elastic hadronic scattering and odderon effects.

\clearpage
\thispagestyle{empty}
\clearpage
{\protect\numberline{}}
\markboth{}{}
\thispagestyle{empty}
\clearpage
{\protect\numberline{}}
\markboth{}{}
\thispagestyle{empty}
\vspace{-1.8cm}

\enlargethispage{1\baselineskip}

\begin{center}
{\bf Odderoncsere a Rugalmas  Proton-Proton és\\ Proton-Antiproton  Szórásban TeV Energiákon}

{ c. doktori értekezésem rövid összefoglalása}

\vspace{2mm}
Szanyi István
\vspace{2mm}

{Témavezető: Csörgő Tamás, az MTA doktora,\\  az Academia Europaea tagja}

{Társ témavezető: Csanád Máté, az MTA doktora}
\end{center}




\noindent A CERN LHC gyorsítóján végzett rugalmas \mbox{proton-proton}~($\rm p p$) és az FNAL Tevatron \mbox{gyorsítóján} végzett rugalmas \mbox{proton-antiproton}~($\rm p\bar p$) ütközési kísérletek TeV-os tömegkö-zépponti energiákon ($\sqrt{s}$) és széles átadott négyesimpulzus-négyzet ($t$) tartományokban új lehetőségeket biztosítottak a hadronok rugalmas ütközéseinek tanulmányozására. \mbox{T.\,Regge} komplex perdületen alapuló elmélete alapján  a nagyenergiás szórási amplitúdót impulzus reprezentációban Regge trajektóriák \mbox{$t$-csatornás} cseréi határozzák meg. A vákuum kvantumszámaival jellemzett pomeron trajektória cserét az 1960-as években vezették be a hadronikus teljes hatáskeresztmetszetek értelmezésére, amely először állandónak bizonyult, de később kiderült, hogy az $\sqrt{s}$ növekedésével növekszik. L.~Lukaszuk és B.~Nicolescu 1973-ban feltételezték az odderonnak, a pomeron negatív paritású és töltésparitású párjának a létezését. Bármely statisztikailag szignifikáns különbség a rugalmas $\rm p p$ és  $\rm p\bar p$ szórás mérhető mennyiségei között TeV energiákon az odderoncsere létezését jelenti. 48~évig nem sikerült döntő bizonyítékot találni az odderoncsere létezésére. Az LHC pp valamint az SPS és Tevatron ${\rm p\bar p}$ rugalmas szórási adatok együttes analízise lehetővé tette a rugalmas pp és ${\rm p\bar p}$ szórás mérhető mennyiségeinek  ugyanabban a kinematikai ($s$,~$t$)~tartományban történő összehasonlítását és a $t$-csatornás odderoncsere létezésének bizonyítását. Az értekezésem részletezi a hozzájárulásomat az odderoncsere felfedezéséhez és az ehhez kapcsolódó tanulmányokhoz.\, Értekezésemben elemeztem a pp és $\rm{p\bar p}$ rugalmas szórási adatokat a fél TeV-tól 13 TeV-ig terjedő $\sqrt{s}$ tartományban a ``soft'' kölcsönhatások szektorában (\mbox{$|t|\lesssim$ 3 GeV$^2$}). Általánosítottam a rugalmas pp szórás ReBB modelljét a rugalmas ${\rm p\bar p}$ szórásra. Megmutattam, hogy a TOTEM által mért pp ${\rm d}\sigma_{\rm el}/{\rm dt}$ extrapolálható $\sqrt{s}=1.96{\rm~TeV}$ energiára és megbecsültem az odderoncsere hatásait $\sqrt{s}=1.96{\rm~TeV}$ \mbox{energián.} A ReBB modell segítségével elvégeztem a kísérleti adatok precíz elemzését és felfedezés szintű odderoncsere jeleket találtam a ReBB modell extrapolációk kísérleti pp és ${\rm p\bar p}$ rugalmas szórási adatokhoz történő hasonlításával. Tanulmányoztam a rugalmas pp szórás ún. H(x) skálázását: interpretáltam a H(x) skálázást, azonosítottam a ReBB  modell H(x) skálázó határesetét és ellenőriztem a H(x) skálázás feltételeinek teljesülését az adatokon. Tanulmányoztam a rugalmas pp szórás differenciális hatáskeresztmetszetének nem exponenciális viselkedését, általánosítottam az alacsony $|t|$ értékekkel járó rugalmas hadronikus szórás Gauss modelljét egy Lévy $\alpha$-stabil modellre. Kidolgoztam a valós résszel bővített  Lévy $\alpha$-stabil általánosított Bialas--Bzdak (LBB) modellt, amely a későbbiekben felhasználható lehet a rugalmas hadronikus szórás és az odderon hatásainak további rész-letes vizsgálatára.


\clearpage

{\protect\numberline{}}
\markboth{}{}
\thispagestyle{empty}
\clearpage

\end{document}